KING'S COLLEGE LONDON
UNIVERSITY OF LONDON

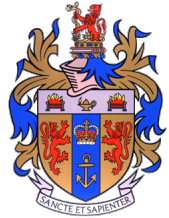

**University of London**

DOCTORAL THESIS

---

# Aspects of quantum gravity

---

*Author:*
Dipl.-Phys. (Univ.) Andreas G. A. PITHIS

*Supervisor:*
Prof. Dr. Mairi SAKELLARIADOU

*Examiners:*
Prof. Dr. Alejandro PEREZ
Prof. Dr. Vincent RIVASSEAU

*A thesis submitted in partial fulfillment of the requirements for the degree of*
*Doctor of Philosophy*

Theoretical Particle Physics & Cosmology Group
Department of Physics
Faculty of Natural & Mathematical Sciences

September 30, 2018

# Declaration of Authorship

The contents of this thesis are the result of the author's own work and of the scientifc collaborations listed below, except where specific reference is made to the work of others. This thesis is based on the following seven research papers, all of which are **published** in peer reviewed journals. They are the outcome of research conducted at King's College London between October 2014 and September 2018.

1. A. G. A. Pithis and H.-Ch. Ruiz Euler, "Anyonic statistics and large horizon diffeomorphisms for Loop Quantum Gravity Black Holes", **published** in Physical Review D 91, 064053 (2015) (22pp), arXiv:1402.2274 [gr-qc];

2. M. de Cesare, A. G. A. Pithis and M. Sakellariadou, "Cosmological implications of interacting Group Field Theory models: cyclic Universe and accelerated expansion", **published** in Physical Review D 94, 064051 (2016) (12pp), arXiv:1606.00352 [gr-qc];

3. A. G. A. Pithis, M. Sakellariadou and Petar Tomov, "Impact of nonlinear effective interactions on GFT quantum gravity condensates", **published** in Physical Review D 94, 064056 (2016) (23pp), arXiv:1607.06662 [gr-qc];

4. A. G. A. Pithis and M. Sakellariadou, "Relational evolution of effectively interacting GFT quantum gravity condensates", **published** in Physical Review D 95, 064004 (2017) (22pp), arXiv:1612.02456 [gr-qc];

5. M. de Cesare, M. Sakellariadou, A. G. A. Pithis and D. Oriti, "Dynamics of anisotropies close to a cosmological bounce in quantum gravity", **published** in Class. Quantum Grav. 35 015014 (2018) (30pp), arXiv:1709.00994 [gr-qc];

6. J. Ben Geloun, A. Kegeles and A. G. A. Pithis, "Minimizers of the equilateral dynamical Boulatov model", **published** in European Physics Journal C (2018) 78: 996, arXiv:1806.09961 [gr-qc];

7. A. G. A. Pithis and J. Thürigen, "Phase transitions in group field theory: The Landau perspective", **published** in Physical Review D 98, 126006 (2018), arXiv:1808.09765 [gr-qc];

My earlier investigation

- A. G. A. Pithis, "Gibbs paradox, Black hole entropy, and the thermodynamics of Isolated Horizons", **published** in Physical Review D 87, 084061 (2013) (7pp), arXiv:1209.2016 [gr-qc];

and the recent letter

- A. C. Jenkins, A. G. A. Pithis and M. Sakellariadou, "Can we detect quantum gravity with compact binary inspirals?", **published** in Physical Review D 98, 104032 (2018), arXiv:1809.06275 [gr-qc];

will not be directly addressed in this thesis.

Andreas G. A. PITHIS

September 2018

KING'S COLLEGE LONDON

UNIVERSITY OF LONDON

# *Abstract*

Theoretical Particle Physics & Cosmology Group

Faculty of Natural & Mathematical Sciences

Department of Physics

Doctor of Philosophy

**Aspects of quantum gravity**

by Dipl.-Phys. (Univ.) Andreas G. A. PITHIS

For more than 80 years theoretical physicists have been trying to develop a theory of quantum gravity which would successfully combine the tenets of Einstein's theory of general relativity (GR) together with those of quantum field theory. At the current stage, there are various competing responses to this challenge under construction. Attacking the problem of quantum gravity from the quantum geometry perspective, where space and spacetime are discrete, the focus of this thesis lies on the application of loop quantum gravity (LQG) and group field theory (GFT). We employ these two closely related non-perturbative approaches to two areas where quantum gravity effects are broadly expected to be relevant: black holes and quantum cosmology.

Concerning black holes, apart from understanding their inner structure, the most pressing issue is to give a microscopic explanation for the phenomenon of black hole entropy in terms of a discrete quantum geometry and relate it to the symmetries of the horizon. Black hole models in LQG are typically constructed via the isolated horizon boundary condition which gives rise to an effective description of the horizon geometry in terms SU(2) Chern-Simons theory. The quantum statistical analysis of this configuration allows to retrieve its entropy which is compatible with the semi-classical Bekenstein-Hawking area law. In this thesis we find a reinterpretation of the statistics of the horizon degrees of freedom as those of a system of non-Abelian anyons.

As regards quantum cosmology, the challenge is to understand how the initial singularity problem of GR can be resolved by means of the discreteness of geometry and how a spacetime continuum can emerge from a large assembly of geometric building blocks. Most recent research in GFT and its condensate cosmology spin-off aims at deriving the effective dynamics for GFT condensate states directly from the microscopic GFT quantum dynamics and subsequently to extract a cosmological interpretation from them. The central conjecture of the condensate cosmology approach is that a possible continuum geometric phase of a particular GFT model is ideally approximated by a condensate state which is considered suitable to describe spatially homogeneous universes. By exploring this idea, new perspectives are revealed for addressing the long-standing question of how to recover the continuum from the collective behaviour of a large set of geometric building blocks in LQG. Remarkably, these efforts have shown that quite naturally a bouncing cosmological solution can be obtained. Its dynamics at late times can be cast into the form of effective Friedmann equations for an isotropic and homogeneous universe.

In this thesis we elaborate on aspects of the above-mentioned conjecture of the condensate phase and study phenomenological consequences of this approach in detail. In particular, we find condensate configurations consisting of many smallest building blocks which may give rise to an effectively continuous emergent geometry in various models. We also explore the cosmological implications of effective interactions between the quantum geometric constituents of the condensate for the first time and show how such interactions can lead to a recollapse or infinite expansion of the emergent universe while preserving the bounce and demonstrate that fine-tuned interactions can lead to an early epoch of accelerated expansion lasting for an arbitrarily large number of e-folds. Finally, we explore the effect of anisotropic perturbations onto GFT condensates and show that these are under control at the bounce and become negligible away from it. This also represents a crucial step towards identifying cosmological anisotropies within this approach.

## *Acknowledgements*

First and foremost, I would like to take the chance and express my gratefulness towards my supervisor Prof. Dr. Mairi Sakellariadou, for giving me the opportunity to come to King's College London and carry on with the things that pleasure my heart the most.

Σ'ευχαριστώ πάρα πολύ Μαίρη! Σ'ενα άλλο σύμπαν σίγουρα θα ξαναχορέψουμε.

Throughout the last years I had the great privilege to have access to and support from outstanding research infrastructure: The work presented in this thesis would not have been possible without the financial support coming from King's College London which covered my fees and living expenses through a generous scholarship. Many thanks also go to the kind people in the Department of Physics for their help, particularly Jean Alexandre, Eugene Lim, Patrick Mesquida and Rowena Peake. I would also like to thank the Max-Planck-Institute for Gravitational Physics, Potsdam (Germany) for its repeated hospitality which allowed me to profit from its cutting-edge research environment. A highlight during my Ph.D. were the two months I could spend at Perimeter Institite, Waterloo (Canada) which were generously supported through its Visiting Graduate Fellowship. I would especially like to thank Sylvain Carrozza, Marc Geiller, Steffen Gielen and Wolfgang Wieland for making this possible and so enjoyable! Special thanks go to Lorenzo Sindoni for all his advice and encouragement! This journey into quantum geometry started in 2011 when I had the chance to spend a fantastic year in the quantum gravity research group at the CPT, Marseille (France) embedded into the beautiful scenery of the Calanques. In all of these places I met marvelous dialogue partners who shaped my perspective. I've never taken it for granted that I was allowed to be there — I was so infinitely lucky!

I am also grateful to all of my collaborators for sharing their knowledge with me and for pushing me further: Joseph Ben Geloun, Marco de Cesare, Alexander Jenkins, Alexander Kegeles, Daniele Oriti, Hans-Christian Ruiz, Mairi Sakellariadou, Johannes Thürigen and Petar Tomov.

I would also like to thank Prof. Dr. Alejandro Perez and Prof. Dr. Vincent Rivasseau for coming to London from France to serve as examiners of my thesis, for the friendly atmosphere during my defense and their kind words.

Now it is time to thank all those remarkable people who had an indirect influence onto this work (and beyond!) by making my life better, happier and funnier: Aniko, Antouanetta, Ben, Dimitris, Fabian, Giorgos, Hans & Petar, Iñigo, Jan, JP (also for your proofreading man!!!), Johann, Johannes, Kameliya, Katrin, Kristian, Lilli, Martin, Matthias, Max, Rita, Ronja, Rosy, Timo, Vasilis, and Yadi. Thank you so much for your friendship!

London. What a city. It took me a while to appreciate what it has to offer and this is simply due to the good friends that I made there: Alix, Amy, Berber, Christophe, Emma, Fabienne, Fedor, Idriss, Lara, Maja, Maria, Oriane and Thomas (1 & 2). Special thanks go to Fedor the fourth. Meeting you had such a positive impact onto my quality of life over there! We shall meet again at 130 bpm (or beyond). Jeder mit jedem ;-)

Finally, my deepest gratitude goes to my family (especially to my sister Anne-Sophie who cooked for me during the last few days before submission and always rescues me when the house is on fire!) for accepting me the way I am, supporting me in any way they can, being there in difficult times and connecting me to the real world. I feel blessed to have you in my life!

One foot on the ground, and the other high up in the skies. Always!

# Ιθάκη

As you set out for Ithaka
hope the voyage is a long one,
full of adventure, full of discovery.
Laistrygonians and Cyclops,
angry Poseidon—don't be afraid of them:
you'll never find things like that on your way
as long as you keep your thoughts raised high,
as long as a rare excitement
stirs your spirit and your body.
Laistrygonians and Cyclops,
wild Poseidon—you won't encounter them
unless you bring them along inside your soul,
unless your soul sets them up in front of you.

Hope the voyage is a long one.
May there be many a summer morning when,
with what pleasure, what joy,
you come into harbors seen for the first time;
may you stop at Phoenician trading stations
to buy fine things,
mother of pearl and coral, amber and ebony,
sensual perfume of every kind—
as many sensual perfumes as you can;
and may you visit many Egyptian cities
to gather stores of knowledge from their scholars.

Keep Ithaka always in your mind.
Arriving there is what you are destined for.
But do not hurry the journey at all.
Better if it lasts for years,
so you are old by the time you reach the island,
wealthy with all you have gained on the way,
not expecting Ithaka to make you rich.

Ithaka gave you the marvelous journey.
Without her you would not have set out.
She has nothing left to give you now.

And if you find her poor, Ithaka won't have fooled you.
Wise as you will have become, so full of experience,
you will have understood by then what these Ithakas mean.

# Preface

## Thesis aim and outline

The goal of this thesis is to apply loop quantum gravity (LQG) to black holes and the related group field theory (GFT) approach to quantum cosmology. In particular, the aim is to study in which way quantum gravitational effects predicted by these theories lead to modifications and improvements as compared to the standard description provided by general relativity (GR).

Specifically, in the case of the application of LQG to black holes the aim is to study if the statistics of the horizon degrees of freedom could be anyonic (as could be naively expected from the dimensionality of the problem) and to inspect potential observational consequences thereof. The result is a reinterpretation of the common reading of the statistics used so far. Moreover, this problem sets the perfect example to gain control over the basics of the loop quantisation which, among many features, is also shared by GFT. This proves highly beneficial for the second part of this thesis. However, it should be noted that the investigation into such exotic statistics relies on the smooth manifold structure (on which LQG is standardly built) since we use that particle exchange can be understood as a parallel transport on such a background. In contrast, in GFT one (fully) dispenses with such a structure and works with combinatorial information only. We would also like to remark that given the limited space, the focus of this first part of the thesis is not to give a detailed motivation of the laws of black hole mechanics, their formulation for isolated horizons and the full details of the derivation of the horizon theory. These can be found in the cited literature.

The second and by far more extensive part of this thesis is concerned with the application of GFT to quantum cosmology. The aim is to understand how the initial singularity problem of GR can be overcome by means of discrete quantum geometry and how a space-time continuum can be recovered from a large number of quanta of geometry. The GFT condensate cosmology approach is motivated by the conjecture that a condensate phase of a suitable GFT model could correspond to a continuum geometry. From a technical point of view, for the investigation of condensate models in the simplest approximation, one has to find solutions to the classical equations of motion of the corresponding GFT model.

Given this setting, the scope of this part of the thesis is to elaborate on aspects of this conjecture and study phenomenological consequences of this approach in detail. The results of this work show, among others, the importance of phenomenology for quantum gravity model building. More specifically, we study various models for $3d$ and $4d$ quantum gravity to this end, investigate the quantum geometric information stored in the solutions and interpret the results against the backdrop of this conjecture. We also examine if a phase transition towards a condensate phase in geometric GFT models can actually be found using Landau's mean field theory. Moreover, we inspect the phenomenological consequences of simplified GFT interactions onto the expansion behaviour of condensates and study the impact of anisotropic perturbations onto models which display bouncing solutions. We see this work as a critical examination of the condensate cosmology approach and the hope is that our results can lead to its improvement and to its corroboration as a new approach to quantum cosmology. It should be noted that we dispensed with a survey of problems of the standard model of cosmology, the inflationary paradigm and non-gravitational bouncing scenarios. Also, a review expostion of canonical quantum cosmology approaches was left out. When referring to them, we direct to the appropriate sources in the literature. A general motivation for studying quantum cosmology and the resolution of the initial singularity are given in the Introduction.

Care was taken to make this thesis as self-contained as possible. Wherever this is not the case due to spatial limitations, this is clearly spelled out and signposts to the literature and primary sources is given (see above). Chapters are mostly interrelated and build on one another. However, the chapters on canonical quantisation and on the LQG black hole model could in principle be read seperately from the rest. Appendices gather either review material or detailed calculations to make the exposition of the main body of this thesis lighter. A comprehensive bibliography is given at the end of this document.

A brief outline is the following: In the Introduction we survey the motivation for the construction of a theory of quantum gravity. In Chapter 2 we review quantum geometro-dynamics and LQG as examples for the canonical quantisation of GR. Chapter 3 applies LQG techniques to black hole physics and analyses the statistics of the horizon quantum geometry. Chapter 4 firstly gives an extensive overview over path integral approaches to quantum gravity. Both the continuum and discrete perspective are discussed. With regard to the discrete approaches, this is done to motivate the GFT approach towards the end this particular chapter. This is also meant to contrast how the attempt at the recovery of the continuum is pursued in these as compared to the way GFT attacks this issue in its condensate cosmology spin-off. Chapter 5 then introduces the GFT condensate cosmology approach. We also investigate solutions to the dynamical Boulatov model as a first application of the formalism and probe the condensate hypothesis using Landau's mean field theory applied to GFT. In Chapter 6 we study consequences for the cosmology of the early and late emergent universe in two more realistic rank-4 models. Finally, in the Conclusion we review our results and discuss further implications of our work.

*To my family.*

# Contents

# List of Figures

# List of Tables

# Notations, conventions and symbols

We mostly use units where Planck's (reduced) constant and the speed of light in vacuum are equal to unity, i.e. $\hbar = c = 1$, unless otherwise stated. Spacetime indices in tensor fields are denoted by Greek letters. Small Latin letters from the beginning of the alphabet refer to their spatial components. Capitalised Latin letters from the second part of the alphabet denote Lorentz indices while small versions thereof denote SU(2) indices. Whenever there is a risk of confusing the meaning of indices this is noted. The signature of the metric used here is $(-, +, +, +)$. Newton's constant is denoted by $G_{\mathrm{N}}$ and $G$ denotes a Lie group.

We may give a list of symbols frequently used throughout this thesis:

| | |
|---|---|
| $g_{\mu\nu}$ | spacetime metric |
| $e_\mu$ | tetrad field |
| $\eta_{\mu\nu}$ | Minkowski metric |
| $\nabla_\mu$ | affine connection |
| $R_{\mu\nu\rho\sigma}$ | Riemann curvature tensor |
| $R_{\mu\nu}$ | Ricci tensor |
| $R$ | Ricci scalar |
| $G_{\mu\nu}$ | Einstein tensor |
| $A$ | connection 1-form |
| $F$ | curvature 2-form of $A$ |
| $\Lambda$ | cosmological constant |
| $T_{\mu\nu}$ | energy-momentum tensor |
| $a$ | scale factor |
| $H$ | Hubble rate |
| $\ell_p$ | Planck length |
| $\gamma$ | Barbero-Immirzi parameter |
| $k_{\mathrm{B}}$ | Boltzmann constant |

A list of repeatedly used abbreviations is:

| | |
|---|---|
| Eq. | an equation |
| Ref. | a reference to the bibliography |
| cf. | compare with |
| GR | general relativity |
| QFT | quantum field theory |
| CS | Chern-Simons |
| MCG | mapping class group |
| RG | renormalisation group |
| FRG | functional renormalisation group |
| IR | infrared |
| UV | ultraviolet |
| LQG | loop quantum gravity |
| LQC | loop quantum cosmology |
| GFT | group field theory |
| RC | Regge calculus |
| QRC | quantum Regge Calculus |
| EDT | Euclidean dynamical triangulations |
| CDT | causal dynamical triangulations |
| TM | tensor models |
| EPRL | Engle-Pereira-Rovelli-Livine |
| BEC | Bose-Einstein condensate |

# Chapter 1

# Introduction

> By doing something a half centimetre high,
> you are more likely to get a sense of the Universe
> than if you try to do the whole sky.
>
> ———————————————————
>
> Alberto Giacometti

Why do we need a theory of quantum gravity at all?

More than a century has passed since quantum mechanics and general relativity (GR) [1] were discovered, essentially revolutionising our understanding of matter, space and time. Research has led to a manifold of quantum theories (QT) and quantum field theories (QFT) [2] on one side, which are used to describe the phenomena of physics on a microscopic scale. On the other side stands GR, which describes the fundamental interaction between matter and spacetime geometry by means of Einstein's field equations and is used to depict the structure of the Universe on a large scale.

Each of these frameworks describes its respective intended domain of physical phenomena to an astounding degree of accuracy, as ever increasing empirical evidence demonstrates. For example, QFT as applied to the Standard Model of particle physics, provides the description of the fundamental interactions of the electromagnetic, weak and strong forces and the classification of all known elementary particles. Precision tests at the LHC have established it as the most precise scientific theory available [3]. Likewise, the recent measurements of gravitational waves produced by inspiralling binary systems [4] have confirmed the predictions of GR with equal fidelity.

However, already from a pragmatic point of view the question arises if an overlapping domain of quantum gravitational phenomena exists and if so, how we could describe and observe it. It is argued that an interface of both frameworks is needed to provide a satisfying description of the microstructure of spacetime together with matter at the so-called Planck scale [1, 5, 6]. This is the natural scale where effects of quantum gravity are expected to occur. Furthermore, a careful analysis of the underlying elementary and universal assumptions of these frameworks leads to the observation that they are mutually incompatible from a conceptual point of view. The need to overcome this confusion in fundamental physics has spurred research on quantum gravity for more than 80 years and has led to the development of a plethora of approaches each with their individual strengths and weaknesses; yet a complete and consistent solution has notheless remained elusive, as all candidates suffer from formal and conceptual problems [6].

It is often invoked that their biggest common problem would ultimately lie in the difficulty to test their predictions experimentally (if these are forthcoming) due to the fact that their relevant scale of application is too far away for current technological capacities. To use the current absence of empirical evidence as an argument against research in this field would however ignore the fact that the precision of cosmological and astrophysical observations has considerably increased over the last decades which might eventually lead to the detection of quantum gravity imprints e.g. onto the spectrum of primordial fluctuations. Such data will provide the needed guidance to dismiss, modify or even construct overall new approaches to the problem.

In the following we want to dwell a little bit more on the mutual incompatibilities as well as conceptual problems which plague the two frameworks and give further incentives to motivate the different strategies developed to escpape this impasse.

The structure of ordinary QFTs assumes the existence of a fixed, non-dynamical background metric living on a fixed, non-dynamical topological and differentiable manifold [7–9]. This framework breaks down when the gravitational field and the manifold structure become dynamical and no fixed background metric is available. In particular, it neglects the backreaction between geometry and matter, i.e. the interwoven co-evolution of the dynamical background structure and the matter fields, as expressed via Einstein's field equations

$$G_{\mu\nu}(g) = 8\pi G_{\mathrm{N}} T_{\mu\nu}(g) \tag{1.1}$$

and becomes fully inadequate for the description of extreme astrophysical and cosmological situations where the metric is expected to fluctuate wildly. Even worse, such a backreaction cannot be consistently described by means of these equations of motion. This is because matter fields are fundamentally quantum mechanical and thus obey probabilistic laws whereas the geometric content is classical and deterministic. Even when ignoring this and promoting the energy-momentum tensor to an operator, the computation of its expectation value would depend on a fixed spacetime background, though the idea of the field equations is to have the metric dynamical in the first place. Again this seems strongly inconsistent, even though it leads to revealing and beautiful insights from QFT on curved spacetime for quantum gravity (see below). Thus, taking both the universal coupling of gravity to all forms of energy and the universality of quantum physics seriously the quantisation of the gravitational field is naturally suggested [1].

Moreover, indications for a breakdown of QFT and GR at very small length scales can be identified. Divergences at large momenta typically render QFTs ill-defined [2] and its expected that if gravity was consistently taken into account, providing a natural ultraviolet cutoff in terms of the Planck length, this limitation could be cured. A theory of quantum gravity should then also be able to say something about the fate of the corresponding infinite vacuum energies which should contribute to the cosmological constant and explain its surprising tininess [5]. On the other hand, Penrose and Hawking have proven that there are inevitable spacetime singularities in the context of gravitational collapse under reasonable conditions on causality and energy [10–12]. This is stated by the singularity theorems, referring to the assumed singularities which govern the internal structure of black holes and to the initial singularity of the cosmos. Hence, in domains of strong gravitational fields, GR loses its predictivity and cannot be a valid theory without restrictions. From a fundamental point of view, such singularities are unphysical and it is expected that quantum effects lead to their resolution [13].

One could try to make progress with the problem of quantum gravity by studying quantum disturbances around a fixed classical background metric. Given the success of the perturbative quantisation recipe for non-gravitational theories at small coupling this could seem as a viable option. In this way, in the low-energy limit particle-like excitations of spin-2 are found, corresponding to gravitions. However, the naive power counting argument entails that this method leads to a non-renormalisable theory exempt of any predictive power in the ultraviolet [14, 15]. It is also not clear if this method yields the same results when physically different backgrounds are perturbed [5]. It was hoped that supersymmetric extensions of gravity could enhance the ultraviolet behaviour by offering a mechanism to cancel perturbative divergences [16] but endeavours along this way proved futile so far [5].

Inspected from a different angle, one might expect the situation to be improved when considering that the continuum should effectively be replaced at very small length scales in favour for some kind of discretum.

Clear physical hints that the smooth spacetime geometry has to give way for a discrete, atomistic and combinatorial picture is provided by the phenomenon of black hole thermodynamics [17]. It became clear that black holes emit radiation at Hawking temperature proportional to their surface gravity leading to their evaporation and that they should be associated an entropy proportional to their area [18–20]. Notice this is obtained using QFT on curved spacetime the descriptive power of which breaks down at the final stages of black hole evaporation (as no backreaction is considered). It is nevertheless understood as a first approximation to a theory of quantum gravity [1] and the result for the entropy calls for an explanation through more fundamental degrees of freedom behind the macroscopic description of the gravitational field as given by the metric.

As outlined above, the difficulty in making progress in this field is rooted to the lack of experiments which have access to the phyiscs at the smallest length scales. It is thus vital to consider quantum gravity in a cosmological context because the highest possible energy scales were reached in the Planck era of the cosmos, i.e. in the vicinity of and shortly after the big bang. One can thus naturally expect that traces of quantum gravity have left a fingerprint on the spectrum of the cosmic microwave background radiation, see e.g. Refs. [21–26]. Moreover, quantum gravity could provide an underpinning or replacement of the standard inflationary scenario [27] the shortcomings of which can be identified, among others, in its inability to clarify the choice of initial conditions used, to resolve the initial singularity [28] and to explain the trans-Planckian mode problem [29]. In fact, a promising alternative to resolve the problems of the standard model of cosmology is provided by bouncing cosmologies [30], in particular those where the bounce is caused by quantum gravity effects. Thus cosmology and quantum cosmology have an important role to play in gaining insight into the problem of quantum gravity.

From a philosophical point of view, there are also more meta-theoretical arguments for constructing a theory of quantum gravity [31, 32]. For instance, it could be expected that a theory which unifies the concepts of GR and QFT should, apart from making original predictions, also be stronger in explaining established facts. A unification of concepts is often also associated with a reduction of complexity by finding a single coherent framework. Aiming at unity could refer to mapping the unity of Nature, i.e. that Nature has a unified structure and one expects that a systematic description of its empirically accessible parts can be given. It could also refer to the unity of scientific method, i.e. that there is a unique way to generate scientific knowledge and that scientific theories are supposed to be unified concerning their terminology, ontology and nomologicality. A historicistic perspective on the evolution of scientific theories seems to support this point of view, at least in the context of physics where convergence between theories over the course of time is observed. These arguments are certainly sufficient to motivate the quest for a theory of quantum gravity. The necessary arguments for the quantisation of gravity, however, are provided by physical arguments, some of which were elaborated above.

In the light of these issues and points, we may roughly group modern approaches to quantum gravity into two classes: perturbative (i.e. background dependent) and non-perturbative (i.e. background independent) approaches [5].

The only known consistent representative in the first class is string theory which attempts to provide a description which unifies all fundamental interactions through more fundamental objects living on a higher-dimensional target space [33]. Essentially, the idea is to increase the amount of symmetries as compared to GR and QFT with the aim to regain perturbative renormalisability. This is strongly inspired by the replacement of the perturbatively non-renormalisable Fermi model for the weak interaction through the renormalisable electroweak theory. Importantly, the mass spectrum of the particle-like excitations of the string contains a massless spin-2 particle which can be understood as the graviton. This is due to the fact that the low-energy effective action of string theory contains the Einstein-Hilbert action up to corrections. However, the introduction of extra background structure comes at a high price: The compactification of the extra dimensions can be done in various ways leading to a vast vacuum degeneracy of the theory and it is unclear if any of these vacua lead to a low-energy effective theory which is in agreement with the standard model of particle physics. It is also unclear how to achieve the spontaneous breaking of supersymmetry which is needed to eliminate not observed superpartners from the spectrum. The problem of the vacuum degeneracy is also not ameliorated via the M-theory interpretation of string theory but rather aggravated leading to the landscape problem. An interesting offspring of string theory understood as an application of the holographic principle is the celebrated AdS/CFT conjecture which argues for a correspondence of string theory on an anti-de Sitter background to a conformal QFT on the boundary of this space. Whether this example for a theory of quantum gravity is realistic can be questioned due to the fact that our Universe is in a de Sitter phase [5].

Non-perturbative approaches, on which this thesis focuses, venerate the role of background independence and general covariance as unveiled by GR and keep them as guiding principles for their construction. Also no additional structure is added to conform to the principle of minimality. In most of these approaches, the spacetime continuum is renunciated and is instead replaced by degrees of freedom of discrete and combinatorial nature. These approaches fall into two subclasses according to whether gravity is quantised by canonical quantisation or covariantly through a discrete version of the path integral. Quantum geometrodynamics [34] and loop quantum gravity (LQG) [5] fall into the former class. Particular representatives of the second class of theories are the closely related covariant loop quantum gravity/spin foam approach [35], group field theory (GFT) [6, 36–38], tensor models (TM) [39–44] and simplicial quantum gravity approaches like quantum Regge calculus (QRC) [45–47] and Euclidean and causal dynamical triangulations (EDT,CDT) [45, 48]. The perturbative expansion of the path integral of these theories then generates a sum over discrete geometries. The most difficult problem for all of these approaches is the recovery of continuum spacetime, diffeomorphism invariance and GR as an effective description for the dynamics of the geometry in an appropriate limit. This thesis will give a detailed exposition of these approaches in the subsequent chapters, with particular reference to quantum gravity (i. e. quantum geometry) from the point of view of LQG applied to black hole physics and GFT applied to quantum cosmology. We would like to remark that the important impact of matter degrees of freedom plays only a supporting role here because matter coupling in this context is a problem on its own and of formidable difficulty.

# Chapter 2

# Canonical quantum gravity

> One instant, one aspect of nature contains it all.
>
> ———————————————————
>
> Claude Monet,
> commenting on what he called his
> *paysages d'eau.*

> No more earth, no more sky, no more limits.
> Now time and space are suspended.
>
> ———————————————————
>
> A journalist,
> commenting on the same.

In the course of this Chapter we review two attempts at the canonical quantisation of general relativity (GR).[1] We first summarise the quantum geometrodynamics perspective and then review elements of loop quantum gravity (LQG). The formulation of both approaches is explicitly background-independent and non-perturbative. We keep the exposition of this material brief and focus onto the introduction of notions which are relevant to the subsequent chapters of this thesis. In particular, Section 2.2 on LQG is indispensable for the application of loop gravity techniques to black holes in Chapter 3 and is also needed to understand the foundations of the group field theory (GFT) approach, as given in Section 4.2.4, and its application to quantum cosmology in Chapters 5 and 6.

---

[1] This is to be contrasted to attempts at the path integral quantisation of gravity either given in the continuum or in the discrete formulation, as reviewed in Chapter 4.

## 2.1 Quantum geometrodynamics in a nutshell

The following section discusses the quantum geometrodynamics approach to the canonical quantisation of GR which was developed by Dirac, Wheeler, De Witt, Arnowitt, Deser and Misner (ADM) [13, 49–53]. To this aim, we first have to find the Hamiltonian formulation of the Einstein-Hilbert (EH) action

$$S_{\text{EH}} = \frac{1}{2\kappa} \int d^4x \sqrt{-g} R^2 \tag{2.1}$$

which necessitates the imposition of a $3 + 1$-splitting of spacetime. In the action $\kappa$ equals $8\pi G_{\text{N}}$, $g = \det(g_{\mu\nu})$ and $R$ denotes the Ricci scalar. The splitting is accomplished by assuming that $(\mathcal{M}, g)$ is a globally hyperbolic spacetime. Then a global causal time function $t \in \mathbb{R}$ can be chosen and $\mathcal{M}$ can be foliated into Cauchy hypersurfaces $\Sigma_t$. This fixes the topology of spacetime to $\mathcal{M} = \mathbb{R} \times \Sigma$, where $\Sigma$ denotes a $3d$ manifold which has arbitrary topology and is of spacelike signature. Importantly, the choice of the global time function does not introduce time as an absolute quantity and does not lead to a preferred foliation as guaranteed by the diffeomorphism invariance of the theory. Hence, all foliations are physically equivalent.

With the foliation and local coordinates $(t, x)$ given, we define the time flow vector field $T^\mu(x)$ for convenience as

$$T^\mu(x) = (1, 0, 0, 0), \tag{2.2}$$

along which the hypersurfaces $\Sigma_t$ are aligned. We decompose it into its normal and tangential parts, such that

$$T^\mu(x) = N(x) n^\mu(x) + N^\mu(x), \tag{2.3}$$

where $n^\mu$ denotes the timelike unit normal vector field to $\Sigma_t$. We may then parameterise it by

$$n^\mu = \left( \frac{1}{N}, -\frac{N^a}{N} \right) \quad \text{and} \quad N^\mu = (0, N^a). \tag{2.4}$$

---

[2]A discussion of the subsequent points including the cosmological constant term and boundary terms is presented in Refs. [34] and [5], respectively.

In these expressions $N$ denotes the lapse function and $N^a$ is the shift vector. Equipped with this, the 4-metric rewrites as

$$g_{\mu\nu} = \begin{pmatrix} (N_a N^a - N^2) & N_b \\ N_a & g_{ab} \end{pmatrix},$$ (2.5)

where $g_{ab}$ with $a, b = 1, 2, 3$ denotes the 3-metric. The latter is not equivalent to the intrinsic metric on $\Sigma_t$, in general. In fact, it is given by the first fundamental form on $\Sigma_t$

$$h_{\mu\nu} = g_{\mu\nu} + n_\mu n_\nu.$$ (2.6)

It induces the tensor calculus on $\Sigma_t$ from the one on $\mathcal{M}$. The second fundamental form is then given by the extrinsic curvature of $\Sigma_t$

$$K_{\mu\nu} = \frac{1}{2} \mathcal{L}_n h_{\mu\nu} = h_\mu^{\mu'} h_\nu^{\nu'} \nabla_{\mu'} n_{\nu'},$$ (2.7)

where $\mathcal{L}$ denotes the Lie-derivative.[3]

With these two objects we can induce the Riemann tensor on $\Sigma_t$, as encoded by the Gauss-Codazzi equation

$$^{(3)}R = \,^{(3)}R_{\mu\nu\rho\sigma} h^{\mu\rho} h^{\nu\sigma} = \left( K_{\mu\nu} K^{\mu\nu} - K^2 \right) + h^{\mu\rho} h^{\nu\sigma} R_{\mu\nu\rho\sigma}$$ (2.8)

and rewrite the Langrangian in Eq. (2.1). This reasoning allows to identify the configuration space variables as $h_{ab}$ with conjugate momenta

$$\pi^{ab} = \frac{\sqrt{h}}{2\kappa} \left( K^{ab} - K h^{ab} \right)$$ (2.9)

and give the Legendre transform of the Einstein-Hilbert action as

$$S_{\text{EH}}[h_{ab}, \pi^{ab}, N, N^a] = \frac{1}{2\kappa} \int \mathrm{d}t \int \mathrm{d}^3 x \left( \pi^{ab} \dot{h}_{ab} - N^a H_a - N H \right)$$ (2.10)

which is the so-called canonical ADM action. Therein we have

$$H_a = -2 h_{ac} \nabla_b \left( \pi^{bc} \right) \quad \text{and} \quad H = 2\kappa G_{abcd} \pi^{ab} \pi^{cd} - \frac{\sqrt{h}}{2\kappa}\,^{(3)}R$$ (2.11)

---

[3]The trace $\theta$ of the extrinsic curvature measures the expansion of a geodesic congruence orthogonal to $\Sigma$ [1].

with $G_{abcd} = \frac{1}{2\sqrt{h}}\left(h_{ac}h_{bd} + h_{ad}h_{bc} - h_{ab}h_{cd}\right)$[4] and $N, N^a$ are identified as Lagrange multipliers. Variation with respect to the latter gives the Hamiltonian/scalar and spatial diffeomorphism/vector constraint, i.e.

$$H = 0 \quad \text{and} \quad H^a = 0 \tag{2.12}$$

to be satisfied by on-shell/physical configurations. Both together define the so-called constraint hypersurface in phase space.[5]

The phase space carries the symplectic structure

$$\left\{\pi^{ab}(t,x), h_{cd}(t,x')\right\} = 2\kappa\delta^a_{(c}\delta^b_{d)}\delta(x,x') \tag{2.13}$$

which can be derived from the symplectic potential, i.e. the first term in Eq. (2.10). Using this symplectic structure together with the smearing of the constraints

$$\vec{H}(\vec{N}) = \int_\Sigma \mathrm{d}^3x H^a(x)N_a(x) \quad \text{and} \quad H(N) = \int_\Sigma \mathrm{d}^3x H(x)N(x) \tag{2.14}$$

we may give the algebra of hypersurface deformations[6]

$$\left\{\vec{H}(\vec{N}), \vec{H}(\vec{N}')\right\} = -2\kappa\vec{H}(\mathcal{L}_{\vec{N}}\vec{N}')$$
$$\left\{\vec{H}(\vec{N}), H(N)\right\} = -2\kappa H(\mathcal{L}_{\vec{N}}N)$$
$$\left\{H(N), H(N')\right\} = -2\kappa\vec{H}\left(\vec{N}(N, N', h)\right) \tag{2.15}$$

the right-hand sides of which vanish on the constraint hypersurface. Hence, the constraints are of first class and generate gauge transformations. More specifically, $H^\mu$ are generators

---

[4]This object is also known as the supermetric due to the fact that it acts as a metric in the space of all metrics. This so-called superspace is defined via $\mathcal{S}(\Sigma) \cong \mathrm{Riem}(\Sigma)/\mathrm{Diff}(\Sigma)$ where $\mathrm{Riem}(\Sigma)$ denotes the set of all 3-metrics of the manifold $\Sigma$ [34, 54].

[5]Notice that the Hamiltonian $H = \frac{1}{2\kappa}\int \mathrm{d}^3x N^\mu H_\mu$ vanishes for physical configurations, implying that there are no dynamics with respect to the parameter $t$. This peculiarity leads to the problem of time in GR and quantum gravity [31, 55, 56].

[6]One refers to this algebra also as the Dirac or Bergmann-Komar algebra. It is in general not a Lie algebra.

of the spacetime diffeomorphism group Diff($\mathcal{M}$), but on-shell only.[78]

When proceeding with the Dirac quantisation [57] of this system, in a nutshell we would have to promote the kinematical phase space to the kinematical Hilbert space $\mathcal{H}_{\text{kin}}$ and the constraints $H^\mu$ to operators $\hat{H}^\mu$ therein. Subsequently, we would have to find the physical Hilbert space $\mathcal{H}_{\text{phys}}$ with its inner product and candidates for observables. The physical Hilbert space consists of the states which solve the constraints, i.e. $\hat{H}^\mu|\psi\rangle = 0$. Disappointingly, already the first step cannot be completed in the quantum geometrodynamics programme since the scalar product on $\mathcal{H}_{\text{kin}}$ cannot be given. This problem is naturally handed down to the construction of $\mathcal{H}_{\text{diff}}$, consisting of states which solve $\hat{H}^a|\psi\rangle = 0$ and, among others, is a problem for the construction of $\mathcal{H}_{\text{phys}}$, found by solving the Wheeler-DeWitt equation

$$\hat{H}|\psi\rangle = 0. \tag{2.16}$$

So far, it has only been possible to make progress in this approach when symmetry-reduced scenarios are considered. In these studies, one imposes a high degree of symmetry in the classical theory (e.g. homogeneity and isotropy) leading to midi- and minisuperspace models and then proceeds with Dirac quantisation. For certain situations, as applications to quantum cosmology show, these models might offer sufficient approximations but they are expected to be inadequate in full generality due to the fact that symmetry reductions violate the uncertainty principle [34].[9]

## 2.2 Elements of Loop Quantum Gravity

Another attempt at the canonical quantisation of gravity is loop quantum gravity. In contrast to quantum geometrodynamics, the kinematical Hilbert space is under full control in this approach. This has allowed the application of LQG methods to study aspects of black hole physics and quantum cosmology. In spite of many efforts, the dynamics of the theory are still not fully understood. To bypass this problem, the spin foam and group field theory approaches to quantum gravity were developed, as reviewed later on in Sections 4.2.3 and 4.2.4, which attack the problem of the dynamics through a functional integral prescription. In the following, we give a brief presentation of the classical theory which underlies LQG and then proceed to elements of its canonical quantisation. To this aim we mostly follow Refs. [5, 58–60]. This is also done in view of the application of LQG to a black hole model based on Chern-Simons theory coupled to curvature defects in Chapter 3 and will also serve as the background for the application of GFT techniques to quantum cosmology in Chapters 5 and 6.

---

[7]If off-shell configurations are also considered (as e.g. in the quantum theory), then the group generated by the Dirac algebra is different to Diff($\mathcal{M}$). This so-called Bergmann-Komar group is a dynamical symmetry group of GR as opposed to Diff($\mathcal{M}$) being the kinematical symmetry group of any generally covariant theory. From a geometric point of view, the diffeomorphism group maps different foliations into one another while the Bergmann-Komar group deforms them. We refer to Refs. [5, 34] for a thorough discussion of these matters.

[8]The Poisson structure defines the $(12 \cdot \infty^3)$-dimensional kinematical phase space of the theory. Reduced by the dimension of the constraint hypersurface $(4 \cdot \infty^3)$ and that of the gauge orbits generated by the gauge transformations $(4 \cdot \infty^3)$, this leads to a $4 \cdot \infty^3$-dimensional physical phase space. Hence, the number of physical degrees of freedom of GR amounts to 2.

[9]Notice that fixing the topology of spacetime at the classical level as implied by global hyperbolicity has the effect that the quantum theory defined via canonical quantisation disallows topology change. We return to this point in Chapter 4 in the context of the path integral quantisation of gravity.

### 2.2.1 Holst action

As recalled above, the formulation of quantum geometrodynamics relies on the Einstein-Hilbert action and thus stresses the importance of the metric field. In constast, the formulation of the classical theory on which LQG is built, puts emphasis on a connection and on a frame field, as already suggested by the Palatini formulation of gravity.

To see this, we rewrite the metric $g$ on $\mathcal{M}$ in terms of 1-forms, the so-called tetrad fields $e^I(x)$ with $I = 0, ..3$, and the flat (internal) metric $\eta_{IJ}$, i.e.

$$g_{\mu\nu}(x)\mathrm{d}x^\mu\mathrm{d}x^\nu = e^I_\mu(x)e^J_\nu(x)\eta_{IJ}\mathrm{d}x^\mu\mathrm{d}x^\nu, \tag{2.17}$$

establishing a local isomorphism between a general reference frame and an inertial one. The index $I$ indicates that the tetrads transform with respect to the Lorentz group, i.e.

$$e^I(x) \to \tilde{e}^I(x) = \Lambda(x)^I_J e^J(x) \tag{2.18}$$

with $\Lambda(x) \in \mathrm{SO}(3,1)$. This group amounts to an additional gauge symmetry of general relativity when the theory is reformulated in terms of tetrads and requires us to introduce an associated connection.[10] In units where $2\kappa = 1$, one may write for the Einstein-Hilbert action

$$S_{\mathrm{EH}}[g] = \int \mathrm{d}^4x\sqrt{-g}R = \int \mathrm{tr}\left[\star\left(e \wedge e\right) \wedge F(\omega(e))\right] = S_{\mathrm{EH}}[e], \tag{2.19}$$

---

[10]Beyond our motivation to introduce tetrads here, notice that this formalism is needed to couple fermions to general relativity.

where

$$F^{IJ} = \mathrm{d}\omega^{IJ} + \omega^I_K \wedge \omega^{KJ} \tag{2.20}$$

denotes the $\mathfrak{so}(3,1)$-valued curvature 2-form of the unique torsionless spin connection $\omega^{IJ}$.[11] We emphasise that apart from the usual diffeomorphism invariance, this action is explicitly invariant under local Lorentz transformations.

One can further rewrite the action into Palatini/first-order form which depends on two independent fields, i.e.

$$S_{\mathrm{P}}[e,\omega] = \int \mathrm{tr}\left[\star\left(e \wedge e\right) \wedge F(\omega)\right]. \tag{2.21}$$

It gives the same equations of motion as the Einstein-Hilbert action if we require the tetrads to be non-degenerate.[12] We may appreciate that in this way $e$ and $\omega$ are ennobled to fundamental fields while the metric $g$ is fully degraded to a derived quantity.

At the heart of LQG lies the Holst action

$$S_{\mathrm{H}}[e,\omega] = \int \mathrm{tr}\left[\left(\star\left(e \wedge e\right) + \frac{1}{\gamma}e \wedge e\right) \wedge F\right], \tag{2.22}$$

where the so-called Holst term, coupled by the Barbero-Immirzi parameter $\gamma \in \mathbb{R} - \{0\}$, was added to Eq. (2.21) and it is compatible with diffeomorphism and local Lorentz invariance. Also this action yields the same equations of motion as the Einstein-Hilbert action.[13] As we will see below, the parameter $\gamma$ and the term it couples play an important role in the quantum theory to be formulated.

In passing, we would like to remark that one could in principle add further (but finitely many) terms to the Palatini action which are compatible with the given symmetries. The

---

[11] In the metric formulation one requires $\Gamma(g)$ to be metric compatible and torsion free, giving the Levi-Civita connection. In the same way, the connection $\omega^{IJ}$ is required to fulfill the tetrad postulate and to be torsion free, yielding the unique torsionless spin connection. With this we may relate the curvature of the connection to the Riemann tensor as $F^{IJ}_{\mu\nu}(\omega(e)) = e^{I\rho}e^{J\sigma}R_{\mu\nu\rho\sigma}(e)$.

[12] The reason for this is that $\delta_\omega S = 0$ yields that the connection is the unique torsionless spin connection while $\delta_e S = 0$ then leads to Einstein's field equations.

[13] The condition of torsionlessness implies that this additional term has no effect onto the equations of motion derived from $\delta_\omega S = 0$ while the first Bianchi identity guarantees that it does not effect the equations of motion derived from $\delta_e S = 0$. Notice, however, that in presence of a source of torsion, the Holst term is of relevance for the classical theory [61, 62].

most general action for pure gravity then takes the form

$$
\begin{aligned}
S[e, \omega] = {} & \alpha_1 \int \operatorname{tr}\left[\star\left(e \wedge e\right) \wedge F(\omega)\right] + \alpha_2 \int \operatorname{tr}\left[\star\left(e \wedge e\right) \wedge e \wedge e\right] \\
& + \alpha_3 \int \operatorname{tr}\left[e \wedge e \wedge F(\omega)\right] + \alpha_4 \int \operatorname{tr}\left[F \wedge F\right] \\
& + \alpha_5 \int \operatorname{tr}\left[\star F \wedge F\right] + \alpha_6 \int \operatorname{tr}\left[\mathrm{d}_\omega e \wedge \mathrm{d}_\omega e - e \wedge e \wedge F\right],
\end{aligned}
\tag{2.23}
$$

where $\alpha_1, \ldots, \alpha_6$ denote appropriate coupling constants.[14] The first term corresponds to the already discussed Einstein-Hilbert-Palatini action. Together with the third term one retrieves the Holst action. The second term captures the cosmological constant term. Finally, terms four to six correspond to the Pontryagin, Gauss-Bonnet-Euler and Nieh-Yan terms which are all topological. Notice that the Nieh-Yan term in a torsion-free theory, i.e. when $T^I = \mathrm{d}_\omega e^I = 0$ holds, immediately reduces to the Holst term. As for the Holst term, the last three terms do not alter the classical equations of motion of gravity.[15] However, they do affect the canonical structrue of the theory (i.e. they produce canonical transformations) and thus might have an impact in the quantum theory. We refer to Refs. [63, 64] and references therein for a thorough discussion of these matters. In the following, we will only focus on the Hamiltonian formulation for the Holst action.

### 2.2.2 Hamiltonian formulation and new variables

As in Section 2.1, when assuming $\mathcal{M}$ to be globally hyperbolic, we can proceed with the Hamiltonian formulation of the theory. Using local coordinates $(t, x)$ on the splitting $\mathcal{M} \cong \mathbb{R} \times \Sigma$[16], this leads to the two canonically conjugated variables

$$
E_i^a = e e_i^a = \frac{1}{2} \epsilon_{ijk} \epsilon^{abc} e_b^j e_c^k,
\tag{2.24}
$$

---

[14]Observe the advantage of the tetrad over the metric formulation: In the metric theory, in principle the theory space is infinite dimensional while the tetrad formulation strongly reduces the space of actions.

[15]It should be noted that the Holst, Pontryagin and Nieh-Yan terms are parity odd in fact.

[16]In fact, the choice of coordinates amounts to a partial gauge fixing of the gauge freedom introduced by the Lorentz symmetry. This fixing to the time-gauge amounts to the reduction of the Lorentz group to its rotation subgroup which leaves invariant the normal to the hypersurface. It aligns the time axis (as given by the co-vector $e_a^0$) to the one of the foliation (as given by the hypersurfce normal). This procedure is naturally adapted to the Hamiltonian formulation of gravity where the $3 + 1$-splitting of spacetime is already available.

called the densitised inverse triad with $e = \det(e) = \sqrt{h}$[17] and the Ashtekar-Barbero connection

$$A_a^i = \gamma K_a^i + \omega_a^i = \gamma K_a^i + \frac{1}{2}\epsilon_{jk}^i \omega_a^{jk} \tag{2.25}$$

where $K_a^i \equiv \omega_a^{0i}$ denotes the extrinsic curvature of $\Sigma$[18] with the fundamental Poisson bracket

$$\{A_a^i(x), E_j^b(y)\} = \gamma \delta_a^b \delta_j^i \delta^3(x, y). \tag{2.26}$$

The indices $a, b$ refer to $\Sigma$ while $i, j$ refer to the internal space. With these variables, one may bring the Holst action in canonical form, giving

$$S_{\mathrm{H}}[A, E, N, N^a] = \frac{1}{\gamma} \int \mathrm{d}t \int_\Sigma \mathrm{d}^3 x \left( \dot{A}_a^i E_i^a - \left( A_0^i G_i + NH + N^a H_a \right) \right). \tag{2.27}$$

The Lagrange multipliers $A_0^i$, $N^a$ and $N$ enforce the Gauss constraint

$$G_i(A, E) = D_a E_i^a = \partial_a E_i^a + \epsilon_{ijk} A_a^j E^{ak}, \tag{2.28}$$

the vector constraint

$$H_a(A, E) = \frac{1}{\gamma} F_{ab}^j E_j^b - \frac{1 + \gamma^2}{\gamma} K_a^i G_i, \tag{2.29}$$

and the scalar/Hamiltonian constraint

$$H(A, E) = \left( F_{ab}^j - (\gamma^2 + 1)\epsilon_{jmn} K_a^m K_b^n \right) \frac{\epsilon_j^{kl} E_k^a E_l^b}{2\sqrt{|\det(E)|}} + \frac{1 + \gamma^2}{\gamma} G^i \partial_a \frac{E_i^a}{\sqrt{|\det(E)|}}, \tag{2.30}$$

respectively.

When smearing the constraints appropriately over the hypersurface $\Sigma$, we have

$$G(\Lambda) = \int_\Sigma \mathrm{d}^3 x \ \Lambda^i(x) G_i(x), \quad V(\vec{N}) = \int_\Sigma \mathrm{d}^3 x \ N^a H_a \quad \text{and} \quad S(N) = \int_\Sigma \mathrm{d}^3 x \ N(x) H(x). \tag{2.31}$$

---

[17]The densitised inverse triad is an $\mathfrak{su}(2)$-valued 2-form which encodes metric information of $\Sigma$.

[18]The Ashtekar-Barbero connection is an SU(2)-connection allowing us to define a notion of parallel transport on the hypersurface $\Sigma$. Notice that it is constructed from the spin connection $\omega^{IJ}$ but it is not its pull back [5, 65]. Only when $\gamma = i$, leading to the formulation called Ashtekar gravity, the present SU(2)-gauge group is indeed the self-dual subgroup of the Lorentz group. For the recovery of GR one then has to impose reality conditions which lead to major obstacles in the formulation of the quantum theory. Due to this fact, LQG has mostly been developed for real-valued Barbero-Immirzi parameter [5].

With this we may give the constraint algebra [5, 58, 59], yielding

$$
\begin{aligned}
\{G(\Lambda), G(\Lambda')\} &= G([\Lambda, \Lambda']), \\
\{G(\Lambda), V(\vec{N})\} &= G(\mathcal{L}_{\vec{N}}\Lambda), \\
\{G(\Lambda), H(N)\} &= 0, \\
\{V(\vec{N}), V(\vec{N'})\} &= V([\vec{N}, \vec{N'}]), \\
\{V(\vec{N}), S(M)\} &= -S(\mathcal{L}_{\vec{N}}M), \\
\{S(N), S(M)\} &= -V\left((N\partial_b M - M\partial_b N)\, h^{ab}\right) \\
&\quad - G\left((N\partial_b M - M\partial_b N)\, h^{ab} A_a\right) \\
&\quad - \frac{1+\gamma^2}{\gamma^2} G\left(\frac{[E^a\partial_a N, E^b\partial_b M]}{|\det(E)|}\right),
\end{aligned}
\tag{2.32}
$$

the right-hand sides of which vanish on the constraint hypersurface. Hence, all constraints are of first class. Similar to Section 2.1, the constraint algebra is not a Lie-algebra because the bracket between two scalar constraints gives a structure function, not a structure constant. One can show that the Gauss constraint generates SU(2)-gauge transformations, reflecting the fact that shifting to the tetrad formalism we introduced an SO(1, 3)-gauge symmetry to the theory. Moreover, on-shell one can show that $H$ and $H^a$ together generate the diffeomorphism group Diff($\mathcal{M}$) [5].

### 2.2.3 Quantum theory

#### 2.2.3.1 Kinematical Hilbert space $\mathcal{H}_{\mathbf{kin}}$

As a preparation for the Dirac quantisation, we introduce the smearing of the canonically conjugated variables to exclude any information regarding the background from the definition of the quantum algebra. Since the densitised inverse triad is a 2-form, it is smeared over a surface $S$ embedded in $\Sigma$ giving the flux

$$
E_i(S) = \int_S \mathrm{d}^2\sigma \; n_a E_i^a,
\tag{2.33}
$$

where $n_a$ is the normal to the surface. In contrast, the connection is smeared over a 1-dimensional path $\gamma$ since $A$ is a 1-form. This is accomplished by introducing the holonomy

$$h_\gamma(A) = \mathcal{P}\mathrm{e}^{\int_\gamma A}, \tag{2.34}$$

where $\mathcal{P}$ denotes the path-ordering.[19]  In this way, the Poisson algebra of the canonically conjugated pair $(A_a^i, E_i^a)$ is turned into the so-called holonomy-flux algebra. In the next step, a representation of this algebra on an auxiliary kinematical Hilbert space has to be found.[20]

Notice that for the definition of such a Hilbert space the explicit knowledge of the inner product is needed. Then the challenge is to find a measure on the space of connections which does not refer to any fixed background metric. The solution to this problem is provided by the introduction of cylindrical functions. These are functionals which depend on the holonomies $h_e(A)$ along a finite set of oriented paths $e$, called links. To define these objects, consider a graph $\Gamma$ in $\Sigma$ consisting of nodes, links and adjacency relations between the nodes by means of links. With this, a cylindrical function is a pair $(\Gamma, f)$ of a graph $\Gamma$ with total number of links $L$ and a smooth function

$$f : \mathrm{SU}(2)^L \to \mathbb{C} \tag{2.36}$$

giving a functional of the connection $A$, i.e.

$$\langle A | \Gamma, f \rangle = \psi_{(\Gamma, f)}(A) = f\left(h_{e_1}(A), \dots, h_{e_L}(A)\right) \in \mathrm{Cyl}_\Gamma. \tag{2.37}$$

Endowed with a scalar product,

$$\langle \psi_{(\Gamma, f)} | \psi_{(\Gamma, g)} \rangle = \int \prod_e \mathrm{d}h_e \, \overline{f\left(h_{e_1}(A), \dots, h_{e_L}(A)\right)} g\left(h_{e_1}(A), \dots, h_{e_L}(A)\right) \tag{2.38}$$

with $\mathrm{d}h$ being the Haar measure on $\mathrm{SU}(2)$, the space $\mathrm{Cyl}_\Gamma$ is turned into the Hilbert space $\mathcal{H}_\Gamma$. The Hilbert space of all cylindrical functions for all graphs is given by the direct sum

---

[19]The holonomy is formally defined as the solution to the differential equation

$$\frac{\mathrm{d}}{\mathrm{d}t} h_\gamma(t) - h_\gamma(t) A(\gamma(t)) = 0, \quad \text{with} \quad h_\gamma(0) = 1. \tag{2.35}$$

[20]Notice that despite the fact that the three-dimensionally smeared triads do Poisson commute with each other, the fluxes do not. If they were assumed to commutate, the Jacobi identity would be violated [66, 67]. This Poisson non-commutativity plays an essential role for the quantum discreteness of geometry [5, 68].

of Hilbert spaces $\mathcal{H}_\Gamma$ associated to a given graph, i.e.

$$\mathcal{H}_{\mathrm{kin}} = \oplus_{\Gamma \subset \Sigma} \mathcal{H}_\Gamma \tag{2.39}$$

and the generalisation of the scalar product to $\mathcal{H}_{kin}$ is immediate. Importantly, it can be shown that this Hilbert space is equivalent to a Hilbert space over connections on $\Sigma$, namely

$$\mathcal{H}_{\mathrm{kin}} = L^2[\mathcal{A}, \mathrm{d}\mu_{\mathrm{AL}}]. \tag{2.40}$$

In this expression $\mathcal{A}$ denotes the space of (generalised) connections and $\mathrm{d}\mu_{\mathrm{AL}}$ is a natural diffeomorphism invariant measure, the so-called Ashtekar-Lewandowski measure [69].

By the Peter-Weyl theorem, a function $f$ in $L^2(\mathrm{SU}(2), \mathrm{d}\mu_{\mathrm{Haar}})$ can be decomposed as

$$f(g) = \sum_j d_j f^j_{mn} D^j_{mn}(g), \tag{2.41}$$

with Fourier coefficients $f^j_{mn}$, $j \in \{0, \frac{1}{2}, 1, \ldots\}$, $m, n \in \{-j, \ldots, j\}$ and where $D^j_{mn}(g)$ are the matrix coefficients of the unitary irreducible representations of dimension $d_j = 2j + 1$ defined by the Wigner matrices $D^j(g)$, see Appendix C.2 for details regarding harmonic analysis on SU(2). This can be directly applied to elements in $\mathcal{H}_\Gamma$, wherefore a cylindrical function $\psi_{(\Gamma, f)}(A)$ can be decomposed as

$$\psi_{(\Gamma, f)}(A) = \sum_{j_e, m_e, n_e} f^{j_1 \ldots j_L}_{m_1 \ldots m_L, n_1 \ldots n_L} \prod_{i=1}^{L} d_{j_i} D^{j_i}_{m_i n_i}(h_{e_i}(A)). \tag{2.42}$$

It has been shown that the holonomy-flux algebra admits a unique representation on the Hilbert space $\mathcal{H}_{\mathrm{kin}}$ [70, 71] which thus defines a kinematical Hilbert space for (quantum) general relativity. In the next step, the constraints Eqs. (2.28), (2.29) and ( 2.30) have to be promoted to operators on this Hilbert space and then to be solved to arrive at the Hilbert space of physical states, as depicted by

$$\mathcal{H}_{\mathrm{kin}} \xrightarrow{\hat{G}_i \psi = 0} \mathcal{H}^0_{\mathrm{kin}} \xrightarrow{\hat{H}^a \psi = 0} \mathcal{H}_{\mathrm{Diff}} \xrightarrow{\hat{H} \psi = 0} \mathcal{H}_{\mathrm{phys}}. \tag{2.43}$$

### 2.2.3.2 Gauge invariant Hilbert space $\mathcal{H}_{\mathrm{kin}}^0$

The imposition of the quantum Gauss constraint yields a new Hilbert space, denoted by $\mathcal{H}_{\mathrm{kin}}^0$, consisting of SU(2)-gauge invariant states. To see this, notice that gauge transformations act on the source and targets of links of the graph $\Gamma$, hence on its nodes $n$. Requiring gauge invariance of states, means that the cylindrical functions have to be invariant under the action of the group at the nodes, i.e.

$$f_0\left(h_1, \ldots, h_L\right) = f_0\left(g_{s_1} h_1 g_{t_1}^{-1}, \ldots, g_{s_L} h_L g_{t_L}^{-1}\right) \tag{2.44}$$

which we implement through group averaging

$$f_0\left(h_1, \ldots, h_L\right) = \int \prod_n \mathrm{d}g_n \; f\left(g_{s_1} h_1 g_{t_1}^{-1}, \ldots, g_{s_L} h_L g_{t_L}^{-1}\right). \tag{2.45}$$

A complete basis of $\mathcal{H}_{\mathrm{kin}}^0$ is defined by so-called spin network states. These are given by the triple $S = (\Gamma, \{j_e\}, \{i_n\})$ with $L$ links and $N$ nodes where $i_n$ denotes an element of the intertwiner space at the node $n$, i.e.

$$i_n \in \mathcal{H}_n = \mathrm{Inv}\left(\otimes_{e \in n} \mathcal{H}^{j_e}\right). \tag{2.46}$$

With this, a spin network state as a functional of the connection $A$ is then schematically given as a linear combination of products of representation matrices contracted with intertwiners, i.e.

$$\psi_S(A) = \langle A | S \rangle = \otimes_n i_n \otimes_e D^{j_e}(h_e).^{21} \tag{2.47}$$

We may then write for the gauge invariant Hilbert space on a graph $\Gamma$

$$\mathcal{H}_\Gamma^0 = L^2[\mathrm{SU}(2)^L / \mathrm{SU}(2)^N, \mathrm{d}\mu_{\mathrm{Haar}}], \tag{2.48}$$

where the quotient with respect to $\mathrm{SU}(2)^N$ expresses SU(2)-gauge invariance at the nodes. When all graphs are taken into account, we have

$$\mathcal{H}_{\mathrm{kin}}^0 = \oplus_{\Gamma \subset \Sigma} \mathcal{H}_\Gamma^0. \tag{2.49}$$

---

[21] In Appendix E we give a concrete example for the computation of intertwiners for a 4-valent open spin network which has the geometric interpretation of a tetrahedron.

**Quantum geometry**

In the following, we will see that spatial geometry is discrete in LQG already at the kinematical level and that the physical interpretation of spin network states lies in their description of quantum geometries. It should be emphasised that this discreteness is a result of the quantum theory and not a built-in discretisation as in lattice quantum gravity approaches.[22]

Geometric operators in LQG can be constructed from the quantum fluxes $\hat{E}_i^a$, quantising lengths, areas, volumes and angles. Let us exemplify this with the case of the area operator. The classical area of a surface is computed by means of the metric on this surface, or equivalently can be rewritten in terms of densitised inverse triads, i.e.

$$A(S) = \int_S \mathrm{d}^2x \ \sqrt{\det(^{(2)}g)} = \int_S \mathrm{d}^2x \ \sqrt{E_i^a E^{bi} n_a n_b},$$ (2.50)

where $n_a$ denotes the normal to the surface $S$ in $\Sigma$. In the quantum theory the area operator can be rigorously defined through its action on spin network states $\psi_\Gamma$. For this, observe that the graph $\Gamma$ intersects the surface at a finite set of points. If we finely decompose this surface into $C$ two-dimensional cells $S_c$, such that each plaquette is punctured by the graph only once, the area operator is then given by

$$\hat{A}(S)\psi_\Gamma = \lim_{C\to\infty} \sum_{c=1}^{C} \sqrt{\hat{E}_i(S_c)\hat{E}^i(S_c)}\psi_\Gamma$$
$$= 8\pi\gamma\ell_p^2 \sum_{p\in S\cup\Gamma} \sqrt{j_p(j_p+1)}\psi_\Gamma$$ (2.51)

and spin network states are its eigenstates. It is a SU(2)-gauge invariant quantity and applicable for any surface in $\mathcal{M}$.[23] We notice that it has discrete eigenvalues proportional to the squared Planck length $\ell_p^2$ which is a strong indication for the intrinsic discreteness of physical space. In addition, the smallest non-vanishing eigenvalue in the spectrum is given through the representation with $j = \frac{1}{2}$, implying that physical space has minimal size at

---

[22]It should be emphasised that the discreteness of geometry holds on the kinematical level and is not shown to hold dynamically in generic situations. This is due to the fact that the geometric operators are not Dirac observables [5]. As we will see in the next chapter, in the context of the isolated horizon framework to describe black holes in LQG the area of the horizon is indeed a Dirac observable and its discreteness thus holds dynamically.

[23]For convenience, we assumed that no node of the graph $\Gamma$ lies on the surface $S$. The general case is discussed in Ref. [72].

the Planck scale [31].[24]

Similarly, the volume operator can be constructed. Classically, the volume of a region $R$ in $\Sigma$ is given by

$$
\begin{aligned}
V(R) &= \int_R \mathrm{d}^3 x \sqrt{\det({}^{(3)}g)} \\
&= \int_R \mathrm{d}^3 x \sqrt{|\det(\mathrm{E})|} = \int_R \mathrm{d}^3 x \sqrt{\left|\frac{1}{3!}\epsilon_{abc}\epsilon^{ijk} E_i^a E_j^b E_k^c\right|}
\end{aligned}
\tag{2.52}
$$

and the quantisation is analogous to the one of the area operator. Note, however, that there are two well-defined but distinct volume operators available in the LQG literature [58]. Both only act on the nodes of a graph and annihilate 3-valent nodes. Non-trivial contributions to the volume are obtained for ($n \geq 4$)-valent nodes. For the special case of valency 4 both operators take the same form.[25] Importantly, precise calculations show that their spectra are discrete and that there is a minimal expectation value proportional to $\ell_p^3$.[26][27] In this way, it is understood that spin network nodes have the interpretation of quantum states of a convex polyhedron [60].

In passing we would like to remark that the discreteness plays a key role in the application of LQG to the problem of the classical singularities in GR at the centre of black holes and at the big bang, see Refs. [68] and [75, 76], respectively, and references therein. We will put the quantisation of area to direct use in the next chapter and will see that the discreteness is also at the heart of the recovery of the black hole entropy from LQG.

### 2.2.3.3 Diffeomorphism invariant Hilbert space $\mathcal{H}_{\mathbf{Diff}}$ and physical Hilbert space $\mathcal{H}_{\mathbf{phys}}$

**Imposition of the vector constraint**

The implementation of the quantum diffeomorphism constraint

$$
\widehat{H}^a \psi(A) = 0
\tag{2.53}
$$

---

[24]Notice that the trivial mode is excluded by cylindrical equivalence, a property specific to the gauge invariant Hilbert space [58].

[25]We discuss properties of $\hat{V}$ for this case in greater detail in Appendix E. This is also done in view of the application of the volume operator in the GFT condensate cosmology framework.

[26]The quantisation of length and angle operators is discussed in Refs. [73] and [74], respectively.

[27]We note that due to the discretness of geometry, LQG is generally expected to be free of ultraviolet divergences [5].

is more difficult. To see this, consider the action of a spatial diffeomorphism $\phi$ on a holonomy supported on a path $\gamma$, i.e.

$$h_\gamma(\phi A) = h_{\phi \circ \gamma}(A), \tag{2.54}$$

which induces an action onto the graph structure of spin network states as

$$\hat{U}_\phi : \mathrm{Cyl}_\Gamma \to \mathrm{Cyl}_{\phi \circ \Gamma}. \tag{2.55}$$

Since the Ashtekar-Lewandowski measure $\mathrm{d}\mu_{\mathrm{AL}}$ is also diffeomorphism invariant, this action is well-defined and unitary. However, $\hat{U}_\phi$ is not weakly continuous and therefore its infinitesimal generators do not exist, making Eq. (2.53) meaningless. Therefore, diffeomorphisms acting on cylindrical functions have to be finite which makes the construction of diffeomorphism invariant states difficult.

It is possible to deal with this issue by means of group averaging techniques. Following Ref. [58], we introduce a graph $\Gamma$ with a cylindrical function (and all labels) and define the group of graph symmetries as

$$GS_\Gamma = \mathrm{Diff}_\Gamma / \mathrm{TDiff}_\Gamma, \tag{2.56}$$

where $\mathrm{TDiff}_\Gamma$ denotes the group of trivial diffeomorphisms and $\mathrm{Diff}_\Gamma$ is the group of diffeomorphisms preserving the labelled graph. This group is finite and acts non-trivially on $\mathcal{H}^0_{kin}$. In the next step, we define a projection map which averages states in $\mathcal{H}^0_{kin}$ with respect to $GS_\Gamma$, i.e.

$$\hat{P}_{\mathrm{Diff},\Gamma} \psi_\Gamma = \frac{1}{n_\Gamma} \sum_{\phi \in GS_\Gamma} \hat{U}_\phi \psi_\Gamma, \tag{2.57}$$

where $n_\Gamma$ denotes the number of the elements of $GS_\Gamma$. The second averaging is now done with respect to diffeormorphisms which move the graph $\Gamma$. For this, we consider the algebraic dual space $\mathcal{H}^{0*}_{kin}$ of $\mathcal{H}^0_{kin}$ consisting of elements $\eta(\psi_\Gamma)$ such that

$$\eta(\psi_\Gamma)[\psi'_{\Gamma'}] = \sum_{\phi \in \mathrm{Diff}(\Sigma)/\mathrm{Diff}_\Gamma} \langle \hat{U}_\phi \hat{P}_{\mathrm{Diff},\Gamma} \psi_\Gamma | \psi'_{\Gamma'} \rangle, \tag{2.58}$$

where $\langle \cdot | \cdot \rangle$ denotes the inner product on the kinematical Hilbert space. Importantly, this dual space is the space of diffeomorphism invariant functionals due to the diffeomorphism invariance of the scalar product on the kinematical Hilbert space. Hence, $\eta$ defines a map

$$\eta : \mathcal{H}_{\text{kin}}^0 \rightarrow \mathcal{H}_{\text{Diff}}^* \tag{2.59}$$

and on $\mathcal{H}_{\text{Diff}}^*$ the Hermitian inner product reads as

$$\langle \eta(\psi_\Gamma) | \eta(\psi'_{\Gamma'}) \rangle = \eta(\psi_\Gamma)[\psi'_{\Gamma'}], \tag{2.60}$$

which completes the construction of the general solution to the diffeomorphism constraint.

Diffeomorphism invariant spin network states are thus defined on equivalence classes $[\Gamma]$ of graphs $\Gamma$ under diffeomorphisms, i.e. $[S] = ([\Gamma], \{j_e\}, \{i_n\})$. In other words, two spin network states are equal if their graphs lie in the same equivalence class. The equivalence classes differ if their underlying graphs are differently knotted. Hence, the diffeomorphism invariant Hilbert space of LQG is spanned by states called knotted spin network states which are labelled by their graphs, knots therein and representation labels.[28] The knotting feature will show up again in the next chapter when indicent spin network links onto a black hole horizon are allowed to braid with one another.

The fact that all information about the embedding in $\Sigma$ has been washed out in this construction, suggests that the smooth manifold structure, on which LQG is originally built, can be replaced by that of a piecewise linear manifold. Spin network graphs are then defined using abstract graphs which are combinatorial objects dual to cellular decompositions [78]. It is expected that diffeomorphism invariance is then recovered only from combinatorial information in the continuum limit. This perspective is central to the spin foam approach for the covariant quantisation of LQG [35].

**Imposition of the Hamiltonian Constraint**

Important for the completion of Dirac's quantisation programme is the imposition of the scalar constraint which leads to states in the physical Hilbert space $\mathcal{H}_{\text{phys}}$. In this work, we will not directly refer to this aspect and will therefore only give a brief list of results and problems related to it. For details we refer to Ref. [5].

It is in principle possible to give a well-defined definition of the Hamiltonian constraint operator, its action is finite and modifies spin network states at their nodes leading to new links. With this it has been shown that an infinite number of solution states exists and that the Dirac algebra is anomaly-free. Moreover, the finiteness is preserved when matter is coupled to gravity. This is an indication that taking the quantum nature of gravity seriously, regulates the infinities encountered in quantum field theories.

---

[28]For a discussion concerning the separability of $\mathcal{H}_{\text{Diff}}$ and the definition of observables thereon, we refer to Refs. [58, 77] and [5], respectively.

In spite of this remarkable progress compared to the quantum geometrodynamics approach, the quantisation programme is only partially completed so far. This is due to the fact, that the full spectrum of the Hamiltonian operator and the characterisation of the physical Hilbert space $\mathcal{H}_{\text{phys}}$ are not known up to now. Moreover, other ways to define $\hat{H}$ have been found.

These issues in defining the dynamics of the quantum theory have led to the development of the so-called master constraint programme [5], where the diffeomorphism and the scalar constraint are implemented together, and to the spin foam approach, which provides a path integral prescription for the quantisation of LQG [35]. We will give an overview of the latter in Section 4.2.3. Notice that understanding the dynamics of the theory is of paramount importance to clarify major outstanding conceptual problems of the theory: It is expected to provide key insight into the issue of the continuum limit of the theory and the fate of Lorentz invariance therein [68].

In the following chapter we will employ LQG techniques to the quantum description of the black hole horizon within the isolated horizon framework. Focus will be given to the role of boundary symmetries therein, particularly the role of so-called large diffeomorphisms will be investigated. LQG techniques are also naturally abundant in GFT and its application to quantum cosmology forming the second thematic unit of this thesis thereafter.

# Chapter 3

# Black hole entropy and large horizon diffeomorphisms in LQG

> The black holes of Nature are the most perfect macroscopic objects there are in the Universe: the only elements in their construction are our concepts of space and time.

<div align="right">Subrahmanyan Chandrasekhar</div>

Einstein's field equations teach us that a star of sufficiently large mass collapses beyond its Schwarzschild radius into a gravitational singularity at the end of its lifetime. A black hole is formed the size of which is determined by the event horizon covering the singularity. This horizon sets the frontier of all events which can be observed by an external observer [1]. An abundance of astrophysical data in support of the existence of these intriguing objects has been collected so far [79].

However, due to the fact that singularities are unphysical divergences of the gravitational field[1] which render spacetime geodesically incomplete, general relativity looses its predictability and cannot be used to describe the interior structure of the black hole. This is implied by the famous singularity theorems [10–12]. The belief that quantum gravitational effects taking place at the Planck scale could lead to a resolution of such singularities [13] has much driven the motivation to attend to a more fundamental account of the gravitational field.

---

[1]An example of a coordinate invariant quantity built from the Riemannian curvature tensor is the Kretschmann scalar. For a spherically symmetric star of mass $M$ it is given by

$$R^{\mu\nu\rho\sigma}R_{\mu\nu\rho\sigma} = \frac{48M^2}{r^6} \tag{3.1}$$

which diverges towards the origin $r = 0$. This case is a particular example of a curvature singularity [1].

Further incentive for such an endeavour is provided by the fact that thermodynamic quantities can be assigned to black holes based on their gravitational properties. This is established by the four laws of black hole mechanics [17] which suggest relations between the horizon area and entropy, the black hole surface gravity and temperature as well as the black hole mass and energy. The correct proportionality factors between these related quantities were clarified through the application of quantum field theory on curved spacetime which predict that black holes radiate like black bodies [19, 20], see Ref. [1] for an overview. In particular, this establishes the Bekenstein-Hawking entropy formula, i.e.

$$ S = k_{\mathrm{B}} \frac{c^3}{\hbar G_{\mathrm{N}}} \frac{A}{4} = k_{\mathrm{B}} \frac{A}{4\ell_p^2} \tag{3.2} $$

which intertwines the geometry of spacetime, gravity, quantum theory and thermodynamics with each other [18–20]. Given that thermodynamics finds its microscopic underpinning through statistical mechanics, it is expected that black hole entropy arises from the microstates of an underlying fundamental theory of the gravitational field which can describe the quantum structure of the horizon geometry.

It has been established for a while, that the black hole as an inner boundary of space can be described in equilibrium locally by the isolated horizon boundary condition [80–82]. The introduction of this notion is justified since the usual definition of a black hole as a spacetime region of no escape is global. This means that it requires the knowledge of the entire spacetime as well as that it is in equilibrium. Consequently, it does not appear to be useful for describing local physics.

Luckily, these conceptual problems are resolved by means of the quasi-local notion of an isolated horizon. From a physical point of view its introduction amounts to having no fluxes of matter and/or gravitational energy across it. From a technical point of view the boundary conditions lead to a surface term for the horizon in the overall action of the gravitational field which in terms of Ashtekar-Barbero variables is proportional to the action of a topological gauge theory, namely Chern-Simons (CS) theory. Furthermore, one can show that this description is fully compatible with the laws of black hole mechanics.

The quantisation of spacetimes with such an isolated horizon by means of LQG techniques leads to the following picture: The quantum geometry of the bulk is given by a spin network the graph of which pierces the horizon surface yielding punctures thereon. These punctures correspond to a gas of topological defects which store the curvature information of the horizon and thus represent the quantum excitations of the gravitational field of the horizon. More precisely, the quantum geometry of the horizon is encoded by an SU(2) CS-theory at level $k$ given on a punctured 2$d$-sphere [83–85].

Equipped with this, one may count the microstates of the corresponding Hilbert space [83–99]. Together with the introduction of proper notions of a quasi-local energy and a local temperature of the isolated horizon its statistical mechanical analysis is facilitated [100–103]. The entropy of this horizon can be retrieved with this and the result is remarkably compatible with the semi-classical Bekenstein-Hawking entropy formula up to a quantum hair correction due to the quantum geometry of the isolated horizon.

Despite these successes in matching the semi-classical results, one can raise the question if the statistics of the horizon degrees of freedom could actually be different from the one assumed in the usual state counting procdures, considering the well-known fact from solid state physics that quantum objects in 2$d$ obey anyonic/braiding statistics. To see that such an exotic type of statistics is indeed present, we employ symplectic geometry tools and show that the punctures lead to global obstructions for symplectic vector fields on the isolated horizon to be Hamiltonian. Upon quantisation, this kinematical ambiguity leads to non-Abelian phases which give rise to non-Abelian anyonic statistics.

The fact that the group of permutations in 2$d$ corresponds to the braid group, allows us to identify these non-Abelian phases as representations of the braid group of the punctured isolated horizon. By demonstrating that this group is equivalent to the group of large diffeomorphisms[2] of the punctured surface, we establish a clear relation between this boundary symmetry group and the statistics of the model. This leads us to the reinterpretation of the quantum isolated horizon model as one which explicitly exhibits non-Abelian anyonic statistics.

To this aim, this Chapter is organised as follows. As a background for our work we assume as given the isolated horizon framework [80–82] and its quantisation à la LQG. Since there one borrows techniques from CS-theory [83–85], for reasons of self-consistency and completeness we will firstly review the symmetries of CS-theory as well as its Hamiltonian formulation in Section 3.1.1. Then we recapitulate properties of the LQG black hole model and its statistics in the following Section 3.1.2. In the core Section 3.2 we elaborate the main and new results. There we will firstly inspect the topological features of the physical phase space and its relation to the braid group, which reveals the anyonic nature of the horizon degrees of freedom in Section 3.2.1. Afterwards we will further investigate the braid group symmetry of the punctured sphere by relating it to the large diffeomorphisms of the horizon and then discuss the effect of the occurring non-Abelian phases in Section 3.2.2. We will then connect the discussion of this property of the horizon degrees of freedom to formal aspects of the theory of non-Abelian anyons known from solid state physics in Section 3.2.3. Since this Chapter suggests that the braiding statistics is suppressed for a large values of the CS-level $k$, we comment on the sensitivity of the entropy to $k$ and give qualitative arguments why the black hole radiance spectrum should display traces of the braiding in Section 3.2.4. Finally, Section 3.3 closes this Chapter with a discussion of the results and comments on possible future investigations. The material presented in this Chapter is largely based on the work of the author in Ref. [104].[3]

---

[2]In general, large gauge and large diffeomorphism transformations are not imposed by constraints in the action. However, they can have an interesting effect on the boundary states of theories at the quantum level, i.e. they can transform under representations of the group of these boundary transformations [57].

## 3.1 Chern-Simons theory and LQG black holes

### 3.1.1 Symmetries of Chern-Simons theory

Within this Section we review essentials of CS-theory with special regard to its symmetry properties, the difference between small and large diffeomorphisms and its Hamiltonian formulation which will be exploited afterwards.

The action of CS-theory on an oriented smooth 3-manifold $M$ is given as

$$S_{\text{CS}}[\tilde{A}] = \frac{k}{4\pi} \int_M \text{tr} \left[ \mathrm{d}\tilde{A} \wedge \tilde{A} + \frac{2}{3} \tilde{A} \wedge \tilde{A} \wedge \tilde{A} \right], \qquad (3.3)$$

for a $G$-valued connection $\tilde{A} = \tilde{A}^i_\mu J_i \mathrm{d}x^\mu$ and $k$ denotes the coupling constant (level). $G$ is a compact, simple and simply connected Lie Group and the generators $\{J_i\}$ with $i = 1, \ldots, \dim G$ form the basis of the corresponding Lie algebra. Stationarity of the action leads to the equation of motion

$$F = \mathrm{d}\tilde{A} + \tilde{A} \wedge \tilde{A} = 0. \qquad (3.4)$$

---

[3]The desciption of black holes from LQG is research in progress and yet to be completed. A detailed overview of the current status of this programme, including a discussion of the problem of the fate of information in black hole evaporation, can be found in Ref. [68].

Inspecting its gauge symmetries, the overall gauge group is given by the semi-direct product of $\mathrm{Diff}_0(M)$ (i.e. the group of small diffeomorphisms) with the infinite dimensional and possibly topologically non-trivial $\mathcal{G} = C^\infty(M, G)$ [105–108].

Let us dwell for a moment on this point and firstly consider transformations which are elements in $\mathcal{G}$. This leads us to the well-known transformation law for the connection

$$\tilde{A} \rightarrow \tilde{A}^g = g\tilde{A}g^{-1} - (\mathrm{d}g)g^{-1}, \tag{3.5}$$

with $g \in \mathcal{G}$. In fact, $\mathcal{G}$ comprises of two parts which are called small and large gauge transformations. We call gauge transformations small if they are connected to the identity and one easily sees that $S_{\mathrm{CS}}$ is invariant with respect to them. Let $g$ be such a transformation given in its finite form as $g = \mathrm{e}^{iJ_i\zeta^i}$ where $\zeta^i$ are the gauge parameters. Infinitesimally, $g \approx 1 - iJ_i\zeta^i$ with $\zeta \ll 1$ and this yields

$$\delta\tilde{A} = \tilde{A}^g - \tilde{A} \approx \mathrm{d}_{\tilde{A}}\zeta. \tag{3.6}$$

Importantly, invariance under small gauge transformations is not enough to guarantee the invariance with respect to finite transformations. This is due to the fact that there are topologically non-trivial finite gauge transformations with homotopy class different from 0. One calls them large gauge transformations. If one demands that the path integral

$$Z_k(M) = \int \mathcal{D}\tilde{A} \; e^{iS_{\mathrm{CS}}[\tilde{A}]} \tag{3.7}$$

is invariant with respect to small and large gauge transformations, it can be shown that for closed $M$ and compact $G$ the coupling constant $k$ must be an integer and thus discrete.

Similar to $\mathcal{G}$, one differentiates two types of diffeomorphisms, namely small and large ones. Small diffeomorphisms are homotopic to the identity, can be infinitesimally generated and form the group $\mathrm{Diff}_0(M)$. Since CS-theory is a TQFT of Schwarz type, its action, equations of motion and observables do not require the existence of a metric. It is thus diffeomorphism invariant [105–107, 109], i.e. invariant with respect to $\mathrm{Diff}_0(M)$.

Large diffeomorphisms on the other hand cannot be obtained from summing up an infinite number of infinitesimal transformations and are not homotopic to the identity. They form a group called mapping class group which is denoted by $\mathrm{MCG}(G) = \mathrm{Diff}(M)/\mathrm{Diff}_0(M)$.

In the context of CS-theory one can show that on-shell, i.e. when Eq. (3.4) is fulfilled,

small diffeomorphisms are equivalent to small gauge transformations. To see this, consider the change of the connection $\tilde{A}$ under an infinitesimal coordinate transformation $x^\mu \rightarrow x^\mu + \xi^\mu$ with $\mu \in \{0, 1, 2\}$. This is expressed as

$$\delta_\xi \tilde{A} = \mathcal{L}_\xi \tilde{A} = (\mathrm{i}_\xi \mathrm{d} + \mathrm{d} \mathrm{i}_\xi)\tilde{A} = \mathrm{i}_\xi F + \mathrm{d}_{\tilde{A}}(\mathrm{i}_\xi \tilde{A}), \tag{3.8}$$

wherein $\mathrm{d}_{\tilde{A}}$ denotes the gauge-covariant exterior derivative and $\xi$ is an infinitesimal generator of small diffeomorphisms. On-shell this expression is just an ordinary infinitesimal gauge transformation as Eq. (3.6) with the gauge parameter $\zeta^i = \xi^\mu \tilde{A}^i_\mu$.

In stark contrast to this, large diffeomorphisms and large gauge transformations are discrete and strictly distinct symmetries of the theory. In the quantum theory one cannot simply demand that states should be invariant under the action of these groups. Instead, they can act as symmetry transformations. As we have just seen, invariance with respect to large gauge transformations leads to a quantisation of the level $k$. As regards the mapping class group, further below we will argue for its importance in the treatment of the quantum isolated horizon framework of LQG and we will relate it to the statistical symmetry giving rise to anyonic/braiding statistics.

All these symmetry considerations of course also hold for the Hamiltonian formulation of CS-theory [105–107, 110–115]. There, the gauge field is split into $\tilde{A} = A_0 \mathrm{d}x^0 + A_a \mathrm{d}x^a$ due to the product structure of $M = \mathbb{R} \times \Sigma$, where $\Sigma$ is an arbitrary orientable surface. Then the spatial components $A = A_a \mathrm{d}x^a$ of the gauge field are considered as the dynamical variables. The appearing $A_0$-component has null conjugate momentum and serves as a Lagrange multiplier in the action

$$S[A, A_0] = \frac{k}{4\pi} \int_{\mathbb{R}} \int_\Sigma \mathrm{tr}\left[-A\partial_0 A + 2A_0 F\right], \tag{3.9}$$

enforcing the first class constraint $F = 0$. From the infinitesimal variation of the action one also obtains a boundary term, which we can identify as the symplectic potential

$$\theta = \frac{k}{4\pi} \int_\Sigma \mathrm{tr}\left[A \wedge \delta A\right] + \delta\rho(A). \tag{3.10}$$

Therein $\rho$ denotes an arbitrary functional of $A$ and $\delta\rho$ expresses the freedom of canonical transformations [2]. The symbol $\delta$ corresponds to the exterior derivative on the space of gauge potentials on $\Sigma$. With this, the symplectic 2-form is obtained by

$$\omega = \delta\theta = \frac{k}{4\pi} \int_\Sigma \text{tr} \left[ \delta A \wedge \delta A \right].$$ (3.11)

If gauge symmetries have not yet been reduced out, $\omega$ is presymplectic and thus has zero modes generating gauge symmetries as discussed above. Upon symplectic reduction one yields the physical or reduced phase space. We consider $\omega$ to be non-degenerate below.

Together with the physical phase space given by the moduli space of flat connections

$$\Gamma = \{A | F = 0\} / \mathcal{G},$$ (3.12)

we have a symplectic manifold $(\Gamma, \omega)$, where $\mathcal{G} = C^\infty(\Sigma, G)$.

### 3.1.2 LQG black hole model and its statistical mechanics

With the following material we review parts of the classical isolated horizon framework and its quantisation [83–85, 90, 91] and summarise essentials of their statistical mechanical analysis as in Ref. [101].

The isolated horizon field theory lives on a 3-manifold $\Delta$, which is a cylinder $\Delta = \mathbb{R} \times S^2$, where $\mathbb{R}$ parameterises the time $t$ and $G$ is SU(2) hereafter. Spherically symmetric isolated horizons can be described as a dynamical system by a presymplectic form $\omega_{\text{horizon}}$, which corresponds to that of an SU(2)-CS-theory. For a proof and a general discussion see [83–85, 90, 91]. Physically this means that the gravitational field of the horizon resides in a topological phase. The overall symplectic structure splits as

$$\omega_{\text{total}} = \omega_{\text{bulk}} + \omega_{\text{horizon}}$$ (3.13)

and field components from bulk and horizon are coupled properly together by the isolated horizon boundary condition which in terms of Ashtekar-Barbero variables reads as

$$F^i(A^i) + \frac{\pi(1-\gamma^2)}{a_{\text{H}}} \Sigma^i = 0,$$ (3.14)

where $a_{\text{H}}$ denotes the classical horizon area and $\gamma$ is the Barbero-Immirzi parameter. $F^i$ is the curvature 2-form of the Ashtekar-Barbero connection $A^i$ being pulled-back to $S^2$ and $\Sigma^i = \epsilon^i_{jk} e^i \wedge e^k$ denotes the solder 2-form of the bulk theory and the internal index $i \in \{1, 2, 3\}$ indicates that the respective object is labelled with an element of $\mathfrak{su}(2)$ in the defining representation.[4]

In the following we will use that in LQG one regularises at the classical level the Poisson algebra of the Ashtekar-Barbero connection and the densitised triad. As we have seen in Section 2.2.3, the resulting smeared algebra is the so-called holonomy-flux algebra [5, 31, 58, 59]. If one embeds a surface of spherical topology such as the one of a classical isolated horizon $\Delta$ into a surrounding spacelike 3-space, then it is pierced by paths $\gamma$ (supporting the bulk holonomies) at the points $\mathcal{P} = \{p_1, ..., p_N\}$, as depicted in Fig. 3.1.

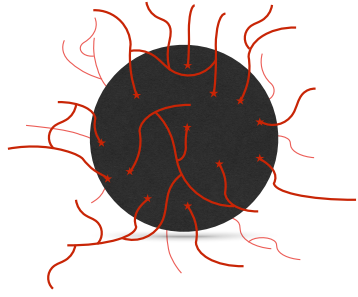

FIGURE 3.1: Bulk holonomies impinging on the horizon surface.

One interprets this set as a distribution of sources on $S^2$, each labelled with an representation $\{\rho_p\}_1^N$ of $\mathfrak{su}(2)$ and $\Sigma^i$ is then

$$\Sigma^i = 16\pi \ell_p^2 \gamma \sum_{p \in \mathcal{P}} J_{\rho_p}^i \delta^2(x, x_p). \tag{3.15}$$

Substituting this into Eq. (3.14) yields

$$F^i + \frac{4\pi}{k} \sum_{p \in \mathcal{P}} J_{\rho_p}^i \delta^2(x, x_p) = 0. \tag{3.16}$$

---

[4]Notice that in the subsequent presentation we have introduced the notion of horizon punctures already at the classical level. In the primary sources, e.g. Refs. [83–85] these are only introduced after quantising the bulk first, giving rise to a surface Hilbert space which is precisely the one of CS-theory in the presence of punctures. Introducing them on the classical level has the advantage that we can more clearly analyse their impact onto the topology of the phase space of the horizon theory in the next subsection. The occurrence of the non-Abelian phases we wish to compute, however, does not depend on this way of reasoning.

The action for the horizon theory thus is

$$S_{\text{horizon}}[A, A_0] = \frac{k}{4\pi} \int_{\mathbb{R}} \int_{\Sigma} \text{tr} \left( -A\partial_0 A + 2A_0 F \right)$$
$$+ \int_{\mathbb{R}} \text{d}t \sum_{p \in \mathcal{P}} \text{tr} \left( J_{\rho_p} A_0(x_p) \right), \tag{3.17}$$

where the level of the CS-theory is now given by $k = \frac{a_H}{4\pi\gamma(1-\gamma^2)\ell_p^2}$. The Euler-Lagrange equations lead to the (induced) constraint

$$G^i \equiv F^i + \frac{4\pi}{k} \sum_{p \in \mathcal{P}} J^i_{\rho_p} \delta^2(x, x_p) = 0, \tag{3.18}$$

delineating that the curvature of the connection on the surface is concentrated at the points of the punctures. This first class constraint generates (small) gauge transformations and (small) diffeomorphisms. More precisely, the horizon part of the smeared Gauss constraint is

$$G[\zeta, A] = \int_{S^2} \zeta_i G^i \approx 0, \tag{3.19}$$

for all $\zeta : \Delta \to \mathfrak{su}(2)$, whereas the diffeomorphism constraint is

$$V[\xi, A] = \int_{S^2} \xi^\mu A_{\mu i} G^i \approx 0, \tag{3.20}$$

for all vectors $\xi$ ($\mu = 1, 2$) which are tangent to the horizon. The form of Eqs. (3.19) and (3.20) implies the on-shell equivalence of small diffeomorphisms and small gauge transformations as in Eqs. (3.6) and (3.8). In addition, for the sources at the points $\{x_p\}_1^N$ one has conjugations

$$J^i_{\rho_p} \to J^{ig}_{\rho_p} = g^{-1} J^i_{\rho_p} g \in \mathcal{C}_p^g \tag{3.21}$$

and the gauge invariance of $F^i$ implies $\mathcal{C}_p = \mathcal{C}_p^g$. The physical phase space of this system is then given as

$$\Gamma = \{\{A | F = 0\} \times \mathcal{C}_1 \times \ldots \times \mathcal{C}_N\} / \{\text{gauge transformations}\} \tag{3.22}$$

as in Refs. [112–115].

The form of the overall symplectic structure Eq. (3.13) motivates to quantise the bulk and horizon degrees of freedom separately. The quantum geometry of the bulk is given by a spin network, the graph of which impinges on the horizon surface yielding the punctures. Hence, for the quantum geometry of the horizon we use the quantum version of Eq. (3.18)

$$\left( \hat{F}^i + \frac{4\pi}{k} \sum_{p\in\mathcal{P}} \delta^2(x, x_p)\hat{J}^i_{\rho_p} \right) \psi_{\text{horizon}} = 0, \tag{3.23}$$

which selects elements of the physical Hilbert space of the horizon theory. Notice that at each puncture $p$ the $\mathfrak{su}(2)$-algebra $[\hat{J}^i_{\rho_p}, \hat{J}^j_{\rho_p}] = \epsilon^{ij}_k \hat{J}^k_{\rho_p}$ holds.

The quantum version of Eq. (3.14) couples bulk and horizon quantum degrees of freedom properly back together.[5] The physical Hilbert space is then given by

$$\mathcal{H}_{\text{phys}} = \left( \bigoplus_{\mathcal{P}} \mathcal{H}^{\mathcal{P}}_{\text{bulk}} \otimes \mathcal{H}^{\mathcal{P}}_{\text{horizon}} \right)/\mathcal{G}_{\text{total}}, \tag{3.24}$$

where $\mathcal{H}^{\mathcal{P}}_{\text{bulk}}$ denotes the bulk space of states. One denotes by $\mathcal{G}_{\text{total}} = \mathcal{G}_{\text{bulk}} \ltimes \mathcal{G}_{\text{horizon}}$ internal SU(2)-transformations, diffeomorphisms which preserve the surface and eventually motions, generated by the Hamiltonian constraint $H$ [83–85]. Since the isolated horizon framework stipulates that the lapse is restricted to vanish on the horizon, the scalar constraint $H$ is only imposed in the bulk. As we have seen, the horizon states satisfying the boundary condition (3.23) are automatically gauge and diffeomorphism invariant. After imposition of the according constraints one has

$$\mathcal{H}_{\text{phys}} = \bigoplus_N \bigoplus_{(j)^N_1} \mathcal{H}^{(j)^N_1}_{\text{bulk,phys}} \otimes \text{Inv}_k(\otimes_p j_p), \tag{3.25}$$

where $\text{Inv}_k(\otimes_p j_p)$ is the CS-Hilbert space on the punctured sphere with $j_p \leq \frac{k}{2}$ and $\mathcal{H}^{(j)^N_1}_{\text{bulk, phys}}$ denotes the physical Hilbert space of the bulk for a corresponding puncture configuration [91, 116, 117].

In order to analyse the thermodynamical properties of the horizon, one computes the total number of (micro-)states available to it, i.e.

$$W(\{\mathcal{P}\}) = \sum_{\mathcal{P}} \dim(\text{Inv}_k(\otimes_p j_p), \tag{3.26}$$

where we constrain ourselves only to those horizon states which are compatible with $a_{\text{H}}$ and $j_p \leq \frac{k}{2}$. In the following, let $n_j$ denote the occupation number of a certain puncture type that is labelled by an irreducible representation $\rho_j$ of $\mathfrak{su}(2)$. Then Eq. (3.26) can be rewritten for a quantum configuration $\{n_j\}$ as

$$W(\{n_j\}) = \frac{N!}{\prod_j n_j!} \; \frac{2}{k+2} \sum_{l=0}^{\frac{k}{2}} \sin^2\left(\frac{(2l+1)\pi}{k+2}\right) \prod_j d_j^{n_j}(l), \qquad (3.27)$$

wherein

$$d_j(l) \equiv \left[ \frac{\sin\left(\frac{(2j+1)(2l+1)\pi}{k+2}\right)}{\sin\left(\frac{(2l+1)\pi}{k+2}\right)} \right], \qquad (3.28)$$

as done in Refs. [84–87, 99]. The total number of punctures is denoted by $N = \sum_j^{\frac{k}{2}} n_j$ and the combinatorial pre-factor indicates that in the purely gravitational case the punctures are considered as distinguishable [83, 118–122]. Since the level $k$ of the theory is proportional to $\frac{a_H}{\ell_p^2}$ it is convenient to consider the limit $k \to \infty$ of Eq. (3.27) giving

$$W(\{n_j\}) = N! \prod_j \frac{(2j+1)^{n_j}}{n_j!}, \qquad (3.29)$$

where we neglected the next-to-leading order term in $k$ which would give rise to a logarithmic correction of the entropy [84–87, 123–125]. The previous expression counts the number of distinct microstates belonging to the distribution set $\{n_j\}$ and is that of a typical Maxwell-Boltzmann statistics for dinstinguishable entities [126–128].[6]

---

[5]In fact, only the exponentiated version of $\hat{F}^i$ is well-defined in the quantum theory [83] but the subsequent discussion will not be altered by this.

The introduction of proper notions of a quasi-local energy and a local temperature (derived from semi-classical input)

$$E = \frac{A}{8\pi\ell}, \quad T = \frac{\ell_p^2}{2\pi\ell} \qquad (3.30)$$

of the isolated horizon associated with a stationary observer at distance $\ell$ from the horizon facilitates its statistical mechanical analysis [100–103].[7] With the horizon area spectrum

---

[6]Choosing the horizon degrees of freedom to be distinguishable (in the purely gravitational case) is well motivated in the LQG literature [83, 116–122]. In fact, the boundary condition Eq. (3.14) demands that incident bulk edges and horizon punctures have to be labelled by the same representations. Effectively, this renders the punctures distinguishable and imposes an ordering relation onto them. This ordering cannot be changed by small diffeomorphisms of the punctured horizon which are elements of $\text{Diff}_0\left(S^2_{\{n_j\}}\right)$. Only a non-trivial permutation can do this, giving rise to a new microstate of the system. From the statistical point of view all microstates accessible to the system in the macrostate $(E, N)$ have to be counted, as reflected by the statistical distributions (3.27) and (3.29), respectively. Below we will see that such permutations are in fact achieved through large diffeomorphisms.

[7]This makes obvious that on the quantum level the horizon Hamiltonian operator commutes with the area operator. Hence, the latter is a Dirac observable and the physical discreteness of geometry holds dynamically.

obtained via

$$\hat{A}|\{n_j\}\rangle = 8\pi\ell_p^2\gamma\sum_j n_j\sqrt{j(j+1)}|\{n_j\}\rangle \tag{3.31}$$

the canonical partition function reads

$$Z(T,N) = \overline{\sum_{\{n_j\}}} W(\{n_j\})e^{-\beta E}, \tag{3.32}$$

where the dashed summation runs over all distribution sets that conform to the restriction $N = \sum_j n_j$. With this, the expression for the entropy is obtained as

$$S = -\beta^2\partial_\beta\left(\frac{\log Z}{\beta}\right) = \frac{A}{4\ell_p^2} + \sigma(\gamma)N, \tag{3.33}$$

with

$$\sigma(\gamma) = \log\left(\sum_j(2j+1)e^{-2\pi\gamma\sqrt{j(j+1)}}\right). \tag{3.34}$$

The entropy function is both extensive in $A$ and $N$ as it should in order to agree with the laws of (phenomenlogical) thermodynamics and black hole mechanics. Remarkably, it is compatible with the semi-classical Bekenstein-Hawking entropy formula [18, 20], since the second summand only expresses a quantum hair correction due to the quantum geometry of the isolated horizon.

## 3.2 Anyonic statistics and LQG black holes

### 3.2.1 Symplectic geometry and anyons

The dimensionality of the problem steers us into a closer investigation of the features associated with the topology of the phase space. This will reveal how the horizon degrees of freedom obey anyonic statistics.

Quantum statistics refers to the phase which arises when two particles of a multi-particle quantum system are exchanged with each other. The purpose of this Section is to explain how this phase arises in the present system. The key idea is that due to the punctures (understood as topological defects), connections on the horizon become elements of the non-trivial first de Rham cohomology group on the phase space (3.22). This group can be related to the fundamental group of the configuration space the representations of which label inequivalent quantisations in the quantum theory. We give a prescription to compute these representations and link them to anyonic statistics. To see this clearly, first we have to analyse features of phase spaces with topological defects using Refs. [129–131].

To this aim, consider a generic symplectic manifold $(\Gamma, \omega)$. One calls a vector field $\eta$ on $\Gamma$ that preserves $\omega$, i.e. $\mathcal{L}_\eta \omega = 0$, a symplectic vector field. Using Cartan's magic formula and the closedness of $\omega$, one has

$$\mathcal{L}_\eta \omega = \mathrm{d} \left( \mathrm{i}_\eta \omega \right) = 0. \tag{3.35}$$

$\eta$ is only symplectic if $\mathrm{i}_\eta \omega$ is closed, whereas it is a Hamiltonian vector field, if additionally $i_\eta \omega$ is exact. It is a fact, that locally on every contractible (i.e. simply connected) open set, symplectic vector fields are Hamiltonian. Additionally, if one has trivial first de Rham cohomology group, i.e $H^1(\Gamma; \mathbb{R}) = 0$, then globally every symplectic vector field is Hamiltonian and we can write $\mathrm{i}_\eta \omega = -\mathrm{d}f$, for some function $f \in C^\infty(\Gamma, \mathbb{R})$. The diffeomorphisms of $\Gamma$, which are generated by Hamiltonian vector fields are known as canonical transformations. However, in case that $H^1(\Gamma; \mathbb{R}) \neq 0$, for some transformations $\eta$ the corresponding $\mathrm{i}_\eta \omega$ is a non-trivial element of $H^1(\Gamma; \mathbb{R})$ and therefore there is no globally defined function $f$ on $\Gamma$ for this transformation. Equivalently, there can be several choices for the canonical 1-form $\theta$ differing by elements of $H^1(\Gamma; \mathbb{R})$, but giving rise to the same symplectic 2-form $\omega$. The ambiguity in $\theta$ has no effect on the classical equations of motions but nevertheless $H^1(\Gamma; \mathbb{R})$ "measures" the obstruction for symplectic vector fields to be Hamiltonian.

Let us apply this to the one-form (3.10) on the phase space without defects (3.12). If $\tilde{\theta}$ is closed, then $\theta$ and $\theta + \tilde{\theta}$ will lead to the same $\omega$. If $\tilde{\theta}$ was closed and exact, we could figure $\tilde{\theta}$ as $\tilde{\theta} = \delta\rho(A)$, where $\rho(A)$ is some globally defined functional and the connection $A$ lives on $M \cong \mathbb{R} \times \Sigma$. The function $\rho(A)$ is a canonical transformation and one can transform $\delta\rho(A)$ to 0, as implied by Poincaré's lemma. On the contrary, in the case of the isolated horizon we have $\Delta \cong \mathbb{R} \times S^2$ with punctures (i.e. topological defects) on it, so we have to consider the phase space (3.22). There $\tilde{\theta}$ is closed but due to the defects not exact. Therefore, $\tilde{\theta}$ is a non-trivial element of the de Rham cohomology $H^1(\Gamma; \mathbb{R})$ and it cannot be transformed to $\tilde{\theta} = 0$ upon canonical transformation. One can only locally write $\tilde{\theta} = \delta\rho(A)$, since $\rho(A)$ is not globally definable. This is of relevance for the quantum theory of generic anyonic systems [2] and also for our problem, as we see below.

To see this, one reparameterises the phase space (3.22) by holonomies, as in Refs. [112–115]. This yields

$$\Gamma = \{\rho \in \text{Hom}\left(\pi_1\left(\mathcal{F}_N(S^2)\right), \text{SU}(2)\right) | \rho(c_p) \in \mathcal{C}_p^G\}/\text{SU}(2). \qquad (3.36)$$

The set $\{c_p\}$ stands for the generators of $\text{Hom}\left(\pi_1\left(\mathcal{F}_N(S^2)\right), \text{SU}(2)\right)$ which agree with non-contractible oriented loops around the punctures $\{p_i\}_1^N$ and $\mathcal{F}_N(S^2)$ denotes the configuration space. For the specific case of distinguishable puncture species $\{n_j\}_{\frac{1}{2}}^{\frac{k}{2}}$ distributed on $S^2$ it is given by

$$\mathcal{F}_N(S^2) = \{(x_1, ..., x_N) \in (S^2)^N | x_p \neq x_{p'} \text{ for } p \neq p'\}. \qquad (3.37)$$

Importantly, the fundamental group $\pi_1$ of this space is the spherical braid group

$$B_{n_{\frac{1}{2}}, ..., n_{\frac{k}{2}}}(S^2) \qquad (3.38)$$

on $N$ strands (cf. Appendices (A.2) and (A.4)).

As a consequence of this, a kinematical ambiguity arises in the quantisation of the classical system on the configuration space $\mathcal{F}_N(S^2)$ and it has to be classified by the set of all irreducible unitary representations of the fundamental group $\pi_1\left(\mathcal{F}_N(S^2)\right)$. In other words, this defines the action of non-Abelian phases

$$\rho \in \text{Hom}\left(B_{n_{\frac{1}{2}}, ..., n_{\frac{k}{2}}}(S^2), \text{SU}(2)\right) \qquad (3.39)$$

onto horizon states. These phases account for the non-Abelian anyonic statistics of the horizon puncture system, in the same way as in ordinary anyonic systems [132–137].[8] We want to emphasise, that for this classification *no knowledge* of the dynamics of the system is needed. If the black hole was still modelled by a topological 2-sphere with punctures but different constraints, such phases would still show up. The account of Γ's topological intricacies thus solely unveils the anyonic nature of the LQG horizon degrees of freedom.

In the following, we will discuss how to compute such non-Abelian phases which actually correspond to a parallel transport of punctures along and around each other. To illustrate this, consider the winding of a puncture around a second one along a loop $C$. This is illustrated in Fig. 3.2, where other punctures are suppressed for convenience.

---

[8]If $A$ was a U(1)-connection, the phase would be Abelian and it is well established that such a phase is needed to describe particles of Abelian anyonic statistics in 2$d$ [132–134, 138].

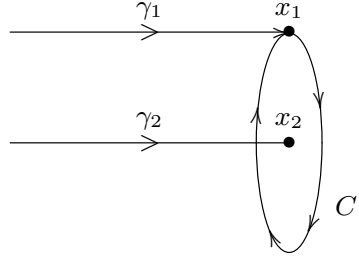

FIGURE 3.2: Parallel transport of a puncture around another one.

A way to compute the phase is to consider the configuration space (3.37) and rewrite it by means of $S^2 \cong \mathbb{C} \cup \{\infty\}$ as

$$\mathcal{F}_N(S^2) \cong \{(z_1, ..., z_N) \in (S^2)^N | z_p \neq z_{p'} \text{ for } p \neq p'\}. \tag{3.40}$$

This is equivalent to

$$\{(S^2)^N - \bigcup_{1 \leq p < p' \leq N} K_{pp'}\}, \tag{3.41}$$

where $K_{pp'} = \{(z_1, ..., z_N) \in (S^2)^N | z_p = z_{p'}\}$. The form of the phase space (3.36), allows us to trade $H^1(\Gamma, \mathbb{R})$ for $H^1(\mathcal{F}_N(S^2), \mathbb{R})$. The closed holomorphic 1-form

$$\omega_{pp'} = \frac{1}{2\pi i} d \log(z_p - z_{p'}) \tag{3.42}$$

on $\mathcal{F}_N(S^2)$ represents the de Rham cohomology class of generators $\omega_{pp'} \in H^1(\mathcal{F}_N(S^2); \mathbb{Z})$ with $1 \leq p < p' \leq N$ [139]. We thus introduce the Knizhnik-Zamolodchikov/Kohno connection to the context of LQG black holes as

$$\hat{A}_K = \frac{4\pi}{k+2} \sum_{1 \leq p < p' \leq N} \hat{J}^i_{\rho_p} \otimes \hat{J}^i_{\rho_{p'}} \; \omega_{pp'}, \tag{3.43}$$

wherein $\otimes$ denotes the Kronecker product [110, 111, 140–142].

Using this, a simultaneous puncture rearrangement can be given using the holonomy operator of Eq. (3.43)

$$\rho(\hat{A}_K, \gamma) = \mathcal{P} \; e^{i \oint_\gamma A_K}, \tag{3.44}$$

where the loop $\gamma$ is taken from the homotopy class $[\gamma] \in B_{n_{\frac{1}{2}}, \ldots, n_{\frac{k}{2}}}(S^2)$.[9]   Hence, by exchanging/moving the horizon degrees of freedom on the punctured 2-sphere, the wave function picks up a non-Abelian phase, namely

$$\psi_{\text{horizon}} \to \rho_{[\gamma]}(\hat{A}_{\text{K}})\psi_{\text{horizon}}. \tag{3.45}$$

This specifies the above-given statement that inequivalent quantisations on the multiply connected configuration space $\mathcal{F}_N(S^2)$ are marked by representations

$$\rho : B_{n_{\frac{1}{2}}, \ldots, n_{\frac{k}{2}}}(S^2) \to \text{SU}(2). \tag{3.46}$$

We have thus clarified, how (non-Abelian) anyonic statistics is encoded in the description of the LQG black hole model based on a puncture system representing the quantum degrees of freedom of the horizon. Below we want to further investigate this exchange behaviour by relating it to the large diffeomorphisms of the punctured horizon.

### 3.2.2   Large diffeomorphisms and the braid group

Motivated by our preliminary discussion of the symmetries of CS-theory and the fact that the horizon puncture system is invariant with respect to small diffeomorphisms, we want to take a closer look onto the action of the large diffeomorphisms on our system.

The large diffeomorphisms of the punctured 2-sphere fall into the mapping class group $M_{n_{\frac{1}{2}}, \ldots, n_{\frac{k}{2}}}(S^2)$, which we discuss in Appendix (A.5). Apriori, horizon states could either be invariant under it or transform by an unitary representation of it [143]. In the former case, large diffeomorphisms would be considered as gauge, whereas in the latter they would be regarded as a symmetry of the theory for which we will argue below.

The action principle does not dictate the transformation properties of the physical states under the diffeomorphisms which are not in the identity component. This is because no constraints are associated to them. Small diffeomorphisms are generated by the constraints encoded in the action (3.20), so only they should apriori be factored out. To demand the invariance under large diffeomorphism transformations would amount to an extra assumption [57]. On the classical level a diffeomorphism of the punctured $S^2$ induces a linear transformation on $H^1(\mathcal{F}_N(S^2), \mathbb{R})$, which in turn is the reason why the latter gives rise to a representation of the mapping class group [105–107].

---

[9]Notice that the contour integral of a meromorphic 1-from, such as $\omega_{pp'}$, along a loop $\gamma$ on $S^2$ that encircles all poles will vanish. The sum over all residues yields 0 because such a loop can always be shrunk to a point on the back of the sphere (cf. Appendix (A.4)).

Interestingly, these diffeomorphisms can be easily connected to the previous discussion of the statistical symmetry of the puncture system since

$$M_{n_{\frac{1}{2}}, \ldots, n_{\frac{k}{2}}}(S^2) \cong B_{n_{\frac{1}{2}}, \ldots, n_{\frac{k}{2}}}(S^2)/\mathbb{Z}_2 \tag{3.47}$$

holds, as recovered in Appendix (A.30).

Let us exemplify this point by considering $N$ punctures on one hemisphere of $S^2$. This would be homeomorphic to a $N$-punctured disc. From algebraic topology one knows, that the mapping class group of the $N$-punctured disc $M_N(D^2)$ is isomorphic to the braid group of the disc $B_N(D^2)$ on $N$ strands which in turn is equivalent to $B_N(\mathbb{R}^2)$. Hence, for this topology the statistical symmetry of the puncture system is exactly given by the large diffeomophisms.

Using the last subsection, we are able to calculate unitary representations of braiding generators e.g. for the setting of 2 labelled punctures. By executing the contour integral in Eq. (3.44) in the case of two punctures, one yields the monodromy operator

$$\hat{M}_{(1,2)}\psi \equiv \rho(\hat{A}_{\mathrm{K}}, \sigma_1^2)\psi = q^{2\hat{J}_{\rho_1}^i \otimes \hat{J}_{\rho_2}^i}\psi, \tag{3.48}$$

where $\sigma_1$ is a generator of the braid group (cf. Appendix (A.2)), $q = e^{i\frac{2\pi}{k+2}}$ is the so-called deformation parameter and we dropped the subscript of the wave function.[10][11] Since $\hat{M}$ represents the case of two consecutive exchanges of puncture one with two, the braiding matrix is

$$\hat{B} \equiv \rho(\hat{A}_{\mathrm{K}}, \sigma_1) = q^{\hat{J}_{\rho_1}^i \otimes \hat{J}_{\rho_2}^i} \; P_{12}, \tag{3.50}$$

where $P_{12}$ is the permutation operator. We depict the effect of $\hat{M}$ and $\hat{B}$ in Fig. 3.3.

---

[10]If we consider e.g. the case where both punctures are labelled with the fundamental representation of $\mathfrak{su}(2)$, we obtain for the monodromy with Eqs. (3.43) and (3.44)

$$\hat{M} = \begin{pmatrix} q^{\frac{1}{2}} & 0 & 0 & 0 \\ 0 & \frac{1}{2}\left(q^{\frac{1}{2}} + q^{\frac{3}{2}}\right) & \frac{1}{2}\left(q^{\frac{1}{2}} - q^{-\frac{3}{2}}\right) & 0 \\ 0 & \frac{1}{2}\left(q^{\frac{1}{2}} - q^{-\frac{3}{2}}\right) & \frac{1}{2}\left(q^{\frac{1}{2}} + q^{\frac{3}{2}}\right) & 0 \\ 0 & 0 & 0 & q^{\frac{1}{2}} \end{pmatrix} \tag{3.49}$$

with eigenvalues $q^{\frac{1}{2}}$ (triplet) and $q^{-\frac{3}{2}}$ (singlet).

[11]Higher powers of $\hat{M}$ correspond to different winding numbers and encirclement of several punctures corresponds to the ordered product of monodromy operators.

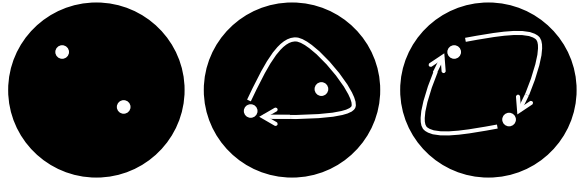

FIGURE 3.3: Two horizon punctures: unbraided vs. upon the application of $\hat{M}$ and $\hat{B}$, respectively.

In the limit of large black holes, i.e. $k \to \infty$, the operator $\hat{M}$ is just the identity. and $\hat{B}$ reduces to (a non-Abelian representation of) the permutation operator $P_{12}$. Pictorially, in the case of large black holes the topological information about what happened along the braid is forgotten. Hence, we infer that in this model the braiding is a quantum effect becoming relevant for small (and smaller getting) black holes.

Since the action of large diffeomorphisms on punctures will also drag the incident bulk edges along, this will cause a braiding of the spin network at least in the vicinity of the horizon as in Fig. 3.4.

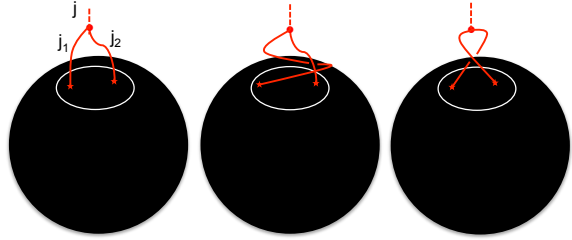

FIGURE 3.4: Two incident bulk edges piercing the horizon: unbraided vs. upon the application of $\hat{M}$ and $\hat{B}$, respectively.

This has observable consequences for a local stationary observer who resides on the node in Fig. 3.4 at proper distance $\ell$ to the horizon. To see this, observe that the field strength $\hat{F}^i$ is an observable which transforms under the action of large diffeomorphisms of the punctured surface as

$$\hat{F}^i \to \hat{F}'^i = \rho^{-1}\hat{F}^i\rho \tag{3.51}$$

because

$$[\hat{F}^i, \rho^n] \neq 0. \tag{3.52}$$

holds. This allows to discern braided from unbraided states.[12]

---

[12]This can be shown by taking the difference of the expectation values

$$\langle \rho^n\psi|\hat{F}^i|\rho^n\psi\rangle - \langle\psi|\hat{F}^i|\psi\rangle, \tag{3.53}$$

### 3.2.3   Aspects of the algebraic theory of $\mathfrak{su}(2)_k$-anyons

It is known from solid state physics that quantum systems in $2d$ exhibit anyonic statistics [132–135, 138]. The question arises, how the anyonic nature of the puncture system, as captured by the non-Abelian phases, affects its statistics and consequently the form of its entropy. This Section illustrates that the Hilbert space and consequently the entropy of the isolated horizon quantum system is completely analogous to the results for a corresponding system of non-Abelian anyons in condensed matter physics. This is demonstrated by direct comparison to the abstract definition of a model of $\mathfrak{su}(2)_k$-anyons [136, 137].

The mathematical formulation of a model of non-Abelian anyons in solid state physics is involved and demands more than the above given braid group description. One actually needs representations of the braid group which are compatible with the notion of fusion. The mathematical structure which consistently captures these features is a modular tensor category, specifically a unitary braided fusion category [136, 137, 144, 145]. Without going into the mathematical intricacies, we will consider a particular set of classes of non-Abelian anyons. These are the $\mathfrak{su}(2)_k$-anyons, which arise in non-Abelian Chern-Simons theory with $G = \mathrm{SU}(2)$ and level $k \geq 2$. A particular class of non-Abelian anyons is therein defined by each value of the level $k$.

For the full specification of the braiding statistics of a system of such anyons one has to give the following data:

(1.) Anyon species/superselection sectors forming a finite set $M$: The different anyons are labelled by anyonic charges $j \in \{0, \frac{1}{2}, 1, \ldots, \frac{k}{2}\}$.

---

where $\rho^n = \left(q^{\hat{J}_{\rho_1}^j \otimes \hat{J}_{\rho_2}^j} P_{pp'}\right)^n$ with $n = 1, 2$. Without loss of generality we set $N = 2$ and when using the Baker-Campbell-Hausdorff formula and its Hadamard lemma one yields

$$\langle \psi | \sum_{m=0}^{\infty} \frac{\left(\frac{n}{i}\frac{2\pi}{k+2}\right)^m}{m!} [\hat{J}_{\rho_1}^j \otimes \hat{J}_{\rho_2}^j, (P_{12}^{-1})^n \hat{F}^i (P_{12})^n]_m - \hat{F}^i | \psi \rangle \tag{3.54}$$

where

$$[\hat{J}_{\rho_1}^j \otimes \hat{J}_{\rho_2}^j, (P_{12}^{-1})^n \hat{F}^i (P_{12})^n]_0 \equiv (P_{12}^{-1})^n \hat{F}^i (P_{12})^n \tag{3.55}$$

and

$$[\hat{J}_{\rho_1}^j \otimes \hat{J}_{\rho_2}^j, (P_{12}^{-1})^n \hat{F}^i (P_{12})^n]_m \equiv [\hat{J}_{\rho_1}^j \otimes \hat{J}_{\rho_2}^j, [\hat{J}_{\rho_1}^j \otimes \hat{J}_{\rho_2}^j, (P_{12}^{-1})^n \hat{F}^i (P_{12})^n]_{m-1}]. \tag{3.56}$$

Multiplying from the left with $\rho^n$ gives Eq. (3.52). For example, when $n = 2$ the commutator yields

$$[\hat{F}^i, \hat{M}] = i\frac{4\pi}{k+2}\frac{4\pi}{k} i\epsilon_{jk}{}^i (\delta^2(x, x_1)\hat{J}_{\rho_1}^j \otimes \hat{J}_{\rho_2}^k + \hat{J}_{\rho_1}^j \otimes \hat{J}_{\rho_2}^k \delta^2(x, x_2)) + \mathcal{O}(k^{-3}). \tag{3.57}$$

When considering large black holes, the effect of the braiding onto the field strength would be negligible but it would become relevant for smaller (and smaller getting) black holes.

Comment: The constituents of the quantum isolated horizon form such a finite set $M$, each puncture is labelled with a spin $j \in \{0, \frac{1}{2}, 1, \ldots, \frac{k}{2}\}$ and gives rise to a quantum of area $a_j = 8\pi\gamma\ell_p^2\sqrt{j(j+1)}$.

(2.) Fusion rules: Similar to ordinary spin systems, the anyon labels are combined by certain fusion rules, determining their collective behaviour. For any combination of anyons $j_1, j_2, j \in M$ there is a fixed finite dimensional Hilbert space $V_j^{j_1 j_2}$ called splitting space, whereas we call $V_{j_1 j_2}^j$ the fusion space. The non-negative integers $N_{j_1 j_2}^j = \dim V_j^{j_1 j_2} = \dim V_{j_1 j_2}^j$ are called fusion multiplicites. $0 \in M$ denotes the vacuum sector. In terms of the fusion matrices $N_j$ the composition rule reads as

$$(j_1) \otimes (j_2) = \bigoplus_{j=|j_1-j_2|}^{\min(j_1+j_2, k-j_1-j_2)} (N_{j_1})_{j_2}^j \ (j).\tag{3.58}$$

The quantum dimension $d_j$ of an anyon with charge $j$ and the fusion matrices are related by $N_j \mathbf{d} = d_j \mathbf{d}$. The components of the vector $\mathbf{d}$ are the quantum dimensions of all anyon species occuring in the model. The total quantum dimension is defined as $\mathcal{D} \equiv \sqrt{\sum_j d_j^2}$. For $\mathfrak{su}(2)_k$-anyons the quantum dimensions are computed iteratively by $d_0 = 1$, $d_{\frac{1}{2}} = 2\cos(\pi/(k+2))$ and $d_j = d_{\frac{1}{2}} d_{j-\frac{1}{2}} - d_{j-1}$ with $j \geq 1$. The Hilbert space of the $N$-punctured sphere with charges/anyons at each puncture is constructed by sewing together a chain of $(N-2)$ 3-punctured spheres, called pants decomposition. non-Abelian anyons have $d_j > 1$, which is generally not an integer. This is characteristic of the non-locality of the Hilbert space which is not simply the tensor product of $d_j$-dimensional Hilbert spaces locally associated to each anyon.

Commment: The constitutents of the quantum isolated horizon are known to obey precisely the same fusion rules. By summing over all possible puncture configurations and equipped with an appropriate combinatorial pre-factor, these rules were used in Refs. [84–87, 99] to obtain for the total number of microstates the expression (3.27).

(3.) The $R$-matrix: This object is used to describe an exchange of two anyons $j_1, j_2$ through braiding after the splitting of anyon $j$. The description of braiding in terms of basic data is specified by the unitary action of $R$ on splitting spaces as $R_j^{j_1 j_2} : V_j^{j_1 j_2} \to V_j^{j_2 j_1}$. Unitarity implies $N_{j_1 j_2}^j = N_{j_2 j_1}^j$ and $R$'s action is diagramatically represented as in Fig. 3.5.

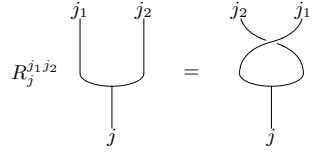

FIGURE 3.5: Graphical representation of the R-matrix.

Comment: In the context of quantum isolated horizons the $R$-matrix showed up in the discussion of the representation theory of the quantum group $U_q(\mathfrak{su}(2))$ in Ref. [99]. The following point will also deal with its relation to the braiding matrix given by Eq. (3.50) and thus with the statistics which we have extensively discussed in the previous subsections.

(4.) The $F$-matrix: The fusion of three anyons is associative and therefore one has two ways to fuse three anyons to a fourth. These two ways are related by a basis change. It is specified by

$$F_{j_1 j_2 j_3}^{j_4} : \bigoplus_j V_{j_1 j_2}^j \otimes V_{j j_3}^{j_4} \to \bigoplus_{j'} V_{j_1 j'}^{j_4} \otimes V_{j_2 j_3}^{j'} \tag{3.59}$$

and its action is diagrammatically represented as in Fig. 3.6.

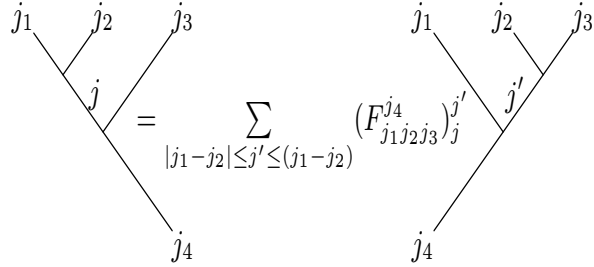

FIGURE 3.6: Graphical representation of the F-matrix.

The $F$-matrix is unitary, obeys the orthogonality relation

$$\sum_l (F_{j_1 j_2 j_3}^{j_4})_j^l (F_{j_4 j_1 j_2}^{j_3})_l^{j'} = \delta_j^{j'} \tag{3.60}$$

and is subject to two further consistency conditions. The first is called the pentagon relation/Biedenharn-Elliott identity,

$$(F^e_{abh})^j_f (F^e_{fcd})^h_g = \sum_k (F^j_{bcd})^h_k (F^e_{akd})^j_g (F^g_{abc})^f_e. \tag{3.61}$$

The second is termed as the hexagon relation,

$$R^g_{ac}(F^d_{bac})^g_e R^e_{ab} = \sum_f (F^d_{bca})^g_f R^d_{af}(F^d_{abc})^f_e. \tag{3.62}$$

Finally, the braiding and the $R$-matrix are related by

$$B_{j_1 j_2} = \sum_j (F^{j_4}_{j_1 j_3 j_2}{}^{-1})^{j'}_j R^j_{j_1 j_2} (F^{j_4}_{j_1 j_3 j_2})^{j'}_j, \tag{3.63}$$

where $B_{j_1 j_2} \in V^{j_2 j_1}_{j_1 j_2} = \bigoplus_j V^{j_2 j_1}_j \otimes V^j_{j_1 j_2}$ and its action is diagrammatically represented as in Fig. 3.7.

FIGURE 3.7: Graphical representation of the B-matrix.

Comment: The $F$-matrix is the analogue of Wigner's ($q$-deformed) $\{6j\}$-symbol from re-coupling theory [146] which is extensively used in LQG [5, 31, 58, 59]. We identify

$$(F^{j_4}_{j_1 j_2 j_3})^{j'}_j = \begin{Bmatrix} j_1 & j_2 & j' \\ j_3 & j_4 & j \end{Bmatrix}_q. \tag{3.64}$$

Although the $R$-matrix has already been discussed in context of the quantum geometry of isolated horizons as in Ref. [99], its relation Eq. (3.63) to the braiding matrix Eq. (3.50) and particularly to the anyonic statistics of the model, as done in the previous subsections, is a novel feature.

(5.) The modular $S$-matrix simultaneously diagonalises all the fusion matrices $\{N_j\}$.

Through the Verlinde formula [140–142, 147–149] it is related to the fusion multiplicities as

$$(N_{j_1})^j_{j_2} = \sum_d \frac{S^d_{j_2} S^d_{j_1} (S^{-1})^j_d}{S^d_0},$$      (3.65)

where $S_{j_1 j_2} = \sqrt{\frac{2}{k+2}} \sin\left(\frac{(2j_1+1)(2j_2+1)\pi}{k+2}\right)$.

Comment: In the comment to point (2.) the Verlinde formula was already implicit to find the dimension of the Hilbert space.

(6.) Topological spin $h_j$ and twist $\theta_j$: The twist $\theta_j$ is related to the topological spin by

$$\theta_j = e^{i2\pi h_j} = R^0_{jj}.$$      (3.66)

Their relation to the chiral central charge $c_- = c - \bar{c}$ is given by

$$\frac{1}{\mathcal{D}} \sum_j d^2_j \theta_j = e^{i\frac{2\pi}{8} c_-}.$$      (3.67)

For the $\mathfrak{su}(2)_k$-WZW-model used in Refs. [86, 87] one has the central charge $c = \frac{3k}{(k+2)}$ and the conformal dimensions $h_j = \frac{j(j+1)}{(k+2)}$. The topological spin feature shows up, if one considers particles in $2 + 1$ dimensions to be of finite extent rather than being point-like. In the context of CS-theory the thickening to a ribbon is called framing and it is needed to preserve general covariance at the quantum level [105–107, 110, 111]. Considering the possibility of a $2\pi$ rotation of a single particle relative to the rest of the system amounts to a change of the quantum wavefunction by a phase $e^{i2\pi\delta}$ with $\delta = h_j$. Their finite extent renders their world lines to ribbons which are twisted by such rotations. Hence, Eq. (3.66) expresses the (topological) spin-statistics connection of anyons. However, notice that $h_j$ should be discriminated from the actual spin of the object, which is related to the transformation properties with respect to the $2d$ rotation group SO(2) [132–134, 138, 150, 151]. Even if the considered system does not exhibit rotational invariance, $h_j$ is of course properly defined.

Comment: When focusing on just one (thickened) puncture out of $N$ distributed on $S^2$, then it cuts out a disc $D^2$ with boundary $S^1$. In the context of anyon models one would consider $S^2$ without $D^2$ as the bulk supporting the system of the remaining $N - 1$ anyons and the $1d$-circle as the edge [136, 137, 144, 145]. Typically, if a $2d$-system supports anyons in the bulk, one also has chiral massless excitations propagating along the $1d$-edge described by a conformal field theory (CFT) and the energy flux of which is proportional to the chiral central charge $c_-$. The anyons in the bulk do not determine $c_-$ completely, hence Eq. (3.67) fixes $c_-$ only modulo 8. Apart from this, to consider the horizon punctures as extended objects giving in turn rise to the twist $\theta_j$ is a new feature. In the large black hole limit, the topological spin $h_j$ vanishes and the twist is equal to 1 rendering the ribbon-like nature unimportant. It should be clarified how these qualitative arguments are related to the recent Ref. [152] which explores the notion of CFT/gravity correspondence in the context of this LQG black hole model.

To summarise, from the points (1.)-(6.) of this definition (1.)-(3.) and (5.) have already been known in the description of quantum isolated horizons in LQG. The latter's description for $G = \mathrm{SU}(2)$ has been accomplished by means of a Wess-Zumino-Witten-CFT on the bounding $S^2$ [86, 87], $\mathrm{SU}(2)_k$ CS-theory on $\mathbb{R} \times S^2$ [84, 85, 90] or the representation theory of the quantum deformed $\mathrm{SU}_q(2)$, with $q$ a non-trivial root of unity [99]. Not surprisingly, all these approaches lead to the same expression for the dimension of the isolated horizon Hilbert space. They agree because the notions of $2d$ modular functor, $3d$ TQFT and modular tensor category are essentially the same [153]. This work adds points (4.) and (6.) to the literature on quantum isolated horizons, especially with regard to the braiding matrix, topological spin and the twist. We want to highlight, that the former are crucial in order to interpret the Hilbert space of the quantum isolated horizon as being analogous to the fusion Hilbert space of non-Abelian anyons. Since the dimension of the fusion Hilbert space is computed in exactly the same manner, considering the horizon punctures as non-Abelian anyons neither changes its dimension nor its entropy.

Nevertheless, apart from the well-known and exploited fact, that the CS-level $k$ serves as a IR cut-off by $j \leq \frac{k}{2}$ [84–87, 99, 154–157], from Eq. (3.50) we have deduced that the strength of the non-local effects due to the braiding is controlled by $k$ and that they disappear for large black holes. Hence, the non-local characteristics of the horizon Hilbert space vanishes when $k \to \infty$ and we are left with the tensor product of $d_j$-dimensional Hilbert spaces that are locally associated to each horizon degree of freedom.

### 3.2.4 $k$-dependance of the entropy and black hole radiance spectrum

In light of the previous subsections, we allow ourselves to add a qualitative discussion of the relevance of the level $k$ for the entropy and the radiance spectrum of the quantum isolated horizon.

Qualitatively, the braiding corresponds to non-local quantum correlations between the horizon degrees of freedom and thus adds order to the collective. Since order reduces entropy, this suggests a reducing effect on the horizon entropy for smaller (and smaller getting) black holes. If we assume without loss of generality that all punctures take $j = \frac{1}{2}$ in (3.27) then for the entropy $S \propto \log W(\{n_j\})$ one has with constant $N$ that

$$S(k_1) < S(k_2) < S(k \to \infty) \tag{3.68}$$

for levels $k_1 < k_2 < \infty$ and $\lim_{k \to \infty} \partial_k (S(k)) = \text{const}$. We attribute this to $k$'s double role as a cut-off and as a parameter controlling the non-local correlations. Interestingly, the analysis of the entropy $S(k)$ in Ref. [125] has shown that

$$S = \lambda A + \alpha \log A, \ \lambda = \text{const.} \tag{3.69}$$

for $k \to \infty$ (with the notorious logarithmic correction with $\alpha = -\frac{3}{2}$ as in Refs. [84–87, 123, 124]) whereas it was found that

$$S = \lambda(k)A(k) \tag{3.70}$$

for finite $k$ (small black holes) holds. Hence, it would be worthwhile to explore if the occurrence of the logarithmic correction is related to the vanishing of the non-local effects, i.e., the collaps of the group of large diffeomorphisms to the permutation group in the large $k$-limit.

Apart from the consequences for the entropy, it could be interesting to see whether there are any traces of the non-trivial statistics of the horizon degrees of freedom in the outgoing radiation. To this aim, we invoke the following qualitative picture for the mechanism responsible for black hole radiance, as given in Refs. [102, 103, 158]. Starting with the microstates given in Section 3.1.2, we assume that the black hole is initially in an eigenstate $|i\rangle$ of the horizon area operator $\hat{A}$. Upon transition to a nearby state $|f\rangle$ with slightly smaller area, radiation of energy $\Delta E_{if}$ is emitted, which in turn leads to a reduction of the black hole energy. Let the emitted quantum be of the gravitational field with energy $\Delta E_{if} = \hbar \omega_{if}$, where $\omega_{if}$ denotes its frequency at infinity. This transition is mediated by the action of the full Hamiltonian operator on a node near the horizon (as on the left in Fig. 3.4), which leads to a change in the spin associated to some of the attached edges. The analysis of the spectrum of this emission process thus yields a discrete set of lines which depend on the matrix elements of the Hamiltonian operator. For the determinition of the intensities of the corresponding spectral lines and the form of the emission spectrum, one then uses the analogy to transitions in atomic physics. By virtue of Fermi's golden rule, this yields for the probablity of a transition $i \to f$:

$$P_{if} = \frac{2\pi}{\hbar} \ |\langle \hat{H}_{if} \rangle|^2 \ \delta(\omega - \omega_{if}) \ \frac{\omega^2 \mathrm{d}\omega \mathrm{d}\Omega}{(2\pi\hbar)^3}. \tag{3.71}$$

The matrix element of the part of the Hamiltonian of the system being responsible for the transition is $\hat{H}_{if}$ and $\mathrm{d}\Omega$ is the differential solid angle. From this one gains the total energy $\mathrm{d}I$ emitted by the system per unit time as

$$\mathrm{d}I_{if} = 2\pi\omega \ p(i) \ |\langle \hat{H}_{if} \rangle|^2 \ \delta(\omega - \omega_{if}) \ \frac{\omega^2 \mathrm{d}\omega \mathrm{d}\Omega}{(2\pi\hbar)^3}, \tag{3.72}$$

where $p(i)$ is the probablity to find the system in the initial state $i$. Due to the fact, that the level spacing between the eigenvalues of $\hat{A}$ decreases exponentially for large areas, the separation of the spectral lines can be rather small, thus justifying the approximation of the spectrum by a continuous profile in accordance with the calculation of the black-body spectrum derived via semi-classical arguments by Hawking [19, 20].

To calculate the intensity distributions, the probability distribution $p(i)$ and the matrix elements $\hat{H}_{if}$ have to be known. These are also the relevant quantities to be inspected when checking if any (perhaps slight) alteration of the spectrum due to the braiding is expectable. Firstly, when assuming for simplicity that all accessible microstates occur with equal probability, one has $p(i) \propto \mathrm{e}^{-S}$ since $S \propto \log W(\{n_j\})$. Knowing that $W(\{n_j\})$ is explicitly $k$-dependent and having identified that the variation of $S$ with respect to $k$ is also due to the non-local effects, i.e. the braiding, one would have $p(i) \propto \mathrm{e}^{-S(k)}$ and the spectrum would indeed be changed by this. Secondly, the matrix elements $\hat{H}_{if} = \langle f | \hat{H} | i \rangle$ could very well be computed with braided states e.g. $|i'\rangle = \hat{B}|i\rangle$ (also $\hat{H}$ does not in general commute with the non-Abelian phases) which would have a non-trivial effect on the spectrum and make it explicitly $k$-dependent. In the limit of large black holes, however, the spectrum would reduce to the one advocated in Ref. [158]. We leave the issue of rigorously quantifying the spectrum in the braided case, also in the improved local setting of Refs. [102, 103], which used the matrix elements computed in Ref. [159], for future investigations.

The rigorous analysis of the emission of non-gravitational quanta would require a more detailed understanding of matter couplings in LQG. Nevertheless, when invoking the semi-classical Parikh-Wilczek tunneling framework [160] which understands the emission of a particle from the black hole as a tunneling process, quantum gravity corrections to the emission spectrum using the entropy-area relation (3.69) were given in Refs. [161–163]. In the tunneling picture the emission probablity is proportional to a phase space factor

$$P_{if} \propto \frac{\mathrm{e}^{S_f}}{\mathrm{e}^{S_i}} = \mathrm{e}^{\Delta S}. \tag{3.73}$$

Using the entropy $S$ as in Eq. (3.69) gives for the emission of a particle of energy $\Delta E$ from a black hole of total energy $E$

$$P_{if} \propto \left(1 - \frac{\Delta E}{E}\right)^{2\alpha} \mathrm{e}^{-8\pi E \Delta E \left(1 - \frac{\Delta E}{E}\right)}, \tag{3.74}$$

which explicitly depends on the log-corrections implied by LQG. For a discussion of the consequences of the first factor, see Refs. [161–163]. However, when using the $k$-dependent entropy as in Eq. (3.70), the first factor drops out and $P_{if}$ becomes explicitly $k$-dependent which could provide traces of the non-trivial braiding and statistics in the outgoing radiation. A detailed analysis of these tentative arguments in the full theory would hinge much on a better understanding of the Hamiltonian operator as well as the matter coupling in LQG and is thus left for future investigations.

## 3.3 Discussion of the results

The purpose of this Chapter was to investigate whether and how the notion of anyonic/braiding statistics has bearing on the current LQG black hole model, based on the isolated horizon framework and its quantisation. The main result is that such a model explicitly displays (non-Abelian) anyonic statistics (as conjectured in Refs. [101, 164–166]) by direct comparison to the definition of a model of $\mathfrak{su}(2)_k$-anyons known from solid state physics. This work also establishes a clear relation in between the boundary symmetry group and the statistics of the model. In this way, we reinterpreted the statistics of the quantum isolated horizon model used so far in the literature. In the following, we discuss further implications and open questions.

As pointed out e.g. in Ref. [167], in LQG one typically considers the large diffeomorphisms to act trivially on the diff-invariant states for practical reasons [5]. However, as we have seen here, the Hamiltonian formulation of CS-theory on the horizon gives rise to a physical Hilbert space on which unitary representations of the mapping class group act non-trivially. This action leads to the anyonic statistics of the punctures and to a braiding of the incident bulk spin network edges which could potentially change their knot class. In light of this result, it could be interesting to re-evaluate the role of large diffeomorphisms in the bulk, too.

This scenario of anyonic statistics should also have bearing on the recent reformulation of the boundary theory in terms of triad fields [168, 169], when other types of boundary surfaces (e.g with different topology than $S^2$ [170–172]) or other boundary conditions are considered, provided that on the kinematical level the basic data is given by a bulk spin-network graph $\Gamma$ piercing the boundary at a set of points. In that case, non-Abelian phases should again show up.

When considering large black holes (given by the limit $k \to \infty$), we observed that the (non-Abelian) phases reduce to permutations. Hence, the large $k$-limit effectively collapses much of the group of large diffeomorphisms to its discrete analogue, the permutation group. To some extent this is counterintuitive since one would actually expect that diffeomorphism symmetry is a derived property in the semi-classical limit (i.e. when $k \to \infty$). It could therefore be interesting to replace the smooth manifold structure in the original formulation of the isolated horizon model by the weaker concept of a piecewise linear manifold and study the role of diffeomorphisms therein.

We want to emphasise that here only gravitational degrees of freedom were considered and that these were treated as distinguishable. Recently, in Ref. [166] it was shown by phenomenological arguments that a holographic degeneracy factor accounting for matter degrees of freedom together with indistinguishable punctures leads to a horizon entropy which is fully consistent with semi-classical treatments. This is not in contradiction to our work here, since we excluded matter degrees of freedom from the very beginning. To render the punctures indistinguishable, one would have to symmetrise the quantum version of the isolated horizon boundary condition. For consistency, this would most likely require the bulk degrees of freedom to be symmetrised, too. This is potentially related to the recent application of GFT (as a second quantised version of LQG) and its condensate states to describe the isolated horizon geometry [173, 174] and deserves to be further investigated.

Finally, we drew a connection to systems of anyons in solid state physics giving rise to topological states of matter like fractional quantum Hall systems [136, 137, 144, 145, 175, 176]. For such topologically ordered 2$d$-solid state systems a universal characterisation of many-particle quantum entanglement was found [177, 178]. In the entanglement entropy of such systems a universal entropy reducing constant occurs. This topological entanglement entropy accounts for the correlations related to the non-local nature underlying the anyonic statistics. Such a notion of entropy is yet to be be studied in the context of LQG black holes.

This finishes the first thematic unit of this thesis in terms of the application canonical quantisation techniques. In the following chapter we will divert our attention to covariant quantisation methods to motivate the GFT approach and its application to quantum cosmology afterwards.

# Chapter 4

# Path integral approaches to quantum gravity

*Mille viae ducunt homines per saecula Romam.*

Alain de Lille,

Liber Parabolarum.

## 4.1   The continuum perspective

The quantum gravitational path integral

$$Z = \int \mathcal{D}g \; e^{iS_{\mathrm{GR}}[g]} \tag{4.1}$$

was first formulated in Ref. [179] and provides the prototype of a background independent covariant quantisation of spacetime geometry. The integration runs over all field histories, given by spacetime metrics $g$ of the 4-manifold $\mathcal{M}$ up to diffeormorphisms thereof. If $\mathcal{M}$ is bounded by two 3-geometries $(\Sigma, h)$ and $(\Sigma', h')$ the transition amplitude is given by

$$\langle h | h' \rangle = \int \mathcal{D}g \; e^{i\left(S_{\mathrm{GR}}[g] + S_{\mathrm{boundary}}[h,h']\right)} \tag{4.2}$$

integrating over all 4-geometries which induce the given boundary geometries and the action is suitably modified by boundary terms. At this point it is worth remembering that path integrals stress that the amplitudes for physical processes are the fundamental objects of the theory.

Beyond its conceptual beauty, it is difficult to make sense of this expression for various reasons. First, topology change is conceivable as it could be induced by strong quantum fluctuations of the geometry. However, it seems impossible to implement a sum over topologies on the right hand side of Eq. (4.1) given the fact that 4-manifolds are not classifiable. Second, among the well-known difficulties one encounters when dealing with measures on infinite dimensional spaces, it is hard to define a probability measure on the space of metrics modulo diffeomorphisms. This problem also applies to the left hand side of Eq. (4.2). As we have seen in Section 2.1, the scalar product cannot be given without a Lebesgue measure on this space and consequently already the notions of kinematical state and kinematical Hilbert space in metric variables is flawed. Moreover, given that the symmetry group of the canonical theory is the Bergmann-Komar group, as reviewed in Section 2.1, while that of the path integral is $\text{Diff}(\mathcal{M})$ the precise relation between the canonical and covariant quantisation procedures is obscured. A related point of criticism then concerns the lack of a clear interpretation of expectation values of observables as expectation values in a rigorously defined physical Hilbert space [5]. Finally, another issue concerns the fact that the Euclidean gravitational action, reached after performing a Wick rotation, is not bounded from below yielding a divergent path integral.[1] A better understanding of these issues would be desirable, given the application of such path integrals e.g. in quantum cosmology research [181–189].

In the absence of a full understanding of Eq. (4.1), it is nevertheless possible to extract information about the quantum nature of gravity from this approach using perturbation theory. To this aim, one introduces the background field method [190, 191] where the metric is expanded in terms of an arbitrary (possibly curved) classical background $\bar{g}$ with perturbations $h$ to be quantised later on, i.e.

$$g = \bar{g} + h. \tag{4.3}$$

Choosing a flat background $\eta$ is already enough to see the merits and issues of the perturbative approach. With Eq. (4.3) one linearises the Einstein-Hilbert action (in the harmonic gauge) leading to a quadratic kinetic term plus higher-order interaction terms. The background metric allows to perform a Wick rotation after which standard quantum field theory techniques can be employed to quantise the perturbations.[2] This leads to the discovery that the (low energy) quanta of the gravitational field, the gravitons, are massless and of spin 2. In analogy with other quantum field theories, scattering amplitudes of gravitons with matter degrees of freedom and with themselves can be computed. An interesting consequence of such computations shows that in the infrared limit (at 1-loop order) the classical Newtonian potential receives corrections, i.e.

---

[1]It has been argued that this so-called conformal-factor problem could be counterbalanced by means of the path integral measure [180].

$$V(r) = -G_\text{N} \frac{m_1 m_2}{r} \left( 1 + \frac{3}{2} \frac{r_\text{S}}{r} + \frac{41}{10\pi} \frac{\ell_p^2}{r^2} \right), \tag{4.4}$$

where the Schwarzschild radius of the system is $r_\text{S} = \frac{2G_\text{N} M}{c^2}$ with $M = m_1 + m_2$ and we restored $\hbar$ and $c$ for convenience. The first departure is the well-kown generally relativistic correction term while the other is a genuine quantum gravity correction [192–194]. Despite its miniscule size, it is a concrete, model-independent quantum gravity prediction, which, if observed, would provide the first experimental evidence for the quantum nature of gravity.[3]

For reasons explained below, this background-dependent perturbative approach to the quantisation of gravity only makes sense as a low energy effective description [201–203]. Intuitively, it has to break down since at higher energies the backreaction of the perturbations onto the background will inevitably increase leading to an invalidation of the linearisation. Concretely, towards higher energies, quantum fluctuations are not under control in the sense that the theory is non-renormalisable. This stands in stark contrast to the electroweak and strong interactions which are described by perturbatively renormalisable quantum field theories [204–210]. As is well-known, the mass dimensionality of the coupling constant for an individual interaction determines the renormalisability of a theory. By dimensional analysis one finds that $G_\text{N}$ is dimensionful and that its mass dimension $[G_\text{N}] = 2 - d$ is negative in $d = 4$. As a consequence, an endless number of counterterms must be introduced to cancel divergences at arbitrary loop orders. Since their couplings are free parameters, an unlimited number of coupling constants has to be experimentally fixed. This implies the loss of predictivity of the theory [14, 15, 211–213]. A possible conclusion from this perturbative non-renormalisability is that the ultraviolet (UV) divergences disappear when gravity is properly quantised in a non-perturbative way.

---

[2] In the weak field expansion one only performs Gaussian integrals, so details about the non-trivial measure are considered unimportant [46].

[3] The observational signatures of this correction to the two-body gravitational potential have already been investigated in the context of solar system dynamics in Refs [195–199]. More recently, its impact onto the gravitational wave emission of inspiralling compact binaries was studied by the author in Ref. [200].

Since gravity is weak at low energies as opposed to high energies, it could also be taken as a hint that the degrees of freedom adequate for describing gravity at low energies are very dissimilar to those encountered at high energies.

An alternative path to quantise gravity within the continuum formulation (in the Euclidean regime) beyond the perturbative ansatz just described, is taken by the the asymptotic safety programme [214–223].[4] To bypass the problems of the perturbative quantisation, this approach assumes the existence of a non-perturbative (i.e. interacting) fixed point for gravity in the UV. Support for this assumption is obtained through the application of functional renormalisation group methods [224]. Intriguingly, close the non-trivial fixed point, one then finds that spacetime is effectively two-dimensional [225]. Since the FRG analyses depends on truncations, their reliability can be questioned. In addition, the asymptotic safety programme also makes use of an auxiliary background metric. Background independence then has to be restored at the level of physical observables which is a difficult issue [223]. In light of these points, it is important to find support for the existence of a non-trivial UV fixed point through other non-perturbative, possibly discrete approaches to quantum gravity, see e.g. Refs. [226, 227].

Beyond such arguments invoking universality, keeping in mind the technical and conceptual issues in the definiton of the gravitational path integral, it is legitimate to ask the question if it can be computed through discretisation techniques and performing the continuum limit at some point and in some way. This bridges the gap to the next section where we present a list of approaches attempting to exactly do this.

---

[4]It should be noted that it is so far unclear how to consistently formulate this programme in the Lorentzian regime.

## 4.2   The discrete perspective

As discussed above, there are several reasons why it is hard to fully clarify the meaning of Eq. (4.1). A possible way to breathe new life into the gravitational path integral is to replace the path integration over geometries, possibly extended by a sum over topologies, by means of a sum over triangulations, together with exchanging the continuum action with its discretised reformulation. To succeed in this endeavor, however, configurations have to be properly weighted in the sum over triangulations and the continuum limit has to be speficied. These points are intimately related. For example, if the recovery of the continuum from discrete building blocks was achieved through a phase transition, the probability weights would determine its details. Similarly, if a semi-classical limit was involved, this could require configurations to peak on some triangulated classical geometry.

In the following we will discuss quantum Regge calculus, Euclidean/causal dynamical triangulations, matrix/tensor models, spin foam models and group field theory (GFT) as examples for different but nevertheless closely related approaches to deal with the discretisation of the path integral. To a varying degree of detail the strategies to recover the continuum are surveyed to provide a contrast to the way this is purported to be done in the condensate cosmology approach of GFT.

By means of the question if the discreteness is considered physical or unphysical in these approaches, one may superficially organise them into two classes. Spin foam models and GFT both share the view that the discreteness of space(time) is real, rooted in the result that the spectra of geometric operators in loop quantum gravity are discrete. However, the strategies to recover classical geometries seem to be different for them. In contrast, in the other approaches a different interpretation is adopted and the discreteness is regarded as a mathematical tool allowing us to rewrite the continuum path integral in a discrete form. The philosophy behind taking the continuum limit is rather similar among them, though they differ in the details of the discretisation and probability densities chosen.

### 4.2.1 Simplicial quantum gravity

In the following, we give a brief exposition of Regge calculus (RC) which is a straightforward discretisation of classical GR [228]. The orginal motivation for the construction of RC was to establish a scheme which would allow to solve Einstein's field equations numerically. Later on, it was understood that it could be made useful for research on quantum gravity following the spirit of lattice gauge theory which had proved useful for the quatisation of non-Abelian gauge theories [229]. Using this, we discuss two non-perturbative approaches to quantum gravity constructed from RC afterwards. These are quantum Regge calculus and Euclidean/causal dynamical triangulations. To this aim, we closely follow Refs. [45–48, 230–233] and refer to the references therein for details.

Simplicial quantum gravity is a broad term which subsumes theories which are based on triangulations $\Delta$[5] of a $d$-dimensional Riemannian manifold $\mathcal{M}$ and the dynamics of which are encoded by the discretised version of the Euclideanised Einstein-Hilbert action, the so-called Regge action [228]. To understand its construction better, we introduce some terminology first.

A $d$-dimensional piecewise linear manifold is defined as a collection of $d$-dimensional (flat) polytopes glued together along their $(d-1)$-dimensional faces so that the topological dimension is preserved. Most generally, it is not possible to embed such objects isometrically into $\mathbb{R}^d$ which leads to curvature defects. As a simplification, we choose the constitutent polytopes to be simplices and thus identify piecewise linear manifolds with those of simplicial type. These building blocks are advantageous because their geometry can be entirely spefioied by their edge lengths.

A $d$-simplex $s$ is a $d$-dimensional object with $d+1$ vertices connected by $d(d+1)/2$ edges or line segments. In general, a $d$-simplex has $\binom{d+1}{d}$ $(d-1)$-simplices in its boundary. A subsimplex of $s$ of dimension $d-1$ is called face $f$, those of dimension $d-2$ are called bones or hinges $h$ and those of dimension 1 are called edges. A simplicial complex is a collection of simplices which are glued along their subsimplices. With this, a simplicial manifold is defined by a simplicial complex in which the neighborhood of any vertex, corresponding to the set of simplices which share the same vertex, is homeomorphic to a $d$-dimensional ball $B_d$ in $\mathbb{R}^d$.

---

[5]A triangulation of a topological space $X$ is a simplicial complex $\mathcal{C}$ which is homeomorphic to $X$ together with a homeomorphism $h : \mathcal{C} \to X$. We may say "$\mathcal{C}$ is a simplicial decomposition of $X$". For example, the simplest triangulation of a 2-sphere $S^2$ is given in terms of a regular tetrahedron. More details concerning this terminology follow below.

As said, the edge lengths $\ell$ fully specify the metric properties, i.e. the shape of the $d$-simplex and are easily constructed using

$$\ell_{ij}^2 = \eta_{\mu\nu}\,(x_i - x_j)^\mu\,(x_i - x_j)^\nu\,, \tag{4.5}$$

where $x_i^\mu$ with $\mu = 1\ldots d$ denote the coordinates of the $i$-th lattice site and $\eta_{\mu\nu}$ is the flat Euclidean metric. In such a piecewise linear space one detects curvature by moving around elementary loops which are dual to a hinge $h$. From the dihedral angle $\theta_{s,h}$, which is the angle in between two faces $f$ in the $d$-simplex meeting at a hinge $h$, the deficit angle $\delta_h$ can be computed. It is given by

$$\delta_h = 2\pi - \sum_{s \supset h} \theta_{s,h}, \tag{4.6}$$

where the sum goes over all $d$-simplices $s$ meeting on the hinge $h$. In other words, a Regge geometry is a special case of a continuum Riemannian manifold, with a flat metric in the interior of its simplices $s$ where curvature is only assigned to its hinges $h$.

Equipped with this technology, it is possible to translate the terms appearing in the Euclidean Einstein-Hilbert action

$$S_{\text{EH,E}} = -\frac{1}{2\kappa} \int_{\mathcal{M}} \mathrm{d}^4 x \sqrt{g}\,(R - 2\Lambda) \tag{4.7}$$

into the Euclideanised Regge action

$$S_{\text{Regge,E}}[\Delta, \{\ell_{ij}\}] = -\frac{1}{\kappa} \sum_{\text{hinges } h} V_h^{(d-2)} \delta_h + \lambda \sum_{\text{simplices } s} V_s^{(d)}, \tag{4.8}$$

with $\kappa = 8\pi G_{\text{N}}$ and $\lambda = \frac{\Lambda}{\kappa}$, see Refs. [46, 230, 231] for details. $V_h^{(d-2)}$ denotes the volume of a hinge which corresponds to an area in four and a length in three dimensions.[6] This procedure can be extended to the case of higher-curvature terms in the action [46, 231] and manifolds with boundary [234].[7] Importantly, all ingredients $V_s^{(d)}$, $V_h^{(d-2)}$ and $\delta_h$ in Eq. (4.8) are functions of the squared length variables $\ell_{ij}^2$ on the triangulation $\Delta$. We observe that these encode the relevant geometric degrees of freedom of the theory. Their advantage lies in the fact that they are completely coordinate-independent, as suggested by the title of Ref. [228]. Varying the action $S_{\text{Regge,E}}$ with respect to the edge legths then leads to the simplicial equivalent to Einstein's field equations.[8] In the continuum limit, obtained by refining the triangulation $\Delta$, the Regge geometry becomes Riemannian and $S_{\text{Regge,E}}$ converges to the Euclideanised Einstein-Hilbert action [235, 236].[9] In this sense, background independence (i.e. independence from the triangulation) is implemented when the continuum limit is taken.

---

[6]At zero cosmological constant, in two dimensions the Regge action gives the discrete analogue of the Gauss-Bonnet theorem, i.e. $\frac{1}{\kappa} \sum_{\text{sites } p} \delta_p = \frac{2\pi\chi}{\kappa}$, where $\chi$ is the Euler characteristic of the surface. Hence, it is a topological invariant which is in agreement with the well-known result from continuum GR in $2d$.

In view of constructing a quantum theory from this classical framework using the path integral approach, we have the freedom either to keep the triangulation fixed while varying the edge lenths or to fix the latter while varying the triangulations or to vary both. The first two options have given rise to the two non-perturbative approaches to quantum gravity named quantum Regge calculus and Euclidean/causal dynamical triangulations which we will showcase in the following.[10] Both turn out to be very closely related to other non-perturbative approaches, namely covariant LQG, GFT and tensor models, as we will see later on. The third option, called random Regge triangulations (cf. Ref. [237]), is much less studied and will not be presented here.

### 4.2.1.1 Quantum Regge calculus

Quantum Regge calculus (QRC) is based on the idea to fix a triangulation $\Delta$ of $\mathcal{M}$ while allowing edge lengths to be dynamical. The quantum dynamics are then encoded by the

---

[7]More precisely, the discrete analogue of the Euclideanised Gibbons-Hawking-York boundary term $S_{\text{GHY}} = -\frac{1}{\kappa} \int_{\partial \mathcal{M}} \mathrm{d}^{d-1}x \sqrt{h} K$ with $h$ being the determinant of the induced metric on the boundary $\partial \mathcal{M}$ and $K$ the associated extrinsic curvature is given by

$$S_{\text{boundary}} = -\frac{1}{\kappa} \sum_{\substack{\text{hinges } h \text{ in} \\ \text{the boundary}}} V_h^{(d-2)} \psi_h, \tag{4.9}$$

with $\psi_h = \pi - \sum_{s \supset h} \delta_h$ denoting the angle between two faces meeting at a hinge in the boundary.

[8]For example, in the absence of any sources and a cosmological constant term, in three dimensions the lattice equation of motion simply is $\delta_h = 0$. Of course, this is in full agreement with the flat space solutions found in continuum GR in $3d$.

[9]This is to be contrasted to the case of lattice gauge theory where the continuum limit is taken by sending the number of lattice sites to infinity and the lattice spacing to zero [229].

[10]It should be kept in mind that approaches resting on the path integral formulation lack a clear physical interpretation of the expectation values of observables in a physical Hilbert space.

Euclidean path integral

$$Z_{\text{QRC}}(\Delta) = \int d\mu(\ell_{ij}^2) e^{-\frac{1}{\hbar} S_{\text{Regge},\text{E}}[\Delta, \{\ell_{ij}\}]}, \tag{4.10}$$

with the path integral measure $d\mu(\{\ell_{ij}^2\})$ such that $\int d\mu$ represents the discrete analogue of the sum over all metrics [238–240]. In general, the partition function is rendered convergent by requiring a cut-off for short and long edge lengths. Local observables are functions of the length variables on $(d-1)$-dimensional slices of $\Delta$ and one is particularly interested in their expectation values evaluated for large simplicial complexes.

It is appropriate to remark that there is still controversy over the precise form of $d\mu$. This is rooted in the difficulties to construct a path integral measure which satisfies the discrete analogue of diffeomorphism invariance of GR in the continuum [47, 241]. Hence, QRC actually consists of a class of models and different representatives therein each make use of a different measure.

There are various applications of this framework, e.g. to quantum cosmology [242–245] and calculations of the graviton propagator in the limit of weak perturbations about flat space [246]. Explorations of the phase structure of the theory by studying the behaviour of long-range correlations and the existence of critical points were conducted in Refs. [247–252] which led to evidence for a transition, potentially of second order, between a region of rough and smooth geometry. Moreover, employing techniques similar to those used in lattice gauge theory facilitated the coupling of Regge gravity to scalar matter [253, 254] and SU(2)-gauge fields [255–259] which appear to only have a small impact onto the phase structure of the gravitational sector [45].[11]

**The Ponzano-Regge model**

Finally, before closing this subsection, it is expedient to discuss the paradigmatic Ponzano-Regge (PR) model [271] which is closely related to QRC and spin foam theory. It is a (heuristic) model for $3d$ Euclidean quantum gravity and can be seen as the first application of Regge calculus to quantum gravity. For this, consider a fixed triangulation $\Delta$ of a Riemannian 3-manifold $\mathcal{M}$ with the additional assumption that the length of each edge of

---

[11]It is possible to formulate a Lorentzian version of Regge calculus and define the corresponding path integral [228, 260]. Likewise, using the Hamiltonian formulation of Regge calculus as the discrete analogue of the Arnowitt-Deser-Misner $3 + 1$-formulation of GR, a discrete form of the Wheeler-DeWitt equation can be studied [261–270].

$\Delta$ can only take discrete values.[12] In other words, we label the edges with an irreducible representation of SU(2), i.e. $j = 0, \frac{1}{2}, 1, \frac{3}{2} \ldots$ and the lengths $\ell$ are proportional to $\hbar(j + \frac{1}{2})$. To each (Euclidean) tetrahedron in the triangulation we assign a Wigner-$\{6j\}$ symbol, depicted by Fig. 4.1. The quantum dynamics of the PR-model are then encoded by the

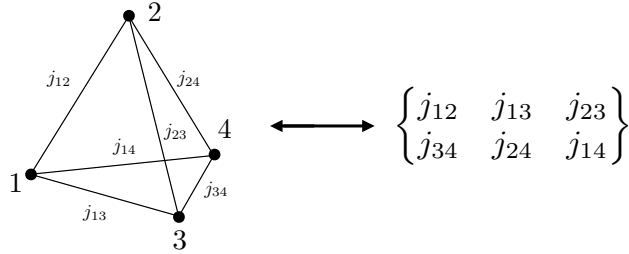

FIGURE 4.1: Tetrahedron in the triangulation $\Delta$ and suitably labelled $\{6j\}$-symbol assigned to it.

state sum

$$Z_{\mathrm{PR}}(\Delta) = \sum_{j_{ij}} \prod_{(ij)} (2j_{ij} + 1) \, (-1)^{\chi(j_{ij})} \prod_{\text{tetrahedra } t \in \Delta} \{6j\}_t, \tag{4.11}$$

where $\chi(j_{ij})$ is some function of the labels $j_{ij}$ [273]. In case that $\mathcal{M}$ has a boundary, $Z_{\mathrm{PR}}$ defines a transition amplitude for a $2d$ Regge geometry.

In general, the partition function is badly divergent. Nevertheless, it is worthwhile to examine its behaviour in the semi-classical limit which is obtained by keeping the edge lengths finite while simultaneously taking the limits $\hbar \to 0$ and $j_{ij} \to \infty$. To this aim, one can utilise that the $\{6j\}$-symbol for large $j_{ij}$ asymptotically behaves like

$$\begin{Bmatrix} j_{12} & j_{13} & j_{23} \\ j_{34} & j_{24} & j_{14} \end{Bmatrix} \sim \frac{1}{\sqrt{12\pi V}} \cos \left( \sum_{(ij)} j_{ij} \theta_{ij} + \frac{\pi}{4} \right), \tag{4.12}$$

where $V$ denotes the volume of the tetrahedron and $\theta_{ij}$ is the exterior dihedral angle about the edge of length $\ell_{ij}$. Using then that for large $j_{ij}$ the sum can be replaced by an integral,

---

[12]This is to be contrasted to the ordinary QRC procedure to take continuous edge lenths. It is possible to give a physical motivation fo the discrete lengths in the PR model from within LQG [272] where, as presented in Section 2.2.3.2, the spectra of geometric operators are quantised.

Ponzano and Regge were able to show that

$$
Z_{\text{PR}}(\Delta) \sim
$$
$$
\int \prod_{(ij)} \mathrm{d}j_{ij} (2j_{ij} + 1) \left( \prod_{\text{tetrahedra } t} \frac{1}{\sqrt{V_t}} \right) \left( e^{\frac{i}{\hbar} S_{\text{Regge,E}}[\Delta, \ell(j_{ij})^2] + i\frac{\pi}{4}} + e^{-\frac{i}{\hbar} S_{\text{Regge,E}}[\Delta, \ell(j_{ij})^2] - i\frac{\pi}{4}} \right).
$$
$$(4.13)$$

This expression bears strong resemblence with Eq. (4.10). In fact, the measure factor can be related to a particular QRC path integral measure and it can be argued that the two complex weights (corresponding to forward and backward propagation in coordinate time) can be identified under parity symmetry, while the $\frac{\pi}{4}$-factors do not affect the classical dynamics. It was observed by Turaev and Viro [274] that the partition function can be rendered finite when employing the representations of the quantum group $\text{SU}(2)_q$ (with $q$ being a root of unity) instead of those of $\text{SU}(2)$. This regularises the PR-model precisely because the number of representations is then finite. We will explore this model and the paradigmatic role it played for the development of the spin foam approach further in Section 4.2.3.

### 4.2.1.2 Euclidean and causal dynamical triangulations

A different way to define a quantum theory based on the Regge action is to fix the edge legths $\ell_{ij}$ to a rigid value, say $a$, while summing over a class of triangulations. This is the so-called Euclidean dynamical triangulations (EDT) approach [45, 48, 232].

The main motivation to proceed complementarily to the QRC approach originates in the aforementioned issues with the path integral measure. Focussing only on one fixed triangulation in the path integral leads to the exclusion of a huge amount of configurations while others are overcounted due to the edge length variation. EDT proposes a solution to these issues by summing over equilateral triangulations only. Hence, it rests on the expectation that its ensemble is more evenly distributed on the space of all geometries compared to the one of QRC on a fixed triangulation $\Delta$ and consequently better approximates smooth Riemannian manifolds [275].[13] Another motivation can be drawn from the fact that the natural and diffeomorphism invariant ultraviolet cut-off $a$ regularises the path integral and facilitates to define a continuum limit procedure.

---

[13] In other words, the use of standardised building blocks in EDT and CDT should not affect the continuum results. Such an universality argument has to be individually checked for different models, in fact.

Due to the restriction to equilaterality, the Euclideanised Regge action Eq. (4.8) takes a particularly simple form, namely

$$S_{\text{Regge,E}}[\Delta, a^2] = -\kappa_{d-2} N_{d-2}(\Delta) + \kappa_d N_d(\Delta) \tag{4.14}$$

where $N_{d-2}$ and $N_d$ denote the numbers of $(d-2)$- and $d$-simplices contained in the simplicial manifold $\Delta$. The coupling constants $\kappa_{d-2}$ and $\kappa_d$ are related to the (bare) gravitational and cosmological constant. The quantum dynamics of the model are then encoded by statistical mechanical ensembles of equilateral triangulations of $d$ dimensional manifolds (typically $S^4$ is used) which are weighted by the Regge action. The partition function is given by the purely combinatorial expression

$$Z_{\text{EDT}}(\kappa_{d-2}, \kappa_d) = \sum_{\Delta \in \{\Delta\}} \frac{1}{C(\Delta)} e^{-\frac{1}{\hbar} S_{\text{Regge,E}}[\Delta, a^2]}, \tag{4.15}$$

wherein $C(\Delta)$ denotes the order of the automorphism group of $\Delta$ which, on the simplical level, encodes the diffeomorphism symmetry of the continuum theory.

The goal is to compute $Z_{\text{EDT}}$ and take the limit $a \to 0$ afterwards with the hope that for particular values of the coupling constants a continuum limit can be obtained allowing for the extraction of a continuum theory (which should be closely related to general relativity). Notice that sending $a$ to zero renders the model background independent in the sense of setting dynamical all the degrees of freedom which were initially rigid.

In two dimensions, this approach is fully successful due to the existence of a closed expression for the partition function at fixed topology [276–281] and is in agreement with the quantisation in the continuum [282–284]. However, when studying the ensemble of configurations for $d > 2$ numerically with Monte-Carlo methods [285–291], one observes two phases corresponding to two distinct macroscopically pathological geometries. Concretely, consider the case there $d = 4$. In the first case, where $\kappa_2$ is small, the building blocks of the geometry are put carelessly together, creating a crumpled space of no extension marked by a high connectivity and large (possibly infinite) Hausdorff dimension $d_{\text{H}}$[14].

---

[14] The Hausdorff dimension $d_{\text{H}}$ of a subset $S$ of a metric space $X$ is defined by the infimum of all real-valued $d$ for which the $d$-dimensional Hausdorff content of $S$ is zero. The $d$-dimensional Hausdorff content of $S$ is defined by $C_H^d(S) \equiv \lim_{\sup_i r_i \to 0} \inf \left\{ \sum_i r_i^d : \text{there exists a cover of } S \text{ by balls with radii } r_i > 0 \right\}$.

Interestingly, when the value of $\kappa_2$ is increased, one encounters a phase transition towards a maximally extended space which is built from one dimensional branched out filaments. Such geometries are called branched polymers or trees [292, 293] and have Hausdorff dimension $d_H \approx 2$. The phase transition connecting the crumpled and branched polymer phase is of first order which means that there is no smooth passage between them [294, 295]. The inclusion of matter degrees of freedom does not seem to have a positive impact on this situation [45]. Thus, no phase corresponding to a smooth $d$-dimensional geometry can be found.

This can be seen as an indication that the continuum limit for the partition function has not been properly identified. First-order phase transitions are associated with a jump in the entropy of the system. When going from the crumpled phase over to the branched polymer phase, one makes a transition from a phase marked by entropy dominance to a phase dominated by the Euclidean action. For EDT it seems impossible to find the correct continuum limit because it is difficult (if not impossible) to balance the two contributions. A possible explanation for this is that one sums over a class of geometries too large which should be restricted instead [48, 232]. An illustration of this is given by EDT in two dimensions. If topology change is explicitly allowed, an initial configuration can branch off generating baby universes glued to a mother universe. Such baby universes dominate the partition function which in the continuum limit leads to a Hausdorff dimension of $d_H \approx 4$ [296, 297]. A way out of this problem is given by adding a restriction onto the partition function which disallows topology change and forbids the generation of baby universes.

Such a restriction can be motivated from within Lorentzian GR. Considering the classical evolution of a connected space-like hypersurface, the topology will remain unchanged. Points of topology change would lead to a degenerate metric, i.e., the light cone structure would become locally degenerate. One would like to exclude such a process from the quantum dynamics. Thus, the idea is to only sum over geometries with Lorentzian signature admitting a global foliation in proper time and that are compatible with (quantum) causality in the sense that topology change is forbidden [298, 299].[15] This insight has led to the causal dynamical triangulations (CDT) approach. The Lorentzian theory is much better behaved and its results, summarised below, may indicate that the correct implementation of a causal structure is essential for having a sensible theory of quantum geometry which produces extended physical geometries in the continuum limit.

---

[15] This view is also adopted when defining a continuum path integral for Lorentzian quantum gravity [300, 301].

In the following, we want to describe the construction of the CDT partition function in some detail, following [48, 232, 233, 302, 303]. To this aim, consider a $d$-dimensional compact spatial hypersurface $\Sigma$ (typically $S^3$) and a triangulation thereof denoted by $\tilde{\Delta}_t(\Sigma)$ which is built entirely from $d$-dimensional simplices $s_d$, assumed to be flat. The spacetime in between two triangulations of this kind at integer times $t$ and $t + 1$ is interpolated by $(d + 1)$-dimensional simplices $s_{d+1}(d - n, n)$, again assumed to be flat, which have a $s_{d-n}$- and a $s_n$-simplex as subsimplices at $t$ and $t + 1$ with $n = 0, 1, \ldots, d$. In CDT one distinguishes in between $d + 1$ different types of $(d + 1)$-simplices. We illustrate this for the case of CDT in $3 + 1$ dimensions in Fig. 4.2.

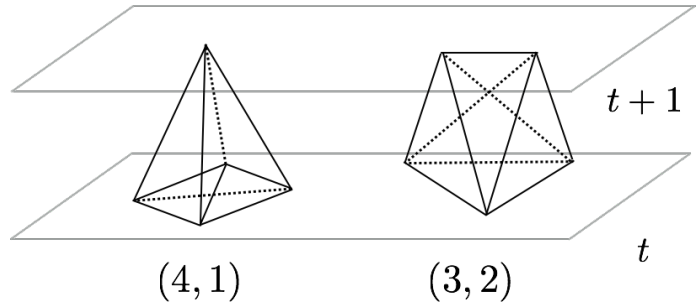

FIGURE 4.2: Two flat 4-simplices of type $(4, 1)$ and $(3, 2)$. The numbers refer to the number of vertices at constant integer time $t$ and $t + 1$. The remaining two building blocks are the time-reversed versions thereof, namely the $(1, 4)$- and the $(2, 3)$-simplices. The number 4 in $(4, 1)$ refers to the four vertices spanning a tetrahedron in the slice at time $t$.

In the $(d + 1)$-dimensional simplices we assign to each time-like edge the length $a_t$ and to each space-like edge the length $a_s$. One then assumes the relation

$$a_t^2 = -\alpha a_s^2 \tag{4.16}$$

to hold in between the edge lengths which allows us to define a map uniquely relating each Lorentzian CDT spacetime triangulation to that of an EDT. Practically, this is achieved by sending $\alpha$ to $-\alpha$ and fixing $a \equiv a_s$, thus changing all time-like lengths to space-like ones. This amounts to a Wick rotation

$$iS_{\text{Regge,L}}[\tilde{\Delta}, \alpha, a^2] = -S_{\text{Regge,E}}[\tilde{\Delta}, -\alpha, a^2] \tag{4.17}$$

which transforms the Lorentzian path integral (based on the Lorentzian Regge action) with complex weights into a Euclidean one (based on the Euclideanised Regge action) with real weights.[16]

For the concrete scenario where $\mathcal{M} \cong [0,1] \times S^3$ the Euclideanised Regge action is then given as

$$
\begin{aligned}
S_{\text{Regge,E}}[\tilde{\Delta}, -\alpha, a^2] = & - \left( \kappa_0 + 6f(\alpha) \right) N_0(\tilde{\Delta}) \\
& + \kappa_4 \left( N_4^{(3,2)}(\tilde{\Delta}) + N_4^{(4,1)}(\tilde{\Delta}) \right) \\
& + f(\alpha) \left( N_4^{(3,2)}(\tilde{\Delta}) + 2N_4^{(4,1)}(\tilde{\Delta}) \right),
\end{aligned}
\tag{4.18}
$$

where $f(\alpha)$ is a function of $\alpha$ (such that for $f(1) = 0$ this action takes the form of the EDT action, Eq. (4.14)), $N_0(\tilde{\Delta})$ is the number of vertices in the triangulation $\tilde{\Delta}$ and $N_4 = N_4^{(4,1)} + N_4^{(3,2)}$ is the number of 4-simplices therein. The coupling constants $\kappa_0$ and $\kappa_4$ are related to the (bare) gravitational and cosmological constant. With this, we may schematically write the CDT partition function as

$$
Z_{\text{CDT}}(\alpha, \kappa_0, \kappa_4) = \sum_{\tilde{\Delta} \in \{\tilde{\Delta}\}} \frac{1}{C(\tilde{\Delta})} e^{-\frac{1}{\hbar} S_{\text{Regge,E}}[\tilde{\Delta}, -\alpha, a^2]}.
\tag{4.19}
$$

Notice that due to the causality conditions imposed, the set of triangulations $\{\tilde{\Delta}\}$ to be summed over in this Euclideanised path integral is smaller than that of a model which is Euclidean from the onset.[17]

The goal is then to compute this sum, remove the regulator $a$ and study the phase structure of the theory. Strikingly, one finds three different phases: A crumpled phase, a branched polymer phase and a phase corresponding to an extended semi-classical continuum geometry [48, 303, 316–319].[18] Analogous results are found for CDT in $(2+1)$-dimensions [323, 324]. One verifies the large-scale features of the latter phase by computing the effective dimension of the geometry it represents and yields for the Hausdorff dimension $d_\mathrm{H} \approx d+1$.[19] In a second step, it is observed that the dynamics of the emergent geometry of this phase can be effectively described by means of an action compatible with the simple minisuperspace action of de Sitter space [303, 325–327].

---

[16]More recently, it has been attempted to formulate CDT without referring to a preferred global foliation by implementing the causal structure only locally. So far this has led to results which are in agreement with those of the standard CDT formulation [304–306]. However, up to now it is unclear whether this modification leads to unitarity violations.

[17]Attempts to modify EDT models by circumventing the imposition of causality conditions with the goal to obtain an interesting continuum theory [307–315] have so far not led to results similar to those of CDT.

[18]The transition in between the first two phases and that in between the first and the last phase is of first order, while the transition from the second to third phase is of second order. For a more refined and more recent account of the phase structure, we refer to Refs. [320–322].

Finally, we would like to mention that the construction of more interesting observables (other than the ones typically studied like the 3- and 4-volume) is a formidable and central challenge in CDT, as for all quantum gravity approaches, and is actively researched, see e.g. Refs. [328, 329]. In passing, it should also be added that coupling matter to CDT is largely uncharted terrain since it is difficult to identify physically interesting matter-gravity observables. So far, only the coupling of a point particle to CDT was investigated, see Refs. [330, 331].

---

[19]Calculations of the spectral dimension for CDT in (3+1)-dimensions show a dynamical reduction of the dimensionality from $d_\mathrm{S} = 4$ on large scales to $d_\mathrm{S} = 2$ on short scales. This result bears strong similarity to what one obtains from within the asymptotic safety programme for quantum gravity [225], see Section 4.1. Notice that the spectral dimension follows from studying a diffusion process on the ensemble of geometries.

## 4.2.2 Matrix and tensor models

Another non-perturbative approach to quantum gravity is presented by tensor models which generalise vector and matrix models to higher dimensions. Tensor models generate random discrete geometries in terms of Feynman diagrams in the perturbative expansion of the corresponding path integral which is intended to provide a discretised version of the path integral of GR. Typically, the problem of the continuum limit is approached by tuning the correlation functions to critical values of the coupling constants. In the following, we will give a brief presentation of the basic aspects of matrix and (traditional as well as modern) tensor models. To this aim, we closely follow Refs. [44, 276, 332–336].[20]

### 4.2.2.1 Matrix models

Matrix models are statistical models for $2d$ random geometries. The fundamental objects of the matrix model approach are given by $N \times N$ matrices $M$ where the entry $M_{ij}$ corresponds to a 1-dimensional object graphically represented by a line with the end points $i$ and $j$. As an example, one can assume $M$ to be Hermitian. As a construction principle for the action $S(M)$ of $M$ one requires an analogue of the locality principle of ordinary field theories to hold. This corresponds to the invariance of $S(M)$ under U$(N)$-transformations of $M$. Consequently, $S(M)$ can only be a sum of products of traces and the simplest non-trivial action is then given by

$$S(M) = \frac{1}{2}\operatorname{tr} M^2 - \frac{\lambda}{\sqrt{N}}\operatorname{tr} M^3 = \frac{1}{2}M^i{}_j M^j{}_i - \frac{g}{\sqrt{N}}M^i{}_j M^j{}_k M^k{}_i. \qquad (4.20)$$

The propagator, retrieved from the kinetic term, is represented by a two-stranded line, a ribbon, where each strand is labelled by one index of the matrix $M$, while the interaction term gives a 3-valent vertex and its combinatorial pattern implements a rerouting of strands. The partition function is then defined by

$$Z_{\text{MM}} = \int \mathrm{d}M e^{-S(M)} \qquad (4.21)$$

---

[20] We will omit the topic of vector models in the following subsection. Suffice it to mention that vector models are rank-1 tensor models which can be solved in the large $N$-limit, where $N$ denotes the dimension of the vector. The geometric interpretation of critical points is that of continuous chains, i.e. branched polymers [337–344].

and generates in perturbative expansion ribbon Feynman diagrams $\Gamma$. With this, the generating function of connected correlators, i.e. the free energy $\ln Z$, is indexed by closed and connected ribbon graphs $\Gamma_c$. Importantly, when identifying the propagator with a transverse line and a 3-vertex with a triangle, where the propagator dictates how to glue triangles along common edges, one sees that ribbon graphs can be given a dual simplicial representation in terms of triangulated surfaces of arbitrary topology. This is depicted by Figs. 4.3 and 4.4.[21]

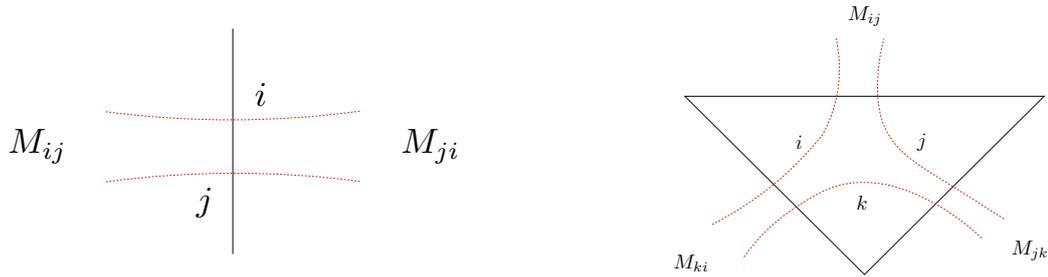

FIGURE 4.3: Graphic representation of the propagator (left) and the cubic interaction (right).

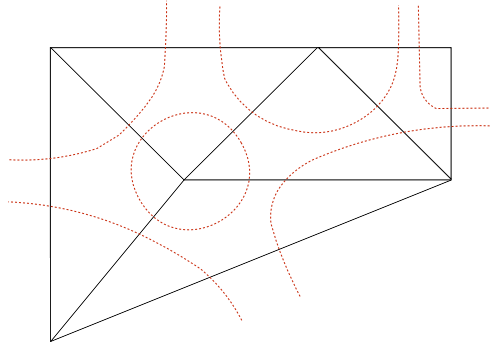

FIGURE 4.4: Snapshot of an open ribbon graph including its dual triangulation. The ribbon graph contains one face.

More concretely, we may write the free energy as

$$\ln Z_{\mathrm{MM}} = \sum_{\Gamma_c} \frac{\lambda^{n(\Gamma_c)}}{\mathrm{sym}(\Gamma_c)} \mathcal{A}_{\Gamma_c}, \tag{4.22}$$

---

[21]To broaden slightly the perspective about the basic ingredients of matrix models and their graphical interpretation, notice that for Hermitian matrices, only oriented triangulations can be generated (and it would be more appropriate to attach arrows in the propagator and vertices in Fig. 4.3), while for real symmetric matrices and $O(N)$-invariant action, orientable and non-orientable triangulations can be obtained. In addition, if a quartic interaction was used instead of a cubic one, the perturbative expansion would generate quadrangulations and the action would possess an additional $\mathbb{Z}_2$-symmetry [332, 333].

where $n(\Gamma_c)$ denotes the number of 3-valent vertices of $\Gamma_c$ and sym is a symmetry factor. The amplitude $\mathcal{A}_{\Gamma_c}$ associated to $\Gamma_c$ is

$$\mathcal{A}_{\Gamma_c} = N^{f(\Gamma_c) - \frac{1}{2}n(\Gamma_c)} \tag{4.23}$$

with $f(\Gamma_c)$ being the number of faces of $\Gamma_c$ and a face is a loop formed by a closed strand. In the following, consider the triangulation $\Delta$ dual to the ribbon graph $\Gamma_c$ with $v$ denoting the number of vertices, $e$ the number of edges and $t$ the number of triangles therein. With this, one may write

$$f(\Gamma_c) - \frac{1}{2}n(\Gamma_c) = v - \frac{1}{2}t \stackrel{3t=2e}{=} v - e + t \stackrel{!}{=} \chi = 2 - 2g \tag{4.24}$$

where $\chi$ is the Euler characteristic of the triangulation $\Delta$ of the closed and orientable surface $\Sigma$ of genus $g$ or equivalently of $\Gamma_c$.[22] Plugging this into Eq. (4.23), one finds

$$\ln Z_{\mathrm{MM}} = \sum_{\Gamma_c} \frac{\lambda^{n(\Gamma_c)}}{\mathrm{sym}(\Gamma_c)} N^{\chi(\Gamma_c)} = \sum_{g \in \mathbb{N}_0} N^\chi \sum_{\Gamma_c, \text{ fixed } g} \frac{\lambda^{n(\Gamma_c)}}{\mathrm{sym}(\Gamma_c)} \equiv \sum_{g \in \mathbb{N}_0} N^\chi F_g(\lambda) \tag{4.25}$$

which shows that the contributions from different $g$ can be separated from each other [345]. Each of these individual contributions encompasses infinitely many orders in the coupling constant $\lambda$ and consequently encodes non-perturbative effects. This is to be highlighted since increasing orders in $\lambda$ correspond to finer triangulations and are thus related to the continuum limit of matrix models.

This becomes transparent, when inspecting the large $N$-limit of this expansion. It is clear that the free energy is then dominated by the contribution of vanishing genus, i.e. $F_0(\lambda)$, hence spherical topologies [276, 346]. It can then be shown that the power series expansion of $F_0(\lambda)$ behaves like

$$F_0(\lambda) \sim \sum_{n(\Gamma_c)} n^{\gamma-3} \left( \frac{\lambda}{\lambda_{\mathrm{crit}}} \right)^n \sim |\lambda - \lambda_{\mathrm{crit}}|^{2-\gamma}, \tag{4.26}$$

with critical exponent $\gamma = -\frac{1}{2}$ where the last approximation holds for a large number of triangles. This expression diverges at the critical value $\lambda = \lambda_{\mathrm{crit}}$ signalling a phase transition [282, 347]. To interpret this divergence as a continuum limit, one computes the expectation value of the area of the surface $\Sigma$ with respect to the ensemble of matrices for large number of triangles, i.e.,

$$\langle A \rangle = a \langle n(\Gamma_c) \rangle = a \partial_\lambda F_0(\lambda) \sim \frac{a}{|\lambda - \lambda_{\text{crit}}|}, \tag{4.27}$$

wherein the triangles in $\Delta$ are assumed to be equilateral and of unit area $a$. Given the validity of this expression for $n$ approaching infinity, if one simultaneously sends $a$ to zero and $\lambda$ to $\lambda_{\text{crit}}$ while keeping $\langle A \rangle$ fixed, the associated triangulation becomes infinitely refined, i.e. continuous. The geometric interpretation of the critical point is then that of the so-called Brownian sphere [348–351].

---

[22]For non-orientable $\Sigma$, $\chi = 2 - g$ holds.

It is possible to generalise this line of reasoning to retain the contributions of non-planar surfaces. To this aim, it is possible to show that for generic $g$

$$F_g(\lambda) \sim \sum_{n(\Gamma_c)} n^{(\gamma-2)\frac{\chi}{2}-1} \left( \frac{\lambda}{\lambda_{\text{crit}}} \right)^n \sim |\lambda - \lambda_{\text{crit}}|^{(2-\gamma)\frac{\chi}{2}} \tag{4.28}$$

holds in the limit of many triangles. One sees that the successive coefficient functions $F_g(\lambda)$ all diverge at the same critical value of the coupling constant $\lambda = \lambda_{\text{crit}}$. The enhanced behaviour of the contributions of higher genus surfaces can counterbalance the large $N$ high genus suppression in Eq. (4.25) if the limits $N$ to infinity and $\lambda$ to $\lambda_{\text{crit}}$ are taken together. In this so-called double scaling limit [352–354] one keeps

$$k^{-1} \equiv N|\lambda - \lambda_{\text{crit}}|^{\frac{(2-\gamma)}{2}} \tag{4.29}$$

fixed, so that in

$$\ln Z_{\text{MM}} \sim \sum_{g \in \mathbb{N}_0} k^{-\chi} c_g \tag{4.30}$$

all genus surfaces contribute coherently, weighted by some constants $c_g$. If the coupling constant is tuned to criticality, infinite ribbon graphs with fixed $g$ will dominate this expansion. Here, the matrix model reaches a phase transition where it describes infinitely refined topological surfaces.

Finally, we are ready to relate this tool to generate $2d$ random geometries to gravity. For this, we use that the Euclideanised Einstein-Hilbert action in $2d$ can be written as

$$S_{\text{EH,E}} = \frac{1}{2\kappa} \int_\Sigma \mathrm{d}^2 x \sqrt{g} \left( 2\Lambda - R \right) = \frac{\Lambda}{\kappa} A(\Sigma) - \frac{2\pi}{\kappa} \chi(\Sigma) \tag{4.31}$$

and employ a minimal choice of discretizing the surface $\Sigma$ in terms of equilateral triangles of area $a$, yielding the Regge equivalent of the action

$$S_{\text{Regge,E}} = \frac{\Lambda}{\kappa} at - \frac{2\pi}{\kappa} \chi(\Delta). \tag{4.32}$$

With this discretisation the simplicial gravity partition function may be written as

$$Z_{\text{SG},2d} = \sum_{\Delta} \frac{1}{C(\Delta)} e^{\frac{2\pi}{\kappa} \chi(\Delta) - \frac{\Lambda}{\kappa} at}, \tag{4.33}$$

where $C(\Delta)$ denotes the order of the automorphism group of $\Delta$ and $\hbar$ is set to unity for convenience. With the identification of $e^{-\frac{\Lambda}{\kappa} a}$ with $\lambda$ and $e^{\frac{2\pi}{\kappa}}$ with $N$[23], one may write

$$\sum_{\Delta} \frac{1}{C(\Delta)} e^{\frac{2\pi}{\kappa} \chi(\Delta) - \frac{\Lambda}{\kappa} at} \overset{!}{=} \sum_{\Gamma_c} \frac{\lambda^{n(\Gamma_c)}}{\text{sym}(\Gamma_c)} N^{\chi(\Gamma_c)} = \ln Z_{\text{MM}}. \tag{4.34}$$

In the continuum limit, this corresponds to a partition function for the Euclideanised Einstein-Hilbert gravity in $2d$, namely

$$Z_{2d \text{ GR}} = \sum_{g \in \mathbb{N}_0} \int \mathcal{D}g \, e^{-S_{\text{EH,E}}}. \tag{4.35}$$

This reasoning relates the path integral for simplicial gravity with that obtained from the simple matrix model we dicussed. It makes transparent that matrix models provide a sum over all possible $2d$ simplicial complexes of all topologies, which is why one says that the matrix model defines a 3rd quantisation of GR in two dimensions. Intuitively, the sum over surfaces of all genera accounts for the fact that large quantum fluctuations may be able to change the genus of a considered surface [332, 333].[24]

To bring it down to a round figure, for matrix models one can establish a clear link to simplicial gravity as their Feynman amplitudes are in correspondence with $2d$ simplicial complexes. In fact, the perturbative sum over these generates simplicial manifolds only. Moreover and strikingly, this sum can be reorganised as a topological expansion indexed by the genus $g$. The large $N$-limit then demonstrated that it is dominated by the simplest topology. If $N$ is sent to infinity while the coupling constant $\lambda$ is tuned to criticality, the contribution of all topologies becomes visible. In this continuum limit, the results are found to be in agreement with those obtained via the quantisation of Liouville gravity. Moreover, the matrix model approach can be generalised to the case where gravity is coupled to conformal matter, leading to multicritical limits, and the results are again in agreement with those of Liouville gravity, as reviewed in Refs. [332, 333].

---

[23]This associates the weak coupling regime of Euclidean gravity in $2d$ to the large $N$ limit of this matrix model.

[24]It should be noted that there exists a matrix model which is able to generate causal/Lorentzian dynamical triangulations in $(1 + 1)$-dimensions. The causality condition of CDT, as reviewed in Section 4.2.1.2, can be imposed onto a matrix model so that it incorporates the space-like and time-like labelling of edges. For details, we refer to Ref. [355].

In passing, we would like to remark that matrix models can be promoted to field theories which are closely related to non-commutative (quantum) field theories. An example is the famous Grosse-Wulkenhaar model which is a $\varphi^4$-theory on non-commutative $\mathbb{R}^4$ [356, 357]. Its relation to matrix models is due to the fact that there exists a matrix representation of this space for which the Moyal-product becomes a product of matrices. It has been shown that this model is asymptotically safe [358] and that it can be solved in the planar sector [359–362]. This is interesting because it demonstrates that supersymmetry (as in the case of $N = 4$ supersymmetric Yang-Mills theory) is not needed to achieve integrability for a quantum field theory in $4d$. The example of the Grosse-Wulkenhaar model also is a major source of inspiration for the development of the tensor track programme for quantum gravity [39–43]. This programme is based on the modern theory of random tensors to which we will devote our attention in the following subsection.

### 4.2.2.2 Tensor models

**Traditional tensor models**

The successes of the matrix model approach quickly prompted the abstraction of the formalism to tensors of rank $d > 2$ with the goal to describe higher dimensional random geometries and relate them to simplicial gravity [363–368]. In analogy with matrix models, in $d$-dimensions the general idea is to graphically represent the entries of a $N^{\times d}$ tensor $T$ as $(d-1)$-simplices. In the minimalistic ansatz for the tensor action, the combinatorics of the interaction term is then supposed to dictate the glueing of $(d+1)$ $(d-1)$-simplices to form a $d$-simplex and the kinetic term ensures that two $d$-simplices are stuck together at common $(d-1)$-subsimplices. Furthermore, the tensors are assumed to be symmetrised. In the following, we shall illustrate that already for the case of $d = 3$ the naive implementation of this idea leads to a set of problems. To this aim, define the action for the tensor

$T_{ijk}$ with $i, j, k = 1 \ldots N$ by

$$S(T) = \frac{1}{2} \operatorname{tr} T^2 - \lambda \operatorname{tr} T^4 = \frac{1}{2} \sum_{i,j,k} T_{ijk} T_{kji} - \lambda \sum_{\substack{i,j,k \\ l,m,n}} T_{ijk} T_{kml} T_{mjn} T_{lni}. \qquad (4.36)$$

The graphical representation of the kinetic and interaction term is illustrated in Fig. 4.5.

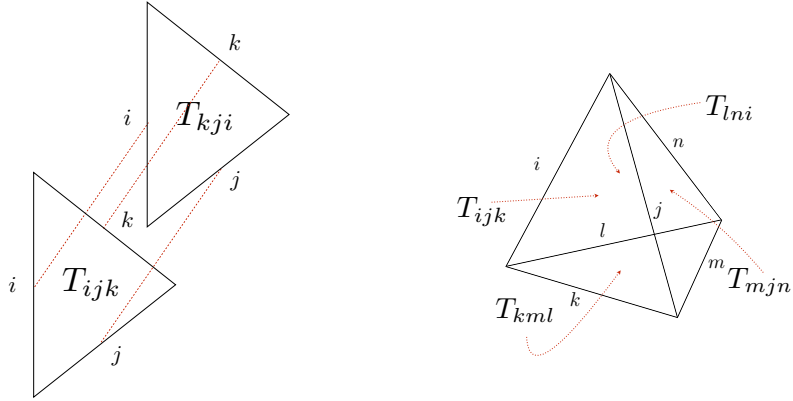

FIGURE 4.5: Graphic representation of the propagator (left) and the quartic interaction (right).

With this, we can formally write the partition function in perturbative expansion

$$Z = \int \mathcal{D}T e^{-S(T)} = \sum_{\Gamma} \frac{\lambda^{n(\Gamma)}}{\operatorname{sym}(\Gamma)} \mathcal{A}_{\Gamma}, \qquad (4.37)$$

where $n(\Gamma)$ is the number of vertices in the Feynman diagram $\Gamma$ and $\mathcal{A}_{\Gamma}$ is the associated amplitude.

Unfortunately, this construction generates Feynman diagrams of which not all correspond to discretisations of topological manifolds. In fact, the Feynman graphs can be associated to general topological spaces, so they correspond to manifolds, pseudomanifolds[25] and even more singular topologies [370–373]. The latter are highly pathological configurations and proliferate in the perturbative expansion. This is in stark contrast to the case of matrix models and is due to the fact that for $d > 2$ no a priori restriction on the glueing of $d$-simplices (the tetrahedra in Fig. 4.5) at common $(d-1)$-subsimplices is imposed. A restriction to an orientation preserving glueing of the $(d-1)$-subsimplices would not be strong enough to make a difference, since their $(d-2)$-subsimplices (and consecutive lower-dimensional subsimplices) would still be arbitrarily identified leading to pathologies. In view of these problems, it is impossible to rewrite the perturbative sum as a $1/N$-expansion as for matrix models and there is no way to reorganise the sum in terms of topological invariants. On the positive side, for those Feynman graphs dual to triangulations of topological manifolds, the relation to simplicial gravity path integral amplitudes is still valid in the same way as for matrix models.

---

[25]A pseudomanifold is topological space $X$ endowed with a triangulation $\Delta$ which is non-branching and strongly connected. From a non-technical point of view, it is a combinatorial realisation of a manifold with singularities [369].

**Modern tensor models**

Importantly, it is possible to give a precise and subtle mathematical prescription to enforce by means of additional combinatorial structure that the Feynman graphs are truely dual to simplicial complexes and all pathologies are excluded. This prescription is provided by coloured tensor models for which a reorganisation of the perturbation series in terms of a $1/N$-expansion can be found [371–376].[26]

The failure of the traditional tensor models is rooted in the symmetrisation of the tensor indices. These issues can be lifted when the tensors are discerned by an additional combinatorial label, termed colour. Following Ref. [335], for the construction of coloured tensor models, one introduces $(d + 1)$ coloured rank-$d$ tensors $T^c_{a_1...a_d}$ and their complex conjugates $\bar{T}^c_{a_1...a_d}$ with the set of colours $c \in \{1, \ldots, d+1\}$. The indices reach from 1 to $N$. The application of complex tensors ensures orientability. Each rank-$d$ tensor of colour $c$ can be graphically represented by $d$ strands of this colour or equivalently, in a collapsed version of this picture, as a coloured half-edge and yet in another way in terms of a coloured $(d-1)$-simplex.

The action of the model is encoded by

$$S(T, \bar{T}) = \sum_{c=1}^{d+1} \sum_{a_1,...,a_d} T^c_{a_1...a_d} \bar{T}^c_{a_1...a_d} - \frac{\lambda}{N^{d(d-1)/4}} \sum_{\{a_{cc'}, \ c<c'\}} \prod_{c=1}^{d+1} T^c_{\mathbf{a}_c} \quad + \quad \text{c.c.,} \quad (4.38)$$

where $\mathbf{a}_c = (a_{cc-1}, \ldots, a_{c1}, a_{cd+1}, \ldots, a_{cc+1})$ and $a_{cc'}$ and $a_{c'c}$ are identified [374–376]. Indices in $\mathbf{a}_c$ run modulo $d+1$ in the set of colours. The combinatorics of the two types of interactions coupled by $\lambda$ and $\bar{\lambda}$ respectively, encode the pairwise glueing of $(d+1)$

---

[26]From a historical viewpoint, the failure of the traditional tensor models spurred the construction of the EDT/CDT approach, as reviewed in Section 4.2.1.2, as well as the development of the GFT program, initiated by the famous Boulatov and Ooguri models [377, 378]. GFTs are tensor models dressed with additional data and were first proposed to overcome the aforementioned problems. The realisation that such pathological configurations [370] could be overcome was first made within the context of GFT, led to the development of coloured GFTs [371–373] and then to coloured tensor models [374–376].

$(d-1)$-simplices along $(d-2)$-subsimplices to form a $d$-simplex. Alternatively, this can be graphically represented by means of white nodes (for tensors $T^c$) and black nodes (for conjugated tensors $\bar{T}^c$) at which $(d+1)$ coloured half-edges meet forming a coloured graph[27]. This pattern of contractions in different graphical representations is illustrated for the case of $d = 3$ in Figs. 4.6 and 4.7. Finally, the kinetic term dictates the glueing of two $d$-simplices along common two $(d-1)$-simplices (more precisely, the contraction of a tensor of colour $c$ with a conjugate tensor of colour $c$ only), where the corresponding subsimplices are identified respecting their orientations.

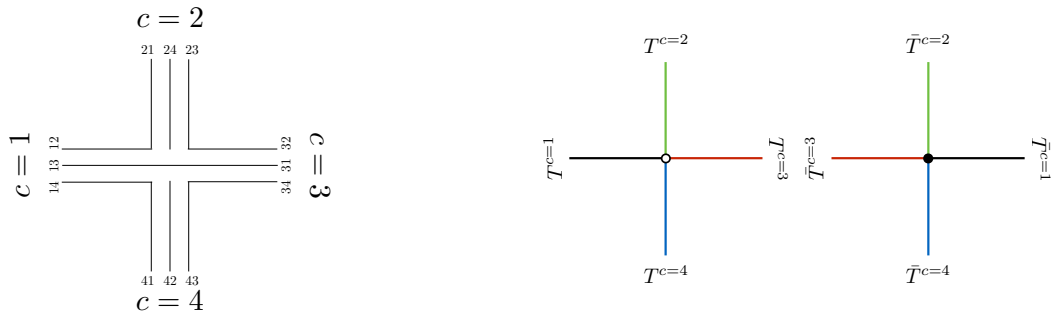

Figure 4.6: Graphic representation of the coloured 4-vertex interaction in $d = 3$. The stranded representation (left) and the coloured graph representation (right) for the tensors $T^c$ and $\bar{T}^c$. A propagator would connect the two coloured graphs (right) through colour $c = 3$ such that their orientation is preserved.

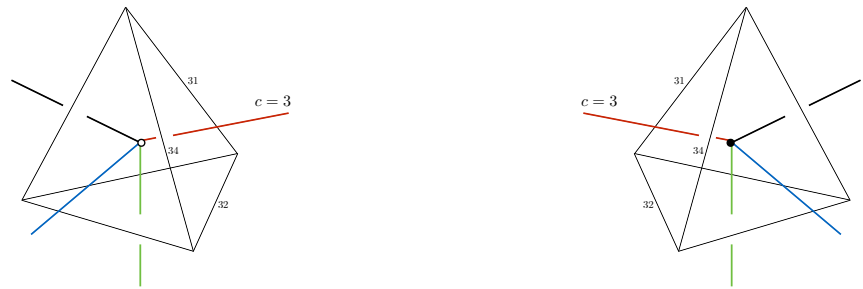

Figure 4.7: Graphic representation of two coloured 4-vertices in terms of two 3-simplices with coloured graphs superimposed.

---

[27] For $n > 2$, a $n$-coloured graph is a bipartite regular $n$-valent graph. Its edges are labelled by colours $c$ from the set $\{1, \ldots, n\}$ and at each vertex $n$ distinctly colourised edges meet [335].

The partition function in perturbative expansion leads to

$$Z = \sum_{\Gamma} \frac{(\lambda\bar{\lambda})^{\frac{n(\Gamma)}{2}}}{\text{sym}(\Gamma)} \mathcal{A}_{\Gamma}, \qquad (4.39)$$

with $\Gamma$ denoting $(d+1)$-coloured graphs and $n(\Gamma)$ corresponds to the number of its nodes. With the number of faces of $\Gamma$ being $f(\Gamma)$, the amplitude $\mathcal{A}_{\Gamma}$ is given by

$$\mathcal{A}_{\Gamma} = N^{f(\Gamma) - n(\Gamma)\frac{d(d-1)}{4}}. \qquad (4.40)$$

To emphasise once again, the labelling in terms of coloured graphs ensures the reduction of the combinatorial complexity of the Feynman diagrams so that no pathological configurations are generated [335].

When introducing the notion of Gurau degree

$$\omega(\Gamma) = \sum_J g_J \qquad (4.41)$$

being the sum of the genera of particular subgraphs of $\Gamma$ called jackets[28], the amplitudes can be rewritten as

$$\mathcal{A}_{\Gamma} = N^{d - \frac{2}{(d-1)!}\omega(\Gamma)}, \qquad (4.42)$$

which allows to reorganise the perturbative series as a $1/N$-expansion wherein the order in $N$ corresponds to packages indexed by the Gurau degree [374–376, 381, 382]. This can be seen as the higher-dimensional analogue of the genus-labelled topological expansion for matrix models. Importantly, however, the Gurau degree is not a topological invariant but encapsulates topological as well as triangulation dependent information.

When the large $N$ limit is performed, this expansion is dominated by graphs of Gurau degree 0, termed as melonic graphs. Such graphs correspond to triangulations of the $d$-dimensional sphere. The most singular part of the free energy for this sector is given by

$$F_{\omega=0}(g) \sim |g_c - g|^{2-\gamma} \qquad (4.43)$$

with $g = |\lambda|^2$ and critical exponent $\gamma = \frac{1}{2}$ [383]. At criticality, a phase transition is observed [381–383] the geometric interpretation of which is that of an infinitely refined topological space corresponding to a branched polymer, as demonstrated in Ref. [384]. A natural way to find emergent geometries which are richer then is to study the double scaling limit of such tensor models [385–388].[29]

---

[28]Within $\Gamma$ one can find specific ribbon subdiagrams called jackets. In the simplicial representation, these are dual to $2d$ discretised orientable surfaces of certain genus which are embedded in the larger simplicial complex corresponding to $\Gamma$. The number of faces of $\Gamma$ is strictly related to $\omega(\Gamma)$. For details, we refer to Refs. [374–376, 379, 380].

Further developments were sparked when it was realised that starting off with a coloured simplicial model, one can integrate out all coloured tensors of colour $c = 2, \ldots, d+1$ in the partition function, yielding an effective action for the remaining tensors of colour $c = 1$ [381, 382]. Hence, this effective action is the action of a single uncoloured tensor. It is a sum of effective interaction terms the combinatorics of which are encoded in terms of coloured graphs (labelled by those colours which were integrated out). One refers to these $d$-coloured graphs as bubbles. In the remaining tensors $T^{c=1}$ and $\bar{T}^{c=1}$, the colours $c = 2, \ldots, d+1$ are understood to label the position of an index. In the bubbles, the pairing of indices dictates that an index in the $c$th position of $T^{c=1}$ can only be connected to an index of $\bar{T}^{c=1}$ in exactly the same position. For example, a bubble of order 4 corresponds to the glueing of four tetrahedra with four external boundary triangles labelled by the colour $c = 1$ (corresponding to the effective tensors) which is encoded by

$$\sum_{a_i, b_i} T_{a_2 a_3 a_4} \bar{T}_{a_2 a_3 b_4} T_{b_2 b_3 b_4} \bar{T}_{b_2 b_3 a_4} \tag{4.44}$$

and depicted on the right of Fig. 4.8.

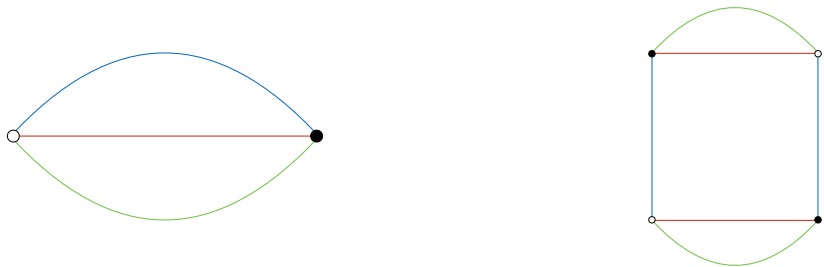

Figure 4.8: Bubble interactions of order 2 (left) and 4 (right).

Remarkably, it can be shown that contractions of this type, i.e. the effective action, are invariant under the external tensor product $\mathrm{U}(N)^{\otimes d}$. This group of transformations

---

acts on the effective tensors (we suppress the $c = 1$ label) like

$$T_{a_1 \ldots a_d} \to T'_{b_1 \ldots b_d} = \sum_{a_1, \ldots, a_d} U^{(1)}_{b_1 a_1} \cdots U^{(d)}_{b_d a_d} T_{a_1 \ldots a_d} \tag{4.45}$$

and analogously for $\bar{T}$. Stepping away from the above construction, more generally, we can then construct actions for uncoloured tensors where the actions are built from the tensor invariants only [381, 382]. Important for this construction is that these uncoloured tensors are assumed to be generic, i.e., they are not symmetrised (as the traditional tensor models) or antisymmetrised. In analogy to matrix models, this $\mathrm{U}(N)^{\otimes d}$-invariance is an analogue of the locality principle for ordinary field theories. In this picture, the above-introduced colours can retroactively be regarded as a consequence of a symmetry principle. For such models, the ideas leading to the $1/N$-expansion can be carried over and in the large $N$ limit, again dominated by melonic graphs, the critical behaviour can be examined [381, 382]. They also possess a double scaling limit [386–388]. We refer to Ref. [44] for further results and a broad overview of this fastly expanding research topic.[30][31][32]

---

[30]Most explored models which support a $1/N$-expansion, generate the above-mentioned branched-polymer geometries. Recent studies have shown that non-melonic interactions can be enhanced leading to a two-phase structure of a branched-polymer and a $2d$ quantum gravity (planar) phase with an intermediate regime of so-called proliferating baby-universes [389–392]. Finding specific models which allow to escape the emergence of lower dimensional geometries at criticality is subject to ongoing research. Another attempt at going beyond such undesired geometries by adding additional structure to the theory, is proposed by GFT where metric information is associated to the Feynman graphs. We turn to this topic in Section 4.2.4.

[31]The notion of tensor invariance can also be applied tensor field theories, e.g. tensorial GFTs. The tensor invariance of tensor field theories is softly broken by their non-trivial propagator which leads to a renormalisation group flow. For details regarding this class of theories, we refer to the reviews Refs. [39–43, 393].

[32]We would like to remark that the domain of applicability of the new tensor models goes well beyond quantum gravity and extends to statistical physics problems [394–398]. Most noteworthily, they can be linked to the Sachdev-Ye-Kitaev model of non-Fermi liquids (and quantum black holes in AdS$_2$), as reviewed in Ref. [399].

### 4.2.3  Spin foam models

As we have recalled above, the Feynman path integral for gravity seeks a description of transition amplitudes between $3d$ boundary configurations $h, h'$ of the gravitational field, formally given by

$$\langle h|h'\rangle = \int \mathcal{D}g \ e^{i\left(S_{\mathrm{GR}}[g]+S_{\mathrm{boundary}}[h,h']\right)}. \tag{4.46}$$

The spin foam approach tries to provide a conretisation of this idea in terms of a path integral formulation for LQG. The driving idea behind this formulation is to construct a framework which is manifestly background independent and non-perturbative in the sense that there is no fixed background geometry on which the path integral quantisation is performed. In LQG, the boundary configurations can be rigorously described in terms of spin network states, encoding the spatial discrete quantum geometry solely by means of combinatorial and algebraic data. Spin foam amplitudes then define the transition in between such states giving rise to a discrete quantum spacetime history turning the sum-over-histories view of the Feynman path integral into a sum-over-quantum spacetime geometries. In this way, the spin foam approach suggests a possible resolution of the previously mentioned functional measure problem of the path integral formulated in the continuum. In principle, it allows to compute expectation values of observables as expectation values in a rigorously defined physical Hilbert space. Moreover, in the absence of a clear understanding of the dynamics in the canonical theory, it is primarily intended to give a clear definition of the quantum dynamics of LQG. The following content introduces the basic notions of spin foam theory closely following Refs. [35, 400].

**Spin foams and spin foam models**

Consider the local gauge group of gravity $G$. A *spin foam* $\sigma$ is formally defined by

1. a 2-complex $\mathcal{C}$, consisting of finite sets of faces, edges and vertices (and their relations to another),

2. a set of representation labels (spins) $\{j_f\}$ of $G$ which are associated with the faces $f \in \mathcal{C}$ and

3. a set of intertwiners $\{i_e\}$ of $G$ associated with the edges $e \in \mathcal{C}$.

A *spin foam model* attributes amplitudes to the spin foams, i.e., $\mathcal{A} : \sigma \to \mathbb{C}$, (in principle) compatible with the composition rule

$$\mathcal{A}[\sigma \circ \sigma'] = \mathcal{A}[\sigma]\mathcal{A}[\sigma'].$$ (4.47)

The amplitude assings a probability for a process that transforms a configuration of grains of space into another one. With this, the partition function is defined as

$$Z = \sum_\sigma w(\sigma)\mathcal{A}[\sigma] = \sum_\sigma w(\sigma) \sum_{j_f, i_e} \prod_v \mathcal{A}_v(j_f, i_e),$$ (4.48)

where $w(\sigma)$ is a weight factor depending on $\sigma$ and the vertex amplitude $\mathcal{A}_v$ is an amplitude attributed to each vertex depending on their adjacent labels $j_f$ and $i_e$. The previous expression can also be written as

$$Z = \sum_\sigma w(\sigma) \sum_{j_f, i_e} \prod_f \mathcal{A}_f(j_f) \prod_e \mathcal{A}_e(j_f, i_e) \prod_v \mathcal{A}_v(j_f, i_e)$$ (4.49)

with the face and edge amplitudes $\mathcal{A}_e$ and $\mathcal{A}_f$. For most models $\mathcal{A}_f(j_f) = \dim(j_f)$ is chosen. Different choices of sets of 2-complexes, group $G$/representation labels $j_f, i_e$ and amplitudes define a specific spin foam model. In particular, different amplitudes roughly correspond to different forms of the Hamiltonian operator in canonical LQG.

If the spin foam is bounded by two spin networks $s$ and $s'$, $\sigma : s \to s'$ then describes a transition from the boundary spin network $s = (\Gamma, \{j_l\}, \{i_n\})$ to $s' = (\Gamma', \{j_{l'}\}, \{i_{n'}\})$. Notice, that intertwiners and spins of the exterior faces and edges are of course equal to the corresponding values given by the boundary spin network states $s$ and $s'$, i.e. at the boundary the labelling reduces to SU(2) representation labels $\{j_{l^{(')}}\}$ and $\{i_{n^{(')}}\}$. This vocabulary is illustrated in Fig. 4.9. Tables 4.1 and 4.2 juxtapose the terminology of the triangulation with that of the dual 2-complex in 3$d$.

Table 4.1: Bulk terminology.

| bulk triangulation $\Delta$ | 2-complex $\mathcal{C}$ |
| --- | --- |
| tetrahedron | vertex |
| triangle | edge |
| segment | face |
| point | |

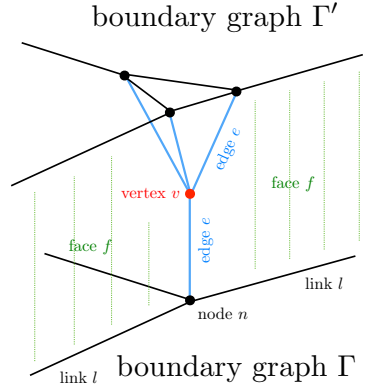

FIGURE 4.9: 2-complex and triangulation terminology in $3d$. The red vertex is dual to a tetrahedron. The figure is understood to be located in a fictitious embedding spacetime and depicts the transition of a single quantum of space to three.

TABLE 4.2: Boundary terminology.

| boundary triangulation $\partial\Delta$ | boundary graph $\Gamma$ |
|---|---|
| triangle | node (boundary vertex) |
| segment | link (boundary edge) |

In this way, one can define transition amplitudes between spin network states as

$$\langle s|s'\rangle_{phys} = \sum_{\sigma:s\to s'} w(\sigma)\mathcal{A}[\sigma]$$

$$= \sum_{\sigma:s\to s'} w(\sigma) \sum_{j_f,i_e} \prod_f \mathcal{A}_f(j_f) \prod_e \mathcal{A}_e(j_f,i_e) \prod_v \mathcal{A}_v(j_f,i_e), \qquad (4.50)$$

where $\langle\cdot|\cdot\rangle_{phys}$ denotes the physical inner product. Notice that only at the vertices the evolution is non-trivial. There, the nodes of the initial spin network branch. Clearly, transition amplitudes obey the superposition principle, since amplitudes are sums of elementary amplitudes and each of these corresponds to a single history. They are (supposed to be) local, in the sense that elementary amplitudes are products of amplitudes associated to local elementary processes and finally, they are required to suffice the property of the local gauge invariance of gravity.

**Ponzano-Regge spin foam model**

In the following, we will exemplify this with the case of $3d$ Euclidean quantum gravity. On the classical level, we choose to work with the first order formalism for GR where the gravitational field on the manifold $\mathcal{M}$ (assumed to be orientable and compact) is represented by a pair of fields $(B, A)$ where the triad $B$ is an $\mathfrak{su}(2)$-valued 1-form and $A$ is SU(2)-valued connection 1-form with curvature 2-form $F(A)$. Their dynamics are encoded by the action

$$S_{3d}[B, A] = \int_{\mathcal{M}} \text{tr} \left( B \wedge F(A) \right), \qquad (4.51)$$

where the trace denotes the Killing form. Variation with respect to both fields leads to the equations of motion

$$F(A) = 0 \quad \text{and} \quad \text{d}_A B = 0 \qquad (4.52)$$

where the former means that the connection $A$ is flat and the latter that it is the unique torsionless spin connection associated to the triad $B$. Physically, the flatness condition means that the vacuum theory is topological in the sense that there are no local degrees of freedom.[33] Both equations together are equivalent to Einstein's field equation in the vacuum.

Formally, the quantum dynamics of this model are encoded by the continuous path integral

$$Z(\mathcal{M}) = \int \mathcal{D}A \mathcal{D}B \; e^{iS_{3d}[B,A]} = \int \mathcal{D}A \; \delta(F(A)), \qquad (4.53)$$

which is ill-defined. However, when suitably discretised, the relation to the Ponzano-Regge model, as given in Section 4.2.1.1, can be unveiled.

To this aim, we introduce a cellular decomposition $\Delta$ of $\mathcal{M}$ and its dual 2-complex $\mathcal{C}$. In the following, we may assume $\Delta$ to be a $3d$ simplicial complex.[34] We also choose an orientation of edges and faces in $\mathcal{C}$. Using this structure, we discretise the continuum fields $B$, $A$ and $F$ using lattice gauge theory techniques.[35] For this let $h_e$ the holonomy of the connection $A$ along an edge $e$ in $\mathcal{C}$, i.e.

$$h_e(A) = \mathcal{P}e^{\int_e A} \in \text{SU}(2). \qquad (4.54)$$

The $\mathfrak{su}(2)$-valued line integral of $B$ along a line segment of $\Delta$ dual a face $f$ in $\mathcal{C}$ shall be

---

[33] In fact, this model is a representative of class of BF-models. For typographic reasons we choose the triad to be denoted by $B$, as opposed to the common notation in terms of $e$.

[34] One could have kept the cellular decomposition $\Delta$ arbitrary but by choosing a simplicial decomposition, we are on a same footing with the (heuristic) Ponzano-Regge model. The generalisation is nevertheless straightforward.

[35] Notice that we discretise the classical theory before quantisation. We do not lose any information along this way due to its topological character.

denoted by $l_f \in \mathbb{R}^3$, the smearing of $B$. In the Wilsonian spirit, the discretised action is then written as

$$S[l_f, h_e] = \sum_f \text{tr}\,(l_f U_f) \tag{4.55}$$

with $U_f$ denoting the product of holonomies around faces (i.e. the plaquette operator)

$$U_f = \prod_{e \in \partial f} h_e^{\epsilon_{fe}} \tag{4.56}$$

where $\epsilon_{ef} = \pm 1$ indicates the relative orientation between the given $e$ and $f$. The product of holonomies $U_f$ relates to the discrete $\mathfrak{su}(2)$-valued curvature $F_f$ around the face $f$ by $U_f = \mathbb{1} + F_f$. With this we may write Eq. (4.53) as

$$Z(\Delta) = \int \prod_f \mathrm{d}l_f \prod_e \mathrm{d}h_e\; \mathrm{e}^{i\,\text{tr}(l_f U_f)} \tag{4.57}$$

with Haar measure $\mathrm{d}h$ on SU(2) and Lebesgue measure $\mathrm{d}l$ on $\mathbb{R}^3$. Integration over $l_f$ yields

$$Z(\Delta) = \prod_e \int \mathrm{d}h_e \prod_f \delta(U_f) \tag{4.58}$$

which is a path integral for a system of flat discrete connections. Using the Peter-Weyl theorem, the delta distribution on SU(2)

$$\delta(g) = \sum_j d_j \chi^j(g) \tag{4.59}$$

with the character $\chi^j(g) = \text{tr}\,D^j(g)$ which allows us to perform the integral. Each holonomy occurs three times and the integral over this package yields a 3-valent intertwiner (i.e. a $3j$-symbol) per edge. The objects obtained in this way can be contracted at each shared vertex in $\mathcal{C}$ leading to a Wigner $6j$-symbol. Altogether, this yields

$$Z(\Delta) = \sum_{j_f} \prod_f (-1)^{2j_f} (2j_f + 1) \prod_v \{6j\}_v \tag{4.60}$$

which is nothing but the partition function of the Ponzano-Regge model introduced in Section 4.2.1.1. When comparing Eq. (4.60) with Eq. (4.48), we see that there is no sum over intertwiners since in the case of a simplicial decomposition we considered here, there is only one intertwiner available. We also observe that the vertex amplitude is equal to the

$6j$-symbol.[36][37]

**Sketching the procedure for gravity in $4d$**

The paradigmatic role of the previous model lies in the fact that a similar correspondence between algebraic expressions and GR also holds in four dimensions. There, according to the currently most favoured procedure which goes under the name of the Engle-Pereira-Rovelli-Livine (EPRL) model [401, 402], one would begin with the Plebanski-Holst (PH) action which rewrites gravity in terms of a BF-theory with $G = \mathrm{Spin}(4)$ or $\mathrm{SL}(2,\mathbb{C})$ subject to constraints. The action reads

$$S_{\mathrm{PH}}[B, A, \lambda] = \frac{1}{2\kappa} \int_{\mathcal{M}} \left[ \left( \star B^{IJ} + \frac{1}{\gamma} B^{IJ} \right) \wedge F_{IJ}(A) + \lambda_{IJKL} B^{IJ} \wedge B^{KL} \right], \qquad (4.61)$$

where $A$ is a $\mathfrak{sl}(2,\mathbb{C})$- or $\mathfrak{spin}(4)$-connection, the bivector $B$ is a Lie algebra-valued 2-form, $\gamma$ is the Barbero-Immirzi parameter and $\kappa = 8\pi G_{\mathrm{N}}$. The Lagrange multiplier $\lambda$ has the property

$$\lambda_{IJKL} = -\lambda_{JIKL} = -\lambda_{IJLK} = \lambda_{KLIJ} \qquad (4.62)$$

and thus $\epsilon^{IJKL}\lambda_{IJKL} = 0$. Variation with respect to $\lambda$ then gives

$$\epsilon^{\mu\nu\rho\sigma} B_{\mu\nu}^{IJ} B_{\rho\sigma}^{KL} = e\epsilon^{IJKL}. \qquad (4.63)$$

With

$$e = \frac{1}{4!} \epsilon_{OPQR} B_{\mu\nu}^{OP} B_{\rho\sigma}^{QR} \epsilon^{\mu\nu\rho\sigma} \qquad (4.64)$$

this is solved by

$$a) \ B = \pm \star (e \wedge e) \quad \text{and} \quad b) \ B = \pm (e \wedge e), \qquad (4.65)$$

---

[36]This expression is divergent due to so-called bubble divergences rooted in redundant $\delta(0)$-factors in Eq. (4.58). These are ultimately related to the topological gauge freedom of the theory. At the discrete level it can be partially gauge fixed to the effect that bubbles are eliminated. Another way to eliminate these divergences is to use the representation theory of the quantum group $\mathrm{SU}(2)_q$ with root of unity instead. Due to the fact that it has only finitely many representations, $Z$ has a finite value. This is the so-called Turaev-Viro model and the quantum deformation to $\mathrm{SU}(2)_q$ can be related to a cosmological constant term in the classical action. For details regarding these points, we refer to Ref. [35].

[37]Notice that the partition function depends on the global topology of $\mathcal{M}$ but is independent from the way it is triangulated. This is a direct consequence of the topological character of gravity in $3d$ which has no local degrees of freedom. Hence, a refinement of the triangulation has no impact on this number. Importantly, triangulation invariance does not hold in the context of gravity in $4d$ which has local degrees of freedom.

where $e$ denotes the tetrad field. If solutions of sector $a$) (the gravitational sector as opposed to the topological sector $b$)) are plugged into the Plebanski-Holst action, one arrives at the Holst action

$$S_{\mathrm{H}}[e, A] = \frac{1}{2\kappa} \int_{\mathcal{M}} \mathrm{tr} \left[ \left( \star (e \wedge e) + \frac{1}{\gamma} e \wedge e \right) \wedge F(A) \right],\qquad(4.66)$$

which lies at the heart of LQG, as we have seen in Section 2.2. The classical equations of motion derived from it agree with those gained from the Einstein-Hilbert action. When introducing a global time foliation, we can extract from this action the so-called simplicity constraints. These constraints are enforced onto the states and amplitudes of the quantum theory to be constructed which in this way is closely connected to canonical LQG. In particular, the boundary states coincide with the kinematical states of the latter.[38]

The previous action looks like the action of a BF-theory when identifying the 2-from in the round brackets (encoding the simplicity constraints) with a *B*-field. The starting point for the construction of the spin foam model is then given by the spin foam quantisation of BF-theory in $4d$. In the $4d$ Euclidean case, following the same steps as in $3d$, for BF-theory for the group $G = \mathrm{Spin}(4) \cong \mathrm{SU}(2) \times \mathrm{SU}(2)$ discretised over the triangulation $\Delta$ of a (compact and orientable) manifold $\mathcal{M}$ this yields

$$Z(\Delta, G) = Z(\Delta, \mathrm{SU}(2))^2 = \sum_{j_f^{\pm}, i_e^{\pm}} \prod_f (2j_f^+ + 1)(2j_f^- + 1) \prod_v \{15j^+\}_v \{15j^-\}_v.\qquad(4.67)$$

In this expression $j_f^{\pm}, i_e^{\pm}$ are half integers labelling left and right representations of $\mathrm{SU}(2)$ and the vertex amplitudes $\{15j^{\pm}\}_v$ arise from the pattern of contraction of intertwiners now reproducing the structure of a four-simplex. To render this into a theory of Euclidean quantum gravity, the discretised version of the simplicity constraints have to be imposed onto the amplitude at the quantum level which loosely speaking leads to a restriction of representation labels. In this way, the space of histories of the BF-theory path integral is constrained to that of gravity.

For the more involved Lorentzian case one proceeds similarly and obtains an amplitude which defines the covariant dynamics of LQG. We refer to Refs. [35, 403] for a detailed presentation of these matters and to Refs. [35, 400, 404] for a discussion of the semi-classical limit. Interesting phenomenological applications of this model to quantum cosmology and quantum black holes were studied in Refs. [405–412] and Refs. [413–416], respectively. With regard to the important point of matter coupling in spin foam models, one is mostly in uncharted waters, however, see Refs. [417–421].[39]

---

[38]A detailed discussion of the discretisation and quantisation ambiguities for various actions as a starting point is given by Ref. [115].

In perspective of the next subsection, the GFT formalism was developed as a covariant QFT formulation of the dynamics of LQG, too. In fact, it can be shown that geometric GFT models provide a generating function for spin foam amplitudes. Importantly, the GFT formalism proves advantageous to study scenarios involving the dynamics of spin network configurations with many links and nodes which represent macroscopic and approximately smooth geometries. In contrast, this is much more difficult to achieve in the language of spin foams and directly relates to the issue of the continuum limit therein.[40] Intuitively, it seems unlikely that a physically realistic theory of quantum gravity has a finite number of degrees of freedom and should be based on a single simplicial complex. If such a configuration described a macroscopic spacetime, it would be very discrete in fact. Although GFTs allow to study an infinite class of simplicial complexes by construction, we will see in the next chapter, that they provide the field theoretic approximation tools to study the physics of many LQG degrees of freedom while bypassing the treatment of highly complicated spin networks.

---

[39]A highly important question for LQG and spin foam models (as well as GFT) touches on the fate of Lorentz invariance (and possible violations thereof due to the quantum gravity granularity of spacetime) at low energies. This point remains to be systematically explored in this context. For a dicussion of this issue we refer to Ref. [35] and references therein.

[40]An interesting take on this problem is presented by the spin foam coarse graining and renormalisation proramme [422–428].

### 4.2.4 Group field theory

GFTs represent a particular class of quantum field theories characterised by their combinatorially non-local interaction terms which aim at generalizing matrix models for $2d$ quantum gravity to higher dimensions. The fields of GFT live on group manifolds $G$ or dually on their associated Lie algebras $\mathfrak{g}$. For quantum gravity intended models, $G$ is interpreted as the local gauge group of gravity.[41]

The perturbative expansion of the GFT path integral is indexed by Feynman diagrams which are dual to cellular complexes because of the particular non-local interactions. Depending on the details of the Feynman amplitudes, the sum over the cellular complexes can be interpreted as a possible discrete definition of the covariant path integral for $4d$ quantum gravity. The reason for this is that beyond the combinatorial details, GFT Feynman graphs can be dressed by group theoretic data of which the function is to encode geometric information corresponding exactly to the elementary variables of LQG [5, 31]. Using this, it can be shown that GFTs provide a formal and complete definition of spin foam models which give a path integral formulation for LQG [35, 403, 429, 430]. Using the dual formulation on the associated Lie algebra $\mathfrak{g}$, it is also possible to manifestly relate their partition functions to (non-commutative) simplicial quantum gravity path integrals [431, 432]. By construction, all data encoded in the fields and their dynamics are solely of combinatorial and algebraic nature thus rendering GFT into a manifestly background independent and generally covariant field theoretic framework [36–38, 433–436].

Recently, a lot of effort has been devoted to the understanding of the non-perturbative sector of GFTs. In particular, studies in GFT condensate cosmology have attracted a fair bit of attention to which a large part of this thesis refers. In the following, we will give a brief account of the basics of (uncoloured) GFTs which are needed in the subsequent chapters of this manuscript. To this aim, we closely follow Refs. [6, 36–38, 437, 438].[42]

---

[41]Typically, one chooses $G = \mathrm{Spin}(4)$, $\mathrm{SL}(2, \mathbb{C})$ or $G = \mathrm{SU}(2)$. The last is the gauge group of Ashtekar-Barbero gravity lying at the heart of canonical LQG.

[42]In fact, GFTs were developed to overcome the issues of traditional tensor models in the first place. To capture the greater complexity of higher-dimensional spacetimes and geometries, the central conviction of the GFT programme is to render the structure and data of the Feynman diagrams richer by working in the context of field theory and by adding additional degrees of freedom in terms of (Lorentz) group theoretic data, as suggested by LQG. While the colourisation of tensor models has led to major progress (suppression of singular configurations in the Feynman expansion and its reorganisation as a $1/N$-expansion, as recalled in Section 4.2.2.2), the metric information encoded by GFTs could be a way to generate richer emergent geometries, as compared to coloured tensor models [44]. Focussing only on this latter point, in the context of the subsequent chapters we work with the traditional formulation of GFTs which lie at the heart of the GFT condensate cosmology approach while keeping in mind that its generalisation to coloured models should be studied in the future.

### 4.2.4.1 Classical theory

The classical field theory is specified by choosing a type of field and an action dictating the dynamics. Most generally, we consider a complex-valued scalar field $\varphi$ living on $d$ copies of the Lie group $G$, i.e.,

$$\varphi(g_{\mathrm{I}}) : G^d \to \mathbb{C} \tag{4.68}$$

with $\mathrm{I} = 1, \ldots, d$. The group elements $g_{\mathrm{I}}$ are parallel transports $\mathcal{P} e^{i \int_{e_{\mathrm{I}}} A}$ associated to $d$ links $e_{\mathrm{I}}$ and $A$ denotes a gravitational connection 1-form. The field is typically assumed to be an $L^2$-function on $G$ with respect to the Haar measure $\mathrm{d}g$.

Importantly, one demands the invariance under the right diagonal action of $G$ on $G^d$, i.e.,

$$\varphi(g_1 h, \ldots, g_d h) = \varphi(g_1 \ldots, g_d), \quad \forall h \in G \tag{4.69}$$

which is a way to guarantee that the parallel transports, emanating from a vertex and terminating at the end point of their respective links $e_{\mathrm{I}}$, only encode gauge invariant data. Its relevance for the geometric interpretation of the fields is clarified below.

For compact $G$ the action is given by

$$S[\varphi, \bar{\varphi}] = \int_G (\mathrm{d}g)^d \int_G (\mathrm{d}g')^d \bar{\varphi}(g_{\mathrm{I}}) \mathcal{K}(g_{\mathrm{I}}, g'_{\mathrm{I}}) \varphi(g'_{\mathrm{I}}) + V. \tag{4.70}$$

The symbol $\mathcal{K}$ denotes the kinetic kernel and $V = V[\varphi, \bar{\varphi}]$ is a non-linear and in general non-local interaction potential. Choices of $\mathcal{K}$, $V$, $d$ and $G$ define a specific model, their geometric relevance is specified below. The classical equation of motion is then given by

$$\int (\mathrm{d}g')^d \, \mathcal{K}(g_{\mathrm{I}}, g'_{\mathrm{I}}) \varphi(g'_{\mathrm{I}}) + \frac{\delta V}{\delta \bar{\varphi}(g_{\mathrm{I}})} = 0. \tag{4.71}$$

In the following, we review in what manner GFT fields encode geometric information dressing simplicial complexes. This makes manifest the identification of rank-$d$ GFT fields with $d$-simplices. As we have recalled above, in the Hamiltonian formulation of Ashtekar-Barbero gravity, where $G = \mathrm{SU}(2)$, the densitised inverse triad is canonically conjugate to the connection [5, 31]. The former represent momentum space variables in which the spatial metric can be written. Through a non-commutative Fourier transform which shifts between configuration and momentum space [67, 439], the GFT formalism can be dually formulated on the latter. Its non-commutative character is rooted in the Poisson non-commutativity of the fluxes, as known from within LQG [66].

To see this, consider the configuration space of the GFT field, i.e., $G^d$ e.g. with $d = 4$ from which the phase space is constructed by the cotangent bundle $T^*G^4 \cong G^4 \times \mathfrak{g}^4$. The non-commutative Fourier transform of a GFT field is then defined by

$$\tilde{\varphi}(B_{\mathrm{I}}) = \int (\mathrm{d}g)^4 \prod_{I=1}^{4} e_{g_{\mathrm{I}}}(B_{\mathrm{I}}) \varphi(g_{\mathrm{I}}), \tag{4.72}$$

where the $B_{\mathrm{I}}$ with $\mathrm{I} = 1, \ldots, 4$ denote the flux variables which parametrise the non-commutative momentum space $\mathfrak{g}^4$. The $e_{g_{\mathrm{I}}}(B_{\mathrm{I}})$ represent a choice of plane waves on $G^4$. Their product is non-commutative, i.e., $e_g(B) \star e_{g'}(B) = e_{gg'}(B)$, indicated by the star product. Notice that we keep the vector arrows above the $B$s suppressed here. The non-commutative delta distribution on the momentum space is given by

$$\delta_{\star}(B) = \int \mathrm{d}g \ e_g(B). \tag{4.73}$$

With this object it is possible to show that the right invariance of the GFT fields corresponds to a closure condition for the fluxes, i.e.,

$$\sum_{\mathrm{I}} B_{\mathrm{I}} = 0. \tag{4.74}$$

It implies the closure of 4 faces dual to the links $e_{\mathrm{I}}$ to constitute a tetrahedron, as depicted in Fig 4.10. Furthermore, it allows for the elimination of one of the $B_{\mathrm{I}}$s when reexpressing the fluxes in terms of discrete triads. The latter is given by $B_i^{ab} = \int_{\triangle_i} e^a \wedge e^b$ with the co-triad field $e^a \in \mathbb{R}^3$ encoding the simplicial geometry. The symbol $\triangle_i$ with $i = 1, 2, 3$ corresponds to the faces associated to the tetrahedron (and we have dropped the fourth due to the closure condition). As shown in Refs. [440, 441], the metric at a given fixed point in the tetrahedron can be reconstructed from this by means of

$$g_{ij} = e_i^a e_j^b \delta_{ab} = \frac{1}{4 \operatorname{tr}(B_1 B_2 B_3)} \epsilon_i^{kl} \epsilon_j^{mn} \tilde{B}_{km} \tilde{B}_{ln}, \tag{4.75}$$

where $\tilde{B}_{ij} \equiv \operatorname{tr}(B_i B_j)$ holds.[43]

---

[43]These arguments are naturally carried over to the general case of rank-$d$ fields.

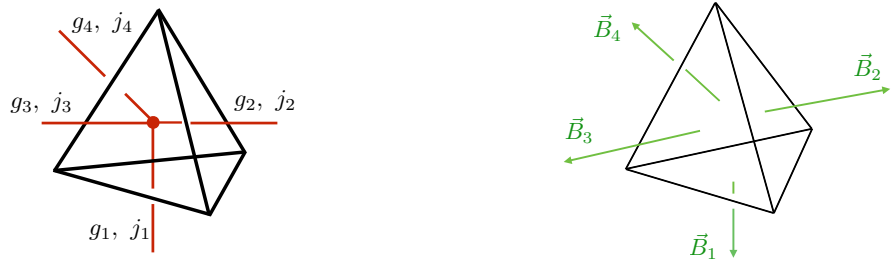

FIGURE 4.10: Graphical representation of a rank-4 GFT field. Group representation (left) and flux representation (right).

#### 4.2.4.2 Path integral quantisation of GFT

The quantum dynamics is defined by the partition function $Z_{GFT}$. If we write a more general interaction term as a sum of polynomials of degree $n$, i.e. $V = \sum_i \lambda_i V_i$, the path integral becomes

$$Z_{\text{GFT}} = \int [\mathcal{D}\varphi][\mathcal{D}\bar{\varphi}] e^{-S[\varphi,\bar{\varphi}]} = \sum_\Gamma \frac{\prod_i \lambda_i^{n_i(\Gamma)}}{\text{Aut}(\Gamma)} \mathcal{A}_\Gamma \qquad (4.76)$$

in the perturbative expansion over the Fock vacuum in terms of the coupling constants $\lambda_i$. The Feynman diagrams are denoted by $\Gamma$, $\text{Aut}(\Gamma)$ is the order of their automorphism group, $n_i(\Gamma)$ denotes the number of interaction vertices of type $i$, and $\mathcal{A}_\Gamma$ is the Feynman amplitude. Crucially, field arguments in $V$ are related to each other in a specific combinatorially non-local pattern which correlates fields among each other just through some of their arguments. This model-specific combinatorial non-locality implies that the GFT Feynman diagrams are dual to cellular complexes of arbitrary topology [36, 38, 433–436].[44]

In view of constructing a partition function for $4d$ quantum gravity, one starts with the GFT path integral quantisation of the so-called Ooguri model [378]. In turn, this defines a quantisation of BF-theory in $4d$ which is a topological field theory. In this model, the (right-invariant) real-valued GFT field is defined over $d = 4$ copies of $G = \text{Spin}(4)$ or $\text{SL}(2, \mathbb{C})$ and corresponds to a quantum tetrahedron or equally a 3-simplex. In the action a simplicial interaction term couples five copies of the field to each other. Their arguments

---

[44]Using the non-commutative Fourier transform as in the previous section, GFTs are turned into non-commutative field theories on Lie algebras. The corresponding Feynman diagrams then explicitly take the shape of simplicial gravity path integrals [432]. This representation also allows to identify the discrete counterpart of continuum diffeomorphisms of GR in the GFT setting [442].

are paired in a particular way to form a 4-simplex, given by

$$V = \frac{\lambda}{5!} \int (\mathrm{d}g)^{10} \; \varphi_{1234}\varphi_{4567}\varphi_{7389}\varphi_{96210}\varphi_{10851} \tag{4.77}$$

with $\varphi(g_1, g_2, g_3, g_4) \equiv \varphi_{1234}$ etc. The ultralocal kinetic term in the action with kernel $\mathcal{K}(g_{\mathrm{I}}, g_{\mathrm{I}}') = \delta(g_{\mathrm{I}}' g_{\mathrm{I}}^{-1})$ is specified by

$$K = \frac{1}{2} \int (\mathrm{d}g)^4 \; \varphi_{1234}^2, \tag{4.78}$$

and dictates how to glue together two such 4-simplices across a shared 3-simplex. In this way, the perturbative expansion generates Feynman diagrams which are dual to simplicial complexes.

The data given so far does not yet permit the reconstruction of a unique geometry for the simplicial complex. In a second step, one has to impose restrictions which reduce the non-geometric topological theory to the gravitational sector.

This can be substantiated by invoking the correspondence between GFT and spin foam models. Indeed, any GFT model defines in its perturbative expansion a spin foam model [36, 38, 430, 433–436]. One can then show, that GFTs based on the Ooguri model, may provide a covariant QFT formulation of the dynamics of LQG. As discussed in Section 4.2.3, boundary spin network states of LQG correspond to discrete quantum 3-geometries and transition amplitudes in between two such boundary states are given by appropriate spin foam amplitudes [5, 31, 35, 403, 429]. A concrete strategy to construct gravitational spin foam models is to start with a spin foam quantisation of topological BF-theory which is equivalent to setting up its discrete path integral. Importantly, it is then turned into a gravitational theory by imposing the so-called *simplicity constraints*, as discussed in Section 4.2.3. These restrict the data dressing the spin foam model such that it becomes equivalent to a discrete path integral for gravity. In particular, we discussed that the constraints allow to establish a link to LQG by restricting the group $G$ to $SU(2)$ on the boundary [401, 443–446].

It is in this way, that each so-constructed spin foam amplitude corresponds to a discrete spacetime history interpolating in between the boundary configurations and thus is identical to a restricted GFT Feynman amplitude. Therefore, the sum over Feynman diagrams given by Eq. (4.76) can be rewritten as a sum over diagrams dual to simplicial complexes decorated with quantum geometric data which clarifies how the GFT partition function can be intuitively understood to encode the sum-over-histories for $4d$ quantum gravity.

Some remarks are in order. Firstly, GFT models based on a non-compact group $G$ are in fact plagued by redundant volume factors in the action which has to be regularised, as discussed in Refs. [440, 441]. It should be noted that the Lorentzian EPRL vertex amplitude in its GFT form has so far not explicitly been put down as a function of its boundary data (i.e. SU(2) representations to match the LQG form of quantum states). This is currently a serious limitation for realistic computations in GFT and GFT condensate cosmology. Secondly, when remaining faithful to simplicial building blocks, one could e.g. consider higher interaction terms which also allow for an interpretation in terms of regular simplicial 4-polytopes but are even-powered. Also, as reviewed in the context of tensor models in Section 4.2.2.2, adding a colouring [373] these complexes are indeed bijective to abstract simplicial pseudo-manifolds [371]. Finally, it is in principle possible to go beyond the choice of simplicial building blocks and define GFTs which are fully compatible with the combinatorics of LQG. There, as seen in Section 2.2, quantum states of the 3-geometry are defined on boundary graphs with vertices of arbitrary valence. These correspond to general polyhedra and not merely to 3-simplices [447].

### 4.2.4.3   GFT as a second quantised formulation

Motivated by the roots of GFT in LQG, it is possible to construct a 2nd quantised Fock space reformulation of the kinematical Hilbert space of LQG of which the states describe discrete quantum 3-geometries. The construction is closely analogous to the one known from ordinary non-relativistic QFTs [437, 438]. In a nutshell, the construction leads to the reinterpretation of spin network vertices as fundamental quanta which are created or annihilated by the field operators of GFT. Pictorially seen, exciting a GFT quantum creates an atom of space (a *choron*) and thus GFTs are not QFTs on space but of space itself.

To start with, the GFT Fock space constitutes itself from a fundamental single-particle Hilbert space $\mathcal{H}_v = L^2(G^d)$ such that

$$\mathcal{F}(\mathcal{H}_v) = \bigoplus_{N=0}^{\infty} \text{sym}\big(\otimes_{i=1}^{N} \mathcal{H}_v^{(i)}\big). \qquad (4.79)$$

The symmetrisation with respect to the permutation group $S_N$ is chosen to account for the choice of bosonic statistics of the field operators and pivotal for the idea of reinterpreting spacetime as a Bose-Einstein condensate (BEC). $\mathcal{H}_v$ is the space of states of a GFT quantum. For $G = \mathrm{SU}(2)$ and the imposition of gauge invariance as in Eq. (4.69), a state represents an open LQG spin network vertex or its dual quantum polyhedron.[45] In the simplicial context, when $d = 4$, a GFT quantum corresponds to a quantum tetrahedron, the Hilbert space of which is

$$\mathcal{H}_v = L^2(G^4/G) \cong \bigoplus_{j_i \in \frac{\mathbb{N}}{2}} \mathrm{Inv}(\otimes_{i=1}^4 \mathcal{H}^{j_i}), \tag{4.80}$$

with $\mathcal{H}^{j_i}$ denoting the Hilbert space of an irreducible unitary representation of $G = \mathrm{SU}(2)$.

In this picture, the no-space state in $\mathcal{F}(\mathcal{H}_v)$ is devoid of any topological and quantum geometric information. It corresponds to the Fock vacuum $|\emptyset\rangle$ defined by

$$\hat{\varphi}(g_{\mathrm{I}})|\emptyset\rangle = 0. \tag{4.81}$$

By convention, it holds that $\langle\emptyset|\emptyset\rangle = 1$. Exciting a one-particle GFT state over the Fock vacuum is expressed by

$$|g_{\mathrm{I}}\rangle = \hat{\varphi}^\dagger(g_{\mathrm{I}})|\emptyset\rangle \tag{4.82}$$

and understood as the creation of a single open 4-valent LQG spin network vertex or of its dual tetrahedron.

The GFT field operators obey the canonical commutation relations (CCR)

$$\left[\hat{\varphi}(g_{\mathrm{I}}), \hat{\varphi}^\dagger(g'_{\mathrm{I}})\right] = \mathbb{1}_G(g_{\mathrm{I}}, g'_{\mathrm{I}}) \text{ and } \left[\hat{\varphi}^{(\dagger)}(g_{\mathrm{I}}), \hat{\varphi}^{(\dagger)}(g'_{\mathrm{I}})\right] = 0. \tag{4.83}$$

The delta distribution $\mathbb{1}_G(g_{\mathrm{I}}, g'_{\mathrm{I}}) = \int_G \mathrm{d}h \prod_{\mathrm{I}} \delta(g_{\mathrm{I}} h g'^{-1}_{\mathrm{I}})$ on the space $G^d/G$ is compatible with the imposition of gauge invariance at the level of the fields, as in Eq. (4.69).[46][47]

---

[45]This also holds true for $G = \mathrm{SL}(2, \mathbb{C})$ and $G = \mathrm{Spin}(4)$ when gauge invariance and simplicity constraints are properly imposed.

[46]References [440, 441] discuss these steps for non-compact group $G$. Notice, however, that on the level of the action volume divergences appear due to the imposed gauge invariance. This then necessitates the regularisation of the action functional.

[47]Importantly, notice that the kinematical Hilbert spaces of LQG and of GFT are not the same. In the construction of the Fock space one takes a direct sum over the number of vertices, in contrast to the set of graphs in LQG. In addition, on the GFT side, one does not impose any cylindrical consistency and equivalence on the quantum states which would be required for a continuum interpretation in terms of a generalised connection $A$. In the abstract, non-embedded context of GFT the exact graph structure is less

Using this, properly symmetrised many particle states can be constructed over the Fock space by

$$|\psi\rangle = \frac{1}{\sqrt{N!}} \sum_{P \in S_N} P \int (\mathrm{d}g)^{dN} \psi(g_I^1, \ldots, g_I^N) \prod_{i=1}^{N} \hat{\varphi}^\dagger(g_I^i)|\emptyset\rangle, \qquad (4.84)$$

with the wave functions $\psi(g_I^1, \ldots, g_I^N) = \langle g_I^1, \ldots, g_I^N | \psi \rangle$. Such states correspond to the excitation of $N$ open disconnected spin network vertices. The contruction of such multi-particle states is needed for the description of extended quantum 3-geometries.

Using this language, one can set up second-quantised Hermitian operators to encode quantum geometric observable data. In particular, an arbitrary one-body operator assumes the form

$$\hat{\mathcal{O}} = \int (\mathrm{d}g)^d \int (\mathrm{d}g')^d \; \hat{\varphi}^\dagger(g_I) \mathcal{O}(g_I, g_I') \hat{\varphi}(g_I'), \qquad (4.85)$$

with $O(g_I, g_I') = \langle g_I | \hat{o} | g_I' \rangle$ given in terms of the matrix elements of the first-quantised operators $\hat{o}$. Due to hermiticity, $\mathcal{O}(g_I, g_I') = (\mathcal{O}(g_I', g_I))^*$ holds. For example, the number operator is given by

$$\hat{N} = \int (\mathrm{d}g)^d \hat{\varphi}^\dagger(g_I) \hat{\varphi}(g_I). \qquad (4.86)$$

Strictly speaking, $N$ exists only when interactions are disregarded which is when all representations of the CCRs are equivalent to the Fock representation, see below.

Another relevant operator encoding geometric information is the vertex volume operator

$$\hat{V} = \int (\mathrm{d}g)^d \int (\mathrm{d}g')^d \; \hat{\varphi}^\dagger(g_I) V(g_I, g_I') \hat{\varphi}(g_I'), \qquad (4.87)$$

wherein $V(g_I, g_I')$ is given in terms of the LQG volume operator between two single-vertex spin networks and an analogous expression holds for the LQG area operator [440, 441, 448, 449]. These operators will play a central role in the next chapter in the study of effective geometries emerging from GFT condensates.

**Fock and non-Fock representations**

Following Ref. [437], for the Fock representation of GFT one defines a set of fundamental operators $\hat{c}_i$ and $\hat{c}_i^\dagger$, with the algebraic relations

$$[\hat{c}_i, \hat{c}_{i'}^\dagger] = \delta_{ii'} \quad \text{and} \quad [\hat{c}_i^{(\dagger)}, \hat{c}_{i'}^{(\dagger)}] = 0$$

satisfying

$$\hat{c}_i|N_i\rangle = \sqrt{N_i}|N_i - 1\rangle \quad \text{and} \quad \hat{c}_i^\dagger|N_i\rangle = \sqrt{N_i + 1}|N_i + 1\rangle,$$

where $N_i$ denote the occupation numbers in the single-vertex state labelled by $i$. The index $i$ encapsulates spin quantum numbers and intertwiner labels such that $i = (\{j\}, \{m\}, \iota)$. The operators $\hat{c}_i$ and $\hat{c}_i^\dagger$ annihilate and create single spin network vertices acting on the Fock vacuum state given by

$$\hat{c}_i|\emptyset\rangle = 0, \quad \forall i.$$

The occupation number operators are then expressed by

$$\hat{N}_i|N_i\rangle = \hat{c}_i^\dagger \hat{c}_i|N_i\rangle = N_i|N_i\rangle$$

with the total number operator $\hat{N} = \sum_i \hat{N}_i$. With this, we may write the bosonic field operators as

$$\hat{\varphi}(g_{\mathrm{I}}) = \sum_i \hat{c}_i \psi_i(g_{\mathrm{I}}), \tag{4.88}$$

where a complete basis of single-vertex wave functions is given by the spin network wave functions for individual spin network vertices

$$\psi_i(g_{\mathrm{I}}) = \langle g_{\mathrm{I}}|i\rangle = C^{j_1 \ldots j_d, \iota}_{n_1 \ldots n_d} \prod_{a=1}^d D^{j_a}_{m_a n_a}(g_a), \tag{4.89}$$

where the $C$s denote (normalised) intertwiners of the group. We refer to Appendix C.2 for details regarding harmonic analysis on SU(2).

Within the context of local QFT [450–456] it is well-known, that in the finite dimensional and non-interacting infinite dimensional cases all irreducible Fock representations are unitarily equivalent and hence there is just one phase associated to the quantum system. However, this is different for interacting fields, models with non-vanishing ground state expectation value and many-body systems in the thermodynamic limit. There the Fock representation is not allowed and $N$ is not a good quantum number for the characterisation of the system since $\hat{N}$ is unbounded from above. In these situations the systems are described by means of non-Fock representations corresponding to inequivalent representations of the commutation relations and thus allow for the occurrence of different phases associated to the considered quantum system. The understanding of these points in GFT, and in particular the notion of phases and their potential relation to continuum geometries, is in its infancy. Inequivalent representations of GFT were first introduced in Ref. [457] and then better understood in Ref. [458]. In Section 6.2 indications for such representations are found when (effective) interactions are considered.

This finishes the introduction of the GFT formalism and its motivation coming from other approaches at the covariant quantisation of gravity. In the following chapter we introduce the condensate cosmology spin-off to GFT where the use of field theory methods (in particular condensate states) plays the key role to extract continuum geometric information, as compared to the other approaches.

# Chapter 5

# Group field theory condensate cosmology

> Most important part of doing physics is the knowledge of approximation.
>
> ———————————————————
>
> Lev Davidovich Landau

## 5.1 Why condensate cosmology?

The most difficult problem for all quantum gravity approaches using discrete and quantum pregeometric structures is the recovery of continuum spacetime, its geometry, diffeomorphism invariance and general relativity as an effective description for the dynamics of the geometry in an appropriate limit. It has been suggested, a possible way of how continuum spacetime and geometry could emerge from a quantum gravity substratum in such theories is by means of at least one phase transition from a discrete pregeometric to a continuum geometric phase. One refers to such a process as "geometrogenesis" [459, 460].

A particular representative in this class of approaches where such a scenario has been proposed is GFT where one tries to identify a continuum geometric phase to a condensate phase of the underlying quantum gravity system [461, 462] with a tentative cosmological interpretation [440, 441, 448].

More specifically, it is conjectured that a phase transition in a GFT system can give rise to a condensate phase which corresponds to a non-perturbative vacuum of the model under consideration. This vacuum is described by a large sample $N = \langle \hat{N} \rangle$ of bosonic GFT quanta which have relaxed into a common ground state (labelled with the same quantum geometric data) that is asymptotically orthogonal to the Fock vacuum, i.e. in the limit where $N \to \infty$ (cf. Refs. [463–465]). Such a state is then considered suitable to model spatially homogeneous geometries the metric of which is the same at every point.

This conjecture is supported by the functional renormalisation group (FRG) [224, 466–469] analyses of specific GFT models which allow to understand their non-perturbative properties. In general, the FRG techniques provide the most powerful theoretical description of thermodynamic phases by means of a coarse graining operation that progressively eliminates short scale fluctuations. Their successful application to matrix models of $2d$ quantum gravity [226, 227, 332, 333, 470] serves as an example for the adaption to GFT models which has recently been very actively pursued [393, 471–476]. In this way, the FRG methods enable one to study the consistency of GFT models, analyse their continuum limit, chart their phase structure and investigate the possible occurrence of phase transitions. In all the models investigated so far, IR fixed points were found, suggesting a phase transition from a symmetric to a broken/condensate phase [393, 471–476].

More precisely, standard FRG methodology has been applied to a couple of models from a class of group field theories called tensorial GFTs [393], for which one requires the fields to possess tensorial properties under a change of basis, see Section 4.2.2.2. The common features of the models analysed so far are a non-trivial kinetic term of Laplacian type and a quartic combinatorially non-local interaction. However, they differ firstly in the size of the rank, secondly in whether gauge invariance is imposed or not, and thirdly in the compactness or non-compactness of the group manifold used as the field domain. Indications for a phase transition separating a symmetric from a broken/condensate phase were found as the mass parameter tends to negative values in the IR limit analogous to a Wilson-Fisher fixed point in the corresponding local QFT. This analogy is illustrated in terms of the phase diagrams in Figs. 5.1 and 5.2. To corroborate the existence of such a phase transition, among others, the theory then has to be studied around the newly assumed ground state by means of a mean field analysis, as noticed in Refs. [393, 471–476]. Amongst other things, such an analysis would then require to investigate solutions to the classical equation of motion obtained via a saddle point approximation of the path integral.

The possible occurrence of a phase transition in such systems is highly interesting, since it could be a realisation of the above-mentioned geometrogenesis scenario. So far, however, the mentioned FRG results for tensorial GFTs can only lend indirect support to the geometrogenesis hypothesis, since a full geometric interpretation of such models is currently lacking. To realise such a hypothesis, one would actually have to proceed towards the analysis of a (potentially coloured) GFT model enriched with additional geometric data and an available simplicial quantum gravity interpretation that is closely linked to LQG. The application of FRG methods to such a model with a combinatorially non-local simplicial interaction term would be needed to give an accurate account of the phase structure of the system. The hope is that studying its renormalisation group flow will reveal an IR fixed point which marks the phase transition into a condensate phase ideally corresponding to a physical continuum geometry. Hence, the aim is to gradually increase the sophistication of the studied toy models to rigorously underpin the GFT condensate assumption and connect it to the geometrogenesis hypothesis [461, 462].

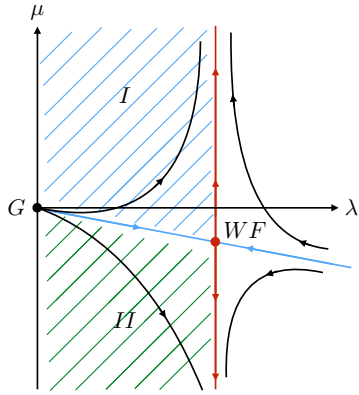

FIGURE 5.1: Phase diagram of a local scalar field theory with quartic interaction on $\mathbb{R}^3$. The mass parameter is denoted by $\mu$ while the interaction couples with $\lambda$. $G$ denotes the Gaussian fixed point and $WF$ the Wilson-Fisher fixed point. In the region hatched in green $\langle \hat{\varphi} \rangle \neq 0$ holds.

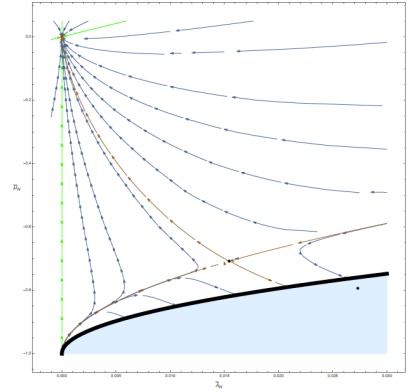

FIGURE 5.2: Exemplary phase diagram of a quartic TGFT on $\mathbb{R}^d$ (cf. Refs. [475, 476]). In the analogue of region $II$ in Fig. 5.1, a non-vanishing expectation value of the field operator is expected to be found.

Given this, the central assumption of the GFT condensate cosmology programme is the existence of a condensate phase for geometric GFT models in three or four dimensions and that it ideally approximates a continuum geometric phase with a cosmological interpretation. The aim then is to derive the effective dynamics for the GFT condensate states directly from the microscopic GFT quantum dynamics using mean field techniques and to extract a cosmological interpretation from them [440, 441, 448]. These techniques are strongly reminiscent of those employed to study the Gross-Pitaevskii equation for, at most, weakly interacting Bose-Einstein condensates [463–465, 477–479]. Indeed, the quickly growing body of studies of effective geometries and their dynamics in Refs. [440, 441, 457, 480–491] admit a description in terms of the one given by general relativity for the corresponding classical geometry. The recovery of the Friedmann dynamics for the emergent homogeneous and isotropic geometry is perhaps the most striking one [486]. This also lends strong support to the idea that GFT condensate states are appropriate for studying the cosmological sector of LQG.[1]

### 5.1.1 Condensate states

In the following, we further discuss the motivation for why GFT condensate states serve as a good ansatz to effectively capture the physics of homogeneous continuum spacetimes following Refs. [440, 441, 448] and review important aspects of their construction.

When considering spatial homogeneity in the continuum, we know that it is possible to reconstruct the geometry from any point as the metric is the same everywhere. This *homogeneity* translates on the level of GFT states to the requirement that all quanta occupy the same quantum geometric state and that in addition to the right-invariance of the GFT field, it is also endowed with an invariance under the diagonal left action of the group, as clarified below. This is the reason for choosing GFT condensate states as the main ingredient for GFT condensate cosmology in close analogy to the theory of real Bose condensates [463–465].[2] Furthermore, for a state to encode in some adequate limit information allowing for the description of a smooth metric geometry in $3d$, one assumes that a large constituent number $N$ will lead to a good *approximation of the continuum*. Moreover, the simplicial building blocks are required to be almost flat. This *near-flatness condition* translates on the level of the states to the requirement that the probability density is concentrated around small values of the curvature.[3] Finally, for a classical cosmological spacetime to emerge from a given quantum state, it should exhibit *semi-classical properties*. Crucially, condensate states automatically fulfill such a desirable feature because they are field coherent states and as such exhibit, in a certain sense, ultraclassical behaviour by saturating the number-phase uncertainty relation and are thus the quantum states which are the closest to classical waves. We will discuss the construction of such states and their properties in the following.

---

[1] We would like to stress that the GFT condensate cosmology approach is not the only one trying to extract the cosmological sector of LQG from a covariant formulation of its dynamics. Spin foam cosmology [405–412] makes use of the spin foam expansion [35] and is thus an expansion in terms of the number of degrees of freedom. Central to this approach is the assumption that the relevant physics is encoded within a fixed number of quanta of geometry, whereas in GFT condensate cosmology there is no a priori restriction on the number of quanta which can be large, in principle.

[2] The procedure to reconstruct the spatial metric by means of the information encoded in the quantum state is discussed in Section 4.2.4.1.

[3] Notice that this is required to identify states with a more straightforward physical interpretation. In fact, at the current stage there is no dynamical principle in GFT available which could lead to a concentration of the probability density at small curvature values.

Using the Fock representation of GFT as introduced in Section 4.2.4.3 with $G = \mathrm{SU}(2)$ coming from the imposition of simplicity constraints on $\mathrm{SL}(2, \mathbb{C})$ data, as in Section 4.2.4.3, we decompose the field operator $\hat{\varphi}(g_\mathrm{I})$ in terms of annihilation operators $\{\hat{c}_i\}$ of single-particle quantum geometry states $\{|i\rangle\}$ yielding

$$\hat{\varphi}(g_\mathrm{I}) = \sum_i \hat{c}_i \psi_i(g_\mathrm{I}). \tag{5.1}$$

Following the logic of the Bogoliubov approximation valid for ultracold, non- to weakly interacting and dilute Bose condensates [463–465, 477–479], if the ground state $i = 0$ has a macroscopic occupation, one separates this expression into a condensate term and one for all the remaining non-condensate components. This yields

$$\hat{\varphi}(g_\mathrm{I}) = c_0 \psi_0(g_\mathrm{I}) + \sum_{i \neq 0} \hat{c}_i \psi_i(g_\mathrm{I}), \tag{5.2}$$

where one replaces the operator $\hat{c}_0$ by the c-number $c_0$ so that the average occupation number of the ground state is given by $N = \langle \hat{c}_0^\dagger \hat{c}_0 \rangle$. In the next step one redefines $\sigma \equiv \sqrt{N} \psi_0$ as well as $\delta\hat{\varphi} \equiv \sum_{i \neq 0} \hat{c}_i \psi_i$ giving rise to

$$\hat{\varphi}(g_\mathrm{I}) = \sigma(g_\mathrm{I}) + \delta\hat{\varphi}(g_\mathrm{I}), \tag{5.3}$$

where $\psi_0$ is normalised to 1. This ansatz is only justified if the ground state is macroscopically occupied, i.e. when $N \gg 1$ and the fluctuations $\delta\hat{\varphi}$ are regarded as small. One calls the classical field $\sigma(g_\mathrm{I})$ the mean field of the condensate which assumes the role of an order parameter. Making use of the particle density $n(g_\mathrm{I}) = |\sigma(g_\mathrm{I})|^2$ and a phase characterising the coherence properties of the condensate, we write the mean field in polar form as

$$\sigma(g_\mathrm{I}) = \sqrt{n} \; \mathrm{e}^{i\theta(g_\mathrm{I})}. \tag{5.4}$$

This illustrates that the order parameter can always be multiplied by an arbitrary phase factor without affecting the physical measurement. This behaviour is identified as a global $\mathrm{U}(1)$-symmetry of the system which is associated with the conservation of the total particle number.

By construction, the Bogoliubov ansatz (5.3) gives rise to a non-zero expectation value of the field operator, i.e., $\langle \hat{\varphi}(g_\mathrm{I}) \rangle \neq 0$, indicating that the condensate state is in, or rather close to, a field coherent state.

A simple multiparticle trial state fulfilling this ansatz is given by

$$|\sigma\rangle = A \, \mathrm{e}^{\hat{\sigma}} |\emptyset\rangle, \quad \hat{\sigma} = \int (\mathrm{d}g)^4 \, \sigma(g_\mathrm{I}) \hat{\varphi}^\dagger(g_\mathrm{I}), \tag{5.5}$$

which is constructed from quantum tetrahedra all encoding the same discrete geometric data.[4] It defines a non-perturbative vacuum over the Fock space. The normalisation factor is given by

$$A = \mathrm{e}^{-\frac{1}{2} \int (\mathrm{d}g)^4 \, |\sigma(g_\mathrm{I})|^2}. \tag{5.6}$$

Such states are field coherent because they are eigenstates of the field operator,

$$\hat{\varphi}(g_\mathrm{I}) |\sigma\rangle = \sigma(g_\mathrm{I}) |\sigma\rangle, \tag{5.7}$$

such that indeed $\langle \hat{\varphi}(g_\mathrm{I}) \rangle = \sigma(g_\mathrm{I}) \neq 0$ holds (as long as $|\sigma\rangle$ is not the Fock vacuum). Due to this property the expectation value of the number operator immediately yields the average particle number

$$N = \int (\mathrm{d}g)^4 \, |\sigma(g_\mathrm{I})|^2 < \infty. \tag{5.8}$$

For practical reasons, it is of course only possible to use such a condensate state for the description of a macroscopic homogeneous universe, if the number of quanta is $N \gg 1$ but finite. If the number operator is well-defined and its expectation value is finite, such states are called Fock coherent states. By construction, such a description is only valid for non-interacting or weakly interacting condensates. Toward the strongly interacting regime, it has to be replaced by one given in terms of non-Fock coherent states, see Section 4.2.4.3 and Appendix B.

Importantly, we require in addition to the right invariance, as in Eq. (4.69), invariance under the diagonal left action of $G$, i.e.,

$$\sigma(k g_\mathrm{I}) = \sigma(g_\mathrm{I}) \tag{5.9}$$

---

[4]More complicated "molecule" states built from the tetrahedra could also be considered, as advocated in Refs. [440, 441, 448].

for all $k \in G$. This property is introduced in order to guarantee that the domain becomes isomorphic to minisuperspace of homogeneous geometries and that the number of dynamical degrees of freedom is thus reduced. To see this, we follow Ref. [480] and use the existence of the natural bijection of quotient spaces $G \backslash G^4 / G$ and $G^3 / \mathrm{Ad}_G$ so that the mean field $\sigma$ can equivalently be seen as a function on the latter space. Its non-commutative Fourier transform $\hat{\sigma}$ then lives on $\mathfrak{g}^{\oplus 3} / \mathrm{Ad}_G$, where $\mathfrak{g}$ denotes the Lie algebra of $G$. Seen as a vector space, $\mathfrak{g}^{\oplus 3}$ is equal to $\mathbb{R}^{\dim(G) \times 3}$. For $G = \mathrm{SU}(2)$ this leads to the quotient space $\mathbb{R}^{3 \times 3} / \mathrm{Ad}_{\mathrm{SU}(2)}$. If we further assume the non-degeneracy condition $\mathrm{tr}\,(B_1 B_2 B_3) \neq 0$ in the reconstructed metric Eq. 4.75, this quotient space is turned into $\mathrm{GL}(3, \mathbb{R}) / \mathrm{Ad}_{\mathrm{SU}(2)}$. Finally, by choosing an orientation through $\mathrm{tr}\,(B_1 B_2 B_3) > 0$ the quotient space turns out to be $\mathrm{GL}(3, \mathbb{R}) / \mathrm{O}(3) \cong \mathrm{SL}(3, \mathbb{R}) / \mathrm{O}(3) \times \mathbb{R} \backslash \{0\}$ which is the space of (non-degenerate) homogeneous 3-metrics [54]. This naturally applies to rank-3 models, too.

### 5.1.2 Effective dynamics

After having discussed the construction of suitable states, we briefly summarise how the effective condensate dynamics can be obtained from the underlying GFT dynamics as in Refs. [440, 441, 448]. To this aim, one uses the infinite tower of Schwinger-Dyson equations

$$0 = \delta_{\bar{\varphi}} \langle \mathcal{O}[\varphi, \bar{\varphi}] \rangle = \left\langle \frac{\delta \mathcal{O}[\varphi, \bar{\varphi}]}{\delta \bar{\varphi}(g_I)} - \mathcal{O}[\varphi, \bar{\varphi}] \frac{\delta S[\varphi, \bar{\varphi}]}{\delta \bar{\varphi}(g_I)} \right\rangle, \qquad (5.10)$$

where $\mathcal{O}$ is a functional of the fields. One extracts an expression for the effective dynamics by setting $\mathcal{O}$ equal to the identity. This leads to

$$\left\langle \frac{\delta S[\varphi, \bar{\varphi}]}{\delta \bar{\varphi}(g_I)} \right\rangle = 0 \qquad (5.11)$$

with the action $S[\varphi, \bar{\varphi}]$ as in Eq. (4.70). When the expectation value is taken with respect to the condensate state $|\sigma\rangle$, one obtains the analogue of the Gross-Pitaevskii (GP) equation for real Bose condensates

$$\int (\mathrm{d}g')^4 \mathcal{K}(g_I, g_I') \sigma(g_I') + \frac{\delta V}{\delta \bar{\sigma}(g_I)} = 0.^5 \qquad (5.12)$$

---

[5] Equivalently, this equation can be obtained via a saddle point approximation of the path integral using the above-introduced field coherent states.

In general, this is a non-linear and non-local equation for the dynamics of the mean field $\sigma$ and is given the interpretation of a quantum cosmology equation. In analogy to the GP equation, it has no direct probabilistic interpretation. These features might appear as a problem when trying to relate the GFT condensate cosmology framework to LQC [75, 76, 492] or Wheeler-DeWitt quantum cosmology [34]. However, they do not pose a problem for the direct extraction of cosmological predictions from the full theory. Based on Refs. [486, 487, 489, 490], in Sections 6.2 and 6.3 we demonstrate how a Friedmann-like evolution equation can be derived from such effective dynamics of specific GFT condensates.

Notice that solutions to the classical equations of motion of GFT are generally poorly understood, especially due to the non-locality of the interaction. Solving these equations would naturally correspond to solving the quantum theory at tree-level. In the following, we will explore solutions to the dynamical Boulatov model under restrictions and try to interpret the results in the context of the condensate cosmology approach. This will be the first application of this formalism in this work and serves as a warm-up to the next chapter. Right after, we will take a closer look at condensate hypothesis and test it using mean field methods.

## 5.2 Solving the dynamical Boulatov model

As we have discussed, a perturbative expansion of the partition function $Z$ for a geometric GFT model of rank $d$ provides a discrete model of $d$-dimensional quantum gravity. The leading thought behind the condensate cosmology approach, however, is the expectation that a description of continuous geometry will require a non-perturbative understanding of the partition function.

As is generally known, the construction of a full non-perturbative quantum field theory is rarely possible, but often it is already enough to construct a perturbation theory around a non-perturbative vacuum [452]. Moreover, if quantum fluctuations are not too strong, a non-perturbative vacuum can be reasonably well approximated by the minimum of the classical action $S$, called the minimiser. In that case, the saddle point or mean field approximation around the minimiser prompts an effective field theory that will provide insights into the non-perturbative regime of the model. For that reason, a study of minimisers of geometric GFT models is an important step towards a better understanding of continuous quantum geometry.

Despite their importance, however, the extrema of such models are poorly understood in the literature. This is mostly due to the fact that their Euler-Lagrange equations are non-linear integro-differential equations which are notoriously difficult to solve, in general. In the Boulatov model, these integro-differential equations can be formulated as integral equations with an integral kernel given by the Wigner $6j$-symbol. A solution to the extremal equations then requires full control of the zeros of the $6j$-symbol, which remains an open problem [493–496].

In addition to this, there seems to be no consensus on the signs of the coupling constants in GFT models. For instance, the convention used in renormalisation analysis [393] is opposite to the one used in the context of the GFT condensate cosmology investigations [440, 441, 457, 480, 486, 487, 489–491]. Despite this discrepancy in the sign convention both analyses rely on the existence of global or at least local minimisers and for that reason require a good understanding of the extrema in GFT.

In the following, we address the minimisers of the Boulatov model [377] augmented by a Laplace-Beltrami operator, hereafter called dynamical Boulatov model [497]. This is a model for Euclidean quantum gravity in $3d$. To make the problem tractable, we look for minimisers in the space of left and right invariant fields corresponding to equilateral triangles. For reasons of clarity, the style of this Section is rather mathematical. We start off with Section 5.2.1 where we give the definition of the model and the space of functions considered hereafter. On this space the Euler-Lagrange equations of the action become solvable, allowing us to provide a full characterisation of solutions in Section 5.2.2.1. We then identify the parameter regimes in which the action admits minima and characterise the minimisers in Section 5.2.2.2. Our main result regarding the extrema is presented in theorem 1 and the subsequent discussion. The characterisation of minimisers is provided in theorem 2. Implications of our results on GFT condensate cosmology are discussed in Section 5.2.3. The proofs of statements in the text are directed to the Appendix to make the presentation lighter. This Section is largely based on the work of the author in Ref. [498].

## 5.2.1 The dynamical Boulatov model

In the following, we recall the construction of the Boulatov model with a focus on mathematical details necessary for the subsequent analysis and set up our notations.

The space of smooth, real-valued functions on the field domain $SU(2)^3$ shall be denoted by $C^\infty(SU(2)^3)$. A subspace $\mathcal{S}$ of it shall be defined by right and cyclic invariant functions, i.e., functions $f$ which are invariant with respect to the diagonal right action of $SU(2)$ and which are invariant with respect to cyclic permutations of its arguments. The right invariance of $f$ ensures a geometric interpretation in terms of triangles and the cyclic relabelling guarantees that the ordering of the field arguments has no physical meaning.[6] The dynamical Boulatov action on $\mathcal{S}$ is given by

$$S_{m,\lambda}[\varphi] = \frac{1}{2} \int (\mathrm{d}g)^3 \varphi_{123} \left(-\Delta + m^2\right) \varphi_{123} + \frac{\lambda}{4!} \int (\mathrm{d}g)^6 \, \varphi_{123}\varphi_{145}\varphi_{256}\varphi_{364}, \qquad (5.13)$$

where $m^2$ and $\lambda$ are real, possibly negative coupling constants, $\mathrm{d}g$ is the Haar measure, $-\Delta$ is the Laplace-Beltrami operator[7], and we abbreviated $\varphi_{123} \equiv \varphi(g_1, g_2, g_3)$ etc. for convenience. The interaction kernel encodes the combinatorics of a tetrahedron. It is symmetric under cyclic permutations of its arguments [377].

---

[6] Notice that by imposing invariance with respect to cyclic permutations of the field arguments, we strictly follow the original definition of the Boulatov model [377]. In later reformulations of the model this property is dropped while an additional combinatorial degree of freedom called colour is attributed to the fields to guarantee that the perturbative expansion of the model is free of topological pathologies [371, 373, 382, 499], as in Section 4.2.2.2.

[7] We include the Laplace-Beltrami operator in the action which is needed for a consistent implementation of a renormalisation scheme [497].

To address the variational problem below, we topologise $\mathcal{S}$ by the family of semi-norms $\|f\|_n \equiv \sup_{(g_1,g_2,g_3) \in \mathrm{SU}(2)^3} |\Delta^n f(g_1, g_2, g_3)|$, with the neighborhood base given by semi-balls [500], $N_{\epsilon,n}(0) = \{\|f\|_n < \epsilon \,|\, f \in \mathcal{S}\}$, for $n \in \mathbb{N}$ and $\epsilon > 0$.

Leading the analysis further, we will restrict the space $\mathcal{S}$ by requiring **(1)** invariance of functions $f$ with respect to the diagonal left action of SU(2): As we have recalled, this left-symmetry is needed to identify the domain space of the fields with the superspace of homogeneous spatial geometries. By the Peter-Weyl theorem, such left and right invariant functions can be written as

$$f(g_1, g_2, g_3) = \sum_{J \in \mathfrak{J}} f^J \, \mathcal{X}^J(g_1, g_2, g_3) \quad \text{with}$$

$$\mathcal{X}^J(g_1, g_2, g_3) \equiv \frac{d_{j_1} d_{j_2} d_{j_3}}{3} \int \mathrm{d}h \sum_{\sigma \in \mathrm{Cycl}} \chi^{j_{\sigma(1)}}(g_1 h) \, \chi^{j_{\sigma(2)}}(g_2 h) \, \chi^{j_{\sigma(3)}}(g_3 h) , \qquad (5.14)$$

where $J = (j_1, j_2, j_3)$ belongs to $\mathfrak{J} \equiv \left(\frac{\mathbb{N}}{2}\right)^3$, Cycl denotes cyclic permutations of the set $\{1,2,3\}$ and $\chi^{j_i}$ denotes the character of an SU(2) representation of dimension $d_j = 2j + 1$ for $j \in \frac{\mathbb{N}}{2}$.[89] Furthermore, for such functions we require **(2)** equilaterality: $f$ is an equilateral function if its non-vanishing Peter-Weyl coefficients are of the form $\left(f^{(j,j,j)}\right)_{j \in \mathbb{N}}$. The condition of equilaterality ensures that the modes correspond to equilateral triangles.[10] We denote the restriction of $\mathcal{S}$ to left invariant equilateral functions by $\mathcal{S}_{\mathrm{EL}}$ and the space of equilateral triples by $\mathfrak{J}_{\mathrm{EL}} = \{(j,j,j) \,|\, j \in \mathbb{N}\}$. Note, that $\mathfrak{J}_{\mathrm{EL}}$ contains only integer multi-indices, since for any half-integer $j$ the matrix coefficients vanish:

$$\mathcal{X}^{(j,j,j)} = 0 \quad \text{with } j = \frac{2n+1}{2} \quad n \in \mathbb{N}.$$

In the following we will sometimes use the notation $f \in \mathcal{S}_{(\mathrm{EL})}$ and $f^J$ with $J \in \mathfrak{J}_{(EL)}$ to signal that the statement holds equally for $\mathcal{S}$ and $\mathcal{S}_{\mathrm{EL}}$ and, correspondingly, with a set of indices belonging to $\mathfrak{J}$ or to $\mathfrak{J}_{\mathrm{EL}}$. For clearer notation we also define the square of the triple $J$ as $J^2 \equiv j_1 (j_1 + 1) + j_2 (j_2 + 1) + j_3 (j_3 + 1)$, and its modulus as $|J| \equiv j_1 + j_2 + j_3$.

---

[8]By [500, theorem 3] the sequence of coefficients $\left(f^J\right)_{J \in \mathfrak{J}}$ is a rapidly decreasing sequence of real numbers and the equality is understood such that the right hand side of the equation converges to $f(g_1, g_2, g_3)$ in the aforementioned topology.

[9]We refer to Appendices C.2 and C.2.2 in particular for details regarding harmonic analysis on SU(2).

[10]In the analysis of specific GFT condensate cosmology models, this last restriction is meant to enforce isotropy. We will come back to this point in the next chapter. Notice that the restriction to equilateral configurations bears strong resemblance to what is done in the closely related contexts of dynamical triangulations [48, 302] and tensor models for quantum gravity [44, 336] where the use of standardized building blocks – by universality arguments – is believed not to affect the continuum results, as discussed in Sections 4.2.1.2 and 4.2.2.2.

Considering the restricted space $\mathcal{S}_{\mathrm{EL}}$ has an important advantage: The action $S_{m,\lambda}$ defines statistical weights of a generating functional by means of a functional integral $Z$. It has been shown that on $\mathcal{S}$ the action $S_{m,\lambda}$ is not bounded from below, regardless of the parameter region [501, 502], thus rendering $Z$ ill-defined. As we will show below, this problem gets resolved on $\mathcal{S}_{\mathrm{EL}}$, since global minimisers of the action exist thereon (at least for some parameter regions). In principle, this allows to perturbatively define the generating functional and to give a well-defined statistical theory.[11]

**Definition 1.** A local minimiser of the action $S_{m,\lambda}$ on $\mathcal{S}_{\mathrm{EL}}$ is a field $\varphi \in \mathcal{S}_{\mathrm{EL}}$, that for some $n \in \mathbb{N}$ and $\epsilon > 0$ satisfies

$$S_{m,\lambda}[\phi] \geq S_{m,\lambda}[\varphi], \tag{5.15}$$

for any $\phi \in N_{\epsilon,n} (\varphi) \cap \mathcal{S}_{\mathrm{EL}}$. If condition Eq. (5.15) is satisfied on the whole space $\mathcal{S}_{\mathrm{EL}}$ we call the minimiser global.

In the following, we will characterise all extrema of the action $S_{m,\lambda}$ on $\mathcal{S}_{\mathrm{EL}}$ for the four different parameter regions

$$(a)\ m^2 < 0 \qquad (b)\ m^2 > 0 \qquad (c)\ m^2 > 0 \qquad (d)\ m^2 < 0$$
$$\lambda < 0 \qquad\qquad \lambda < 0 \qquad\qquad \lambda > 0 \qquad\qquad \lambda > 0$$

and identify which of the extrema are minimisers.[12]

### 5.2.2 Extrema and minimisers

Let $I \subset \mathbb{R}$ denote an interval containing zero; for $t \in I$ and $\varphi, f \in \mathcal{S}_{\mathrm{EL}}$ a necessary condition for $\varphi$ to be a local minimiser on $\mathcal{S}_{\mathrm{EL}}$ is given by

$$S'_{m,\lambda}[\varphi, f] \equiv \partial_t S_{m,\lambda}[\varphi + tf]|_0 = 0, \tag{5.16}$$

$$S''_{m,\lambda}[\varphi, f] \equiv \partial_t^2 S_{m,\lambda}[\varphi + tf]|_0 \geq 0, \tag{5.17}$$

---

[11]One can bound the Boulatov action by adding a so-called pillow term to the action [501]. We leave the impact of such a modification onto the ensuing analysis to future investigations.

[12]Notice that for the subsequent analysis of extrema on $\mathcal{S}_{\mathrm{EL}}$ the cyclicity property mentioned above has no impact and could in principle be lifted from the outset.

for any $f \in \mathcal{S}_{\text{EL}}$.

In the following, we will investigate the extremal condition Eq. (5.16) for the model Eq. (5.13). We will then check if some solutions are minimal and thus fulfill Eq. (5.17) and the condition in definition 1.

**Proposition 1.** *$\varphi \in \mathcal{S}_{\text{(EL)}}$ is an extremum of $S$ if and only if the Peter-Weyl coefficients of $\varphi$ — denoted by $\varphi^J$ — satisfy for any $J \in \mathfrak{J}_{\text{(EL)}}$,*

$$(J^2 + m^2)\varphi^J + \frac{\lambda}{3!} \sum_{K \in \mathfrak{J}_{(EL)}} \varphi^{j_1 k_2 k_3} \varphi^{j_2 k_3 k_1} \varphi^{j_3 k_1 k_2} \begin{Bmatrix} j_1 & j_2 & j_3 \\ k_1 & k_2 & k_3 \end{Bmatrix}^2 = 0, \qquad (5.18)$$

*where $K = (k_1, k_2, k_3) \in \mathfrak{J}_{(EL)}$.*

*Proof.* See Appendix D.1.1. □

The extremal condition Eq. (5.18) is a non-linear tensor equation with an integral kernel given by the $6j$-symbol squared. To this issue adds the fact that the non-trivial zeros of the $6j$-symbol are still under investigation [493–496], making Eq. (5.18) inherently difficult to solve in full generality. Some specific solutions for the case without the Laplace-Beltrami operator and $\lambda < 0$ have been introduced in Ref. [503], but a systematic analysis of extrema was not performed therein.

Although the extremal condition Eq. (5.18) is difficult to solve on $\mathcal{S}$, it turns out to be solvable on $\mathcal{S}_{\text{EL}}$, because in this case the $6j$-symbol significantly simplifies.

### 5.2.2.1 Extrema

In the following we will denote the Wigner $6j$-symbol for $J \in \mathfrak{J}_{\text{(EL)}}$ by

$$\{6j\} \equiv \begin{Bmatrix} j_1 & j_2 & j_3 \\ j_1 & j_2 & j_3 \end{Bmatrix}, \qquad (5.19)$$

and define the space $\mathfrak{J}_{\text{(EL)}}^S$ of $J$'s such that $\mathfrak{J}_{\text{(EL)}}^S = \left\{ J \in \mathfrak{J}_{\text{(EL)}} \mid \text{with } \{6j\} \neq 0 \right\}$. In order to characterise the extrema of the action, we define the space of extremal sequences. Let $C = \left( C^J \right)_{J \in \mathfrak{J}_{\text{(EL)}}}$ denote the sequence of (possibly complex) numbers such that for $J \in \mathfrak{J}_{\text{(EL)}}^S$

$$C^J \in \left\{ 0, \pm \frac{1}{|\{6j\}|} \sqrt{-\frac{3!}{\lambda} (J^2 + m^2)} \right\} \qquad (5.20)$$

and for $J \in \mathfrak{J}_{(\mathrm{EL})}/\mathfrak{J}_{(\mathrm{EL})}^S$

$$C^J = \begin{cases} r \in \mathbb{R} & \text{if } J^2 = -m^2 \\ 0 & \text{otherwise} \end{cases} \tag{5.21}$$

Since $J^2 > 0$, the first case in Eq. (5.21) can only be reached when $m^2$ is negative and for $J \in \mathfrak{J}_{\mathrm{EL}}$, $m^2$ has to be an even integer. For simplicity, we will exclude this case in the following analysis, because it requires a strong fine-tuning on the parameter $m^2$. It is convenient to define the length $\ell$ of the sequence $C$ such that

$$\ell(C) = \sum_{J \in \mathfrak{J}_{\mathrm{EL}}} \left| \mathrm{sgn}\left(C^J\right) \right|, \tag{5.22}$$

with the convention $\mathrm{sgn}(0) = 0$.

**Definition 2.** We define the space of extremal sequences as

$$\mathcal{E}_{m,\lambda} = \left\{ C = \left(C^J\right)_{J \in \mathfrak{J}_{\mathrm{EL}}} \mid C^J \in \mathbb{R}, \, \ell(C) < \infty \right\},$$

where the coefficients of each sequence are of the form Eq. (5.20).

This space of course depends on the values of $m^2$ and $\lambda$, since different choices of these parameters may violate the reality condition $C^J \in \mathbb{R}$. $\mathcal{E}_{m,\lambda}$ fully characterises the space of extrema of the action as stated by the following theorem.[13]

**Theorem 1.** *For any $C \in \mathcal{E}_{m,\lambda}$ the field $\varphi \in \mathcal{S}_{EL}$*

$$\varphi(g_1, g_2, g_3) = \sum_{J \in \mathfrak{J}_{EL}} C^J \, \mathcal{X}^J(g_1, g_2, g_3) \tag{5.23}$$

*is an extremum of the action $S_{m,\lambda}$. Moreover, every equilateral extremum of $S_{m,\lambda}$ is of the above form.*

*Proof.* See Appendix D.1.2.                                                         □

We denote the space of extremal functions by $\tilde{\mathcal{E}}_{m,\lambda}$. It is worth mentioning that, in spite of the non-linearity of the Euler-Lagrange equations, its solutions form a vector space over $(\mathbb{Z}_3, +, \cdot)$, see Appendix D.1.2.

---

[13]It has to be stressed that we are only looking for solutions with sequences of finite length in the remainder which excludes distributional solutions.

We now discuss the space of extremal sequences according to different parameter regions, the major difference of which is captured by the sign of the radicand in Eq. (5.20). We obtain the four cases:

(a) $m^2 < 0, \ \lambda < 0$ the radicand is positive only if

$$J^2 - |m^2| = 3j(j+1) - |m^2| \geq 0, \tag{5.24}$$

which is the case when $j$ satisfies

$$j_{\min} = \left\lceil \frac{1}{6} \left( \sqrt{9 + 12|m^2|} - 3 \right) \right\rceil \leq j, \tag{5.25}$$

where $\lceil \cdot \rceil$ denotes the ceiling function. The space of extremal sequences contains infinitely many sequences of the form

$$\left( 0, \ldots, 0, C^{J_{\min}}, C^{J_{\min}+1}, \ldots \right),$$

where we used the notation $J_{\min} + n \equiv (j_{\min} + n, j_{\min} + n, j_{\min} + n)$ for $n \in \mathbb{N}$, with finitely many non-zero elements $C^J$.

(b) $m^2 > 0, \ \lambda < 0$ all coefficients $C^J$ are real. The space of extremal sequences can be written as

$$\mathcal{E}_{m,\lambda} = \left\{ \left( C^{(0,0,0)}, C^{(1,1,1)}, \ldots \right) \mid \ell(C) < \infty \right\}.$$

(c) $m^2 > 0, \ \lambda > 0$ the reality condition $C^J \in \mathbb{R}$ then requires $C^J = 0$ for all $J \in \mathfrak{J}_{\mathrm{EL}}$. The space of extremal sequences contains a single zero-sequence

$$\mathcal{E}_{m,\lambda} = \{ (0, 0, 0, \ldots) \}.$$

(d) $m^2 < 0, \ \lambda > 0$ the radicand is positive only if

$$3j(j+1) - |m^2| \leq 0, \tag{5.26}$$

or equivalently for $j$ satisfying,

$$0 \leq j \leq \left\lfloor \frac{1}{6} \left( \sqrt{9 + 12 \left| m^2 \right|} - 3 \right) \right\rfloor = j_{\max} \tag{5.27}$$

where $\lfloor \cdot \rfloor$ denotes the floor function. In this case $\mathcal{E}_{m,\lambda}$ contains finitely many sequences of the form

$$\left( C^{(0,0,0)}, \dots, C^{J_{\max}}, 0, 0, \dots \right),$$

where $J_{\max} = (j_{\max}, j_{\max}, j_{\max}) \in \mathfrak{J}_{\text{EL}}$.

At this point, a few comments are in order: according to the geometrical interpretation in the previous subsection, each Fourier mode can be interpreted as a triangle with the edge length given by $j$. The area of the triangle is then measured in terms of $J^2$. In the parameter regime (*d*) relation Eq. (5.26) provides an upper bound on the possible $j$'s for the extrema of the action. Hence, in this case $\left| m^2 \right|$ can be interpreted as the bound on the area of the triangles determined by the extremal solutions. This is an interesting geometrical fact that deserves further investigation.[14]

### 5.2.2.2 Minimisers

We now seek the minimisers of the action and show that only two parameter regions admit global minimisers.

First, notice that in the case, $m^2 < 0$, $\lambda > 0$, the value of $\left| m^2 \right|$ can determine, whether or not the action $S_{m,\lambda}$ is bounded from below. To agree with this, assume the first non-trivial zero of the $6j$-symbol to be at $J_0 \in \mathfrak{J}_{\text{EL}}$ and choose a function $f(g_1, g_2, g_3) \equiv f^{J_0} \mathcal{X}^{J_0}(g_1, g_2, g_3)$ with $f^{J_0} \in \mathbb{R}$. Then, for $\left| m^2 \right| > J_0^2$ the action evaluated at $f$ yields

$$S_{m,\lambda}[f] = \left( f^{J_0} \right)^2 \, \left( J_0^2 - \left| m^2 \right| \right) < 0. \tag{5.28}$$

Hence, the action can become arbitrarily negative and thus is unbounded from below. On the other hand, for $\left| m^2 \right| < J_0^2$ the action has a global minimum as we will show in the following.

---

[14]A second remark is that the method of resolution restricting to equilateral configurations used to tackle Eq. (5.18) certainly exports to GFT models on higher dimensional manifolds $M = G^{\times D}$ with $G = \text{SU}(2)$, $\text{SO}(4)$ and $D \in \mathbb{N}$. We expect that a similar results as in Eq. (5.20) will hold if we replace the $6j$-symbol by the appropriate Wigner symbol and replace the square root by the $D - 2$ root.

In order to give a general classification of solutions we need to exclude cases when the $6j$-symbol vanishes. A quick numerical analysis shows that for $\left|m^2\right| \leq 10^9$, the space of non-trivial zeros of the $6j$-symbol with $J^2 \leq \left|m^2\right|$ is empty, Therefore, theorem 2 captures all possible solutions up to this order. In fact, we conjecture that for equilateral configurations, $\mathfrak{J}_{\mathrm{EL}}/\mathfrak{J}_{\mathrm{EL}}^S = \emptyset$, and our theorem holds for any value of $\left|m^2\right|$.

**Theorem 2.** *Let $\left|m^2\right|$ be such that for $j \leq j_{\max}$ every $J \in \mathfrak{J}_{EL}^S$ and such that there is no $J \in \mathfrak{J}_{EL}/\mathfrak{J}_{EL}^S$ such that $J^2 - \left|m^2\right| = 0$. Then the equilateral extrema of the dynamical Boulatov action are of the following type:*

(a) *For $m^2 < 0$, $\lambda < 0$, all extrema are saddle points.*

(b) *For $m^2 > 0$, $\lambda < 0$, all non-trivial extrema are saddle points and the trivial extremum, $\varphi = 0$, is a local minimiser on $\mathcal{S}_{EL}$.*

(c) *For $m^2 > 0$, $\lambda > 0$ the unique trivial extremum is a global minimiser on $\mathcal{S}_{EL}$.*

(d) *For $m^2 < 0$, $\lambda > 0$ there exist $2^{j_{\max}}$ global minimisers on $\mathcal{S}_{EL}$ given by extremal sequences $C \in \mathcal{E}_{m,\lambda}$ with maximal length, $\ell(C) = j_{\max}$. Any other extremum of length $\ell(C) < j_{\max}$ is a saddle point.*

*Proof.* See Appendix D.1.3. □

### 5.2.3 Discussion of the results

In view of the analysis of different GFT condensate cosmology models in the next chapter, we investigated the minimisers of the dynamical Boulatov action in four different parameter regions of the coupling constants. Our analysis is restricted to the space of smooth, equilateral, left and right invariant functions, also invariant under cyclic permutations of its variables, $\mathcal{S}_{\mathrm{EL}}$. This restriction ensures that the action is bounded from below for some parameter regions. It appears that the very same restrictions allow us to solve the Euler-Lagrange equations for the dynamical Boulatov action and lead to a complete characterisation of minimisers on the restricted space. These results directly apply to the condensate cosmology context when interpreting the field $\varphi$ as the mean field $\sigma$.

In the most interesting parameter region $(d)$, the non-vanishing Fourier modes of extremal solutions are bounded by the coupling constant $m^2$, which suggests a connection between $m^2$ and the area of the triangle of the largest Peter-Weyl mode of the GFT field.

In this region the action has $2^{j_{\max}}$ degenerate (non-trivial) global minimisers, where $j_{\max}$ is a function of the coupling constant $m^2$. The rich structure of global minima makes this region most interesting for further investigations, especially for the statistical theory and GFT condensate cosmology. Case $(c)$ admits a single global minimiser $\varphi = 0$. Perturbation theory around this minimiser defines the perturbation theory in the coupling constant $\lambda$ and can be used to draw a connection to the spin-foam expansion. Hence, our analysis would suggest that this regime is suitable for such relation. However, such a configuration would be rather uninteresting for cosmology studies in GFT since the mean particle number would vanish.

Case $(d)$, on the other hand, may suggest more structure for the quantum theory: a degenerate global minimum could lead to instantons or symmetry breaking in the corresponding statistical field theory. As regards symmetry breaking, this mechanism happens when the classical action admits degenerate global minimisers — related by a symmetry of the classical action — but the tunneling probability between them vanishes. The tunneling probability in ordinary field theory is proportional to the volume of the base manifold. On a manifold with finite volume the tunneling probability is therefore finite. This often pertains to the statement that spontaneous symmetry breaking can not occur in quantum field theories in a box. This realisation, however, contains further assumptions that are satisfied in ordinary field theories but do not hold for GFT. In fact, it has recently been shown that even on the compact base manifold, $M = \mathrm{SU}(2)^d$ the tunneling between different perturbative minima can vanish [458], leading to a phenomenon similar to symmetry breaking. In order to talk about symmetry breaking, we need to identify the symmetry, which in our case, is given by a flip of the sign of at least one of the modes in the Peter-Weyl decomposition of the minimiser (this can be modeled as a $\mathbb{Z}_2$-symmetry). Since the action is of even power in the fields, such a flip will not affect the value of the action and will correspond to a discrete symmetry. For this reason it is possible that the global minimisers of the action provoke the breaking of sign-flip symmetry. This needs to be investigated more rigorously in future work.

For ordinary local quantum field theories, a symmetry breaking mechanism can sometimes be related to a phase transition and the formation of a condensate. In particular, this could be the signal of a Bose-Einstein condensation just as expected for quantum cosmology studies in GFT. A closer look at the solutions found for sector $(d)$ shows that these might bear intriguing perspectives. Indeed, the "particle number", used in the condensate cosmology context, is computable in terms of the $L^2$-norm of the minimiser, i.e.

$$N = \|\varphi\|_{L^2}^2 \leq \frac{3! \, |m^2|}{\lambda} \, j_{\max} \, C_{\max}, \tag{5.29}$$

with $C_{\max} = \max_{J \in \mathfrak{J}_{\mathrm{EL}}^S} \left( \{6j\}^{-2} \right)$. In principle, this number can become large depending on the coupling constants. A large particle number would be desirable from the point of view of the condensate cosmology approach because such configurations could then be interpreted as to define non-trivial homogeneous and isotropic background geometries in $3d$ with Euclidean signature. This point deserves further attention.

We should mention here that minimisers with a divergent $L^2$-norm are not captured by our analysis (dealing only with integrable functions). One necessary modification would be to relax the smoothness condition of the minimisers and use the space of tempered distributions instead. This could be particularly interesting for GFT models without Laplace-Beltrami operator, which correspond to a topological BF-theory. Due to the distributional nature of its minimisers [503] their $L^2$-norm will diverge (which should relate them to non-Fock representations, see Section 4.2.4.3 and Appendix B) making them potentially interesting for quantum cosmological studies [457, 458] and continuum limit studies in general. The solutions to this case must be differently addressed but certainly deserve to be explored. For a more thorough discussion of the results of this Section, we refer to the original material in Ref. [498].

Given this analysis, we conclude that it is indeed possible to extract non-trivial solutions to the classical equation of motion from geometric GFTs which, in the context of the condensate cosmology framework as a mean field approach, may be put to use to model homogeneous and macroscopic geometries. We will come back to this point with the analysis of more realistic models in $4d$ in the next chapter.

## 5.3 Probing the condensate hypothesis with Landau's mean field theory

As recalled above, the GFT condensate cosmology approach rests on the conjecture that a continuum geometric phase could emerge through a transition to a condensate phase in an appropriate geometric GFT model [462]. This conjecture finds indirect support through the functional renormalisation group (FRG) analysis of related tensorial GFTs [393, 471–476]. Despite these successes, the analysis of the phase structure of GFT models with a proper simplicial gravity interpretation, either in the Euclidean or Lorentzian sector, remains an open problem. In particular, it would be important to settle the question whether a condensate phase indeed exists in such models.

In this Section we tackle this issue in terms of mean field techniques. Inspired by the phenomenological perspective of Laudau-Ginzburg mean field theory designed to describe second-order phase transitions, we explore if hints for a phase transition can be found without going through a non-perturbative analysis like the functional renormalisation group methodology. We apply Landau-Ginzburg theory to the case of the GFT model for three-dimensional Euclidean quantum gravity, augmented by a Laplace-Beltrami operator, hereafter called dynamical Boulatov model [377, 497] (as in Section 5.2), and then proceed to a conjugation-invariant rank-1 GFT model on $SL(2, \mathbb{R})$ with local interaction. We check the validity of this mean field approach in the supposed critical region by means of the Ginzburg criterion and find that it provides a trustable description of a phase transition for the non-compact sector of the model on $SL(2, \mathbb{R})$. In the case of the dynamical Boulatov model on compact domain, it proves insufficient to that end. There, non-perturbative methods remain necessary to analyse if a phase transition can occur or not. In this way, our work can be seen as a first step towards the same analysis of the Lorentzian version of the dynamical Boulatov model and, furthermore, as a precursor to a more involved treatment of these models by means of the functional renormalisation group [224, 466, 469, 504].

To this aim, this study is organised as follows: In Section 5.3.1 we first recapitulate Landau-Ginzburg mean field theory and the essence of the Gaussian approximation. This will serve as a template for the analysis for GFT models in the remainder of our work. We then discuss some relevant pecularities of phase transitions in GFT. In Section 5.3.2 we apply the mean field method to GFT models on a compact domain. Firstly, we analyse the relevance of gauge symmetry in the case of a rank-3 model on $SU(2)^3$ with quartic local interaction subject to right, left and right, as well as conjugation-invariance in Section 5.3.2.1. In a next step, we study the effect of non-locality in Section 5.3.2.2 in the case of a rank-1 toy model on $SU(2)$ with a convolution-type of interaction and conjugation invariance, which serves as a warm-up for the case of the dynamical Boulatov model right afterwards. There, we again discuss the cases where the field is subject to right, left and right, as well as conjugation invariance to demonstrate the independence of the results from the symmetries imposed. In Section 5.3.3 we then treat a rank-1 model on the non-compact group $SL(2, \mathbb{R})$ with conjugation invariance and quartic local interaction. We conclude in Section 5.3.4 with a summary and discussion of our results. Relevant details of harmonic analysis on the Lie groups $SU(2)$ and $SL(2, \mathbb{R})$ are supplemented in Appendix C. The work presented in this Section is largely based on the work of the author in Ref. [505].

### 5.3.1 Landau theory for group field theory

The aim of Section 5.3 is to understand phase transitions in GFT in terms of the Gaussian approximation as pioneered by Landau and Ginzburg [506]. Thus, we first recapitulate the general scheme, then remind on the peculiarities of GFT, and finally discuss the physical meaning of the application of Gaussian approximation to GFT.

#### 5.3.1.1 Landau's theory of phase transitions and the Gaussian approximation

In the following, we recapitulate the statistical properties of a scalar field $\varphi(\vec{x})$ on $\mathbb{R}^D$ at mean field level. We then introduce the Ginzburg criterion which allows us to test the validity of the mean field description when studying the critical behaviour of the system [504, 507–509].

The generating functional of all correlation functions

$$Z[J] \equiv e^{W[J]} = \int \mathcal{D}\varphi e^{-S[\varphi] + \int d^D x \ J\varphi} \tag{5.30}$$

with external source $J$ defines the statistical field theory of $\varphi$. $W[J]$ is the generating functional for the connected correlation functions and

$$S = \frac{1}{2} \int d^D x \ \varphi(\vec{x}) \left(-\Delta + m^2\right) \varphi(\vec{x}) + \frac{\lambda}{4!} \int d^D x \ \varphi(\vec{x})^4 \tag{5.31}$$

denotes the bare action. The connected (two-point) correlation function $C$ is given by

$$C(\vec{x} - \vec{x}') = \frac{\delta^2 W[J]}{\delta J(\vec{x})\delta J(\vec{x}')}\bigg|_{J=0} \tag{5.32}$$

depending only on the relative coordinate $\vec{x} - \vec{x}'$ due to translation invariance.

One arrives at *Landau's mean field approximation* when estimating the functional integral $Z[0]$ through the saddle point method for a uniform field configuration $\varphi_0$. It minimises the bare action, that is, solves the classical equations of motion obtained from $S[\varphi_0]$ without source:

$$\varphi_0 = 0 \text{ if } m^2 > 0 \text{ and } \varphi_0 = \pm\sqrt{-\frac{m^2}{\lambda/3!}} \text{ if } m^2 < 0. \tag{5.33}$$

In the *Gaussian approximation* quadratic fluctuations around the saddle point are retained in $S[\varphi]$. Their correlation function $C$ is given by the inverse of

$$\delta_\varphi^2 S\big|_{\varphi_0} = -\Delta + m^2 \text{ if } m^2 > 0 \text{ and } -\Delta - 2m^2 \text{ if } m^2 < 0. \tag{5.34}$$

Equivalently, one can obtain the correlation function from the classical equation of motion with source term in terms of the linearisation $\varphi(\vec{x}) \to \varphi_0 + \delta\varphi(\vec{x})$ and $J(\vec{x}) \to J(\vec{x}) + \delta J(\vec{x})$ [507]. This leads to the differential equation

$$\left(-\Delta + m^2\right)\delta\varphi(\vec{x}) + \frac{\lambda\varphi_0^2}{2}\delta\varphi(\vec{x}) = \delta J(\vec{x}), \tag{5.35}$$

which we may solve by means of the Green's function method. Using the response relation

$$\delta\varphi(\vec{x}) = \int \mathrm{d}^D x' \; C(\vec{x} - \vec{x}')\delta J(\vec{x}'), \tag{5.36}$$

the equation of motion Eq. (5.35) rewrites as

$$\left(-\Delta + m^2 + \frac{\lambda\varphi_0^2}{2}\right)C(\vec{x}) = \delta(\vec{x}), \tag{5.37}$$

which can be solved in Fourier space and leads to an exponentially decaying function in position space.

Given the structure of the effective propagator, the correlation length $\xi$ is defined by

$$\xi^{-2} = m^2 + \frac{\lambda \varphi_0^2}{2} \overset{\text{Eq. (5.33)}}{=} \begin{cases} m^2 & , \ m^2 > 0 \\ -2m^2 & , \ m^2 < 0 \end{cases} \tag{5.38}$$

setting the scale beyond which the exponential decay in $||\vec{x} - \vec{x}'||$ sets in. At the second-order phase transition $\xi \to \infty$ and the correlation function obeys a power-law behaviour.

A test for the validity of the description of the phase transition in terms of the Gaussian approximation is to quantify the strength of the field fluctuations relative to the mean field value in terms of the quantity

$$Q = \frac{\int_\xi d^D x \ C(\vec{x} - \vec{x}')}{\int_\xi d^D x \ \varphi_0^2}, \tag{5.39}$$

defined in the supposedly broken phase. In general, the approximation is self-consistent and deemed trustworthy if $Q \ll 1$ for large $\xi$, that means fluctuations are small along all scales. This condition is the so-called *Ginzburg criterion* [504]. In contrast, the approximation breaks down if fluctuations are large, i.e. $Q \gg 1$, necessitating a non-perturbative treatment instead. On flat space $\mathbb{R}^D$, the asymptotic behaviour for large $\xi$ is

$$Q \sim \lambda \xi^{4-D} \tag{5.40}$$

from which one deduces the breakdown of the Gaussian approximation for the description of the phase transition below the critical dimension $D_c = 4$.

### 5.3.1.2   Phase transition in group field theory

Our goal now is to analyse the effect of the various peculiar properties of GFT, that is the Lie group $G$, the rank $d$, the group symmetry as well as the combinatorial non-locality, on phase transitions in the Gaussian approximation and check its validity via the Ginzburg criterion closely following the exposition of Section 5.3.1.1.

Before we start, we briefly discuss the physical meaning of such a phase transition in GFT which is a research question in its own. Technically, it is rather straightforward to apply Landau theory as outlined in Section 5.3.1.1 to GFT. However, the original meaning of the scalar field effectively describing degrees of freedom on physical space in condensed-matter physics does not apply here. This poses a challenge in particular to the concept of correlation length.

In GFT, spacetime itself is generated as the superposition of geometric configurations in correspondence to discrete geometries in terms of the perturbative expansion of the path integral. As known from matrix models, physically, the most relevant aspect of a phase transition then is that it may describe the critical subspace in coupling space at which an infinite number of such configurations contribute. Approaching the point of phase transition then has the meaning of a limit to continuum spacetime.[15] Complementarily, if different phases exist on the critical subspace itself, there should also be phase transitions between these as for example in (causal) dynamical triangulations [233]), matrix models [511] and tensor models [389, 392], reviewed in the previous chapter.

In GFT the meaning of 'correlation length' is completely different to the usual notion in condensed-matter physics. There, the correlation length $\xi$ is the scale beyond which correlation functions $C(\vec{x} - \vec{x}')$ on space $\vec{x}, \vec{x}' \in \mathbb{R}^D$ decay exponentially. Contrary, the GFT configuration space $G^{\times d}$ is related to parallel transports of the gravitational field through the $d$ boundaries of a simplicial building block of $d$-dimensional spacetime. Parallel transports capture the curvature of spacetime geometry. Thus, a distance on this space describes, roughly speaking, a difference in local curvature. The correlation length then describes the difference of modes with respect to local curvature.

Applying Landau theory to GFT, we here consider phase transitions characterised by arbitrary large correlation length on group space. At this point, fluctuations of arbitrary different group variables, that is parallel transports of the gravitational field, contribute equally to the dynamics. This is the same physical setting as investigated by functional renormalisation group techniques [393, 471–476]. However, the relation to the discrete-to-continuum limit of GFT or tensor models is not obvious. From the physical perspective of the discrete geometries, another possibility is that such a phase transition should be characterised by arbitrary large fluctuations in GFT momentum space given by group representations since these are the eigenmodes of length, area or volume operators of such geometries [512]. While this has been explored in spin-foam models [513], the usual GFT propagator does not allow for such a notion of correlation length.

---

[15] In tensor models there are examples where such a discrete-to-continuum phase transition can be made precise and related to the spontaneous breaking of unitary symmetry [510]. To this end, a description of the tensor model in the intermediate-field representation as a multi-matrix model is used and perturbations around the non-trivial matrix vacuum are studied.

Even with a correlation length on GFT configuration space there remain some ambiguities. Throughout this work we use Eq. (5.38) as a definition for $\xi$ as we consider GFT with the standard kinetic term with Laplacian and mass contribution. However, for a compact group $G$ with compactness scale $a$, correlations can only decay for geodesic distances between $\xi$ and $a$ such that quantities like Ginzburg's measure for fluctuations Eq. (5.39) applied to GFT,

$$Q = \frac{\prod_{i=1}^d \int_{\xi_i} \mathrm{d}g_i \; C(g_1, ..., g_d)}{\prod_{i=1}^d \int_{\xi_i} \mathrm{d}g_i \varphi_0^2} \tag{5.41}$$

are meaningful only for $\xi$ large but smaller than $a$ (cf. [509]). Furthermore, we integrate all single copies $i = 1, ..., d$ of $G$ up to $\xi_i = \xi$ (like on $\mathbb{R}^D$ one integrates over a $D$-cube with edge length $\xi$ [504]). While only a full physical theory of phase transitions in GFT can justify these choices eventually, our Landau analysis already clarifies for the first time various aspects of such transitions through the very necessity to consider the notion of correlation length in GFT.

### 5.3.2   GFTs on a compact domain in the Gaussian approximation

In this Section, we firstly discuss a rank-3 GFT on $\mathrm{SU}(2)^3$ with an ordinary quartic local interaction with right, left and right as well as conjugation invariance in the Gaussian approximation. Right invariance is the standard symmetry in GFT, as explained in Section 4.2.4. If an additional left invariance is imposed, the field domain can be related to the space of homogeneous 2-geometries [480], as follows from the discussion in Section 5.1.1. The fixing to conjugation invariance then gives a special case of this scenario.

Secondly, we explore the effect of non-locality of interactions in two cases in Section 5.3.2.2. The first is a rank-1 toy model on $\mathrm{SU}(2)$ endowed with conjugation invariance with a convolution-type interaction.[16] This model shows already the essential features of non-locality. In this way, it sets the stage for the analysis of the more relevant dynamical Boulatov model right afterwards.

#### 5.3.2.1   A rank-3 model with a quartic local interaction

We start off with a local GFT model for a real-valued field $\varphi$ living on three copies of the Lie group $G = \mathrm{SU}(2)$ subject to different types of invariance as defined below. The model

---

[16] We would like to thank E. Livine for suggesting to us to study this exemplary toy model prior to the more involved case of the dynamical Boulatov model.

has a quartic local interaction given by

$$S[\varphi] = \frac{1}{2} \int (\mathrm{d}g)^3 \varphi(g_1, g_2, g_3) \left(-\Delta + m^2\right) \varphi(g_1, g_2, g_3) + \frac{\lambda}{4!} \int (\mathrm{d}g)^3 \varphi(g_1, g_2, g_3)^4. \quad (5.42)$$

Minimisation of this functional leads to

$$(-\Delta + m^2)\varphi(g_1, g_2, g_3) + \frac{\lambda}{3!}\varphi(g_1, g_2, g_3)^3 = 0. \quad (5.43)$$

In the mean field approximation, for uniform field configurations it is simply solved by

$$\varphi_0 = 0 \text{ if } m^2 > 0 \text{ and } \varphi_0 = \pm\sqrt{-\frac{m^2}{\lambda/3!}} \text{ for } m^2 < 0. \quad (5.44)$$

In the Gaussian approximation, one considers fluctuations around this background. Inserting $\varphi \to \varphi_0 + \delta\varphi$ and $J \to J + \delta J$ in Eq. (5.43) with additional source $J$ and keeping terms to linear order in $\delta\varphi$, we find

$$\left(-\Delta + m^2 + \frac{\lambda\varphi_0^2}{2!}\right)\delta\varphi(g_1, g_2, g_3) = \delta J(g_1, g_2, g_3). \quad (5.45)$$

We solve this equation using the Green's function method. To this aim, we introduce the response relation for the group field

$$\delta\varphi(g_1, g_2, g_3) = \int (\mathrm{d}h)^3 \; C(g_1 h_1^{-1}, g_2 h_2^{-1}, g_3 h_3^{-1})\delta J(h_1, h_2, h_3), \quad (5.46)$$

This leads to

$$\left(-\Delta + m^2 + \frac{\lambda}{2!}\varphi_0^2\right) C(g_1, g_2, g_3) = \delta(g_1, g_2, g_3), \quad (5.47)$$

which we solve in the spin representation in the next subsections. For this, we exploit the fact that the $\delta$-function can be expanded in terms of group characters $\chi^j$ for each representation labelled by half integers $j \in \mathbb{N}/2$, i.e.,

$$\delta(g_1, g_2, g_3) = \sum_{j_1, j_2, j_3} \prod_{i=1}^{3} d_{j_i} \chi^{j_i}(g_i). \quad (5.48)$$

For details regarding the Fourier decomposition on SU(2), we refer to Appendix C.2.

● **Mono-invariance**

At first, we consider GFT with invariance under the right diagonal action of the group, i.e.

$$\varphi(g_1, g_2, g_3) = \varphi(g_1 r, g_2 r, g_3 r), \ \forall g_i, r \in \mathrm{SU}(2) \tag{5.49}$$

which is imposed via group averaging. Hence, the field may be decomposed as

$$\varphi(g_1, g_2, g_3) = \sum_{m_i, n_i, j_i} \varphi^{j_1 j_2 j_3}_{m_1 n_1 m_2 n_2 m_3 n_3} \int \mathrm{d}r \prod_{i=1}^{3} d_{j_i} D^{j_i}_{m_i n_i}(g_i r)$$

$$= \sum_{m_i, \alpha_i, j_i} \varphi^{j_1 j_2 j_3}_{m_1 m_2 m_3} \begin{pmatrix} j_1 & j_2 & j_3 \\ \alpha_1 & \alpha_2 & \alpha_3 \end{pmatrix} \prod_{i=1}^{3} d_{j_i} D^{j_i}_{m_i \alpha_i}(g_i), \tag{5.50}$$

in terms of 3$j$-symbols $\begin{pmatrix} j_1 & j_2 & j_3 \\ \alpha_1 & \alpha_2 & \alpha_3 \end{pmatrix}$ and with modes

$$\varphi^{j_1 j_2 j_3}_{m_1 m_2 m_3} = \sum_{n_1, n_2, n_3} \varphi^{j_1 j_2 j_3}_{m_1 n_1 m_2 n_2 m_3 n_3} \overline{\begin{pmatrix} j_1 & j_2 & j_3 \\ n_1 & n_2 & n_3 \end{pmatrix}}. \tag{5.51}$$

With this symmetry imposed, for $m^2 < 0$ the solution to Eq. (5.47) reads as

$$C(g_1, g_2, g_3) = \sum_{m_i, \alpha_i, j_i} C^{j_1 j_2 j_3}_{m_1 m_2 m_3} \begin{pmatrix} j_1 & j_2 & j_3 \\ \alpha_1 & \alpha_2 & \alpha_3 \end{pmatrix} \prod_{i=1}^{3} d_{j_i} D^{j_i}_{m_i \alpha_i}(g_i) \tag{5.52}$$

where the Fourier coefficients are given by

$$C^{j_1 j_2 j_3}_{m_1 m_2 m_3} = \frac{\overline{\begin{pmatrix} j_1 & j_2 & j_3 \\ m_1 & m_2 & m_3 \end{pmatrix}}}{\sum_{i=1}^{3} j_i(j_i + 1) - 2m^2}. \tag{5.53}$$

To evaluate the strength of the fluctuations relative to the average field in the supposed region of criticality, we have to compute Eq. (5.41) for large $\xi$. However, due to the compactness of SU(2) it does not make sense to consider $\xi > \pi$ as $a = \pi$ is the maximal possible geodesic distance. Thus, we are interested in the "asymptotic" behaviour of $Q$ for large $\xi < a$, that is $\xi$ close to $\pi$. For this reason it is sufficient to compute the integrals at first simply over the entire SU(2)$^3$-domain. If we integrate Eq. (5.52) in this way and use the orthogonality relation of the Wigner matrices for each SU(2)-integration, we observe that only the zero-mode $C_{000}^{000} = 1/2|m^2|$ will contribute to $Q$. Since all modes $j > 0$ yield continuous oscillations which are zero at $\xi = \pi$, the part of the zero mode is indeed dominant for $\xi$ close to $\pi$. For the zero mode we can then perform the integration up to $\xi < \pi$ exactly and find

$$Q \sim \frac{-1/m^2}{\varphi_0^2} \sim \lambda \xi^4. \tag{5.54}$$

For large (but smaller than $a$) correlation lengths $\xi^2 = -\frac{1}{2m^2}$ this expression becomes large, indicating the invalidation of the Gaussian approximation in the region of the expected phase transition.

In fact, for given $\xi < a$ there are always bare couplings $\lambda \ll a^{-4}$ such that $Q \ll 1$ despite being a power function in $\xi$. For $a = \pi$ this is the case for $\xi \ll 10^{-2}$, and this value becomes even smaller for larger compactness scales $a$. Of course, the actual value of the coupling could only be determined by experiment. However, the very concept of second-order phase transitions relies on the possibility of correlation lengths $\xi$ to become very large in a physical sense (though described mathematically by asymptotics, physically it is sufficient if they are much larger than the fluctuations around the ground states of the different phases). Thus, if there is such a phase transition on a compact space, then $\xi$ and as a consequence $Q$ becomes very large indicating the breakdown of the Gaussian approximation.

● **Bi-invariance**

As a second case, we impose invariance with respect to left and right diagonal action of the group on the group field, i.e.

$$\varphi(g_1, g_2, g_3) = \varphi(lg_1r, lg_2r, lg_3r), \ \forall g_i, l, r \in \text{SU}(2), \tag{5.55}$$

implemented via group averaging. Hence, the field may be decomposed as

$$
\begin{aligned}
\varphi(g_1, g_2, g_3) &= \sum_{m_i, n_i, j_i} \varphi_{m_1 n_1 m_2 n_2 m_3 n_3}^{j_1 j_2 j_3} \int \text{d}l \int \text{d}r \prod_{i=1}^{3} d_{j_i} D_{m_i n_i}^{j_i}(lg_i r) \\
&= \sum_{\alpha_i, \beta_i, j_i} \varphi^{j_1 j_2 j_3} \overline{\begin{pmatrix} j_1 & j_2 & j_3 \\ \alpha_1 & \alpha_2 & \alpha_3 \end{pmatrix}} \begin{pmatrix} j_1 & j_2 & j_3 \\ \beta_1 & \beta_2 & \beta_3 \end{pmatrix} \prod_{i=1}^{3} d_{j_i} D_{\alpha_i \beta_i}^{j_i}(g_i) \\
&= \sum_{j_1, j_2, j_3} \varphi^{j_1 j_2 j_3} \int \text{d}h \prod_{i=1}^{3} d_{j_i} \chi^{j_i}(g_i h),
\end{aligned}
\tag{5.56}
$$

where

$$\varphi^{j_1 j_2 j_3} = \varphi^{j_1 j_2 j_3}_{m_1 n_1 m_2 n_2 m_3 n_3} \begin{pmatrix} j_1 & j_2 & j_3 \\ m_1 & m_2 & m_3 \end{pmatrix} \overline{\begin{pmatrix} j_1 & j_2 & j_3 \\ n_1 & n_2 & n_3 \end{pmatrix}}. \tag{5.57}$$

With this symmetry imposed, for the sector $m^2 < 0$ the solution to Eq. (5.47) reads as

$$C(g_1, g_2, g_3) = \sum_{j_1, j_2, j_3} C^{j_1 j_2 j_3} \int \mathrm{d}h \prod_{i=1}^{3} d_{j_i} \chi^{j_i}(g_i h), \tag{5.58}$$

where the Fourier coefficients are given by

$$C^{j_1 j_2 j_3} = \frac{1}{\sum_{i=1}^{3} j_i(j_i + 1) - 2m^2}. \tag{5.59}$$

To evaluate $Q$ in this case, we use the same argument as in the previous subsection. The only difference is that we employ the orthogonality of the characters for each SU(2)-integration to find again that only the zero-mode will contribute when integrating Eq. (5.58) over SU(2)$^3$. Again we find that the zero-mode thus dominates for $\xi$ large (but smaller than $\pi$) where we obtain

$$Q \sim \lambda \xi^4, \tag{5.60}$$

indicating the invalidation of the Gaussian approximation in this region.

● **Conjugation invariance**

Now we consider the case where the field is subject to conjugation invariance, i.e.

$$\varphi(g_1, g_2, g_3) = \varphi(k g_1 k^{-1}, k g_2 k^{-1}, k g_3 k^{-1}), \tag{5.61}$$

which holds for all $g_i$ and $k$ in SU(2). Hence, $\varphi$ is a central function on the domain and can be decomposed in terms of characters, so we write

$$\varphi(g_1, g_2, g_3) = \sum_{j_1, j_2, j_3} \varphi^{j_1 j_2 j_3} \prod_{i=1}^{3} d_{j_i} \chi^{j_i}(g_i). \tag{5.62}$$

With this symmetry, for $m^2 < 0$ the solution to Eq. (5.47) is

$$C(g_1, g_2, g_3) = \sum_{j_1, j_2, j_3} C^{j_1 j_2 j_3} \prod_{i=1}^{3} d_{j_i} \chi^{j_i}(g_i), \tag{5.63}$$

where the Fourier coefficients are again

$$C^{j_1 j_2 j_3} = \frac{1}{\sum_{i=3}^{3} j_i(j_i + 1) - 2m^2}. \tag{5.64}$$

The computation of $Q$ follows along the lines of the previous subsections, leading to

$$Q \sim \lambda \xi^4. \tag{5.65}$$

Again, this entails $Q \gg 1$ in the supposedly critical region.

We conclude that the Gaussian approximation does not provide a trustable description of a phase transition for the present model subject to the different symmetries. Furthermore, though we have chosen rank $d = 3$ here, it is obvious from the calculations that the result generalises to arbitrary rank $d$.

The peculiar form of $Q$ is similar to that found for a scalar field with a quartic local interaction on $S^d$ in Ref. [509]. There it is furthermore demonstrated through a functional renormalisation group analysis that the $\mathbb{Z}_2$-symmetry is always restored in the IR and no phase transition takes place. Such a result might also be found for the models considered here. However, their full non-perturbative analysis is beyond the scope of this Section and will be treated elsewhere.

### 5.3.2.2 Models with a quartic non-local interaction

Now we explore the effect of combinatorial non-locality on the validity of the Gaussian approximation. To this end, we consider two models with non-local quartic interactions, first a rank-1 toy model and second the dynamical Boulatov model. We find similar results for the Ginzburg criterion as for the local model in the preceding Section 5.3.2.1.

### (A) A rank-1 toy model

It is possible to mimic the non-local pairing of field arguments in GFT already for a field with single argument in terms of a non-commutative convolution product. Thus, we consider a real-valued field $\varphi$ on one copy of $G = \mathrm{SU}(2)$ which is subject to conjugation invariance and has dynamics given by the action

$$S[\varphi] = \frac{1}{2} \int \mathrm{d}g \varphi(g) \left( -\Delta + m^2 \right) \varphi(g) + \frac{\lambda}{4!} \int \mathrm{d}g [\varphi \star \varphi \star \varphi \star \varphi](g) \tag{5.66}$$

wherein the convolution product $\star$ is defined via

$$[\varphi \star \varphi](g) = \int \mathrm{d}h \; \varphi(h)\varphi(gh^{-1}) \tag{5.67}$$

such that the quartic convolution expands into

$$[\varphi \star \varphi \star \varphi \star \varphi](g) = \int \mathrm{d}h \int \mathrm{d}k \int \mathrm{d}l \; \varphi(h)\varphi(kh^{-1})\varphi(lk^{-1})\varphi(gl^{-1}). \tag{5.68}$$

Such an interaction already captures the essential aspects of combinatorial non-locality.

In a first step, we again compute the equation of motion, given by

$$
\begin{aligned}
0 = (-\Delta + m^2)\varphi(g) + \frac{\lambda}{4!} \int \mathrm{d}h \int \mathrm{d}k \int \mathrm{d}l \bigg( & \varphi(kh^{-1})\varphi(lk^{-1})\varphi(gl^{-1}) \\
+ \; & \varphi(h^{-1}k)\varphi(lk^{-1})\varphi(gl^{-1}) + \varphi(h^{-1}k)\varphi(k^{-1}l)\varphi(gl^{-1}) \\
+ \; & \varphi(h^{-1}k)\varphi(k^{-1}l)\varphi(l^{-1}g) \bigg).
\end{aligned}
\tag{5.69}
$$

For uniform field configurations $\varphi_0$ the non-locality is washed away and the solution is the same as in the local case,

$$\varphi_0 = 0 \text{ if } m^2 > 0 \text{ and } \varphi_0 = \pm\sqrt{-\frac{m^2}{\lambda/3!}} \text{ for } m^2 < 0. \tag{5.70}$$

In the Gaussian approximation, however, the non-locality is retained to a certain degree. To show this, we linearise Eq. (5.69) with additional source $J$ via the insertion $\varphi \to \varphi_0 + \delta\varphi$ and $J \to J + \delta J$ while only keeping terms up to linear order in $\delta\varphi$. For $m^2 < 0$, this leads to the following integro-differential equation

$$(-\Delta + m^2)\delta\varphi(g) - 3m^2 \int \mathrm{d}l \; \delta\varphi(gl^{-1}) = \delta J(g), \tag{5.71}$$

where the integral term is actually constant due to the properties of the Haar measure. We tackle it using the Green's function method and to this aim introduce the response relation

$$\delta\varphi(g) = \int \mathrm{d}h \; C(gh^{-1})\delta J(h). \tag{5.72}$$

With this we obtain

$$(-\Delta + m^2)C(g) - 3m^2 \int \mathrm{d}l \; C(gl^{-1}) = \delta(g),$$ (5.73)

which we solve in Fourier space. Because of conjugation invariance the correlation function decomposes into

$$C(g) = \sum_j C^j d_j \chi^j(g).$$ (5.74)

Using this and the orthogonality relation for the characters (see Appendix C.2), the integral in Eq. (5.73) simply contributes a zero-mode $C^0$ such that the solution is

$$C^j = \frac{1}{j(j+1) + m^2 - 3m^2 \frac{\delta_{j0}}{d_j}}.$$ (5.75)

Hence, in the Gaussian approximation the correlation function obtains a mild modification due to the non-locality of the interaction. Comparing to the local case Eq. (5.64), the zero mode is the same while for modes $j > 0$ there is a mass term $m^2$ instead of $-2m^2 = 2|m^2|$. Still, the argument for the dominance of the zero mode applies such that we again find the large-$\xi$ behaviour

$$Q \sim \lambda \xi^4.$$ (5.76)

Oscillations are stronger by a factor 2 and have opposite sign as compared to the local case, but they remain irrelevant at large $\xi$.

### (B) The dynamical Boulatov model

The dynamical Boulatov model [377, 497] is a GFT with real-valued field $\varphi$ on three copies of SU(2) with a simplicial quartic interaction

$$S[\varphi] = \frac{1}{2} \int (\mathrm{d}g)^3 \varphi_{123} \left( -\Delta + m^2 \right) \varphi_{123} + \frac{\lambda}{4!} \int (\mathrm{d}g)^6 \; \varphi_{123}\varphi_{145}\varphi_{256}\varphi_{364}$$ (5.77)

where we abbreviate $\varphi_{123} \equiv \varphi(g_1, g_2, g_3)$ etc. from now on. As in the previous section, the action is endowed with an invariance with respect to the right diagonal action of the group SU(2).[17] The action is constructed such that the perturbative expansion of the generating functional around the Fock vacuum is equivalent to the Ponzano-Regge spin foam model [35, 271] which provides a discrete version of the path integral for three-dimensional Euclidean quantum gravity.

---

[17]In the original definition of the Boulatov model there furthermore is an invariance of the field with respect to cyclic permutations of its arguments [377]. This symmetry has no bearing on the ensueing arguments.

As in the other cases, we first compute the equation of motion,

$$(-\Delta + m^2)\varphi_{123} + \frac{\lambda}{3!} \int \mathrm{d}g_4 \mathrm{d}g_5 \mathrm{d}g_6 \varphi_{146}\varphi_{526}\varphi_{543} = 0. \tag{5.78}$$

The projection onto uniform field configurations $\varphi_0$ is not sensitive to the combinatorial non-locality. Thus, it is solved again by

$$\varphi_0 = 0 \text{ if } m^2 > 0 \text{ and } \varphi_0 = \pm\sqrt{-\frac{m^2}{\lambda/3!}} \text{ for } m^2 < 0. \tag{5.79}$$

Turning to the Gaussian approximation, the effect of the non-locality appears for small deviations around this constant background. To see this, we linearise Eq. (5.78) with additional source $J$ via inserting $\varphi \to \varphi_0 + \delta\varphi$ and $J \to J + \delta J$ yielding

$$(-\Delta + m^2)\delta\varphi_{123} + \frac{\lambda}{3!}\varphi_0^2 \int \mathrm{d}g_4 \mathrm{d}g_5 \mathrm{d}g_6 (\delta\varphi_{146} + \delta\varphi_{526} + \delta\varphi_{543}) = \delta J_{123}. \tag{5.80}$$

We solve this integro-differential equation using again the response relation Eq. (5.46) such that

$$\left[ -\Delta + m^2 + \frac{\lambda}{3!}\varphi_0^2 \left( \int \mathrm{d}g_2 \mathrm{d}g_3 + \int \mathrm{d}g_1 \mathrm{d}g_3 + \int \mathrm{d}g_1 \mathrm{d}g_2 \right) \right] C(g_1 h_1^{-1}, g_2 h_2^{-1}, g_3 h_3^{-1})$$

$$= \prod_{i=1}^{3} \delta(g_i h_i^{-1}). \tag{5.81}$$

In the following, we solve this equation in Fourier space for three types of invariance. To deal with the integral kernel, we use the orthogonality relation of the Wigner matrices and characters, see Appendix C.2, in the same way as for the toy model just treated.

● **Mono-invariance**

For field configurations simply endowed with the invariance with respect to the right diagonal action, the correlation function is

$$C(g_1, g_2, g_3) = \sum_{m_i, \alpha_i, j_i} C_{m_1 m_2 m_3}^{j_1 j_2 j_3} \begin{pmatrix} j_1 & j_2 & j_3 \\ \alpha_1 & \alpha_2 & \alpha_3 \end{pmatrix} \prod_{i=1}^{3} d_{j_i} D_{m_i, \alpha_i}^{j_i}(g_i). \qquad (5.82)$$

The Fourier coefficients are given by

$$C_{m_1 m_2 m_3}^{j_1 j_2 j_3} = \frac{\overline{\begin{pmatrix} j_1 & j_2 & j_3 \\ m_1 & m_2 & m_3 \end{pmatrix}}}{\sum_{i=1}^{3} j_i(j_i+1) + m^2 - 3m^2 A}, \qquad (5.83)$$

with

$$A = \sum_{i<k} \frac{\delta_{j_i 0}}{d_{j_i}} \delta_{m_i 0} \delta_{\alpha_i 0} \frac{\delta_{j_k 0}}{d_{j_k}} \delta_{m_k 0} \delta_{\alpha_k 0} = \delta_{j_2 0} \delta_{j_3 0} + \delta_{j_1 0} \delta_{j_2 0} + \delta_{j_1 0} \delta_{j_3 0}. \qquad (5.84)$$

This term leads to a mild modification of the correlation function in the Gaussian approximation similar to the toy model above which is due to the non-locality of the interaction. Hence, we also find the same result for the relative fluctuations $Q$ up to the numerical factors given by $3A$ which slightly modify the amplitude of $j_i > 0$ mode oscillations but do not influence the large-$\xi$ behaviour $Q \sim \lambda \xi^4$ due to the zero mode.

● **Bi-invariance**

For field configurations endowed with the invariance with respect to the left and right diagonal action, the correlator is given by

$$C(g_1, g_2, g_3) = \sum_{j_1, j_2, j_3} C^{j_1 j_2 j_3} \int \mathrm{d}h \prod_{i=1}^{3} d_{j_i} \chi^{j_i}(g_i h). \qquad (5.85)$$

Its Fourier coefficients are

$$C^{j_1 j_2 j_3} = \frac{1}{\sum_{i=1}^{3} j_i(j_i+1) + m^2 - 3m^2 B}, \qquad (5.86)$$

with

$$B = 3 \frac{\delta_{j_1 0}}{d_{j_1}} \frac{\delta_{j_2 0}}{d_{j_2}} \frac{\delta_{j_3 0}}{d_{j_3}}. \qquad (5.87)$$

Due to the non-locality of the interaction, the last term again gives a mild modification of the correlation function in the Gaussian approximation. Its particular form varies from that of the previous case due to the different symmetry imposed onto the field. Still the qualitative behaviour is the same and $Q$ takes the form of Eq. (5.76).

● **Conjugation invariance**

For field configurations subject to conjugation invariance, the solution to Eq. (5.81) expands as

$$C(g_1, g_2, g_3) = \sum_{j_1, j_2, j_3} C^{j_1 j_2 j_3} \prod_{i=1}^{3} d_{j_i} \chi^{j_i}(g_i),$$  (5.88)

with Fourier coefficients given by

$$C^{j_1 j_2 j_3} = \frac{1}{\sum_{i=3}^{3} j_i(j_i + 1) + m^2 - 3m^2 C},$$  (5.89)

and

$$C = \left( \frac{\delta_{j_2 0}}{d_{j_2}} \frac{\delta_{j_3 0}}{d_{j_3}} + \frac{\delta_{j_1 0}}{d_{j_1}} \frac{\delta_{j_2 0}}{d_{j_2}} + \frac{\delta_{j_1 0}}{d_{j_1}} \frac{\delta_{j_3 0}}{d_{j_3}} \right).$$  (5.90)

Again, we find a mild modification of the correlation function in the Gaussian approximation due to the non-locality of the interaction. Qualitatively, it yields the same result as in the other cases.

We may conclude that, following Landau's strategy, the non-local interactions treated here have no relevant effect on the singular behaviour of $Q$. Hence, the Gaussian approximation cannot be trusted to give a valid description of a phase transition for these models. Non-perturbative methods have to be applied to settle the question if a phase transition can take place for these.

This result generalises not only to simplicial interactions of different rank but also to other types of non-locality such as tensor-invariant interactions (as studied for example in Refs. [393, 471–476]). The reason is that they all lead to integro-differential equations of the type of Eq. 5.80 differing only in the specific structure of integrations. For tensor-invariant interactions one has for example terms with a different number of integrations. But the result is only a different specific form of $\delta_{j0}$ terms in the modification of the representation-space propagator (like the terms $A$, $B$ and $C$ above). These are only responsible for the slight modification of higher-mode oscillations but do not alter the dominant zero-mode contribution.

Concluding this Section on GFT on a compact group, it is important to emphasise once more that it is solely the zero mode of fields on compact manifolds which causes the breakdown of the Gaussian approximation as a description for the theory at phase transition. From the perspective of loop quantum gravity one might alternatively be interested in a modified GFT excluding these zero modes. This is because due to cylindrical equivalence edges with variable $j = 0$ are equivalent to no edge at all in the construction of the kinematical Hilbert space in terms of cylindrical functions on embedded graphs.[18] For such a modified GFT, the result of Landau-Ginzburg theory is possibly completely the opposite, that is, the Gaussian approximation could be valid.

### 5.3.3 GFT on a non-compact domain in the Gaussian approximation

To overcome the issue of large Gaussian fluctuations in GFT on compact configuration space the natural consequence is to consider GFTs with non-compact groups. From a quantum gravity perspective they are also more interesting since they provide models with Lorentzian signature. However, the application of Landau theory to the Lorentzian dynamical Boulatov model is not straightforward. A geometric GFT model for Lorentzian spacetimes in $3d$ has to be based on three copies of the Lie group $SL(2,\mathbb{R})$. Due to non-compactness already the bare GFT action in Lorentzian signature is only well-defined upon regularisation. This is because the imposition of the right invariance yields spurious integrations over at least one copy of $SL(2,\mathbb{R})$ leading to group volume divergences [441]. Next to the increased degree of difficulty due to the intricacies of the representation theory of $SL(2,\mathbb{R})$, not to mention the handling of the tensor product decomposition for a rank-3 model, the volume divergences are the main reason why we devote our attention to a simplified scenario here.

In the following, we discuss a rank-1 toy model on $SL(2,\mathbb{R})$ with a local quartic interaction in the Gaussian approximation. We also restrict our analysis to conjugation-invariant fields, which simplifies the harmonic analysis. This model has no obvious geometric interpretation but, due to the locality of the interaction, it is free of the aforementioned divergences. In this way, the following work serves as a first step towards the analysis of geometric models, focusing on the influence of the non-compact domain onto the critical behaviour.

---

[18] We refer to Ref. [437] where subtle differences between the kinematical Hilbert spaces of LQG and GFT are discussed in detail.

### 5.3.3.1 A rank-1 toy model on $\mathrm{SL}(2,\mathbb{R})$ with a quartic local interaction

In the following, we consider a GFT model for a real-valued field $\varphi$ living on one copy of $G = \mathrm{SL}(2,\mathbb{R})$ which is subject to conjugation invariance. Its dynamics are defined by

$$S[\varphi] = \frac{1}{2} \int \mathrm{d}g \varphi(g) \left(-\Delta + m^2\right) \varphi(g) + \frac{\lambda}{4!} \int \mathrm{d}g \, \varphi(g)^4, \tag{5.91}$$

wherein $\mathrm{d}g$ denotes the Haar measure on $\mathrm{SL}(2,\mathbb{R})$. In a first step, we again compute the equation of motion, given by

$$\left(-\Delta + m^2\right) \varphi(g) + \frac{\lambda}{3!} \varphi(g)^3 = 0. \tag{5.92}$$

In the mean field approximation, for uniform field configurations it is solved by

$$\varphi_0 = 0 \text{ if } m^2 > 0 \text{ and } \varphi_0 = \pm \sqrt{-\frac{m^2}{\lambda/3!}} \text{ for } m^2 < 0. \tag{5.93}$$

We arrive at the Gaussian approximation by linearising Eq. (5.92) with additional source $J$ via the insertion $\varphi \to \varphi_0 + \delta\varphi$ and $J \to J + \delta J$ while only keeping terms up to linear order in $\delta\varphi$. This leads to the differential equation

$$\left(-\Delta + m^2\right) \delta\varphi(g) + \frac{\lambda}{2!} \varphi_0^2 \delta\varphi(g) = \delta J(g), \tag{5.94}$$

which we once again solve via the Green's function method leading to

$$\left(-\Delta + m^2 + \frac{\lambda}{2!} \varphi_0^2\right) C(g) = \delta(g) \tag{5.95}$$

to be tackled in representation space.

To solve this differential equation, we explain briefly the relevant features of $\mathrm{SL}(2,\mathbb{R})$ as well as harmonic analysis thereon and refer to Appendix C.3 for further details. The group $SL(2, \mathbb{R})$ has two Cartan subgroups, a compact one corresponding to rotations

$$H_0 = \left\{ u_\theta = \begin{pmatrix} \cos\theta & \sin\theta \\ -\sin\theta & \cos\theta \end{pmatrix}, \ 0 \leq \theta \leq 2\pi \right\}. \tag{5.96}$$

and a non-compact one corresponding to boosts,

$$H_1 = \left\{ \pm a_t = \begin{pmatrix} \pm e^t & 0 \\ 0 & \pm e^{-t} \end{pmatrix}, \ t \in \mathbb{R} \right\}. \tag{5.97}$$

A regular group element can be conjugated to either one or the other. Using group averaging arguments [514–517], it follows that a conjugation-invariant field defined on $SL(2, \mathbb{R})$ is either supported on the classes of group elements conjugated to $H_0$ or to $H_1$.[19] We call these conjugation classes $G_0$ and $G_\pm$. Upon averaging over them, the field depends only on an angle $\theta \in [0, 2\pi]$ (parameterising the compact domain) or $t \in \mathbb{R}$ (parameterising the non-compact domain), as explained in detail in Appendix C.3.1. Then we have to analyse the Gaussian approximation for such averaged objects separately.

**(A) Gaussian approximation for fields averaged over compact subgroup**

To solve Eq. (5.95) for fields averaged over $G_0$, we use their decomposition and that of the $\delta$-distribution as explained in Appendix C.3.2. The Green's function decomposes for $m^2 < 0$ as

$$C(\theta) = \sum_{n=1}^{\infty} \frac{n}{4\pi} \left( C^+(n) \chi_n^+(\theta) + C^-(n) \chi_n^-(\theta) \right), \tag{5.98}$$

where the Fourier coefficients for $n = 1, 2, \dots$ are

$$C(n) \equiv C^\pm(n) = \frac{1}{\frac{1-n^2}{4} - 2m^2}. \tag{5.99}$$

Due to the restriction to the compact direction, the evaluation of the strength of fluctuations in terms of $Q$, given in this case by

$$Q = \frac{\int_\xi \mathrm{d}\theta \sin^2\theta \, C(\theta)}{\int_\xi \mathrm{d}\theta \sin^2\theta \, \varphi_0^2}, \tag{5.100}$$

---

[19]These two sectors cannot be mapped into one another which can be interpreted as a superselection rule [515–517].

closely follows our observations for the local GFT models on a compact domain constructed from SU(2) in Section 5.3.2.1. Again, due to the compactness of the domain it does not make sense to evaluate the integrals therein for $\xi \to \infty$ but instead only up to $\xi < \pi$. We find the same behaviour as before for SU(2), namely

$$Q \sim \frac{1}{-2m^2} \frac{1}{\varphi_0^2} \sim \lambda \xi^4, \tag{5.101}$$

leading to the invalidation of the Gaussian approximation for large $\xi$. The dominant behaviour here stems from the modes for $n = 1$ which play the role of the zero-mode contribution, as discussed in Section 5.3.2.1. Indeed the two modes labelled by $n = 1$ are zero modes in the sense that they have zero eigenvalue with respect to the Laplacian. The contributions for all the other modes can be neglected. We may also note that due to the structure and resemblance of the characters $\chi_n^\pm(\theta)$ to those of SU(2), the analysis of a model on the former with a non-local convolution type of interaction will reproduce the same result for $Q$ as for the latter, Eq. (5.76).

**(B) Gaussian approximation for fields averaged over non-compact subgroups**
For fields which are averaged over $G_\pm$, we use the decomposition and that of the $\delta$-distribution expatiated on in Appendix C.3.2 to solve Eq. (5.95). The Green's function decomposes in the sector $m^2 < 0$ as

$$
\begin{aligned}
C(t) &= \int_0^\infty \frac{\mathrm{d}s}{4\pi} \frac{s}{2} C(s) \left( \tanh \frac{\pi s}{2} \chi_s^+(t) + \coth \frac{\pi s}{2} \chi_s^-(t) \right) \\
&\quad + \sum_{n=1}^\infty \frac{n}{4\pi} C(n) \left( \chi_n^+(t) + \chi_n^-(t) \right)
\end{aligned} \tag{5.102}
$$

where the Fourier coefficients $C(n)$ are as in Eq. (5.99) and the coefficients of the continuous series are

$$C(s) \equiv C^\pm(s) = \frac{1}{\frac{1+s^2}{4} - 2m^2}. \tag{5.103}$$

It is possible to obtain exact expressions for the different contributions to the Green's function and thus quantify the behaviour of field fluctuations via $Q$.

To compute the part of $C(t)$ stemming from the continuous series, we use the expression for the $\delta$-distribution in Appendix C.3.2 and compute the contributions of the positive and negative branches, separately. To this end, we use the series expansions of tanh and coth

$$\frac{\pi s}{2} \tanh \frac{\pi s}{2} = \sum_{n \in 2\mathbb{Z}+1} \frac{s^2}{s^2 + n^2} , \tag{5.104}$$

$$\frac{\pi s}{2} \coth \frac{\pi s}{2} = \sum_{n \in 2\mathbb{Z}} \frac{s^2}{s^2 + n^2} \tag{5.105}$$

and apply the residue theorem to compute the integrals.[20] This yields

$$\begin{aligned}
C_{\text{cont}}^+(t) &= \frac{1}{4\pi} \int_0^\infty \mathrm{d}s \frac{s}{2} C(s) \tanh \frac{\pi s}{2} \chi_s^+(t) \\
&= \sqrt{\frac{\pi}{2}} \frac{1}{|\sinh t|} \\
&\quad \times \left( -\frac{\pi}{2} e^{-|t|\sqrt{1-8m^2}} \tan\left(\frac{\pi}{2}\sqrt{1-8m^2}\right) - \sum_{n \in 2\mathbb{Z}+1} \frac{|n| e^{-|n||t|}}{1 - n^2 - 8m^2} \right)
\end{aligned} \tag{5.106}$$

and

$$\begin{aligned}
C_{\text{cont}}^-(t) &= \frac{1}{4\pi} \int_0^\infty \mathrm{d}s \frac{s}{2} C(s) \coth \frac{\pi s}{2} \chi_s^-(t) \\
&= \sqrt{\frac{\pi}{2}} \frac{\text{sgn}(\lambda_{\pm a_t})}{|\sinh t|} \\
&\quad \times \left( \frac{\pi}{2} e^{-|t|\sqrt{1-8m^2}} \cot\left(\frac{\pi}{2}\sqrt{1-8m^2}\right) - \sum_{n \in 2\mathbb{Z}\setminus\{0\}} \frac{|n| e^{-|n||t|}}{1 - n^2 - 8m^2} \right).
\end{aligned} \tag{5.107}$$

To compute the part of $C(t)$ originating from the discrete series on the non-compact direction, we can proceed as in the previous subsection and write

$$C_{\text{disc}}^+(t) = \frac{1}{2|\sinh t|} \sum_{n=1}^\infty \frac{e^{-n|t|}}{1 - n^2 - 8m^2} \tag{5.108}$$

and

$$C_{\text{disc}}^-(t) = \frac{\text{sgn}(\pm a_t)}{2|\sinh t|} \sum_{n=1}^\infty \frac{e^{-n|t|}}{1 - n^2 - 8m^2} . \tag{5.109}$$

The sums over $n$ appearing in each of these expressions converge to sums of hypergeometric functions $_2F_1$ the details of which are not relevant here.

---

[20]Notice that for each tanh / coth-branch we obtain per summand in $n$ altogether 4 poles of order 1. The contour integration has to be performed for the poles with positive and negative imaginary parts separately for $t > 0$ and $t < 0$, respectively. The final result can be put together wherefore yielding expressions with arguments in $|t|$. Similar arguments hold for the discrete series below, too.

We are now ready to quantify the strength of fluctuations by evaluating

$$Q = \frac{\int_\xi dt \ \sinh^2 t C(t)}{\int_\xi dt \ \sinh^2 t \varphi_0^2}. \tag{5.110}$$

We proceed step by step and compute this expression for the individual contributions to the Green's function. To this aim, it is sufficient to estimate $Q$ by looking at the asymptotic behaviour of the integrand in the numerator for $t \to \infty$ and integrating it for $\xi \to \infty$ since modes beyond $\xi$ are anyways exponentially suppressed [504, 509]. In the denominator we have to integrate up to finite $\xi$. In this way, we obtain for large $\xi$

$$Q \sim \lambda \xi^4 e^{-2\xi}. \tag{5.111}$$

Hence, the Gaussian approximation is valid at large $\xi$ where $Q \ll 1$. Thus, it provides a trustworthy description of a phase transition at which $\xi \to \infty$.

This result is in agreement with the one obtained for a scalar field with quartic local interaction on the $3d$ hyperboloid $\mathbb{H}^3$ [509]. We may understand the similarity of the results from the fact that $\mathrm{SL}(2, \mathbb{R}) \cong \mathrm{AdS}^3$ which in turn is diffeomorphic to $\mathbb{H}^{1,2}$.

The form of $Q$ is in stark contrast to the results of the previous sections for fields living on compact domains constructed from $\mathrm{SU}(2)$ and suggests that for a phase transition to occur in the GFT context (and to be visible already at the mean field level), the non-compactness of the domain is a decisive prerequisite.

### 5.3.4 Discussion of the results

The purpose of this Section was to investigate the critical behaviour of various GFT models with and without geometric interpretation in the Gaussian approximation. This encompassed the analysis of the validity of the mean field techniques employed to this end. In the following, we list the different models and the respective results.

(1) With the example of a rank-3 model on $\mathrm{SU}(2)^3$ with a quartic local interaction subject to right, left and right as well as conjugation invariance, we showed that the mean field techniques break down at large correlation length $\xi$, irrespective of the symmetries imposed onto the field.

(2) The case of a rank-1 model on $\mathrm{SU}(2)$ with a quartic non-local interaction of convolution-type subject to conjugation invariance showed that the mean field techniques seize to be valid at large $\xi$. Non-locality does effect higher field modes but without changing the order of magnitude. The dominant zero mode responsible for the breakdown of the Gaussian approximation is not altered by non-locality.

(3) For the dynamical Boulatov model we found the same result as for the non-local toy model, only the exact prefactors of higher modes depend on the specific type of non-locality. In this case, we checked also right, left and right as well as conjugation invariance to demonstrate the independence of the result from the symmetries imposed. We attribute the failure of the mean field techniques to the compactness of the field domain used and expect the result to generalise to other non-local interactions such as simplicial interactions for different rank or tensor-invariant interactions.

Finally, in (4) we analysed the critical behaviour in the case of a rank-1 GFT model on $\mathrm{SL}(2, \mathbb{R})$ with a quartic local interaction subject to conjugation invariance. To our best knowledge, in spite of its toy model nature, this is the first time a GFT model with Lorentzian signature has been studied in some detail in the literature. We employed group averaging arguments to separately analyse the validity of the mean field approach for fields averaged over the conjugation classes of the two Cartan subgroups. For the compact direction, we obtained results analogous to the ones found in case (1), whereas for the non-compact direction mean field techniques continue to be valid in the critical region and can serve as a trustable description of a phase transition. This is ultimately rooted in the non-compactness of the field domain.

In the following, we want to comment on the limitations and possible extensions of our discussion.

Given the breakdown of the mean field techniques towards the supposedly critical region for the cases (1)-(3), the impact of higher order fluctuations should be investigated by means of non-perturbative techniques as for example the functional renormalisation group. With these it should be possible to decide whether or not a phase transition can occur. In this sense, our work can also be seen as a motivation to extend the successful functional renormalisation group methodology developed for tensorial GFTs [393, 471–476] to the realm of simplicial GFT.

Before non-perturbative methods are applied, it could be instructive to go beyond the particular realisation of Landau mean field theory with the Gaussian approximation by relaxing one the main assumptions of this approach, namely the projection onto uniform field configurations. A starting point of such a study could be the non-trivial (and not uniform) global minima of the dynamical Boulatov model for right and left invariant as well as equilateral field configurations found in Ref. [498] and extensively discussed in the previous Section 5.2.

Finally, in view of the last part of our work, it would be important to extend the analysis for the locally interacting rank-1 toy model on $\mathrm{SL}(2,\mathbb{R})$ to the rank-3 case where only right invariance is imposed. For this, Ref. [518] could be useful which collects a variety of facts on the representation theory of $\mathrm{SU}(1,1)$ (which is diffeomorphic to $\mathrm{SL}(2,\mathbb{R})$). In a second step, a regularisation scheme should be introduced to tackle the volume divergences for models with a non-local interaction possibly of simplicial type. As an intermediate pedagogical step, a rank-1 toy model with a convolution-type of interaction and conjugation invariance should be studied to this end. At the level of the correlator, this would already indicate possible modifications which could be expected for the case of the full-blown rank-3 model with simplicial interaction and Lorentzian signature, similar as for the case of the dynamical Boulatov model. The goal of such considerations would of course be to understand if phase transitions and different phases can actually exist for such a model and whether these are related to $(2+1)$-dimensional Lorentzian continuum geometries at all. From a larger perspective, this would also allow us to establish contact and compare with the existing literature on $(2+1)$-dimensional Lorentzian loop quantum gravity and spin-foam models [519, 520]. Finally, our results could be a hint that the non-compactness of the field domain in Lorentzian models will play a crucial role for producing a phase transition for a realistic model for $4d$ spacetime. This deserves to be examined in detail in the future.

Notice that, despite no clear evidence for a condensate phase transition in geometric GFT models as provided by functional methods at the moment, we will adopt the condensate formation as a working hypothesis. This view is further backed by the recent findings of inequivalent condensate representations of the canonical commutation relations in GFT, corresponding to states which are sharply peaked on a given value of the connection [458]. In the following, we will use the simple trial states (field coherent states) introduced in this Chapter as non-perturbative states which, together with their effective dynamics, prove to be very useful to describe realistic cosmological scenarios.

# Chapter 6

# Analysis of two condensate cosmology models

> The whole is more than the sum of the parts.
>
> Aristotle, Metaphysica.

## 6.1 Introduction and overview for the analysis

As we have reviewed in the previous chapter, the effective cosmological dynamics for geometric GFT models follows from a mean field approximation of the full quantum theory. Essentially, the dynamics are captured by the classical equations of motion of the specific GFT model, subject to a few additional restrictions. The simplest way to obtain such equations from the microscopic quantum dynamics is to consider operator equations of motion evaluated in mean value on simple field coherent states, which model condensate states. Generally, the resulting equations are non-linear equations for the condensate wavefunctions.

In the following, we proceed with the analysis of two different GFT condensate cosmology models where the field lives on $d = 4$ copies of $G = \mathrm{SU}(2)$, coming from the imposition of simplicity constraints on $\mathrm{SL}(2, \mathbb{C})$ data. As discussed in Section 5.1.1, these condensate states are given a geometric interpretation as homogeneous continuum spatial geometries if the condensate wave function, in addition to the ordinary right invariance, is also invariant under the diagonal left action of the group. In this way, the domain of the field is isomorphic to minisuperspace of homogeneous cosmologies and the number of dynamical degrees of freedom is reduced to guarantee the above interpretation. The major difference in between the two models analysed, lies in whether the field is real-valued or complex-valued, leading to a distinct phenomenology.

In the first model, we start to work in the group representation and will further reduce symmetries on this level which will allow us to obtain nearly-flat isotropic configurations. The second model is treated in the spin representation throughout and a symmetry reduction will lead to a somewhat different but related notion of isotropy.

To extract information about the evolution of such condensate configurations the GFT condensate field can be coupled to a massless and free real-valued scalar field $\phi$, i.e.

$$\sigma : G^4 \times \mathbb{R} \to \mathbb{R} \text{ or } \mathbb{C}, \tag{6.1}$$

as described in detail in Refs. [486, 487, 521]. Just as for the discrete geometric data encoded in the group elements (and their conjugate variables), the interpretation of the real variable $\phi$ as a discretised matter field is grounded in the expression of the (Feynman) amplitudes of the model corresponding to the action $S$, which take the form of lattice gravity path integrals for gravity coupled to a massless scalar field. As a consequence, the expectation values of the relevant observables studied in the remainder will depend on the relational clock $\phi$.

For such a condensate field, when the relational clock is included, the action takes the general form

$$S[\sigma, \bar{\sigma}] = \int (\mathrm{d}g)^d (\mathrm{d}g')^d \mathrm{d}\phi \mathrm{d}\phi' \bar{\sigma}(g_\mathrm{I}, \phi) \mathcal{K}(g_\mathrm{I}, g'_\mathrm{I}, \phi, \phi') \sigma(g'_\mathrm{I}, \phi') + V[\sigma, \bar{\sigma}]. \tag{6.2}$$

Therein, the local kinetic operator we consider here is of the type

$$\mathcal{K} = \delta(g'_\mathrm{I} g_\mathrm{I}^{-1}) \delta(\phi' - \phi) \left[ -\left( \tau \partial_\phi^2 + \sum_{\mathrm{I}=1}^d \Delta_{g_\mathrm{I}} \right) + m^2 \right] \quad \text{with} \quad \tau, m^2 \in \mathbb{R}. \tag{6.3}$$

The Laplacian on the group manifold is motivated by the renormalisation group (RG) analysis of GFT models where it is shown to be indispensible in order to regulate the ultraviolet behaviour of the theory (cf. Ref. [393]). The "mass term" is related to the GFT/spin foam correspondence, as it corresponds to the spin foam edge weights and $\tau$ is an adjustable parameter.

The larger scope of this Chapter is to study the effective geometries and their dynamics as encoded by the two different condensate models and in this way to scrutinise the condensate hypothesis. For the sake of a better overview, we give a brief snapshot of the content studied for the two models as follows:

- In Section 6.2 we investigate the dynamics of the real-valued mean field $\sigma$. At first, we consider the field to be independent from the relational clock. For such static configurations, we explore properties of a free field in an isotropic restriction in Section 6.2.1 and then consider the impact of phenomenologically motivated effective interactions in Section 6.2.2. This is followed by the investigation of the same model under relational evolution in Sections 6.2.4, 6.2.5 and finally 6.2.6 where anisotropic configurations are studied.

- In Section 6.3 we study a complex-valued condensate field subject to the dynamics of the Lorentzian Engle-Pereira-Rovelli-Livine (EPRL) GFT model when the relational clock is turned on. The complexity of the field leads to a different phenomenology as compared to that of a real-valued field, i.e. they generically give rise to bouncing solutions. We first review the dynamics of an isotropic background in Section 6.3.2 and then investigate the behaviour of anisotropic perturbations over it in the vicinity of the bounce in Sections 6.3.3 and 6.3.4. This motivates the study of the impact of effective interactions onto the isotropic background after the bounce in Section 6.3.5.

## 6.2   Model 1: Real-valued symmetry reduced rank-4 GFT

In the first part of this Section, we study and elaborate on a free model of rank-4, the analysis of which was started in an isotropic restriction in Ref. [480]. We note that no additional massless scalar field is added to study the evolution of the system in relational terms first. We extensively discuss the geometric interpretation of such solutions by analysing their curvature properties and by computing the expectation values of the volume and area operators imported from LQG. We then redo this analysis in Section 6.2.2 after introducing combinatorially local interaction terms to the system.

For a particular choice of the signs of the free parameters in the GFT action, we find in the isotropic restriction solutions which (i) are consistent with the condensate ansatz, (ii) are normalisable with respect to the Fock space measure in the weakly non-linear regime, and (iii) obey a specific condition of which the fulfillment is required for the interpretation in terms of continuous manifolds. We generally find for both the free and interacting cases that the expectation values of the geometric operators are dominated by low-spin modes. In this sense, such solutions may be interpreted as giving rise to an effectively continuous geometry. Moreover, we discuss the consequences of the interactions in the strongly non-linear regime, where solutions generally lose their normalisability with respect to the Fock space measure and thus can be interpreted as corresponding to non-Fock representations of the canonical commutation relations. These results are largely based on the work of the author in Ref. [457].

The second part of this Section extends these results and studies the relational evolution of such condensate systems. To do so, we incorporate the concept of a relational clock allowing us to extract information about the dynamics of the emergent 3-geometries. We start off with studying the relational evolution in the free case in Section 6.2.4. In particular, we find in Section 6.2.4.1 and 6.2.4.2 that the corresponding quantum geometry quickly settles into the lowest non-trivial configuration available to the system going in hand with its classicalisation. In addition, the emergent geometry obeys Friedmann-like dynamics which is akin to the results shown in Refs. [486–488]. We investigate the influence of effective interactions onto the condensate system in Section 6.2.5. Most importantly, we demonstrate in Section 6.2.5.1 that for a particular choice of interaction terms one can accommodate for an era of inflationary expansion, however, coming at the cost of fine-tuning their coupling constants. Towards the end, in Section 6.2.6 we lift the isotropic restriction and study more general, i.e., anisotropic free and effectively interacting condensates. This will allow for a more systematic exploration of anisotropic condensate models and their relation to those models studied for instance by loop quantum cosmology (LQC) [75, 76, 492] in the future. In particular, we show in Section 6.2.6.1, how such condensate systems dynamically isotropise. Conversely, it is shown that anisotropic contributions to the condensate dominate towards small volumes but are under control. These findings are largely based on the author's work published in Ref. [490].

## 6.2.1 Static case of a free isotropic GFT condensate

Throughout this subsection, we exhaustively analyse the properties of a "static" and non-interacting condensate configuration with particular regard to its geometric interpretation in terms of the geometric operators imported from LQG.

### 6.2.1.1 The free model and general properties of its solutions

Using Eqs. (5.12) and (6.3) for a static field with $V = 0$ one obtains

$$\left[ -\sum_{I=1}^{4} \Delta_{g_I} + m^2 \right] \sigma(g_I) = 0. \tag{6.4}$$

To find exact solutions to this equation, first we introduce coordinates on the SU(2) group manifold, use invariance properties of $\sigma(g_I)$ and then apply symmetry reductions. We closely follow the results of Ref. [480] and elaborate them to extract a geometric interpretation from them.

To this aim, assume that the connection in the holonomy $g = \mathcal{P} e^{i \int_e A}$ remains approximately constant along the link $e$ with length $\ell_0$ in the $x$-direction, which yields $g \approx e^{i \ell_0 A_x}$. In the polar decomposition, this gives

$$g = \cos(\ell_0 ||\vec{A}_x||) \mathbb{1} + i \vec{\sigma} \frac{\vec{A}_x}{||\vec{A}_x||} \sin(\ell_0 ||\vec{A}_x||), \tag{6.5}$$

with the $\mathfrak{su}(2)$-connection $A_x = \vec{A}_x \cdot \vec{\sigma}$ and the Pauli matrices $\{\sigma_i\}_{i=1,\dots,3}$. In the next step, we introduce the coordinates $(\pi_0, \dots, \pi_3)$ together with $\pi_0^2 + \dots + \pi_3^2 = 1$ which specifies an embedding of SU(2) $\cong S^3$ into $\mathbb{R}^4$. Due to the isomorphism SO(3) $\cong$ SU(2)/$\mathbb{Z}_2$, the choice of sign in $\pi_0 = \pm\sqrt{1 - \vec{\pi}^2}$ corresponds to working on one hemisphere of $S^3$. With

the identification

$$\vec{\pi} = \frac{\vec{A}_x}{||\vec{A}_x||} \sin(\ell_0 ||\vec{A}_x||), \tag{6.6}$$

we can parametrise the holonomies as

$$g(\vec{\pi}) = \sqrt{1 - \vec{\pi}^2} \mathbb{1} + i\vec{\sigma} \cdot \vec{\pi}, \quad ||\vec{\pi}|| \leq 1, \tag{6.7}$$

where $||\vec{\pi}|| = 0$ corresponds to the pole of the hemisphere and $||\vec{\pi}|| = 1$ marks the equator. In these coordinates the Haar measure becomes

$$\mathrm{d}g = \frac{\mathrm{d}\vec{\pi}}{\sqrt{1 - \vec{\pi}^2}}. \tag{6.8}$$

Using the Lie derivative on the group manifold acting on a function $f$, one has for the Lie algebra elements

$$\vec{B} f(g) \equiv i \frac{\mathrm{d}}{\mathrm{d}t} f(\mathrm{e}^{\frac{i}{2} \vec{\sigma} t} g)|_{t=0} \frac{i}{2} \left[ \sqrt{1 - \vec{\pi}^2} \vec{\nabla} + \vec{\pi} \times \vec{\nabla} \right] f. \tag{6.9}$$

With this, the Laplace-Beltrami operator $\vec{B}^2 = -\Delta_g$ in terms of the coordinates $\vec{\pi}$ on SU(2) is given by

$$-\Delta_g f(g) = -[(\delta^{ij} - \pi^i \pi^j)\partial_i \partial_j - 3\pi^i \partial_i] f(\vec{\pi}). \tag{6.10}$$

This applies to all group elements $g_\mathrm{I}$, $I = 1, \ldots, 4$ dressing the spin network vertex dual to the quantum tetrahedron. Thus, the Laplacian part in the equation of motion (6.4) is given by

$$-\sum_\mathrm{I} \Delta_{g_\mathrm{I}} = \sum_\mathrm{I} \vec{B}_\mathrm{I} \cdot \vec{B}_\mathrm{I}. \tag{6.11}$$

Using the invariance of the mean field under the right diagonal action of $G$, which corresponds to the closure condition for the fluxes

$$\sum_\mathrm{I} \vec{B}_\mathrm{I} = 0, \tag{6.12}$$

as detailed in Section 4.2.4.1, Eq. (6.11) rewrites as

$$-\sum_\mathrm{I} \Delta_{g_\mathrm{I}} = 2 \left( \sum_{i=1}^{3} \vec{B}_i \cdot \vec{B}_i + \sum_{i \neq j} \vec{B}_i \cdot \vec{B}_j \right). \tag{6.13}$$

In the most general case, the left and right invariant $\sigma(g_I)$ is defined on the domain $SU(2)\backslash SU(2)^4/SU(2)$ which can be parametrised by six independent coordinates $\pi_{ij} = \vec{\pi}_i \cdot \vec{\pi}_j$, with $i, j = 1, 2, 3$ and $0 \leq |\pi_{ij}| \leq 1$. Hence, when using the above considerations, the action of Eq. (6.13) in the equation of motion (6.4) gives rise to a rather complicated partial differential equation, see Ref. [480].

A strategy to extract solutions, as outlined in Refs. [480, 522], is to impose in a first step a symmetry reduction by considering functions $\sigma$ which only depend on the diagonal components $\pi_{ii}$. In this way, Eq. (6.13) acting on $\sigma(\pi_{ii})$ leads to

$$-\sum_I \Delta_{g_I}\sigma(\pi_{ii}) = -\Bigg[\sum_i 8\pi_{ii}(1-\pi_{ii})\frac{\partial^2}{\partial\pi_{ii}^2} + 4(3-4\pi_{ii})\frac{\partial}{\partial\pi_{ii}} +$$
$$4\sum_{i\neq j}\sqrt{1-\pi_{ii}}\sqrt{1-\pi_{jj}}\pi_{(ij)}\frac{\partial^2}{\partial\pi_{ii}\partial\pi_{jj}}\Bigg]\sigma(\pi_{ii}). \quad (6.14)$$

The second sum stems from the corresponding term in Eq. (6.13). To find particular solutions, it is then advantageous to get rid of the latter which leads to a decoupling of the terms in the different $\pi_{ii}$. To this aim, one can employ a summation ansatz

$$\sigma(\pi_{ii}) = \sigma(\pi_{11}) + \ldots + \sigma(\pi_{33}). \quad (6.15)$$

Finally, assuming all $\pi_{ii}$ to be equal to a variable $p$, i.e.

$$p \equiv \pi_{ii} = \sin^2(\ell_0||\vec{A}_x||), \quad (6.16)$$

Eq. (6.4) can be rewritten as

$$-\Bigg[2p(1-p)\frac{\mathrm{d}^2}{\mathrm{d}p^2} + (3-4p)\frac{\mathrm{d}}{\mathrm{d}p}\Bigg]\sigma(p) + \mu\sigma(p) = 0, \quad (6.17)$$

with $\mu \equiv \frac{m^2}{12}$ and $p \in [0, 1]$. The general solution is given by

$$\sigma(p) = \sqrt[4]{\frac{1-p}{p}}\left[a\,P^{\frac{1}{2}}_{\frac{1}{2}(\sqrt{1-2\mu}-1)}(2p-1) + b\,Q^{\frac{1}{2}}_{\frac{1}{2}(\sqrt{1-2\mu}-1)}(2p-1)\right], \quad (6.18)$$

with $a, b \in \mathbb{C}$ and $P, Q$ are associated Legendre functions of the first and second kinds, respectively [480, 522]. With respect to the measure induced from the full Fock space, one yields for the average particle number

$$N = \int (\mathrm{d}g)^4 |\sigma(g_\mathrm{I})|^2 = 2\pi \int dp \sqrt{\frac{p}{1-p}} |\sigma(p)|^2 < \infty. \tag{6.19}$$

In retrospection, it is sensible to refer to such a reduction, i.e. to just one variable $p$, as an isotropisation. This can be seen when substituting

$$p = \pi_{ii} \equiv \sin^2(\psi) \tag{6.20}$$

into Eq. (6.17) which leads to

$$-\left[ \frac{\mathrm{d}^2}{\mathrm{d}\psi^2} + 2\cot(\psi)\frac{\mathrm{d}}{\mathrm{d}\psi} \right] \sigma(\psi) + 2\mu\sigma(\psi) = 0, \quad \psi \in [0, \frac{\pi}{2}]. \tag{6.21}$$

Comparing this expression to the Laplacian on one hemisphere of $S^3$ which acts on a function $\sigma(\phi, \theta, \psi)$, one has

$$-\Delta\sigma(\phi, \theta, \psi) = -\frac{1}{\sin^2(\psi)}\left[ \frac{\partial}{\partial\psi}\left( \sin^2(\psi)\frac{\partial}{\partial\psi}\sigma \right) + \Delta_{S^2}\sigma \right], \tag{6.22}$$

with $\phi \in [0, 2\pi]$, $\theta \in [0, \pi]$ and $\psi \in [0, \frac{\pi}{2}]$. The function $\sigma$ is called hyperspherically symmetric, isotropic or zonal if it is independent of $\phi$ and $\theta$ [523]. These are spherically symmetric eigenfunctions of $-\Delta_{S^2}$ for which Eq. (6.21) is just equal to Eq. (6.22). Hence, this symmetry reduction is an explicit restriction of the rather general class of condensates to a particular representative with a clearer geometric interpretation. In addition, we note that this reduction should not be confused with the symmetry reduction employed in Wheeler-DeWitt quantum cosmology [34] or LQC [75, 76, 492], since it is applied after quantisation onto the quantum state and not beforehand.

### 6.2.1.2 Nearly-flat solutions

In the following, we want to specify the possible values of $\mu$ in the symmetry reduced case by means of discussing the spectrum of the operator $-\sum_\mathrm{I}\Delta_{g_\mathrm{I}}$. Its self-adjointness and positivity imply that its eigenvalues $\{m^2\}$ lie in $\mathbb{R}_0^+$. The compactness of the domain space $\mathrm{SU}(2)\backslash\mathrm{SU}(2)^4/\mathrm{SU}(2)$ entails that the spectrum is discrete and the respective eigenspaces are finite-dimensional. This also holds for the symmetry reduced case. To concretise the spectrum, we have to introduce boundary conditions, which we infer from physical assumptions. We request from the solutions to the equation of motion to admit an interpretation in terms of smooth metric 3-geometries. Hence, they should obey the above-mentioned near-flatness condition, see Section 5.1.1. In the group representation, this condition translates into demanding that the group elements decorating the quantum tetrahedra to be close to the unit element so that the character gives $\chi(\mathbb{1}_{j_i}) = 2j_i + 1$, see Refs. [440, 441, 448]. On the level of the mean field this leads to the requirement that the (probability) density is concentrated around small values of the connection or its curvature.[1] In the symmetry reduced case this condition holds for $\sigma(p)$ if the (probability) density $|\sigma(p)|^2$ is concentrated around small values of the variable $p$ and tends to zero at the equator traced out at $p = 1$. The latter translates into a Dirichlet boundary condition on the equator,

$$\sigma(p)|_{p=1} = 0, \tag{6.23}$$

which is only obeyed by the $Q$-branch of the general solution, Eq. (6.18).[2] Using this, the spectrum of the Dirichlet Laplacian is given by

$$\mu = -2n(n+1) \quad \text{with} \quad n \in \frac{2\mathbb{N}_0 + 1}{2}.^3 \tag{6.25}$$

Equivalently, these solutions correspond to the eigensolutions of Eq. (6.21) obeying the boundary condition $\sigma(\frac{\pi}{2}) = 0$. They are given by

$$\sigma_j(\psi) = \frac{\sin((2j+1)\psi)}{\sin(\psi)}, \quad \psi \in [0, \frac{\pi}{2}] \tag{6.26}$$

with $j \in \frac{2\mathbb{N}_0 + 1}{2}$ corresponding to the eigenvalues $\mu = -2j(j+1)$. On the interval $[0, \frac{\pi}{2}]$ these solutions are exactly equal to those hyperspherically symmetric eigenfunctions of the

---

[1] It is advantageous to work in the coordinate representation exactly because this simplifies the construction of nearly-flat solutions. This is less obvious when working in the spin representation from the outset, as in Section 6.3.

[2] Due to the linear character of the free problem, the solutions have a rescaling invariance with respect to the chosen boundary conditions. This means that two solutions for different boundary conditions $\sigma'(1)$ can be rescaled into one another according to

$$\frac{N[\sigma'_1(1)]}{N[\sigma'_2(1)]} = \frac{|\sigma'_1(1)|^2}{|\sigma'_2(1)|^2}, \tag{6.24}$$

which obscures the interpretation of the quantity $N$ and other observables in the free case. This rescaling property is lost once (strong) non-linear interactions are considered as in Section 6.2.2.

[3] Importantly, consistency with the flatness condition demands that the only eigenfunction of the Dirichlet Laplacian to the eigenvalue $\mu = 0$ can be given in terms of the trivial function. We emphasise that the zero-mode is excluded from the spectrum due to the flatness condition. This is to be contrasted to the removal of this mode through cylindrical equivalence in the context of LQG and LQC [5].

Laplacian on $S^3$ which vanish on the equator. Furthermore, observe that these would just be the characters $\chi_j(\psi)$ of the respective representation for $j$ if $\sigma_j(\psi)$ was not set to zero on $[\frac{\pi}{2}, \pi]$.

In view of the geometric interpretation of these solutions, we want to illustrate and then discuss the behaviour of the first few eigensolutions by plotting their (probability) density $|\sigma(p)|^2$ in Fig. 6.1 or $|\sigma(\psi)|^2$ in Fig. 6.2, respectively. The plot illustrates that the density is concentrated around small values of the variable $p$ or $\psi$, respectively. In general, eigensolutions remain finitely peaked around $p = 0$ or $\psi = 0$. Solutions for slightly perturbed eigenvalues $\mu$ are infinitely peaked as $\lim_{p \to 0} |\sigma(p)|^2 \sim 1/p$.

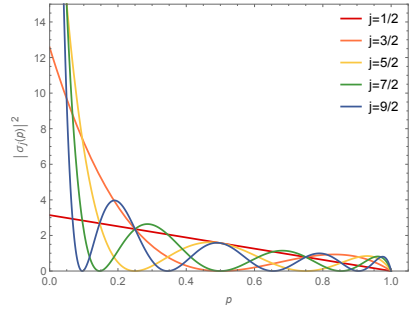

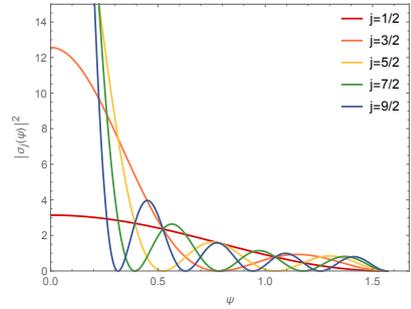

FIGURE 6.1: Probability density of the free mean field over $p$.

FIGURE 6.2: Probability density of the free mean field over $\psi$.

A concentration of the (probability) density around small $p$ corresponds to a concentration around small curvature values. This is because small $p$, itself directly proportional to the gravitational connection $A$, implies small field strength via $F = \mathrm{d}_A A$. Naively, this leads to a small 3-curvature $R$, as is known from the first-order formalism for gravity, see Section 2.2.1. This is important for consistency matters, meaning that the building blocks of the geometry are indeed almost flat which is needed to approximate a smooth 3-space. Around $p = 1$ or $\psi = \frac{\pi}{2}$, tracing out the equator of $S^3$, the solutions vanish. The occurrence of the finite number of oscillatory maxima does not a priori pose a problem to the fulfillment of the near-flatness condition since the eigensolutions are indeed concentrated around small values of $p$ or angles $\psi$. Our solutions obey the above-introduced near-flatness condition since in Eq. (6.26) $\lim_{\psi \to 0} \sigma_j(\psi)$ exactly yields $2j + 1$. In this light, using the solutions $\sigma_j(\psi)$, we can compute the expectation value of the field strength[4] $F^i \sim p$ given by

$$\langle \hat{F}^i \rangle \sim \int_0^{\frac{\pi}{2}} \mathrm{d}\psi \sin^2(\psi) \; |\sigma_j(\psi)|^2 \; \hat{F}^i > 0, \tag{6.28}$$

which is illustrated in Fig. 6.3. From this follows that the average curvature $\langle \hat{F}^i \rangle / N$ is positive and possibly tiny, depending on $N$. The dots indicate the discrete contributions to the field strength for a particular $j$-mean field and show a dominance of the $\frac{1}{2}$-eigensolution over the others.

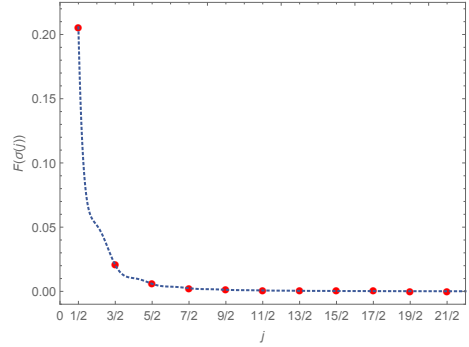

FIGURE 6.3: Un-normalised spectrum of the field strength with respect to the eigensolutions $\sigma_j(\psi)$ in arbitrary units.

In light of the previous discussion, it may seem a bit surprising that the expectation value of the field strength is non-zero despite the fact that $p = 0$ is the most probable value of the corresponding mean field. However, the extended tail of the (probability) density with the finite oscillatory maxima accounts for the average being bigger than the most probable value. The finite value indicates that the space described by the condensate is of finite size. We will come back to this point at the end of this Section.[5]

In the last step, we want to transform our nearly-flat solutions to the spin representation which most directly facilitates the extraction of information about the LQG volume and area operators and is crucial for the geometric interpretation of the solutions. To this aim, notice that due to the left- and right invariance of $\sigma(g_{\mathrm{I}})$, under the given symmetry

---

[4] Relating $p$ to the field strength is justified when considering a plaquette $\square$ in a face of a tetrahedron so that we can make use of the well-known expression

$$F_{ab}^k(A) = \frac{1}{\mathrm{Tr}(\tau^k \tau^k)} \lim_{\mathrm{Area}_\square \to 0} \mathrm{Tr}_j \left( \tau^k \frac{\mathrm{hol}_{\square_{ij}}(A) - 1}{\mathrm{Area}_\square} \right) \delta_a^i \delta_b^j, \tag{6.27}$$

where $a, b \in \{1, 2\}$ and for the $\mathfrak{su}(2)$-algebra elements $\tau_k = -\frac{i}{2}\sigma_k$ the relation $\mathrm{Tr}(\tau^k \tau^k) = -\frac{1}{3}j(j+1)(2j+1)$ holds [75, 76, 492]. This yields $F^k \sim \sin^2(\psi) = p$.

[5] The last word on the flatness behaviour of such solutions also in the interacting case, however, lies with the analysis (of the expectation value) of a currently lacking GFT-curvature operator, as already noticed in Ref. [448].

reduction the mean field becomes a central function on the domain. In this case isotropy coincides with the notion of centrality. Using the Fourier series of a central function on SU(2) [523], the Fourier series for the mean field in the angle parameterisation is given by

$$\sigma_j(\psi) = \sum_{m \in \frac{\mathbb{N}_0}{2}} (2m+1) \, \chi_m(\psi) \, \sigma_{j;m}, \tag{6.29}$$

with the "plane waves" given by the characters $\chi_m(\psi) = \frac{\sin((2m+1)\psi)}{\sin(\psi)}$.[6] The Fourier coefficients are then obtained via

$$\sigma_{j;m} = \frac{2}{\pi} \frac{1}{2m+1} \int_0^{\frac{\pi}{2}} \mathrm{d}\psi \sin^2(\psi) \, \chi_m(\psi) \, \sigma_j(\psi), \tag{6.30}$$

and $m \in \frac{\mathbb{N}_0}{2}$.[7] Using this, the Fourier coefficients of the solutions $\sigma_j(\psi)$ (cf. Eq. (6.26)) yield

$$\sigma_{j;m} = \frac{1}{2\pi} \frac{1}{2m+1} \frac{(-1)^{\frac{2j-1}{2}} \, (2j+1) \, \cos(m\pi)}{(m-j) \, (m+j+1)}, \tag{6.31}$$

with $j \in \frac{2\mathbb{N}_0+1}{2}$. Notice that the occurrence of the two indices $m$ and $j$ is rooted in the chosen boundary conditions.

### 6.2.1.3 Expectation value of geometric operators

The expectation value of the volume and area operators is central to the geometric interpretation of solutions to the equation of motion. In the spin representation, this quantity can be computed with respect to the mean field as

$$\langle \hat{V} \rangle \equiv V = V_0 \sum_{m \in \frac{\mathbb{N}_0}{2}} |\sigma_{j;m}|^2 V_m \quad \text{with} \quad V_m \sim m^{\frac{3}{2}} \tag{6.32}$$

and $V_0 \sim \ell_p^3$. The normalised volume $V/V_0$ is shown in Fig. 6.4 for different values of $j$. Therein, the dots indicate the discrete contributions to the volume for a particular $j$. Eigensolutions for smaller $j$ or $|\mu|$ have a larger volume in comparison to those with larger $j$, especially the $j = \frac{3}{2}$ eigensolution has the relatively largest volume. Importantly, the volume is finite for all $j$ indicating that the space which the condensate approximates must be of finite size. Hence, a general solution which can be decomposed in terms of eigensolutions, describes a finitely sized space of which the largest contributions arise from low-spin modes.

---

[6]The symbol $m$ in the Fourier expansion is not to be confused with the mass term $m^2$ in the kinetic operator.

[7]The double index in $\sigma_{j;m}$ accounts for the fact that $\sigma_j(\psi)$ is non-zero on the interval $[0, \frac{\pi}{2})$ and vanishes on $[\frac{\pi}{2}, \pi]$.

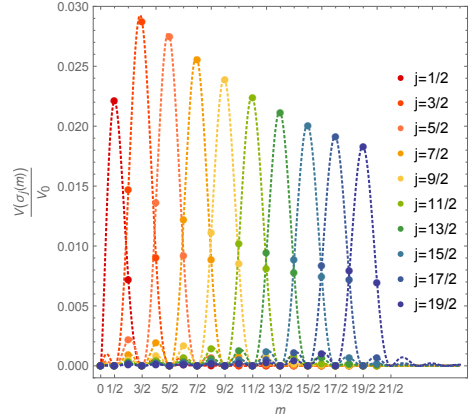

FIGURE 6.4: Normalised spectrum of the volume operator with respect to the eigensolutions $\sigma_j(\psi)$ in arbitrary units.

Finally, Figs. 6.5 and 6.6 illustrate the uncertainty and the relative standard deviation of the volume contributions $V_j$ for the solutions Eq. (6.26) of the volume operator, which are monotonously increasing in $j$ (for $j > \frac{1}{2}$ in the case of the relative standard deviation) and indicate that the expectation value assumes a sharper value if the condensate resides in lower $j$-modes. In Section 6.2.4 we will reconsider the relative standard deviation in the context of the relational evolution of the total volume. We will then show that the relative uncertainty vanishes at late times which indicates the classicalisation of the quantum geometry emerging from the condensate state.

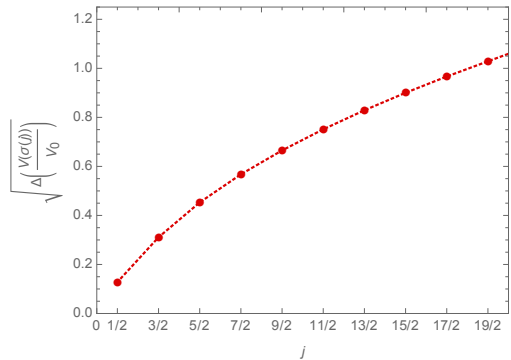

FIGURE 6.5: Standard deviation of the volume operator over $j$.

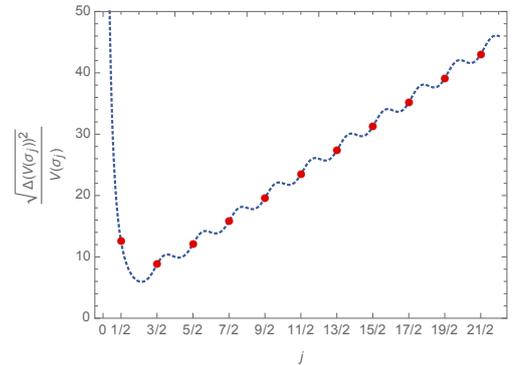

FIGURE 6.6: Relative standard deviation of the volume operator over $j$.

Analogously, the expectation value of the area operator for an individual face of a quantum tetrahedron in the condensate is given as

$$\langle \hat{A} \rangle \equiv A = A_0 \sum_{m \in \frac{\aleph_0}{2}} |\sigma_{j;m}|^2 A_m \tag{6.33}$$

with $A_m \sim (m(m+1))^{\frac{1}{2}}$ and $A_0 \sim \ell_p^2$. Depending on the solution $\sigma_j$, the spectrum of the normalised area $A/A_0$ is illustrated in Fig. 6.7 showing a dominance of the $\frac{1}{2}$-representation and otherwise with a similar interpretation as in the case of the volume operator.

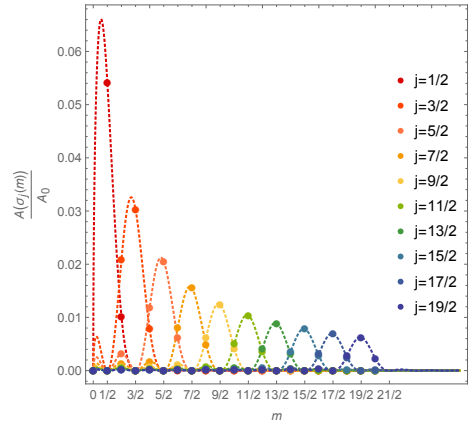

FIGURE 6.7: Normalised spectrum of the area operator with respect to the eigensolutions $\sigma_j(\psi)$ in arbitrary units.

From the above, it is not clear whether a certain eigensolution could be dynamically preferred over others. The near-flatness condition seems to be better fulfilled by lower eigenmodes, that means for those solutions with a lower number of oscillatory maxima. These are the solutions which are mostly concentrated around small connection or curvature values. In this light, it is striking that the computation of the expectation values of the volume, area and the field strength operators all display the dominance of low $j$-modes. This seems to be in favor of the condensate picture where the field quanta are conjectured to condense into the same simple quantum geometric state. Below we explore the case of interacting models which is pivotal for the geometric interpretation of the solutions and the extraction of phenomenology.

We want to make a final remark about restricting our attention solely to the those solutions obeying the near-flatness condition. Of course one could consider more general solutions to Eq. (6.17) which are not necessarily peaked around $p = 0$. Despite the fact, that such solutions cannot be immediately related to continuous 3-geometries according to the near-flatness condition [440, 441], their properties could nevertheless be studied in a similar manner.

### 6.2.1.4 Non-commutative Fourier transform of the solutions $\sigma$

It is possible to retrieve from the free (and nearly-flat) solutions the flux-information with the non-commutative Fourier transform $\mathcal{F}$ on SU(2), as discussed in Section 4.2.4.1. In principle, this facilitates to relate the geometric content of the quantum tetrahedra to a metric.

$\mathcal{F}$ defines an isometric mapping between the space $L^2(\mathrm{SO}(3), \mathrm{d}\mu_H)$ with Haar measure $\mathrm{d}\mu_H$ and the space $L^2_\star(\mathbb{R}^3, d\mu)$ of functions on $\mathfrak{su}(2) \sim \mathbb{R}^3$ with a non-commutative $\star$-product and standard Lebesgue measure $\mathrm{d}\mu$, as expounded in greater detail in Refs. [67, 439]. This transform allows us to shift in between the group and the dual flux representation of the mean field $\sigma$. From the momentum space representation of the condensate field it is in principle possible to reconstruct the metric at a given point in one of the quantum tetrahedra constituting the condensate [440, 441, 448].

For the computation of the non-commutative Fourier transform of the free solutions, we use the coordinates introduced in Section 6.2.1 to parameterise the group manifold SU(2). In particular, since we are working on SO(3) which can be identified to the upper hemisphere of SU(2) $\cong S^3$, we adapt our coordinates to $\vec{\pi} = \sin(\psi)\vec{n}$ with $\psi \in [0, \frac{\pi}{2}]$ and $\vec{n} \in S^2$. In this parametrisation the Haar measure is given by $\mathrm{d}g = \frac{1}{\pi}\sin^2(\psi)\mathrm{d}\psi \mathrm{d}^2\vec{n}$, where $\mathrm{d}^2\vec{n} = \sin(\theta)\mathrm{d}\phi\mathrm{d}\theta$ is the normalised measure on the unit 2-sphere. With this one can recast the transform given by Eq. (4.72) into a standard $\mathbb{R}^3$-Fourier transform, leading to the integral formula

$$\mathcal{F}[\sigma](x) = \tilde{\sigma}(x) = \frac{1}{\pi}\int_{||\vec{\pi}||\leq 1}\frac{\mathrm{d}^3\vec{\pi}}{\sqrt{1-\vec{\pi}^2}}\,\sigma(g(\vec{\pi}))\,\mathrm{e}^{i\vec{\pi}\cdot\vec{B}}. \tag{6.34}$$

Taking advantage of the symmetry of the problem in the isotropic restriction, together with $x \equiv ||\vec{B}||$ and $||\vec{n}|| = 1$, one finds that $\vec{\pi}\cdot\vec{B} = x\sin(\psi)\cos(\theta)$. In a next step one integrates over $\theta$ which leads with $p \equiv \vec{\pi}^2$ to the analogue of the Fourier-Bessel transformation given by

$$\tilde{\sigma}(x) = 2\int_0^1 \frac{\mathrm{d}p}{\sqrt{1-p}}\frac{\sin(\sqrt{p}\,x)}{x}\,\sigma\left(\sqrt{p}\right). \tag{6.35}$$

In a final step, we use $p \equiv \sin^2(\psi)$ to arrive at an expression in terms of the angles $\psi$ given by

$$\tilde{\sigma}(x) = 4 \int_0^{\frac{\pi}{2}} \mathrm{d}\psi \ \sin(\psi) \frac{\sin(\sin(\psi)x)}{x} \ \sigma(\psi). \tag{6.36}$$

Using this, we can compute the non-commutative Fourier transform $\tilde{\sigma}_j(x)$ of the solution to the free equation given by Eq. (6.26). The results for different $j$ are illustrated in Fig. 6.8.

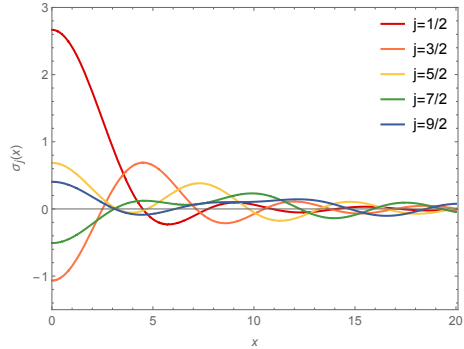

FIGURE 6.8: Non-commutative Fourier transform of the free isotropic solution $\sigma_j(\psi)$.

The reconstruction of the metric $g_{ij}$, encoded by Eq. (4.75), however, is somewhat obstructed, e.g., due to ordering ambiguities in the corresponding operator version $\hat{g}_{ij}$ stemming from the non-commutativity of the fluxes, as noticed in Ref. [448]. Instead, it is much simpler to reconstruct the metric in the spin representation from Eq. (4.75). Given that $\vec{B}_i \sim \vec{J}_i$ under the imposed symmetry restrictions, it can be easily seen that the metric is proportional to $\mathbb{1}_3$, as expected from an isotropic configuration.

## 6.2.2 Static case of an effectively interacting isotropic GFT condensate

In this subsection, we consider the impact of local interaction terms, i.e. pseudopotentials, onto the condensate system. In particular, we analyse their effect onto the behaviour of the solutions and the expectation values of relevant geometric operators.

From the point of view of mean field theory, it is not uncommon to study simplified types of interactions which gloss over the actual microscopic details. They have a practical utility as simplified versions of more complicated ones, and bring us nearer to the physics which we want to probe. Here we speculate that such simplified interactions between the condensate constituents are only relevant in a continuum and large scale limit, where the true combinatorial non-locality of the fundamental theory could be effectively hidden. This phenomenologcal perspective can be motivated by speculating that while the occurrence of ultraviolet fixed points in tensorial GFTs is deeply rooted in their combinatorial non-locality (cf. Refs. [471–476, 524]), the occurrence of infrared fixed points, akin to Wilson-Fisher fixed points in the corresponding local QFTs, seems to be unaffected by this feature. Ultimately, rigorous RG arguments will have the decisive word on whether combinatorially local interaction terms may be derived from the fundamental theory. In this way, studying the effect of pseudopotentials and trying to extract physics from the solutions can be useful to clarify the map between the microscopic and effective macroscopic dynamics of the theory and is instructive to gain experience for the treatment of the corresponding non-local terms which have a clearer discrete geometric interpretation. In this light, we will consider two types of local interactions, mimicking the so-called tensorial and the above-introduced simplicial interactions.

### 6.2.2.1   General setup for effectively interacting condensates

The models on which we build this analysis involve static condensate fields, a Laplacian-type kinetic term and two types of simplified interactions. The first of these mimics so-called tensorial interactions[8] and is given by the pseudopotential

$$V_T[\varphi] = \sum_{n \geq 2} \frac{\kappa_n}{n} \int (\mathrm{d}g)^4 (|\varphi(g_\mathrm{I})|^2)^n. \tag{6.37}$$

It is even powered in the modulus of the field. The equation of motion of the mean field then is

$$\left[ -\sum_{\mathrm{I}=1}^{4} \Delta_{g_\mathrm{I}} + m^2 \right] \sigma(g_\mathrm{I}) + \sigma(g_\mathrm{I}) \sum_{n=2} \kappa_n (|\sigma(g_\mathrm{I})|^2)^{n-1} = 0. \tag{6.38}$$

Observe that the locality of the interaction implies that we do not make use of any non-trivial pairing pattern for the fields. Applying the same symmetry assumptions as above, one has $\sigma(g_1, g_2, g_3, g_4) = \sigma(g, g, g, g) = \sigma(p)$. Considering only one summand for the interaction, we yield

$$-\left[ 2p(1-p)\frac{\mathrm{d}^2}{\mathrm{d}p^2} + (3-4p)\frac{\mathrm{d}}{\mathrm{d}p} \right] \sigma(p) + \mu\sigma(p) + \kappa\sigma(p)(|\sigma(p)|^2)^{n-1} = 0, \tag{6.39}$$

with $n = 2, 3, 4, \ldots$.

---

[8]Interactions of tensorial type are briefly discussed in Section 4.2.2.2 on modern tensor models.

In the following, we focus on the case of real-valued GFT fields and set $n = 2$, for which the equation of motion reads

$$-\left[2p(1-p)\frac{\mathrm{d}^2}{\mathrm{d}p^2} + (3-4p)\frac{\mathrm{d}}{\mathrm{d}p}\right]\sigma(p) + \mu\sigma(p) + \kappa\sigma(p)^3 = 0, \qquad (6.40)$$

with the effective potential

$$V_{\text{eff}}[\sigma] = \frac{\mu}{2}\sigma^2 + \frac{\kappa}{4}\sigma^4. \qquad (6.41)$$

The signs of the coupling constants determine the structure of the ground state of the theory. For appropriately chosen signs of $\mu$ and $\kappa$ the potential, and thus the spectrum of the theory, is bounded from below. However, only for $\mu < 0$ and $\kappa > 0$ one can have a non-trivial (non-perturbative) vacuum with

$$\langle\sigma\rangle \neq 0, \qquad (6.42)$$

which is needed to be in agreement with the condensate state ansatz. The two distinct minima of the potential are located at $\langle\sigma_0\rangle = \pm\sqrt{-\frac{\mu}{\kappa}}$ where the potential has strength $V_{\text{eff}}(\sigma_0) = -\mu^2/4\kappa$, which is lower than the value for the excited configuration $\sigma = 0$. The system would thus settle into one of the minima as its equilibrium configuration and could be used to describe a condensate.[9] This potential is illustrated in Fig. 6.9 and contrasted to the case where $\mu > 0$ for which the potential is a convex function of $\sigma$ with minimum at $\langle\sigma\rangle = 0$. The latter setting cannot be used to describe a condensate with $N \neq 0$. For other choices of signs, the equilibrium configuration $\langle\sigma\rangle = 0$ is unstable or metastable and should be dismissed. The upshot of this discussion is that if the effective action is to represent a stable system and a condensate of GFT quanta, one must choose the signs of the coupling constants accordingly.[10]

---

[9]A sign change of the driving parameter $\mu$ from positive to negative values induces a spontaneous symmetry breaking of the global $\mathbb{Z}_2$-symmetry of the action with interaction as in Eq. (6.37). This symmetry would have guaranteed the conservation of oddness or evenness of the number of GFT quanta as it corresponds to the conserved discrete quantity $(-1)^N$. For complex-valued GFT fields the analoguous situation would correspond to the spontaneous breaking of the global U(1)-symmetry of the action which would have guaranteed the conservation of the particle number $N$.

[10]For real BECs [463–465], $\kappa < 0$ gives an attractive interaction and only a large enough kinetic term can prevent the condensate from collapsing. In the opposite case where $\kappa > 0$, the interaction is repulsive and if it dominates over the kinetic term the condensate is well described in terms of the so-called Thomas-Fermi approximation [463–465].

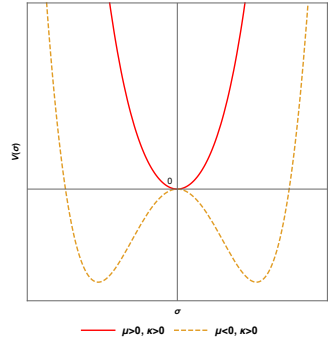

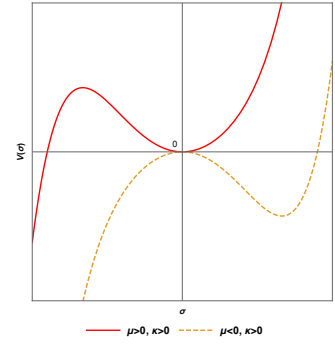

FIGURE 6.9: Plot of the effective potential $V_{\text{eff}}[\sigma] = \frac{\mu}{2}\sigma^2 + \frac{\kappa}{4}\sigma^4$.

FIGURE 6.10: Plot of the effective potential $V_{\text{eff}}[\sigma] = \frac{\mu}{2}\sigma^2 + \frac{\kappa}{5}\sigma^5$.

The second interaction term mimics a simplicial interaction for real-valued GFT fields. It is given by the pseudopotential

$$V_S[\varphi] = \frac{\kappa}{5} \int (\mathrm{d}g)^4 \varphi(g_{\mathrm{I}})^5,\tag{6.43}$$

one has

$$-\left[2p(1-p)\frac{\mathrm{d}^2}{\mathrm{d}p^2} + (3-4p)\frac{\mathrm{d}}{\mathrm{d}p}\right]\sigma(p) + \mu\sigma(p) + \kappa\sigma(p)^4 = 0.\tag{6.44}$$

For such a model the effective potential reads

$$V_{\text{eff}}[\sigma] = \frac{\mu}{2}\sigma^2 + \frac{\kappa}{5}\sigma^5.\tag{6.45}$$

Here we ignore that this potential is unbounded from below to one side.[11] Only for $(\mu < 0, \kappa > 0)$ or $(\mu < 0, \kappa < 0)$ one can have a non-trivial (non-perturbative) vacuum in agreement with the condensate state ansatz and the discussion of the choice of signs is similar to the potential considered first. Classically, the corresponding minima of the potential are then located at $\sigma_0 = \pm\sqrt[3]{\mp\frac{\mu}{\kappa}}$ where the potential has strength $V_{\text{eff}}(\sigma_0) = (\mp\frac{\mu}{\kappa})^{2/3}(3\mu/10)$. This is illustrated in Fig. 6.10.

### 6.2.2.2 Effective interactions as perturbations of the free case

In a first step, we consider the interaction term as a perturbation of the free case discussed in Section 6.2.1 using the same boundary conditions $\sigma(1) = 0$ and different $\sigma'(1)$ to numerically solve the non-linear differential equations (6.40) and (6.44), respectively. By closely following the procedure adopted in the free case, we compute the effect of perturbations onto the (probability) densities and the spectra of geometric operators. In this way, we obtain a clear qualitative picture of the effect of such interactions by comparing with the results for the free case.

---

[11]Notice that when using four arguments in the group field $\varphi$, higher simplicial interaction terms known to be e.g. of power 16 or 500 would lead in the local point of view, adopted here, to bounded effective potentials $V_{\text{eff}}$ like (6.41) and the discussion of their effects would be rather analogous.

In the following, we discuss the behaviour of solutions for the pseudotensorial potential (6.41) with $\mu < 0$ and $\kappa > 0$ and where the qualitative results differ also for the pseudosimplicial potential (6.45) with $\mu < 0$ and $\kappa > 0$ (or $\kappa < 0$) so that the potentials would possess non-trivial minima. The effect of weak non-linearities in the equation of motion onto the solutions is respectively illustrated in Fig. 6.11 and Fig. 6.12 in the $p$- and $\psi$-parameterisations and is contrasted to the behaviour of the free solutions of Section 6.2.1. In general, the finiteness of the free solutions at the origin is lost due to the interactions. Crucially, the concentration of the densities around the origin can still be maintained giving rise to nearly-flat solutions, as long as $|\kappa|$ does not become too big. For larger $j$, i.e. larger $|\mu|$, one sees that the departure from the free solutions is less pronounced because the $\mu$-term of the potential dominates longer over the $\kappa$-term.

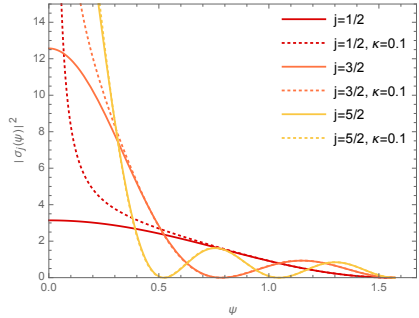

FIGURE 6.11: Probability density of the interacting mean field over $\psi$ for $V_{\text{eff}}[\sigma] = \frac{\mu}{2}\sigma^2 + \frac{\kappa}{4}\sigma^4$.

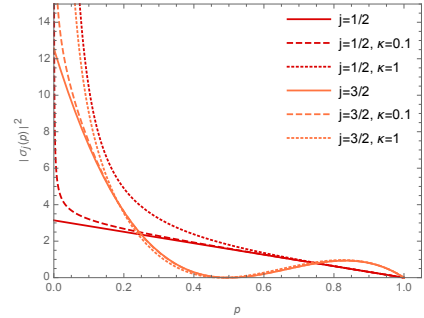

FIGURE 6.12: Probability density of the interacting mean field over $p$ for $V_{\text{eff}}[\sigma] = \frac{\mu}{2}\sigma^2 + \frac{\kappa}{4}\sigma^4$.

When $|\kappa|$ and $|\sigma'(1)|$ are small, solutions will remain normalisable with respect to the Fock space measure, i.e.

$$N = \int (\mathrm{d}g)^4 |\sigma(g_\mathrm{I})|^2 < \infty. \tag{6.46}$$

However, when gearing up toward the strongly non-linear regime, i.e. $\kappa \gtrsim \mathcal{O}(1)$, this feature is lost as $N$ grows and eventually one finds $N \to \infty$.[12] The loss of normalisability of $\sigma$ with respect to the Fock space measure in the strongly non-linear regime goes in hand with the breaking of the rescaling invariance expressed by Eq. (6.24) and signals the breakdown of the ansatz used here. Such behaviour is not surprising, as it is well-known within the context of local QFTs that the proper treatment of interactions necessitates the use of non-Fock representations for which $N$ is infinite, see Section B and [450–456]). We will get back to this point below.

---

[12] It should be noted that the precise values of $\kappa$ and/or $\sigma'(1)$ for which $N \to \infty$ depend on the numerical accuracy of the used solver. In this sense the observation of such behaviour is a qualitative result.

With regard to the average of the field strength, one observes that $\kappa > 0$ increases $\langle \hat{F}^i \rangle / N$ for some $j$ in comparison to the free case, whereas for negative $\kappa$ the expectation value decreases. This behaviour is reminiscent of the effect of similar interactions onto the effective curvature of the space described by the condensate found by the author in Refs. [489, 490], presented here in Sections 6.2.5 and 6.3.5. There it is shown that a bounded interaction potential generically leads to recollapsing GFT condensates.

By means of the numerically computed solutions, one can obtain their corresponding Fourier components and with these one yields in close analogy to the free case the modified spectra of the volume and area operators, illustrated in Fig. 6.13 and Fig. 6.14.

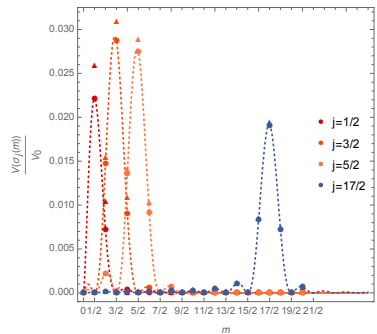

FIGURE 6.13: Normalised spectrum of the volume operator with respect to the interacting mean field $\sigma_j(\psi)$ for $\kappa = 0.22$ (triangles) compared to the respective free solutions (fat dots).

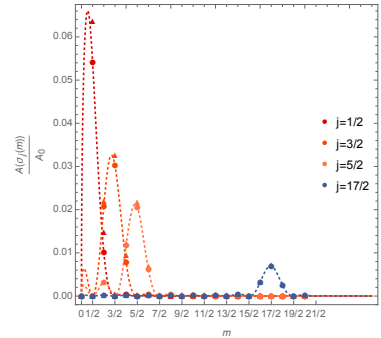

FIGURE 6.14: Normalised spectrum of the area operator with respect to the interacting mean field $\sigma_j(\psi)$ for $\kappa = 0.22$ (triangles) compared to the respective free solutions (fat dots).

The plots clearly indicate that perturbations for $\kappa > 0$ increase both the volume and the area, however, in the weakly non-linear regime they remain finite. More specifically, one observes that the effect of the perturbations upon the spectra of the volume and area are more pronounced for small $j$, i.e. small $|\mu|$, since for these the non-linearity dominates quickly over the $\mu$-term of the potential. Moreover, one notices that when pushing $\kappa$ to larger values as a consequence the volume $V$ and area $A$ quickly blow up in the same way as $N$ does, whereas $\langle \hat{F}^i \rangle / N$, $V/N$ and $A/N$ remain finite.

For the pseudosimplicial potential one obtains qualitatively analogous results with the differences to the free solutions being more emphasised since the non-linearity is stronger.

### 6.2.2.3 Analysis of the condensate close to non-trivial minima

To chart the condensate phase and understand its properties, it is necessary to numerically study the solutions to the non-linear differential equation (6.40) around the non-trivial minima. To this aim, we choose the coupling constants in Eq. (6.41) in such a manner that the potential forms a Mexican hat, as in Fig. 6.9. We select the position of the minimum $\sigma_0$ as well as $\sigma'(1)$ as the boundary conditions in order to find solutions numerically. Since the chart is only defined within $p \in [0, 1]$, we begin solving from $p = 1$ backwards.

Without any loss of generality, we use the same values for $\mu$ as in the previous subsections. Apart from the requirement that they assume negative values they could be completely arbitrary since here we do not study eigensolutions to the Dirichlet Laplacian as in Section 6.2.1.

Figure 6.15 shows the resulting (probability) density and the potential over $p$ and $\psi$ computed for an exemplary choice for the values of the free parameters. Depending on the sign of $\sigma'(1)$ or $\sigma'(\frac{\pi}{2})$, respectively, the solution either climbs over the local maximum at $\sigma = 0$, then reaches the other minimum after which it ascends the left branch of the potential or directly climbs up the right branch shown in Fig. 6.9. For the choice of parameters leading to Fig. 6.15, the solutions are normalisable. In general, for small $\sigma'(1)$ the solutions crawl slowly out of the minima and if $\sigma'(1)$ is almost zero, the solutions remain almost constant up to $p = 0$, where the regular singularity of the differential equation finally kicks in. The contribution of the Laplacian term is less pronounced for smaller $\mu$ than for larger ones, as the right hand side of Fig. 6.15 in the $\psi$-parameterisation illustrates. Similar results are obtained when $\mu$ is kept fixed while decreasing $|\sigma'(\frac{\pi}{2})|$.

It is clear that as long as for the boundary condition $\sigma' \approx 0$ holds, this is equivalent to neglecting the Laplacian part of the kinetic term $\mathcal{K}$ in the equation of motion. The solutions, exemplified by Fig. 6.15, show that the properties of the non-trivial ground state are then defined by the ultralocal action. Solving the equation of motion starting at the minima of the effective potentials gives rise to almost constant, i.e. homogeneous, functions

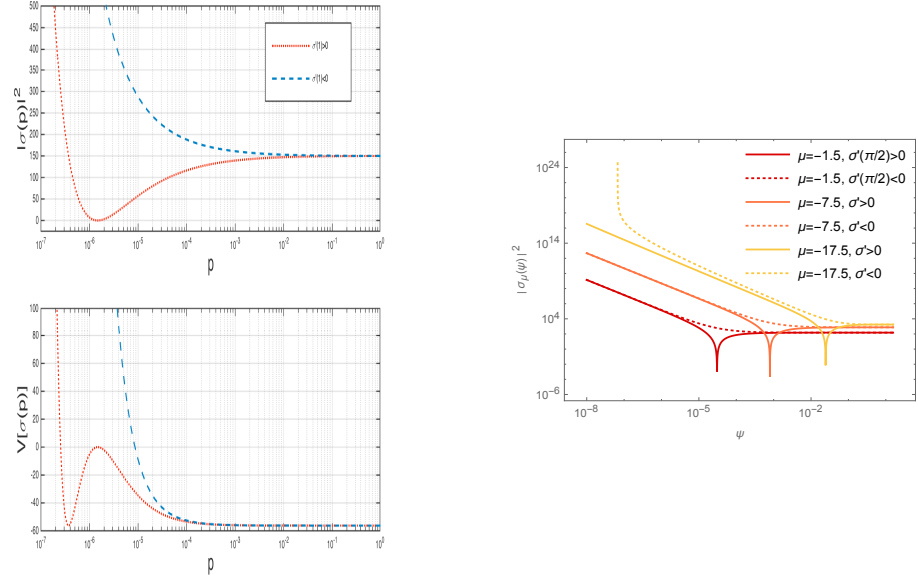

FIGURE 6.15: Left: Semilog plot of the probability density and potential for solutions $\sigma(p)$ with $\mu = -1.5$, $\kappa = 0.01$, $\sigma(1) = 12.2474$ and $\sigma'(1) = \pm 100$ for the potential $V_{\text{eff}}[\sigma] = \frac{\mu}{2}\sigma^2 + \frac{\kappa}{4}\sigma^4$. Solutions were computed by means of MATLAB's ODE45 solver which is based on an explicit Runge-Kutta $(4, 5)$ formula. Output was generated for $10^5$ points on the interval $[0, 1]$ while making use of highly stringent error tolerances. Right: Double-log plot of the probability density for solutions $\sigma(\psi)$ for different $\mu$, with the same $\kappa = 0.01$, and the same $|\sigma'(\frac{\pi}{2})|$ at the respective minima $\sigma(\frac{\pi}{2})$ for the potential $V_{\text{eff}}[\sigma] = \frac{\mu}{2}\sigma^2 + \frac{\kappa}{4}\sigma^4$.

on the domain.[13]

The geometric interpretation of such solutions which "sit" in the equilibrium position is slightly obstructed. This is due to the fact that the above-used near-flatness condition cannot be straightforwardly applied to such solutions. Despite the fact that the (probability) density can be tuned to be concentrated around low curvature values, it is finite close to the equator at $p = 1$, while in other cases it simply remains constant on the whole interval, as Fig. 6.15 shows. This calls for a more differentiated formulation of this condition, perhaps by means of a well defined GFT-operator capturing the average curvature of the 3-space described by means of the condensate state.

In spite of the current lack of such an operator, it is possible to obtain from exemplary numerical solutions the spectrum of the volume and area operators, as illustrated in

---

[13]Completely neglecting the Laplacian from the onset, is only justified when the interaction is dominant which corresponds to the regime of large ground state condensate "density", i.e. $\kappa N \gg 1$. In the context of real BECs this is known as the Thomas-Fermi approximation [463–465].

Figs. 6.16 and 6.17. Solutions which are computed around the non-trivial minima give rise to a different qualitative form of the spectrum of the volume and area as compared to the ones obtained in Sections 6.2.1.3 and 6.2.2.2; nevertheless we emphasise again the relevance of low-spin modes. In general, for different $\mu$ the dominant contribution to the volume $V$ and area $A$ comes from the Fourier coefficients with $m = \frac{1}{2}$, whereas in the previous cases the predominant contribution comes from the Fourier coefficients with $m = j$. This is due to the fact that $\sigma$ remains mostly constant and is thus best approximated by the simplest non-trivial modes for $m = 0, \frac{1}{2}$. In particular, one can check that the contributions to $V$ and $A$ coming from the modes with $m > \frac{1}{2}$, are exponentially suppressed when $\sigma'(1) \approx 0$.[14]

Moreover, the volume and area remain finite in the weakly non-linear case and when the boundary condition $\sigma'(1)$ is relatively small. The use of weak interactions is thus instructive in order to understand the qualitative behaviour of the solutions in particular with regard to the expectation values of the geometric operators. Since the size of $\kappa$ only has a quantitative impact on the spectrum, as the right hand side of Fig. 6.16 suggests, an analogous form of the spectra can also be expected in the strongly non-linear regime. Furthermore, for bigger values of $|\sigma'(1)|$ and/or strongly non-linear interaction terms, the volume and area, as well as the expectation value of the number operator $\hat{N}$ blow quickly up. As noticed above, this signals the breakdown of the simple condensate state ansatz used here and suggests the need for non-Fock coherent states once the strongly correlated regime is explored, see Appendix B.[15]

Such solutions yield for all choices of $\mu < 0$ and $\kappa > 0$ for the averaged observables

$$\frac{\langle \hat{\mathcal{O}} \rangle}{N} \approx const., \tag{6.47}$$

since $\sigma$ is approximately constant. This naturally applies to the averaged field strength $\langle \hat{F}^i \rangle / N$ which is larger than in the corresponding free case. This indicates that the chosen effective GFT interactions have the effect of positively curving the effective geometry described by the condensate state. This is again reminiscent of similar findings by the author in Refs. [489, 490], presented in Sections 6.2.5 and 6.3.5, where it is shown that relationally evolving and effectively interacting GFT condensate cosmology models display recollapsing solutions when the interaction potential is bounded from below.

---

[14]Obviously, the contributions to the volume and area operators stemming from the zero-mode are vanishing, while its occupation number is non-zero here. Notice again that this is due to the fact that the Hilbert space of GFT (in contrast to the one of LQG) includes this mode [437], unless it is exluded e.g. by boundary conditions as above. We refer to Ref. [488] for a discussion of this issue.

[15]The right hand side of Fig. 6.16 also seems to suggest that the volume $V$ is ever increasing for $\kappa \to 0$. However, in such a limit, it is more appropriate to treat the system as in the free case, as discussed in Section 6.2.1.

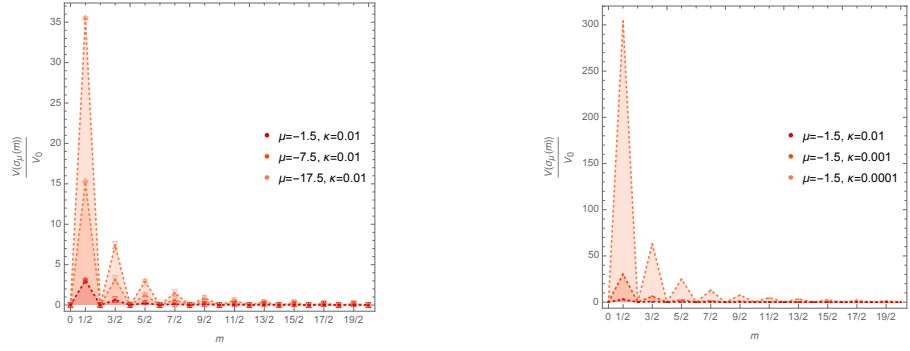

FIGURE 6.16: Left: Normalised discrete spectrum of the volume operator (in arbitrary units) with respect to the interacting mean field $\sigma_\mu(\psi)$: Solutions $\sigma_\mu(\psi)$ were obtained with $\kappa = 0.01$ but boundary conditions differ for each $\mu$ to solve around a non-trivial minimum of the respective potential $V_{\text{eff}}[\sigma] = \frac{\mu}{2}\sigma^2 + \frac{\kappa}{4}\sigma^4$. Right: Normalised discrete spectrum of the volume operator (in arbitrary units) with respect to the interacting mean field $\sigma_\mu(\psi)$: Solutions $\sigma_\mu(\psi)$ were obtained for $\mu = -1.5$, different $\kappa$ and the same boundary conditions $\sigma'(\frac{\pi}{2})$ to solve around the respective non-trivial minima of the potential $V_{\text{eff}}[\sigma] = \frac{\mu}{2}\sigma^2 + \frac{\kappa}{4}\sigma^4$.

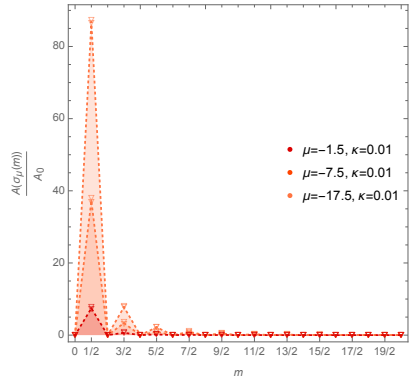

FIGURE 6.17: Normalised discrete spectrum of the area operator (in arbitrary units) with respect to the interacting mean field $\sigma_\mu(\psi)$: Solutions $\sigma_\mu(\psi)$ were obtained with $\kappa = 0.01$ but boundary conditions differ for each $\mu$ to solve around a non-trivial minimum of the potential $V_{\text{eff}}[\sigma] = \frac{\mu}{2}\sigma^2 + \frac{\kappa}{4}\sigma^4$.

Analogously, such a discussion can be repeated for the pseudosimplicial potential, where the solutions to the non-linear equation of motion are illustrated in Fig. 6.18. The resulting behaviour of the relevant operators is similar and will not be repeated here, though it should be kept in mind that only such interaction terms can be more closely related to models with a simplicial quantum gravity interpretation.

To summarise the main points of this subsection: We computed static condensate solutions around the non-trivial minima of the interaction potentials of which the essential features can be defined by means of the ultralocal action. We found that the condensate consists of many GFT quanta residing in the low-spin modes $m = 0, \frac{1}{2}$. This is indicated by the analysis of the discrete spectra of the geometric operators, dominated by the lowest non-trivial mode $m = \frac{1}{2}$. Such low-spins actually correspond to the infrared regime of the theory. Hence, these results fit well into the picture suggested by the above-mentioned functional renormalisation group (FRG) analyses which find infrared fixed points in all GFT models considered so far marking the formation of a condensate phase the main features of which are supposed to be captured by means of the employed condensate state. In this sense, the condensate phase may describe an effectively continuous homogeneous and isotropic 3-space built from many small building blocks of the quantum geometry. Moreover, these results are also interesting from the point of view of LQC where one typically assumes that the quantum geometry resides in the lowest non-trivial configuration without giving any deeper explanation for this. Our results indicate that the condensate cosmology approach may be able to underpin such a basic assumption.

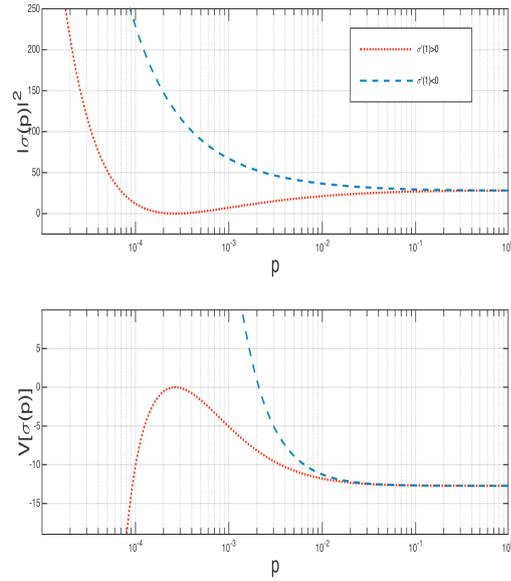

FIGURE 6.18: Semilog plot of the probability density and potential for solutions $\sigma(p)$ with $\mu = -1.5$, $\kappa = 0.01$, $\sigma(1) = 5.3132$ and $\sigma'(1) = \pm 100$ for the potential $V_{\text{eff}}[\sigma] = \frac{\mu}{2}\sigma^2 + \frac{\kappa}{5}\sigma^5$. Solutions were computed by means of MATLAB's ODE113 procedure which is a variable order "Adams-Bashforth-Moulton predictor-corrector" solver. Output was generated for $10^5$ points on the interval $[0,1]$ while making use of highly stringent error tolerances.

### 6.2.3 Discussion of the results of the statics of model 1

In the first part of our analysis of model 1 we investigated the impact of simplified interactions onto static GFT quantum gravity condensate systems describing effective 3-geometries with a tentative cosmological interpretation. To this aim, we extensively examined the geometric properties of a free system in an isotropic restriction by studying the spectra of the volume and area operators imported from LQG and comparing the results to the perturbed case. In a last step, we studied the features of the GFT condensate when the system sits in the non-trivial minima of the effective interaction potentials. The main result of this study is then that the condensate consists of many discrete (non-degenerate) building blocks predominantly of the smallest non-trivial size encoded by the quantum number $m = \frac{1}{2}$ – which supports the idea that an effectively continuous geometry can emerge from the collective behaviour of a discrete pregeometric GFT substratum [461, 462]. In this sense, our results also strengthen the connection with LQC where the typically used quantum states are constructed from the assumption that the quanta of the geometry all reside in the same lowest non-trivial configuration [75, 76, 492]. This lends strong support to the idea that condensate states are appropriate for studying the cosmological sector of LQG.

The results can also be seen as a support of the idea proposed in Ref. [525]: The Laplacian in the kinetic operator $\mathcal{K}$, originally motivated by field theoretic arguments to guarantee the consistent implementation of a renormalisation scheme, might only be a property of the ultraviolet-completed GFT without a significant physical effect in the effectively continuous region which is expected to correspond to the small spin (infrared) regime together with many building blocks of the quantum geometry. In this regime, the kinetic term is then suggested to become ultralocal, thus allowing for a straightforward interpretation of the GFT amplitudes in terms of spin foam amplitudes for quantum gravity. The numerical analysis done here indeed suggests that from the ultralocal action alone one can find that the condensate consists of many GFT quanta residing in the low-spin configuration.

In the following we want to comment on the limitations of our discussion. We implicitly assumed that the condensate ansatz is trustworthy for any $\mu \leq 0$, where $\mu = 0$ marks the critical value at which the phase transition from the unbroken into the condensate phase is supposed to take place [393, 475, 476]. With respect to these findings, our analysis should be complemented by investigating whether indications for a phase transition into a condensate phase can be observed and whether their possible absence might be related to the expectation that true phase transitions are only realised for GFTs on non-compact manifolds, like Lorentzian quantum gravity models, as noticed in Ref. [475, 476]. Studying GFTs with Landau's mean field theory and the Gaussian approximation, as done in Section 5.3, is a step into this direction, but our results there indicate that non-perturbative techniques should be employed to settle these issues.

In this light, it is worth noting that in the context of weakly interacting, diluted and ultracold non-relativistic BECs [463–465] it is well understood that Bogoliubov's mean field and perturbation theory [477–479] becomes invalid and breaks down in the vicinity of the critical point of the phase transition because quantum fluctuations become important. Of course, as is generally known today, mean field approaches only accurately work as effective descriptions of thermodynamic phases well away from critical points. A satisfactory description for such systems which systematically extends Bogoliubov theory and cures its infrared problems has been given in terms of FRG techniques [224, 526–531]. The example of real BECs suggests that the analogue of the Bogoliubov ansatz for quantum gravity condensates should be similarly extended by means of FRG methods at the critical point.

In addition, it is also well understood that Bogoliubov theory for real BECs breaks down, when considering condensates with rather strongly interacting constituents. Likewise, FRG techniques can systematically implement non-perturbative extensions to Bogoliubov's approximation. These suggest that for Bose condensates with approximately pointlike interactions like in superfluid $^4$He, it is only possible to realise a strongly interacting regime for a very dense condensate [526–531]. This example could indicate a similar failure of the quantum gravity condensate ansatz when considering the strongly interacting regime. Indeed, when increasing the coupling constant $\kappa$ in this sector, the average particle number $N$ grows. If $\kappa$ is too large, we find that solutions are generally not normalisable with respect to the Fock space measure. The regime of large number of quanta $N$ and the eventual failure of $|\sigma\rangle$ to be normalisable in this sector certainly mark the breakdown of the Gross-Pitaevskii approximation to the dynamics (5.12) for the simple condensate state constructed with Bogoliubov's ansatz (5.3). In this regime, quantum fluctuations and correlations among the condensate quanta become relevant and only solutions to the full quantum dynamics together with FRG techniques would be capable of capturing adequately their impact. This entails that the approximation used here should only be trusted in a mesoscopic regime where $N$ is not too large, as noticed in Ref. [486]. Nevertheless, the finding of solutions corresponding to non-Fock representations gives a forecast on what should be found when considering non-perturbative extensions of the techniques used here.

In fact, the loss of normalisability is not too surprising because it is a generic feature of massless or interacting (local) QFTs according to Haag's theorem which require the use of non-Fock representations [450–453]. However, finding such solutions is first of all intriguing as a matter of consistency because non-Fock representations are also required in order to describe many particle systems in the thermodynamic limit. It is only in this limit that inequivalent irreducible representations of the CCRs become available which is a prerequisite for the occurrence of non-unique equilibrium states, in turn essential to consistently describing phase transitions [450–453]. It is also interesting for a second reason, since in the context of quantum optics it was understood that such non-Fock coherent states with an infinite number of (soft) photons can be described in terms of a classical radiation field [532]. Hence, the occurrence of non-Fock coherent state solutions in our context might also play a role in the classicalisation of the system and could be important to consistently capture continuum macroscopic information of the GFT system. This would intuitively make sense, because one would expect to look for the physics of continuum spacetimes in the regime far from the perturbative Fock vacuum corresponding to the no-space state.

To fully extract the geometric information encoded by such solutions, it would then also be necessary to go beyond the use of the simplified local interactions and explore the effect of the proper combinatorially non-local interactions encountered in the GFT literature (e.g. on the expectation values of the geometric operators) in order to compare it to the results obtained here. Additionally, it is worth mentioning that only for proper simplicial interaction terms the quantum geometric interpretation is rather straightforward while for the others a full geometric interpretation is currently lacking. This connects to the analysis of solutions to the dynamical Boulatov model in Section 5.2 and that of perturbations for the EPRL GFT model in Section 6.3 under symmetry restrictions, however, more general scenarios should be investigated.

Finally, we remarked in our analysis that the notion of near-flatness used in the previous subsections should be reconsidered for the condensate solutions around the non-trivial minima since it cannot be straightforwardly applied then. Statements regarding the flatness property of such solutions can only be satisfactorily made if the spectrum of a currently unavailable GFT curvature operator is studied. Potentially, this notion does not make sense for such situations anymore, since the curvature of the emergent geometry is curved. This deserves further scrutiny.

In a next step, we will study the time evolution of the condensate with respect to a relational clock. This will allow for the extraction of further phenomenological consequences from our model.

### 6.2.4 Relational dynamics of a free isotropic GFT condensate

This Section investigates the relational evolution of a free GFT condensate system in the above-introduced isotropic restriction. At first, we show that such a system quickly settles into a low-spin configuration in the free case. Using this, we then demonstrate how Friedmann-like dynamics emerge in the semi-classical limit. Finally, we show that the expansion of the emergent space is accelerated, though not strong enough to supplant the inflationary mechanism [27].

#### 6.2.4.1 Emergence of a low-spin phase

In Ref. [488] a mechanism for the dynamical relaxation of a GFT condensate system into a low-spin phase was proposed. We employ this idea, however, in contrast to the strategy followed there, we do not work from the beginning in the spin representation. We also do not make use of the notion of isotropy applied there which was introduced in Ref. [486]. Instead, we work in the coordinate representation, which together with the above-discussed symmetry reductions, lead us to a qualitatively equivalent result. In addition, we illustrate the entire domain of values of the parameters for which exponentially dominating low-spin configurations can be found.

To this aim, we start with the equation of motion for the free case,

$$\left[ \left( \tau \partial_\phi^2 - \sum_{I=1}^4 \Delta_{g_I} \right) + m^2 \right] \sigma(g_I, \phi) = 0, \tag{6.48}$$

which takes account of the evolution with respect to the relational clock $\phi$. With the above-discussed isotropic restriction, this yields the partial differential equation (PDE)

$$\left[ \left( 2\tau \partial_\phi^2 - \frac{\mathrm{d}^2}{\mathrm{d}\psi^2} - 2\cot(\psi)\frac{\mathrm{d}}{\mathrm{d}\psi} \right) + 2\mu \right] \sigma(\psi, \phi) = 0, \quad \psi \in [0, \frac{\pi}{2}]. \tag{6.49}$$

General solutions to this PDE can be obtained by employing a separation ansatz which is justified because the terms in $\phi$ and $\psi$ completely decouple. With $\sigma(\psi, \phi) = \xi(\psi)T(\phi)$ one yields,

$$2\tau \frac{\partial_\phi^2 T(\phi)}{T(\phi)} = \frac{(\frac{\mathrm{d}^2}{\mathrm{d}\psi^2} + 2\cot(\psi)\frac{\mathrm{d}}{\mathrm{d}\psi} - 2\mu)\xi(\psi)}{\xi(\psi)} \equiv \omega = \mathrm{const.} \tag{6.50}$$

The general solution to

$$\partial_\phi^2 T(\phi) = \frac{\omega}{2\tau} \, T(\phi) \ , \tag{6.51}$$

is given by

$$T(\phi) = \left( \mathrm{a}_1 \, \mathrm{e}^{\sqrt{\frac{\omega}{2\tau}}\phi} + \mathrm{a}_2 \, \mathrm{e}^{-\sqrt{\frac{\omega}{2\tau}}\phi} \right), \tag{6.52}$$

with the constants $\mathrm{a}_1, \mathrm{a}_2 \in \mathbb{C}$.

As in Section 6.2.1, the spectrum of the differential equation depending solely on $\psi$ is concretised by imposing boundary conditions. In accordance with the near-flatness condition one chooses the Dirichlet boundary condition $\xi(\psi = \frac{\pi}{2}) = 0$ so that the eigensolutions of the $\psi$-part are given by

$$\xi_j(\psi) = \frac{\sin((2j+1)\psi)}{\sin(\psi)}, \ \ \psi \in [0, \frac{\pi}{2}] \tag{6.53}$$

with eigenvalues $\omega + 2\mu = -4j(j+1)$ and $j \in \frac{2\mathbb{N}_0+1}{2}$.[16]

In the following, we want to study the evolution of the solutions $\sigma_j(\psi, \phi) = \xi_j(\psi)T_j(\phi)$ with respect to the relational clock $\phi$ with particular regard to their behaviour for $\phi \to 0$ and $\phi \to \pm\infty$. Using the form of the spectrum, we can substitute $\omega$ into Eq. (6.52) such that

$$T_j(\phi) = \left( \mathrm{a}_1 \, \mathrm{e}^{\sqrt{\frac{\mu+2j(j+1)}{-\tau}}\phi} + \mathrm{a}_2 \, \mathrm{e}^{-\sqrt{\frac{\mu+2j(j+1)}{-\tau}}\phi} \right) \tag{6.54}$$

the behaviour of which critically depends on the sign of $\frac{\mu+2j(j+1)}{-\tau}$ with $\mu < 0$ and $\tau > 0$. For two possible initial conditions $T_j(0) = 0$ and $T_j'(0) = 0$ the solutions are

$$T_j(\phi) = 2\mathrm{a}_1 \sinh\left( \sqrt{\frac{\mu+2j(j+1)}{-\tau}}\phi \right) \tag{6.55}$$

and

$$T_j(\phi) = 2\mathrm{a}_1 \cosh\left( \sqrt{\frac{\mu+2j(j+1)}{-\tau}}\phi \right), \tag{6.56}$$

---

[16]Notice that $\xi_j(\psi)$ is zero on the interval $[\frac{\pi}{2}, \pi]$ which leads to a double index in $\sigma_{j;m}(\phi)$ from Eq. (6.61) onward as remarked for the static case in Section 6.2.1 .

respectively. Importantly, only in the latter case $\lim_{\phi \to 0} T_j(\phi) \neq 0$ holds leading to a non-vanishing volume, as we shall see below. Hence, for such solutions the singularity problem is avoided.[17][18][19]

Solutions for which

$$\frac{1}{2} \leq j < -\frac{1}{2} + \frac{1}{2}\sqrt{1 - 2\mu} \quad \text{with} \quad \mu < -\frac{3}{2} \tag{6.57}$$

grow exponentially for $\phi \to \pm\infty$ whereas solutions with

$$j \geq -\frac{1}{2} + \frac{1}{2}\sqrt{1 - 2\mu} \quad \text{with} \quad \mu \leq -\frac{3}{2} \tag{6.58}$$

or

$$j \geq \frac{1}{2} \quad \text{with} \quad \mu > -\frac{3}{2} \tag{6.59}$$

display an oscillating behaviour. This is illustrated by means of the differently colourised sectors in Fig. 6.19. Using this, we see that the condensate will quickly be dominated by the lowest representation $j = \frac{1}{2}$ and all others are exponentially suppressed.

To extract further information about the behaviour of the total volume of a general solution

$$\sigma(\psi, \phi) = \sum_{j \in \frac{2\mathbb{N}_0 + 1}{2}} \xi_j(\psi) T_j(\phi), \tag{6.60}$$

we compute the Fourier components $\xi_{j;m}$ of $\xi_j(\psi)$ as in Section 6.2.1. Using this, the expectation value of the volume operator with respect to the mean field $\sigma$ is decomposed as

$$\langle \hat{V}(\phi) \rangle \equiv V(\phi) = V_0 \sum_{j \in \frac{2\mathbb{N}_0 + 1}{2}} \sum_{m \in \frac{\mathbb{N}_0}{2}} |\sigma_{j;m}(\phi)|^2 \, V_m \tag{6.61}$$

---

[17]In principle, constructing solutions which vanish at $\phi = 0$ seems to be arbitrary. Instead, one could use that Eq. (6.54) vanishes for $\phi = \sqrt{\frac{2\tau}{\omega}} \frac{\log(-a_1/a_2)}{2}$ but this would not change the essence of our argument.

[18]We want to remark that the resolution of the singularity problem for real-valued GFT fields thus depends on the choice of initial conditions. This is clearly different to the situation found for complex-valued GFT fields in Refs. [486, 487, 489, 491]. The occurrence of a bounce in these works is deeply rooted in the global U(1)-symmetry of the complex-valued GFT field to which a conserved charge $Q$ can be associated, preventing the field from becoming zero for all values of the relational clock $\phi$ as long as $Q$ is non-zero. This mechanism does not depend on the initial conditions assigned to the mean field. This becomes transparent in Section 6.3 where the results of Refs. [489, 491] are presented.

[19]Notice that "(curvature) singularity avoidance" assumes that non-vanishing volume is related to a finite value of curvature. In the absence of a GFT curvature operator, this should be taken with a grain of salt as it presupposes terminology from general relativity [1], which should in fact be derived from a fundamental framework that GFT asserts to be.

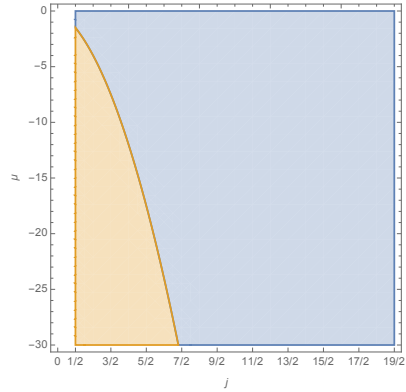

FIGURE 6.19: Two sectors representing exponentially growing (orange) and oscillating (blue) solutions.

with $V_m \sim m^{\frac{3}{2}}$ and $V_0 \sim \ell_p^3$.

From the relational evolution of the volume we deduce that only when $\lim_{\phi \to 0} T_j(\phi) \neq 0$ holds, the volume will be non-vanishing which indicates that the singularity problem can be avoided.

At large $\phi$, the dominant contribution to the total volume is then given by

$$\lim_{\phi \to \pm\infty} V(\phi) = V_0 \; V_{\frac{1}{2}} \; |\xi_{\frac{1}{2},\frac{1}{2}}|^2 \; |a_{1,2}|^2 \; e^{\pm 2\sqrt{\frac{\mu+\frac{3}{2}}{-\tau}}\phi}. \qquad (6.62)$$

Bearing the slight differences in the sign conventions for $\mu$ and $\tau$ as compared to Ref. [488] in mind, this result is, up to a factor of $\frac{1}{2}$ in the argument of the exponential function, just identical to the one obtained there. This tiny difference seems to reflect the differences in the imposition of the notion of isotropy and it is rather interesting that both also quantitatively nearly lead to the same behaviour at late times. In addition, we want to remark that one also here obtains the direct proportionality between the asymptotic total number of quanta $N$ and the volume $V$ which in the context of LQC is needed for the so-called improved dynamics scheme [75, 76, 492, 533], as pointed out in Ref. [488].

A last remark is in order. When computing the relative standard deviation or uncertainty of the volume operator with respect to the coherent state $\sigma$, given by

$$\epsilon = \frac{\sqrt{\langle \hat{V}^2 \rangle_\sigma - \langle \hat{V} \rangle_\sigma^2}}{\langle \hat{V} \rangle_\sigma}, \qquad (6.63)$$

one can show that it is dominated by the contributions of the $\frac{1}{2}$-mode at late times. In particular, one finds that

$$\lim_{\phi \to \pm \infty} \epsilon \ \sim \ |\xi_{\frac{1}{2};\frac{1}{2}}|^{-1} \ \mathrm{e}^{\mp \sqrt{\frac{\mu+\frac{3}{2}}{-\tau}}\phi} \to 0, \tag{6.64}$$

which shows that measured values of the volume are tightly clustered around its mean. This suggests that the quantum geometry classicalises at late times if the condensate system relaxes into a low-spin configuration.

### 6.2.4.2 Emergent Friedmann dynamics and accelerated expansion

#### (A) Emergent Friedmann dynamics

In the following, we make use of the fact that the equation of motion Eq. (5.12) together with the kinetic kernel Eq. (6.3) in the free case lead to the conserved quantity,

$$E = \frac{1}{2} \int (\mathrm{d}g)^4 \left[ \tau(\sigma')^2 - \sigma \Delta \sigma + m^2 \sigma^2 \right], \tag{6.65}$$

termed "GFT energy" in the literature [486]. From a fundamental point of view, its physical meaning has yet to be clarified. In the remainder, we will simply refer to it as conserved quantity $E$.

When working in the above-introduced isotropic restriction with $\mu \equiv \frac{m^2}{12}$ and employing the Fourier decomposition of $\sigma(\psi, \phi)$ this expression can be rewritten into

$$E = \sum_{j \in \frac{2\mathbb{N}_0+1}{2}} \sum_{m \in \frac{\mathbb{N}_0}{2}} E_{j;m} \tag{6.66}$$

where

$$E_{j;m} = \frac{1}{2} \left[ \tau \sigma_{j;m}'^2(\phi) + [2(m(m+1)) + \mu]\sigma_{j;m}^2(\phi) \right] \ . \tag{6.67}$$

Using the introduction of the relational clock, the first Friedmann equation can be written as

$$H^2 = \left( \frac{V'}{3V} \right)^2 \left( \frac{d\phi}{dt} \right)^2 \tag{6.68}$$

where $t$ denotes the proper time. Together with the expectation value of the volume operator Eq. (6.61) one obtains

$$\left(\frac{V'}{3V}\right)^2 = \left(\frac{2\sum\limits_{j,m} V_m \sigma_{j;m}\sqrt{\frac{(2E_{j;m}-[2m(m+1)+\mu])\sigma_{j;m}^2}{\tau}}}{3\sum\limits_{j,m} V_m \sigma_{j;m}^2}\right)^2 \tag{6.69}$$

where Eq. (6.67) was used and the argument in $\sigma_{j;m}(\phi)$ is suppressed from hereon. Given that $\mu < 0$, in the semi-classical limit, where $\sigma_{j;m}^2$ is large (cf. Ref. [486]), one finds

$$\left(\frac{V'}{3V}\right)^2 \approx \left(\frac{2\sum\limits_{j,m} V_m \sigma_{j;m}\sqrt{\frac{[|\mu|-2m(m+1)]\sigma_{j;m}^2}{\tau}}}{3\sum\limits_{j,m} V_m \sigma_{j;m}^2}\right)^2. \tag{6.70}$$

Finally, exploiting the fact that quickly the mean field dynamically settles into the $j = \frac{1}{2}$ configuration, as obtained in the previous subsection, with the identification[20] $3\pi G = \frac{1}{\tau}[|\mu| - \frac{3}{2}]$ one recovers the classical Friedmann equations

$$\left(\frac{V'}{3V}\right)^2 = \frac{4\pi G}{3} \tag{6.71}$$

and

$$\frac{V''}{V} = 12\pi G, \tag{6.72}$$

both given in terms of the relational clock $\phi$. We note that here we neglect the quantum gravity corrections to these equations stemming from the term proportional to $E_{\frac{1}{2};\frac{1}{2}}$ which is of the order of $\sqrt{\hbar G}$ as in Ref. [486].

Some remarks are in order. Firstly, the isotropic restriction on the coordinates of the group manifold used here differs somewhat to the one employed in the spin representation in Ref. [486]. Interestingly, the results do only slightly vary from a quantitative point of view suggesting that they describe (almost) the same quantum geometric configuration. Second, in contrast to Ref. [488] the free condensates used here exclude the occurrence of excitations with $j = 0$. In our context these would correspond to the mean field solving the equation of motion for $\mu = 0$, which is identical to zero (cf. Ref. [457]). Also, the mean field configurations considered here obey the near-flatness condition, that means the quantum geometric building blocks are almost flat as naively required for the emergence of an effectively continuous 3-geometry from them [448, 457]. We also do not restrict ourselves *a priori* to a configuration with just one $j$. Rather, we keep the analysis as general as possible and only use in a final step the relaxation mechanism into the $j = \frac{1}{2}$ configuration to recover the Friedmann equations.

---

[20]More precisely, in Ref. [487] it was shown that such an identification holds asymptotically when the value of the relational clock grows.

## (B) Accelerated expansion

Finally, we want to investigate the question whether in the case of the free model it is possible to obtain an era of accelerated expansion of the space built from the GFT quanta which lasts long enough to replace the standard inflationary scenario by relying entirely on quantum geometric arguments. We do this by closely following the techniques developed by the author in Ref. [489] with some slight deviations to account for the characteristics of the model studied here.

A necessary condition to call a possible era of accelerated expansion an inflationary era is that the number of e-folds

$$N = \frac{1}{3} \log \left( \frac{V_{\text{end}}}{V_{\text{beg}}} \right) \tag{6.73}$$

must be $N \gtrsim 60$ where $V_{\text{beg}}$ is the volume of the universe at the beginning of the phase of accelerated expansion and $V_{\text{end}}$ denotes its volume at the end of it.[21] To simplify the following calculations, we then assume that the volume receives its major contribution at both points from a single $j$-mode. Motivated by the results of Section 6.2.4.1, we choose $j$ to be equal to $\frac{1}{2}$ and thus set $\sigma_{\frac{1}{2}; \frac{1}{2}}(\phi) \equiv \sigma$. Hence Eq. (6.73) is recast into

$$N = \frac{2}{3} \log \left( \frac{\sigma_{\text{end}}}{\sigma_{\text{beg}}} \right). \tag{6.74}$$

In a next step, we have to introduce a physically sensible definition of the notion of acceleration given in terms of the relational clock $\phi$. This is motivated via the Raychaudhuri equation for the acceleration in standard cosmology in terms of proper time $t$ which is

$$\frac{\ddot{a}}{a} = \frac{1}{3} \left[ \frac{\ddot{V}}{V} - \frac{2}{3} \left( \frac{\dot{V}}{V} \right) \right]. \tag{6.75}$$

---

[21]In models with a bounce, it is in principle sufficient to have a smaller number of e-folds [30].

Using that the momentum conjugate to the scalar field $\phi$ is given by $\pi_\phi = V\dot{\phi}$, together with $\dot{V} = \partial_\phi V \left(\frac{\pi_\phi}{V}\right)$, this expression can be rewritten as

$$\frac{\ddot{a}}{a} = \frac{1}{3}\left(\frac{\pi_\phi}{V}\right)^2\left[\frac{\partial_\phi^2 V}{V} - \frac{5}{3}\left(\frac{\partial_\phi V}{V}\right)^2\right] \equiv \frac{1}{3}\left(\frac{\pi_\phi}{V}\right)^2 \mathfrak{a}(\sigma).^{22} \tag{6.76}$$

Using that the expression for the conserved quantity $E$, Eq. (6.67), for such a configuration is

$$E_{\frac{1}{2};\frac{1}{2}} \equiv E = \frac{1}{2}\left[\tau\sigma'^2 + \left(\frac{3}{2} + \mu\right)\sigma^2\right], \tag{6.77}$$

and setting $\sigma' = 0$ we yield

$$\sigma_{\text{beg}}^2 = \frac{2E}{\mu + \frac{3}{2}}, \tag{6.78}$$

for which the acceleration condition $\mathfrak{a}(\sigma) > 0$ is fulfilled. The acceleration phase comes to an end when the expression for $\mathfrak{a}(\sigma)$ vanishes which gives

$$\sigma_{\text{end}}^2 = \frac{7}{2}\frac{E}{\mu + \frac{3}{2}}. \tag{6.79}$$

For these one obtains that the number of e-folds is $N \approx 0.186$. This value agrees with the upper bound found for the complex-valued model considered in Ref. [489]. There is no lower bound on $N$ due to the absence of the conserved charge $Q$ that is only present in complex-valued models, as in clarified in Section 6.3. Also the effect of the Laplacian has no influence on the value of $N$. The small value of $N$ shows that the epoch of accelerated expansion for the case of the free model does not last long enough to offer an alternative to the standard inflationary paradigm. However, with the same techniques we show in Section 6.2.5.1 that for two interaction terms with fine-tuned coupling constants an era of inflation driven by quantum geometric effects can in principle be realised.

We want to remark that the analytic result for $N$ is only realisable if just one mode for the condensate is considered. Taking into account all modes, would require a numerical analysis. However, we do not expect the value of $N$ to grow significantly then since the biggest contribution to $V_{\text{end}}$ and $V_{\text{beg}}$ will always stem from the fastest growing condensate components, i.e. the low-spin modes.

---

[22]We want to remark that using Eq. (6.75) from classical cosmology as an input to motivate Eq. (6.76) should be taken with a grain of salt: At the current stage of the GFT condensate programme it is not possible to give an intrinsic derivation for the acceleration from within GFT due to the absence of more interesting observables apart from the number, area and volume operators.

### 6.2.5 Relational dynamics of an effectively interacting isotropic GFT condensate

This Section studies the dynamics of effectively interacting GFT condensates in the isotropic restriction introduced above. To this aim, we work in the equivalent to the so-called Thomas-Fermi approximation for real BECs. This is the regime in which the effect of the Laplacian operator in the kinetic term is considered to be much smaller compared to any interaction. In the case of real BECs this is a typical simplification of the system when the density of the ground state is very large [463–465, 534]. We start off by analysing this regime first by performing a formal stability analysis of the non-linear dynamical system around its fixed points and second by numerically finding solutions to the equation of motion giving rise to an effective Friedmann equation. We then investigate the acceleration behaviour of the expansion of such systems. We show that for two interaction terms one can in principle find a sufficiently strong acceleration to replace the inflationary paradigm, however, coming at the cost of fine-tuning their coupling constants. Finally, we investigate the formal stability properties of the full dynamical system including the Laplace-Beltrami operator and an interaction term. This allows us to determine which modes, depending on the mass parameter, are (in the language of dynamical systems) stable or unstable which allows us to understand the emergence of a low-spin phase from a different angle.

To start with, for a system with one interaction term, the equation of motion

$$\left[ (\tau \partial_\phi^2 - \sum_{I=1}^{4} \Delta_{g_I}) + m^2 \right] \sigma(g_I, \phi) + \kappa \sigma(g_I, \phi)^{n-1} = 0, \qquad (6.80)$$

with $n = 4, 5, \ldots$ under the above-discussed isotropic restriction leads to

$$\left[ (2\tau \partial_\phi^2) - (\frac{\mathrm{d}^2}{\mathrm{d}\psi^2} + 2\cot(\psi)\frac{\mathrm{d}}{\mathrm{d}\psi}) + 2\mu \right] \sigma(\psi, \phi) + 2\kappa \sigma(\psi, \phi)^{n-1} = 0. \qquad (6.81)$$

When considering the analogue of the Thomas-Fermi approximation for real BECs, the contribution of the Laplace-Beltrami operator is suppressed which implies that $\sigma(\psi, \phi) = \xi(\psi)T(\phi) = c\,T(\phi)$ with some constant $c$. Based on the results of Ref. [457], it was shown in Sections 6.2.1 and 6.2.2 that the Fourier components of such constant functions on the domain are dominated by the lowest non-trivial and non-degenerate mode. This implies that the condensate consists of many smallest possible and discrete building blocks.

Its dynamics are then captured by the non-linear dynamical system

$$(\tau \partial_\phi^2 + \mu)T(\phi) + \kappa T(\phi)^{n-1} = 0 \tag{6.82}$$

where factors of $c$ were absorbed into $\kappa$. With $T(\phi) \equiv x$ and $T'(\phi) \equiv y$ and $\tau \equiv 1$ this can be rewritten as

$$\frac{\mathrm{d}}{\mathrm{d}\phi} \begin{pmatrix} x \\ y \end{pmatrix} = \begin{pmatrix} y \\ -\mu x - \kappa x^{n-1} \end{pmatrix}. \tag{6.83}$$

For even-valued $n$, the formal linear stability analysis of its fixed points $(x,y)_* = \{(0,0), (\pm \sqrt[n-2]{-\frac{\mu}{\kappa}}, 0)\}$ shows that for $\mu < 0$ and $\kappa > 0$ the first is a saddle and, the others are center fixed points. For odd-valued $n$, only for $\mu < 0$, $\kappa > 0$ or $\mu < 0$, $\kappa < 0$ one can have a non-trivial (non-perturbative) vacuum in agreement with the condensate state ansatz, so the fixed points of the dynamical system are given by $(x,y)_* = \{(0,0), (\pm \sqrt[n-2]{\mp\frac{\mu}{\kappa}}, 0)\}$.

Using this, we can give the general solutions close to the fixed points. For example, in the case for an even-valued $n$, close to the saddle point, the solution is given by

$$\begin{pmatrix} T \\ T' \end{pmatrix} = \frac{a_1 e^{-\sqrt{|\mu|}\phi}}{\sqrt{1 + \frac{1}{|\mu|}}} \begin{pmatrix} -\frac{1}{\sqrt{|\mu|}} \\ 1 \end{pmatrix} + \frac{a_2 e^{\sqrt{|\mu|}\phi}}{\sqrt{1 + \frac{1}{|\mu|}}} \begin{pmatrix} \frac{1}{\sqrt{|\mu|}} \\ 1 \end{pmatrix}, \tag{6.84}$$

where $a_1$ and $a_2$ are determined by using the initial conditions. Around the two center fixed points, the linearised solution is given by

$$\begin{pmatrix} T \\ T' \end{pmatrix} = \begin{pmatrix} \pm \sqrt[n-2]{|\mu|/\kappa} \\ 0 \end{pmatrix} + \frac{a_1 e^{-i\sqrt{(n-2)|\mu|}\phi}}{\sqrt{1 + \frac{1}{((n-2)|\mu|)^2}}} \begin{pmatrix} \frac{i}{\sqrt{(n-2)|\mu|}} \\ 1 \end{pmatrix} + \frac{a_2 e^{i\sqrt{(n-2)|\mu|}\phi}}{\sqrt{1 + \frac{1}{((n-2)|\mu|)^2}}} \begin{pmatrix} -\frac{i}{\sqrt{(n-2)|\mu|}} \\ 1 \end{pmatrix}. \tag{6.85}$$

The full numerical solutions obtained by employing an implementation of the Runge-Kutta method at small step size are illustrated in the phase portraits given by Figs. 6.20 and 6.21 which resemble those of a classical point particle in the corresponding effective potential. From the phase portraits we also deduce that depending on the initial conditions one can find configurations the total volume of which vanishes at some point and others for which the volume never becomes zero.

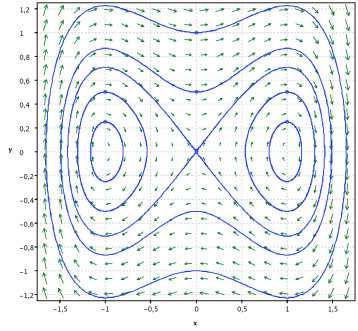

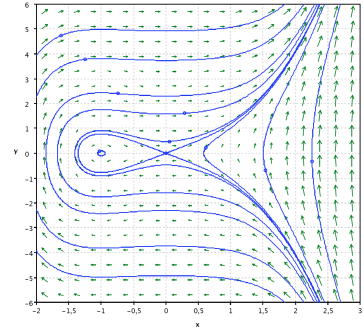

FIGURE 6.20: Phase plane portrait of the dynamical system given by Eq. (6.83) for $\mu = -1$, $\kappa = 1$, $n = 4$ and for different initial conditions where $x = T(\phi)$ and $y = T'(\phi)$.

FIGURE 6.21: Phase plane portrait of the dynamical system given by Eq. (6.83) for $\mu = -1$, $\kappa = -1$, $n = 5$ and for different initial conditions where $x = T(\phi)$ and $y = T'(\phi)$.

When comparing Eq. (6.84) to Eq. (6.52) we also notice that close to the saddle point, i.e. where the non-linearities are neglected, the solutions show a similar evolution behaviour with respect to the relational clock. It is thus obvious that only in the neighborhood of this point the effective Friedmann equation

$$\left(\frac{V'}{3V}\right)^2 = \frac{4\pi G}{3},\tag{6.86}$$

can be recovered when using Eq. (6.84) together with the identification $3\pi G = \frac{|\mu|}{\tau}$. In the latter identification of course no contribution stemming from the Laplacian is taken into account since we work in the Thomas-Fermi regime. Another difference worth to be emphasised then lies in the fact that to recover the effective Friedmann equations in Section 6.2.4.2 the quick relaxation of the system into a low-spin phase was needed whereas here the system is dominated by the $\frac{1}{2}$-configuration from the onset.

Away from the fixed point, however, the effective interactions will modify the evolution process due to the non-linear interactions which is clearly visible from the phase portraits. For instance, for even-powered potentials the evolution will be cyclic in general whereas for odd-powered potentials cyclic evolution competes with an open evolution depending on the initial conditions, as Fig. 6.21 illustrates.

More specifically, we can give the dynamical equation in the effectively interacting case for the volume which takes the form of a modified Friedmann equation in relational terms. For this, we can exploit that $\sigma$ is constant on the domain in the Thomas-Fermi regime, which simplifies the expression for the quantity $E$ in Eq. (6.65) to

$$E = \frac{1}{2} \left[ \sigma'^2 + \mu \sigma^2 + 2 \frac{\kappa}{n} \sigma^n \right],$$ (6.87)

where we included the interaction term. Since the constancy of the mean field on the domain implies that its dominant non-degenerate Fourier component is $\sigma_{\frac{1}{2};\frac{1}{2}}$, the volume can be expressed as $V = V_{\frac{1}{2}} \sigma_{\frac{1}{2};\frac{1}{2}} (\phi)^2$. Thus the first Friedmann equation in terms of the relational clock is given by

$$\left( \frac{V'}{3V} \right)^2 = \frac{4\pi G}{3} + \frac{4}{9} \left[ \frac{2 E V_{\frac{1}{2}}}{V} - \frac{2\kappa}{n} \left( \frac{V_{\frac{1}{2}}}{V} \right)^{1-n/2} \right],$$ (6.88)

where we used the identification $|\mu| = 3\pi G$. As stated above, dimensional analysis suggests that the first term in the parenthesis is of the order of $\sqrt{\hbar G}$ and is thus a quantum gravity correction (cf. Ref. [486]). Since the units of GFT coupling constants are not clear from a fundamental point of view and have to be consistent with (semi-)classical results, similarly, since $V_{\frac{1}{2}} \sim (\hbar G)^{\frac{3}{2}}$ then only if $\kappa$ is of the order of $(\hbar G)^x$ with $x > \frac{3}{4} n - \frac{3}{2}$, the second term stemming from the GFT interaction term can be understood as a quantum gravity correction which vanishes in the limit $\ell_p \to 0$.

### 6.2.5.1 Geometric inflation

We now demonstrate that within the context of the effectively interacting GFT condensate models it is in principle possible to have an era of accelerated expansion lasting long enough to represent an alternative to the standard inflationary scenario [27]. This, however, comes at the cost of using two interaction terms and a fine-tuning of the respective coupling constants. The origin of these results is entirely quantum geometric since the massless scalar field $\phi$ used throughout serves only as a relational clock and does not play the role of the inflaton.

To this aim, we will employ the techniques to extract information about the acceleration encoded by a model with a complex-valued GFT field in the spin representation that were developed by the author in Ref. [489]. Despite the fact that here we work with a real-valued model starting in the group representation and in principle different effective interactions, within the Thomas-Fermi regime we are able to fully recover the same results as in Ref. [489]. They will be more thoroughly displayed in Section 6.3.5 which is the reason why we keep the subsequent exposition as brief as possible.

As argued in Section 6.2.5, in the Thomas-Fermi regime the expectation value of the volume operator can be written as $V = V_{\frac{1}{2}} \sigma_{\frac{1}{2}, \frac{1}{2}}(\phi)$ and we drop indices of the mean field hereafter. Including the "mass term" and two interaction terms, the effective potential reads

$$V_{\text{eff}} = \frac{\mu}{2}\sigma^2 + \frac{\kappa}{n}\sigma^n + \frac{\lambda}{n'}\sigma^{n'} \tag{6.89}$$

wherein for stability purposes we choose $\lambda > 0$ and $n'$ to be even. For the conserved quantity $E$ we can then write

$$E = \frac{1}{2}[\sigma'^2 + 2V_{\text{eff}}]. \tag{6.90}$$

With this we can cast the expression for the "acceleration" $\mathfrak{a}(\sigma)$ given by Eq. (6.76) first into

$$\mathfrak{a}(\sigma) = -\frac{2}{\sigma^2}\left[\frac{(\partial_\phi V_{\text{eff}})\sigma}{\sigma'} + \frac{14}{3}(E - V_{\text{eff}})\right] \tag{6.91}$$

which is then rewritten as

$$\mathfrak{a}(\sigma) = -\frac{2}{3\sigma^4}\left[-4\mu\sigma^4 + 14E\sigma^2 + \alpha\sigma^{n+2} + \beta\sigma^{n'+2}\right] \tag{6.92}$$

with $\alpha = (3 - 14/n)\kappa$ and $\beta = (3 - 14/n')\lambda$.

As stated above, in order to accommodate for an era of inflationary expansion we assume the hierarchy $|\kappa| \gg \lambda$ hereafter which amounts to a fine-tuning. The beginning of the acceleration phase starts when $E = V_{\text{eff}}$ where the acceleration condition $\mathfrak{a}(\sigma_{\text{beg}}) > 0$ must hold. This leads to the condition $\partial_\phi V_{\text{eff}} \leq 0$, which together with the hierarchy implies $|\mu| \geq \kappa$. Furthermore, we have to demand that $\alpha < 0$ so that a large enough number of e-folds can be generated in this model, since otherwise $\sigma_{\text{end}}$ is even smaller than in the free case, treated in Section 6.2.4.2.

At the end of the accelerated expansion phase the acceleration has to vanish, i.e. $\mathfrak{a}(\sigma) = 0$, from which we obtain $\sigma_{\text{end}}$. To do so, we can expect that $\sigma_{\text{end}} \gg \sigma_{\text{beg}}$, which allows us to solve for the root of $\mathfrak{a}$ by solely considering the highest powers in $\sigma$ therein. With this one can argue that in order to guarantee that $\sigma_{\text{end}}$ is the only root of $\mathfrak{a}$ in between $\sigma_{\text{beg}}$ and $\sigma_{\text{end}}$ one needs $\kappa < 0$. Together with $\sigma_{\text{end}} = \sigma_{\text{beg}} \, e^{\frac{3}{2}N}$ and the expressions for $\alpha$ and $\beta$ this yields

$$\beta = -\alpha \, e^{-\frac{3}{2}N(n'-n)} > 0. \tag{6.93}$$

From this, the value of $N$ can be obtained when the values of $\kappa$, $n$ and $\lambda$, $n'$ are given. It also implies that $n' > n \geq 5$.

This is interesting because for $n = 5$ the corresponding local interaction term would mimic a proper combinatorially non-local interaction term which is typically needed to formulate a GFT with a simplicial quantum gravity interpretation. Since $n'$ has to be even for stability reasons, the corresponding even-powered interaction terms are reminiscent of so-called tensorial interactions. Typically, these occur in models where the GFT field is endowed with a specific tensorial transformation property, see Ref. [438].[23] In Section 6.3 we give further phenomenological arguments based on Ref. [489] which indicate that only $n' = 6$ is physically reasonable.

We emphasise that the above-given strategy to extract the acceleration behaviour of the emergent space as described by the mean field $\sigma$ is within our context only applicable to the case of the Thomas-Fermi regime in which the mean field is constant on the domain. To go beyond this and study less restricted scenarios, one would have to employ numerical techniques which can control the regular singularities of the Laplacian arising due to the used coordinate system. In view of Section 6.3, it is interesting that the same acceleration pattern is generated by means of real- and complex-valued models where the powers of the effective interaction terms agree whereas their microscopic details are rather different.

### 6.2.5.2 Stability analysis of the effectively interacting isotropic system

In a final step we want to study the stability of the condensate system close to the fixed points computed in Section 6.2.5 when the effect of the Laplace-Beltrami operator onto the system is also included. To this aim, we employ a linearisation of the non-linear PDE about these points which is the standard procedure used in the context of the stability theory of general dynamical systems. In a first step, without loss of generality, we set $\tau \equiv 1$ and cast the original system into first order form

$$\frac{\mathrm{d}}{\mathrm{d}\phi}\vec{\Sigma} \equiv \frac{\mathrm{d}}{\mathrm{d}\phi}\begin{pmatrix} \sigma \\ \sigma' \end{pmatrix} = \begin{pmatrix} \sigma' \\ \Delta\sigma - \mu\sigma - \kappa\sigma^{n-1} \end{pmatrix} \tag{6.94}$$

with stationary solutions

$$\vec{\Sigma}_*(\psi) = \begin{pmatrix} \sigma(\psi) \\ 0 \end{pmatrix}. \tag{6.95}$$

---

[23]Tensorial interactions are briefly discussed in Section 4.2.2.2 in the context of the new tensor models.

Plugging the ansatz $\vec{\Sigma} = \vec{\Sigma}_* + \vec{\Omega}$ into the original non-linear equation, where $\vec{\Omega}$ represents a small perturbation about the fixed points, one finds the linearisation of the non-linear PDE at the solution $\vec{\Sigma}_*$ to be given by

$$\frac{\mathrm{d}}{\mathrm{d}\phi}\vec{\Omega} = J_{\vec{\Sigma}_*}\vec{\Omega} + \mathcal{O}(\vec{\Omega}^2), \tag{6.96}$$

where $J$ denotes the Jacobian. This gives

$$\frac{\mathrm{d}}{\mathrm{d}\phi}\vec{\Omega} = \begin{pmatrix} 0 & 1 \\ \Delta - \mu - \kappa(n-1)\sigma(\psi)^{n-2} & 0 \end{pmatrix}\vec{\Omega}. \tag{6.97}$$

For any initial condition its solution is given by

$$\vec{\Omega} = \mathrm{e}^{J_{\vec{\Sigma}_*}\phi}\vec{\Omega}_0. \tag{6.98}$$

The solution is called stable if for its eigenvalues $\lambda_i$ with $i \in \{1, 2\}$ of $J_{\vec{\Sigma}_*}$ one has $\mathrm{Re}(\lambda_i) \leq 0$. Otherwise it is called unstable.

For the saddle point $\vec{\Sigma}_* = (0, 0)$ the eigenvalues of the Jacobian read

$$\lambda_{1,2} = \pm\sqrt{-2j(j+1) + |\mu|}. \tag{6.99}$$

Stable solutions are only found if and only if these eigenvalues are purely imaginary. Due to the linearisation, we can see when this happens by directly importing the results of Section 6.2.4.1. This allows us to reinterpret the blue sector in Fig. 6.19 as only comprising of stable solutions whereas the orange sector represents unstable ones. The striking fact that small $j$-modes are more unstable than all the others, leading to their exponential growth, has a simple explanation from the point of view of the formal stability analysis. For a given $\mu$, only for large enough $j$ the contributions stemming from the Laplacian and the negative mass term will be positive altogether. Hence, for such modes, the equation of motion resembles that of a particle in a potential which is bounded from below. When $j$ is too small to compensate for the negative mass term, the latter would appear as unbounded, wherefore these modes are unstable. These are the physically most relevant ones, as they lead to a quick expansion of the emergent space.

In contrast, for the center fixed points $\vec{\Sigma}_* = (\pm \sqrt[n-2]{-\frac{\mu}{\kappa}}, 0)$ the eigenvalues of the Jacobian are

$$\lambda_{1,2} = \pm\sqrt{-2j(j+1) + (2-n)|\mu|} \qquad (6.100)$$

which for $n = 4, 6, \ldots$ always leads to stable solutions.

Supposing that one chooses the vicinity of the saddle point as the starting point of the evolution of the condensate, the latter will roll down the effective potential towards the local minimum and will dynamically settle into a low-spin configuration. Depending on the value of $\mu$ this leads to a condensate configuration where each building block of the quantum geometry is only characterised by the mode for which $j$ is equal to $\frac{1}{2}$.[24]

### 6.2.6 Relational dynamics of an anisotropic GFT condensate

In the following, we lift the isotropic restriction introduced in Section 6.2.1, to pave the way to study more general, i.e. anisotropic GFT condensate configurations from the point of view of the group representation. The purpose of the following exposition is to serve as a starting point for a more systematic analysis of anisotropic GFT condensate configurations, their dynamics, comparison to the study of anisotropic models in LQC [75, 76, 492] as well as spin foam cosmology [405–410] and the possible emergence of generalised Friedmann equations [535] in an appropriate limit.

After explaining how such configurations are obtained, we study their behaviour in the small and large volume regimes. We show that anisotropies are dominant in the former and play a negligible role in the latter regime. We conclude by performing a formal stability analysis of the corresponding effectively interacting GFT condensate system which explores and explains the reasons for such a behaviour. To this aim, we again use techniques used in the context of the stability theory of general dynamical systems.

#### 6.2.6.1 Dynamical isotropisation

In the present subsection, we refrain from the full symmetry reduction employed in Section 6.2.1 and retain some of the anisotropic information which is stored in the mean field. Using such a particular anisotropic configuration, we show that by dynamically settling into a low-spin phase, the condensate isotropises over relational time.

---

[24]Similar results hold for odd-powered potentials, only that these exhibit just one center fixed point apart from the saddle point.

To this aim, we come back to the expression of the full Laplace-Beltrami operator, i.e. Eq. (6.13), and consider a particular geometry of the quantum tetrahedron. In general, the fluxes $\vec{B}_i$ associated to the tetrahedron are perpendicular to its faces. Bearing this in mind, assume that we are working with a trirectangular tetrahedron which has three pairwisely orthogonal faces. Hence, for such a building block of the quantum geometry

$$\sum_{i \neq j} \vec{B}_i \cdot \vec{B}_j = 0 \tag{6.101}$$

holds. Imposing such a constraint, for a mean field which depends only on the diagonal components $\pi_{ii}$ (thus $p_1, p_2, p_3$) and the relational clock $\phi$, this frees us from the second sum in Eq. (6.14) and allows us to decouple all remaining coordinates from one another.[25]

Concretely, using Eq. (6.13) and (6.101) the resulting equation of motion leads to the following PDE for $\sigma(p_1, p_2, p_3, \phi)$

$$\left[ \tau \partial_\phi^2 - \sum_{i=1}^3 \left( 2p_i(1-p_i)\frac{\mathrm{d}^2}{\mathrm{d}p_i^2} + (3-4p_i)\frac{\mathrm{d}}{\mathrm{d}p_i} \right) + \mu \right] \sigma(p_1, p_2, p_3, \phi) = 0. \tag{6.102}$$

Since our assumptions decoupled all the terms in the $p_i$'s and $\phi$ from each other, we can employ a separation ansatz to solve Eq. (6.102). Together with the substitution $p_i = \sin^2(\psi_i)$ the product ansatz

$$\sigma(\psi_1, \psi_2, \psi_3, \phi) = \xi_1(\psi_1)\xi_2(\psi_2)\xi_3(\psi_3)T(\phi) \tag{6.103}$$

yields

$$2\tau \frac{\partial_\phi^2 T(\phi)}{T(\phi)} = \sum_i \frac{\left( \frac{\mathrm{d}^2}{\mathrm{d}\psi_i^2} + 2\cot(\psi_i)\frac{\mathrm{d}}{\mathrm{d}\psi_i} \right)\xi_i(\psi_i)}{\xi_i(\psi_i)} - 2\mu \equiv \omega = \mathrm{const.} \tag{6.104}$$

The general solution to the equation for $T(\phi)$ is again given by

$$T(\phi) = \left( \mathrm{a}_1 \; \mathrm{e}^{\sqrt{\frac{\omega}{2\tau}}\phi} + \mathrm{a}_2 \; \mathrm{e}^{-\sqrt{\frac{\omega}{2\tau}}\phi} \right), \tag{6.105}$$

with the constants $\mathrm{a}_1, \mathrm{a}_2 \in \mathbb{C}$.

---

[25]We notice that Eq. (6.101) implies the vanishing of the off-diagonal components in the reconstructed metric, Eq. (4.75). This suggests that the anisotropic condensate states to be constructed below are possibly related to the Bianchi models of class A, see Refs. [448, 535].

Using $\mu = \sum_{i=1}^{3} \mu_i$ and $\omega = \sum_{i=1}^{3} \omega_i$ one solves in each direction the one-dimensional problem

$$\left[ -\left( \frac{\mathrm{d}^2}{\mathrm{d}\psi_i^2} + 2\cot(\psi_i)\frac{\mathrm{d}}{\mathrm{d}\psi_i} \right) + 2\mu_i + \omega_i \right] \xi_i(\psi_i) = 0, \quad \psi_i \in [0, \frac{\pi}{2}], \tag{6.106}$$

separately. To solve these differential equations, we impose, as above, Dirichlet boundary conditions $\xi_i(\psi_i = \frac{\pi}{2}) = 0$. The eigensolutions are then given by

$$\xi_{i_{j_i}}(\psi_i) = \frac{\sin\left((2j_i + 1)\psi_i\right)}{\sin(\psi_i)}, \quad \psi_i \in [0, \frac{\pi}{2}] \tag{6.107}$$

for the eigenvalues $\omega_i + 2\mu_i = -4j_i(j_i + 1)$ with $j \in \frac{2\mathbb{N}_0 + 1}{2}$. Using the above, we get

$$\omega = -2\mu - 4\sum_i j_i(j_i + 1) \tag{6.108}$$

which in Eq. (6.105) yields

$$T_{j_1,j_2,j_3}(\phi) = \left( \mathrm{a}_1\ \mathrm{e}^{\sqrt{\frac{\mu + 2\sum_i j_i(j_i+1)}{-\tau}}\phi} + \mathrm{a}_2\ \mathrm{e}^{-\sqrt{\frac{\mu + 2\sum_i j_i(j_i+1)}{-\tau}}\phi} \right). \tag{6.109}$$

As in the previous section, its behaviour critically depends on the sign of $\frac{\mu + 2\sum_i j_i(j_i+1)}{-\tau}$ where $\mu < 0$ and $\tau > 0$. Assuming now that $\mu_i < 0$, solutions for which

$$\frac{1}{2} \leq j_i < -\frac{1}{2} + \frac{1}{2}\sqrt{1 - 2\mu_i}\ \text{ with }\ \mu_i < -\frac{3}{2} \tag{6.110}$$

grow exponentially as $\phi \to \pm\infty$ for $i = 1, 2, 3$ simultaneously. Strikingly, the condensate will quickly be dominated by the lowest representation $j_i = \frac{1}{2}$ for all $i = 1, 2, 3$ and all other $j$-contributions are suppressed. This directly implies that toward $\phi \to \pm\infty$ the fully anisotropic configuration

$$\sigma(\psi_1, \psi_2, \psi_3, \phi) = \sum_{j_1,j_2,j_3 \in \frac{2\mathbb{N}_0 + 1}{2}} \xi_{1_{j_1}}(\psi_1)\xi_{2_{j_2}}(\psi_2)\xi_{3_{j_3}}(\psi_3)T_{j_1,j_2,j_3}(\phi) \tag{6.111}$$

will dynamically isotropise, as indicated by the late time behaviour of the expectation value of the volume operator

$$\lim_{\phi \to \pm\infty} V(\phi) = V_0\ V_{\frac{1}{2}}\ \left( |\xi_{1_{\frac{1}{2};\frac{1}{2}}}|^2 \right)^3\ |\mathrm{a}_{1,2}|^2\ \mathrm{e}^{\pm 2\sqrt{\frac{\mu + \frac{9}{2}}{-\tau}}\phi}. \tag{6.112}$$

Consequently, a generalised Friedmann equation (in relational terms) which would take into account the effect of the anisotropies, will reduce at late times to the one for the isotropic configuration, Eq. (6.71).

In a last step, we investigate whether the contributions stemming from the anisotropies become important at small volumes. To this aim, we consider without loss of generality the initial condition $T_{j_1,j_2,j_3}(0) = 0$.[26] For this one has

$$T_{j_1,j_2,j_3}(\phi) = 2\mathrm{a}_1 \sinh\left(\sqrt{\frac{\mu + 2\sum_i j_i(j_i+1)}{-\tau}}\phi\right). \qquad (6.113)$$

The differences between the dynamical behaviour of the isotropic ($j_1 = j_2 = j_3$) and the anisotropic part of the mean field are illustrated by Figs. 6.22 and 6.23. These show that anisotropies only play an important role at small values of the relational clock, i.e. small volumes, whereas at late times the isotropic mode for $j_1 = j_2 = j_3 = \frac{1}{2}$ will clearly dominate. Certainly, this behaviour is qualitatively the same for the other branch of solutions (with initial conditions $T'_{j_1,j_2,j_3}(0) = 0$) where the singularity problem is avoided since $\lim_{\phi\to 0} T_{j_1,j_2,j_3}(\phi) \neq 0$. We also want to remark that in spite of the surge of anisotropies for small volumes, towards $\phi \to 0$ such a behaviour cannot turn a solution corresponding to a finite volume into one for which the volume vanishes, and vice versa.

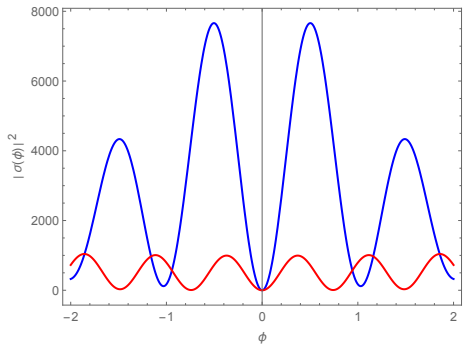
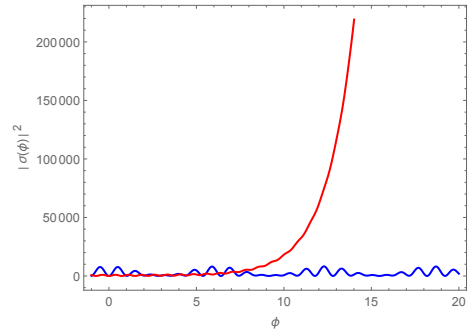

FIGURE 6.22:  Probability density of $\sigma$ for the isotropic (red) and the anisotropic (blue) parts for small values of the relational clock $|\phi|$ for $\mu = -4.6$ and $j_{\max} = \frac{3}{2}$.  Setting $j_{\max}$ to higher values does not qualitatively change the result.

FIGURE 6.23:  Probability density of $\sigma$ for the isotropic (red) and the anisotropic (blue) parts for larger values of the relational clock $\phi$ for $\mu = -4.6$ and $j_{\max} = \frac{3}{2}$.  Setting $j_{\max}$ to higher values does not qualitatively change the result.

From a physical point of view, such a behaviour is certainly very interesting calling for an explanation. In the ensuing subsection, this will be given by means of a formal stability analysis of the corresponding dynamical system around its fixed points which shows that the solutions for isotropic modes with $j_1 = j_2 = j_3 = \frac{1}{2}$ are more unstable as compared to others. These are the physically most relevant modes since they lead to an isotropic expansion of the emergent geometry.

---

[26]More generally, one could also use that Eq. (6.105) vanishes for $\phi = \sqrt{\frac{2\tau}{\omega}}\frac{\log(-\mathrm{a}_1/\mathrm{a}_2)}{2}$. However, this does not change the subsequent discussion qualitatively.

### 6.2.6.2 Stability analysis of the effectively interacting anisotropic system

This subsection analyses the formal stability properties of the anisotropic and effectively interacting condensate system in the vicinity of the fixed points of the dynamics. Due to the analogous treatment for the isotropic configuration done above, the subsequent discussion is kept as short as possible.

Again, we employ a linearisation of the non-linear PDE about the fixed points, however, keeping in mind that the mean field now depends on three coordinates on the domain, as well as on the relational clock, i.e. $\sigma = \sigma(\psi_1, \psi_2, \psi_3, \phi)$. Without loss of generality, we set $\tau \equiv 1$ and rewrite the original equation of motion into first order form

$$\frac{\mathrm{d}}{\mathrm{d}\phi}\vec{\Sigma} \equiv \frac{\mathrm{d}}{\mathrm{d}\phi}\begin{pmatrix} \sigma \\ \sigma' \end{pmatrix} = \begin{pmatrix} \sigma' \\ \Delta\sigma - \mu\sigma - \kappa\sigma^{n-1} \end{pmatrix} \tag{6.114}$$

which has the stationary solutions

$$\vec{\Sigma}_*(\psi_1, \psi_2, \psi_3) = \begin{pmatrix} \sigma(\psi_1, \psi_2, \psi_3) \\ 0 \end{pmatrix}. \tag{6.115}$$

With the ansatz $\vec{\Sigma} = \vec{\Sigma}_* + \vec{\Omega}$ one finds the linearisation of the non-linear PDE for the anisotropic system at the stationary solutions $\vec{\Sigma}_*$ to be given by

$$\frac{\mathrm{d}}{\mathrm{d}\phi}\vec{\Omega} = J_{\vec{\Sigma}_*}\vec{\Omega} + \mathcal{O}(\vec{\Omega}^2), \tag{6.116}$$

wherein $J$ denotes the Jacobian.

The eigenvalues of the Jacobian at the saddle point $\vec{\Sigma}_* = (0, 0)$ are given by

$$\lambda_{1,2} = \pm\sqrt{-2\sum_i j_i(j_i + 1) + |\mu|}. \tag{6.117}$$

Only if the eigenvalues are purely imaginary, the corresponding solutions are stable. From this expression we see that this is the case if $j_1, j_2$ and $j_3$ exceed a certain value for a given $\mu$. For such modes the equation of motion resembles that of a particle in a potential

$$\bar{V}_{j_1,j_2,j_3}[\sigma] = \left[2\sum_i j_i(j_i+1) + \mu\right]\sigma^2 \qquad (6.118)$$

which is bounded from below. When the $j_i$ are too small to compensate for the negative "mass term", the latter would appear as unbounded, wherefore these modes are considered as unstable from the point of view of the formal stability analysis. This leads to the exponential growth of the corresponding modes. In Fig. 6.24 we illustrate the form of $\bar{V}$ for a particular value for $\mu$.

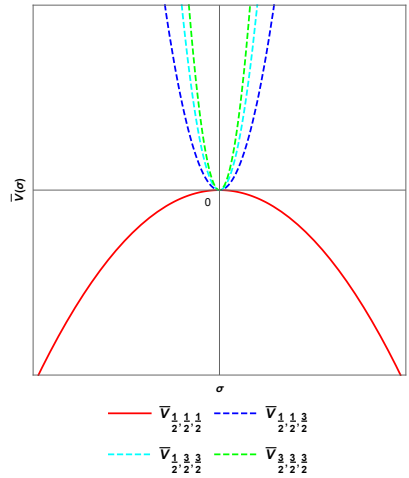

FIGURE 6.24: Form of the potentials $\bar{V}_{j_1,j_2,j_3}[\sigma]$ for $\mu = -5$.

To conclude, using the terminology of dynamical systems, the specific configuration with $j_1 = j_2 = j_3 = \frac{1}{2}$ will be the most unstable one compared to all the others for a given $\mu$, which implies that an initially anisotropic GFT condensate will quickly isotropise by settling into the low-spin phase. On the other hand, anisotropies are more exposed for this reason at small volumes.

Given that our discussion rests on a particular anisotropic building block, in future research it will be important to lift this restriction, to consider even more general configurations and to investigate whether the isotropisation is also realised there. This will also entail a systematic comparison of such condensates to anisotropic quantum cosmological models explored in LQC [75, 76, 492]. In Section 6.3 we will return to the study of anisotropic perturbations close to a bounce in the context of a model based on complex-valued fields.

### 6.2.7 Discussion of the results of the dynamics of model 1

In the second part of our analysis of model 1 we investigated the relational evolution of effectively interacting GFT quantum gravity condensate systems describing the dynamics of effective 3-geometries and to study the tentative quantum cosmological consequences.

To this aim, we worked with real-valued GFT fields in the group representation and thus adopted a different point of view as compared to Refs. [486, 487, 489, 491, 536] and Section 6.3 which rest on working in the spin representation with complex fields. We studied the evolution of such a configurations with respect to a relational clock. In close analogy to and expanding on the results presented in Ref. [488], we showed for our model that it quickly settles into a lowest-spin configuration of the quantum geometry. We then demonstrated that this goes hand in hand with the accelerated and exponential expansion of its volume. The dynamics of the latter can be cast into the form of the classical Friedmann equations given in terms of the relational clock, as in Ref. [486]. In particular, we showed that the relative uncertainty of the volume operator vanishes for such a configuration at late times which can be seen as an indication for the classicalisation of the quantum geometry.

Solutions which avoid the singularity problem and exponentially grow after the bounce can only be found for a negative conserved quatity $E$ in case of real-valued condensate fields. This is to be contrasted to the case of complex-valued fields where the mechanism responsible for the resolution of the initial singularity rests on the global U(1)-symmetry of the field. This is discussed in the next Section.

We then moved on to study the formal stability properties of the evolving isotropic system when effective interactions are included. This was done using tools from the stability theory of general dynamical systems. In particular, we investigated the properties of solutions in the Thomas-Fermi regime. There, the system is dominated by the lowest-spin configuration of the quantum geometry and the dynamics of the volume can be cast into the form of modified Friedmann equations. Depending on whether the effective interaction potentials are bounded from below or not, the evolution of the emergent 3-geometries will lead to a recollapse or to an infinite expansion, respectively. When studying in this regime the effect of two fine-tuned interaction terms mimicking those found in the GFT literature, by applying the techniques developed in Ref. [489], it is also possible to accommodate for an era of accelerated expansion which is strong enough to represent an alternative to the standard inflationary scenario. The study of the influence of effective interactions onto isotropic configurations was then concluded by performing a formal stability analysis of the full system including the effect of the Laplace-Beltrami operator. This allowed us to put the above-given results into a larger context by showing that for a given value of the "GFT mass", at small volumes low-spin modes are highly unstable, thus leading to their exponential growth over time. These are the physically most important modes, as they lead to an expansion of the emergent space.

Together with the findings on the statics of this model, these results show that the condensate models considered here can give rise to effectively continuous, homogeneous and isotropic 3-geometries built from many smallest and almost flat building blocks of the quantum geometry. Their dynamics lead to a rich phenomenology. In particular, it can be shown that the classical Friedmann equations may be recovered in an appropriate limit.

In a final step, we lifted the restriction of isotropy to study more general GFT condensate systems, thus opening an avenue to study anisotropies in the context of GFT condensate cosmology for the first time. The study of anisotropic perturbations in Section 6.3 is a direct consequence of this analysis. Employing again formal stability analysis techniques known in the context of general dynamical systems, we showed for effectively interacting anisotropic configurations that the isotropic contribution with $j_1 = j_2 = j_3 = 1/2$ is highly unstable at small values of the relational clock, i.e. at small volumes. This quickly leads to the isotropisation of the system for increasing values of the relational clock while the emergent geometry expands and thus suggests that within the GFT condensate cosmology framework a natural mechanism for smoothing out anisotropies may be realised. On the other hand, toward small volumes, we show that anisotropies surge. However, nothing like the anisotropy problem known in the literature on bouncing cosmologies occurs, where a regular bounce may be transformed into a singularity due to anisotropies, as reviewed in Ref. [30]. We note that in our context the feature of singularity avoidance of solutions is not altered by the occurrence of anisotropies.

In the following, we want to comment on the limitations and possible extensions of our discussion. Beyond the limitations discussed in Section 6.2.3, the work presented here is based on using real-valued GFT fields which to some extent leads to a different phenomenology compared to complex-valued fields since e.g. the singularity problem can be avoided in different ways. To clarify this difference, a better understanding of the conserved quantities $E$ and $Q$ would be desirable. Related to this is the question of how models exhibiting a bounce can be reconciled with the phase transition picture as a possible realisation of the geometrogenesis scenario [461, 462]. In the latter, the mean field is supposed to vanish at the critical point which cannot happen in models which exhibit bouncing solutions.

Concerning the acceleration behaviour of interacting GFT models, numerical techniques should be developed to generalise the results for more general configurations including anisotropies as well as for models going beyond the Thomas-Fermi regime as used here. In particular, the robustness of the geometric inflation picture should be checked when going beyond the use of effective interactions and studying the impact of proper combinatorially non-local interactions.

With regard to anisotropies, it is pressing to more systematically explore whether and how the dynamics of the volume can be cast into the form of generalised Friedmann equations for anisotropic cosmologies [535]. At the current stage, it seems that the simple second-order character of the evolution equation together with the lack of appropriate observables to capture cosmological anisotropies do not straightforwardly allow for dynamics which are similar to those of Bianchi class A models with *directional* Hubble rates. Be this as it may, we will return to the analysis of anisotropies in the next Section and thoroughly investigate their impact on the bouncing mechanism found for complex-valued GFT fields in Refs. [486, 487].

## 6.3   Model 2: Complex-valued Lorentzian EPRL GFT model

It was first shown that GFT allows to derive an effective Friedmann equation from the evolution of simple condensate states in Refs. [486, 487], within a generalised GFT formulation of the Lorentzian EPRL spin foam model (cf. Ref. [401]) for a complex-valued GFT field. As a general feature, the solutions of this model exhibit a quantum bounce which resolves the initial singularity, provided that the mean field approximation to the dynamics of GFT holds. The cosmologies that have been considered so far are, with the exception of the partial analysis of the author in Ref. [490] (as presented in Section 6.2.6.1), isotropic and are thus governed by a single global degree of freedom corresponding to the total volume (or scale factor) of the universe.

The first goal of this Section is to take a step beyond this isotropic case, and study GFT perturbations involving anisotropic degrees of freedom around isotropic background configurations, focusing on the cosmological evolution around the bouncing region, and remaining within the mean field approximation of the full quantum dynamics. The dynamics of anisotropies is an extremely important issue in fundamental cosmology, and especially in bouncing scenarios, where the growth of anisotropies close to the bounce is a problematic aspect if no mechanism exists to keep it under control [30]. It is also a difficult one to tackle in full generality and in concrete physical terms in GFT condensate cosmology mostly because of the current absence of suitable GFT observables with a clear and compelling geometric interpretation which capture anisotropic degrees of freedom of continuum geometries, and allow to extract their effective cosmological dynamics, in the same way in which it has been done in the isotropic case.

As a first step forward, we sidestep this difficulty by focusing on GFT perturbations of exact isotropic condensate wavefunctions. For any given definition of isotropy, anisotropies are going to be encoded by construction in such perturbations. A detailed study in terms of physical observables will be needed to explore how, exactly, but the fact itself is not in question. In particular, if one is only interested in the specific issue of whether anisotropies become dominant during the effective cosmological evolution or remain subdominant compared to the isotropic background, it is enough to study the dynamics of such GFT perturbations and their relative amplitude compared to the background condensate wavefunction.

This is the issue we focus on.[27]

Another main difficulty we have to face (which affects the whole GFT cosmology programme, like all attempts to extract physics from GFT or spin foam models) is purely technical, and lies in the complication of the analytic expression of the interaction kernel of the EPRL model. The Lorentzian EPRL vertex amplitude, in fact, has an analytic expression which can be variously expressed in integral form (where the integrals are over the Lorentz group and encode the covariant properties of the model) [402], but has not been explicitly put down as a function of its boundary data, which are usually given in terms of SU(2) representations (to match the LQG form of quantum states, as in Section 2.2). This is a serious limitation for the computation of transition amplitudes in the full theory, as well as for the solution of the classical dynamics of the corresponding GFT model, which would be our concern here. This is true both for the background dynamics and for that of the perturbations over it, so it is not sidestepped by our focus on the latter.

What allows us to make some progress on the study of anisotropies in this Section, despite this technical difficulty, is another general fact about the effective cosmological evolution of GFT condensates. GFT interactions become subdominant, compared to the kinetic term, at low densities, thus for small values of the GFT field (or perturbations), beside for small values of the GFT coupling constants, of course. In turn, this is exactly what happens close to the cosmological bounce, i.e. for low enough values of the universe volume[28]. This allows to study, as a first step, the dynamics and relative strength of anisotropic (non-monochromatic) perturbations on an isotropic (monochromatic) background close to the cosmological bounce, before the universe grows in size and occupation number enough that the GFT interactions have to be included in the analysis. Luckily, as we mentioned above, the behaviour of anisotropies close to the bounce, and their relative suppression compared to the isotropic background geometry away from it, where a Friedmann dynamics are expected to be a good description, is also the key question from the point of view of cosmological bouncing scenarios. The simple case we first consider in the first part of this Section, therefore, is also of direct physical interest.

---

[27]In this Section we will often prefer the label "monochromatic" over "isotropic" and "non-monochromatic" over "anisotropic" when referring to the fundamental building blocks, in agreement with the definitions of Ref. [537]. The motivation for this is to withstand the temptation of identifying microscopic and cosmological anisotropies from the outset.

[28]This is true, of course, provided one remains within the mean field approximation, which is not so obvious, since low densities also mean small average number of atoms in the condensate and mean field theory would then be expected to break down.

Regarding the current absence of an explicit closed expression of the simplicial GFT interaction term in its EPRL form, the remaining option is to adopt a more phenomenological approach and model the GFT interactions by simple functions that capture some of their essential features. The second part of this Section is thus devoted to adopting this point of view and to studying its consequences.

Here, we show that interactions play a substantial role in determining the ultimate fate of the emergent universe and lead to its recollapse. Together with the bounce this implies a cyclic evolution. Aiming at further bridging the gap between the quantum gravity era and the standard classical cosmological model, we study the acceleration behaviour of the emergent geometry. Despite the undeniable success of the inflationary paradigm in providing an explanation for structure growth and solving cosmological puzzles, its *ad hoc* underlying assumptions do not find support in a more fundamental theory. More specifically, since the onset of inflation is supposed to take place at Planckian times, the dynamics of the Universe at this stage should find a more suitable formulation so as to take quantum gravitational effects into account. Under this backdrop, we show that in the model considered here it is possible to achieve an arbitrarily large number of e-folds by imposing a hierarchy between two interaction coefficients. This is highly interesting and may indicate that the quantum dynamics of the gravitational field itself could effectively give rise to dynamical features similar to those of inflationary models, without the need to introduce a new hypothetical field (the inflaton) with an *ad hoc* potential.

To this aim, we first review elements of the Lorentzian EPRL GFT model based on SU(2) states for quantum gravity [401] coupled to a massless scalar field [521] in Section 6.3.1 and then discuss how an effective cosmological dynamics can be extracted from it for the simple GFT condensate states employed in this Chapter in Section 6.3.2. In particular, we discuss the symmetry assumptions on the condensate states and the isotropic reduction, which allow to extract an effective (modified) Friedmann dynamics and a cosmic bounce when interactions are neglected. Notice that we work in the spin representation from the outset here, as opposed to the first model treated in the previous Section 6.2.

We then obtain the dynamics of non-monochromatic perturbations to first order in Section 6.3.3. The dynamics are given by a system of four coupled linear differential equations. The terms arising from the linearisation of the interaction have non-constant coefficients depending on the background. Within this approximation, only one of the four faces of the tetrahedra can be perturbed. The others must match the spin of the background, due to a constraint given by the EPRL vertex. This is done in full generality (for EPRL-like models with a broad class of kinetic kernels). The dynamics can then be recast in a simpler and more compact form in the particular case of a local kinetic kernel and when considering a background with only spin $j = \frac{1}{2}$ being excited. Considering this specific case, in Section 6.3.4 we study the dynamics of non-monochromatic components at the bounce, where interactions are negligible. Hence, both the background and the perturbations satisfy linear equations of motion. In this regime, there is no need to impose the condition that perturbations are much smaller than the background. Thus, the perturbed geometry of the emergent spacetime can in principle be quite different from the one given by the background. An important result we obtain is the determination of a region of parameter space such that perturbations are bounded at all times while the background field grows unbounded. It is thus justified to neglect microscopic anisotropies after the bounce, when the non-linear regime is entered. However, around the bounce the magnitude of the perturbations can be of the same order as the background, leading to interesting consequences. To illustrate this point, we compute geometric quantities to characterise the non-trivial corrections to the "mean geometry" of the elementary isotropic constituents around the cosmological bounce. Our results do not depend on where we set the initial conditions for the non-monochromatic perturbations, i.e. whether their amplitude is maximal at the bounce or in the contracting/expanding phase. These results are largely based on the work of the author in Ref. [491].

In the second part of this unit we adopt the more phenomenological point of view and study the impact of effective interactions onto the condensate dynamics in Section 6.3.5. In particular, in Section 6.3.5.1 we show how the higher power interaction term induces a recollapse and we discuss the cyclicity of solutions of the model. Section 6.3.5.2 is divided in two parts. In the first part, we discuss the case of a non-interacting GFT model and show that it does not support a reasonably large number of e-folds. In the second one, we show how this is made possible by considering suitable interactions terms. In Section 6.3.5.3, we discuss how the different terms in the GFT effective potential can be reinterpreted, from the point of view of an effective Friedmann equation, as sources corresponding to effective fluids with peculiar equations of state. These findings are largely based on the work of the author in Ref. [489]. To a certain extent, this latter part of the analysis also applies to real-valued models examined by the author in Ref. [490], as discussed in Section 6.2.5.

### 6.3.1 Introduction of the model

In the following, we will review details of the group field theory formulation of the Lorentzian EPRL model, as given in Ref. [486]. This model is based on a complex-valued GFT field living $SU(2)^4 \times \mathbb{R}$, that is

$$\varphi = \varphi(g_1, g_2, g_3, g_4; \phi) = \varphi(g_I; \phi), \tag{6.119}$$

where the $\phi$ is a free and massless scalar field used as a relational clock [486, 487, 521], as in Section 6.1. Its introduction is crucial for the cosmological applications of GFT. The first part of the domain is chosen to be $SU(2)^4$, coming from the imposition of simplicity constraints on $SL(2, \mathbb{C})$ data, like in the EPRL model.

The field is invariant under the diagonal right action of $SU(2)$

$$\varphi(g_I h; \phi) = \varphi(g_I; \phi), \quad \forall h \in SU(2), \tag{6.120}$$

again understood as the closure of the tetrahedra dual to the vertices of the same spin network states.

Since GFT fields are $L^2$-functions with respect to the Haar measure on $SU(2)$, by the Peter-Weyl theorem, we can expand them in a basis of functions labelled by the irreducible representations of the same group (see Appendix C.2). For quantum states, this is the decomposition in terms of spin network states. For the GFT field used in this Section, this decomposition is

$$\varphi(g_I; \phi) = \sum_{j_I, m_I, n_I, \iota} \varphi^{j_1 j_2 j_3 j_4, \iota}_{m_1 m_2 m_3 m_4}(\phi) \, C^{j_1 j_2 j_3 j_4, \iota}_{n_1 n_2 n_3 n_4} \prod_{i=1}^{4} d_{j_i} D^{j_i}_{m_i n_i}(g_i). \tag{6.121}$$

The right invariance leads to the Hilbert space $\mathcal{H} = L^2 \left( SU(2)^4/SU(2), (\mathrm{d}g)^4 \right)$. This is the intertwiner space of a four-valent open spin network vertex, and also the Hilbert space of states for a single tetrahedron, a basis for which is given by the intertwiners $C^{j_1 j_2 j_3 j_4, \iota}_{n_1 n_2 n_3 n_4}$, which are elements in

$$\mathcal{H}_{\mathrm{kin},4} = \mathrm{Inv}_{SU(2)} \left[ \mathcal{H}^{j_1} \otimes \mathcal{H}^{j_2} \otimes \mathcal{H}^{j_3} \otimes \mathcal{H}^{j_4} \right]. \tag{6.122}$$

The index $\iota$ labels elements in a basis in $\mathcal{H}_{\text{kin},4}$, and represents an additional degree of freedom in the kinematical description of the GFT field and of its quantum states.

For example, $\iota$ can be chosen so as to label eigenstates of the volume operator for a single tetrahedron. With this choice, the volume operator diagonally acts on a wavefunction for a single tetrahedron $\varphi$ (which we indicate with the same symbol as the classical GFT field, since they are functionally analogous) decomposed as in Eq. (6.121)

$$\hat{V}\varphi(g_{\mathrm{I}}) = \sum_{j_{\mathrm{I}},m_{\mathrm{I}},n_{\mathrm{I}},\iota} V^{j_1,j_2,j_3,j_4,\iota} \varphi^{j_1 j_2 j_3 j_4,\iota}_{m_1 m_2 m_3 m_4}(\phi)\, C^{j_1 j_2 j_3 j_4,\iota}_{n_1 n_2 n_3 n_4} \prod_{i=1}^{4} d_{j_i} D^{j_i}_{m_i n_i}(g_i). \qquad (6.123)$$

This action of the volume operator is the same as in LQG, where the volume eigenvalues for four-valent vertices have been studied extensively, see e.g. Ref. [537]. More details on the quantum geometry of GFT states are found in Appendix E.2.

Like any GFT model, the action is decomposed into the sum of a kinetic and an interaction term

$$S = K + V_5 + \overline{V}_5\,. \qquad (6.124)$$

The most general (local) kinetic term for an SU(2)-based GFT field of rank-4 is

$$K = \int \mathrm{d}\phi \sum_{j^a_{\mathrm{I}},m^a_{\mathrm{I}},\iota_a} \overline{\varphi}^{j_1^1 j_2^1 j_3^1 j_4^1\, \iota_1}_{m_1^1 m_2^1 m_3^1 m_4^1} \mathcal{K}^{j_1^1 j_2^1 j_3^1 j_4^1\, \iota_1}_{m_1^1 m_2^1 m_3^1 m_4^1} \varphi^{j_1^2 j_2^2 j_3^2 j_4^2\, \iota_2}_{m_1^2 m_2^2 m_3^2 m_4^2}$$
$$\times \delta^{j_1^1 j_1^2} \delta_{m_1^1 m_1^2} \delta^{j_2^1 j_2^2} \delta_{m_2^1 m_2^2} \delta^{j_3^1 j_3^2} \delta_{m_3^1 m_3^2} \delta^{j_4^1 j_4^2} \delta_{m_4^1 m_4^2} \delta^{\iota_1 \iota_2}, \qquad (6.125)$$

and one has the interaction term corresponding to simplicial combinatorial structures given by

$$V_5 = \frac{1}{5} \int \mathrm{d}\phi$$
$$\times \sum_{j_i,m_i,\iota_a} \varphi^{j_1 j_2 j_3 j_4 \iota_1}_{m_1 m_2 m_3 m_4} \varphi^{j_4 j_5 j_6 j_7 \iota_2}_{-m_4 m_5 m_6 m_7} \varphi^{j_7 j_3 j_8 j_9 \iota_3}_{-m_7 -m_3 m_8 m_9} \varphi^{j_9 j_6 j_2 j_{10} \iota_4}_{-m_9 -m_6 -m_2 m_{10}} \varphi^{j_{10} j_8 j_5 j_1 \iota_5}_{-m_{10} -m_8 -m_5 -m_1}$$
$$\times \prod_{i=1}^{10} (-1)^{j_i - m_i} \; \mathcal{V}_5(j_1,\ldots,j_{10}; \iota_1,\ldots \iota_5), \qquad (6.126)$$

where we suppressed the clock variable $\phi$ in the field for convenience. The details of the EPRL model would be encoded in the choice of kernels $\mathcal{K}$ and $\mathcal{V}_5$, and it is the interaction kernel that encodes the Lorentzian embedding of the theory and its full covariance, and what goes usually under the name of "spin foam vertex amplitude", here with boundary

SU(2)-states. The explicit expression for such interaction kernel can be found in Ref. [486] and, in more details in Ref. [402]. Here, we do not need to be explicit about the functional form of the interaction kernel (in fact it has not been spelled out to date for the Lorentzian case), while we will say more about the kinetic term in the following. Some discrete symmetries of the interaction kernel will however be relevant for what follows. In fact, the coefficients $\mathcal{V}_5$ are invariant under permutations of the spins and of the intertwiners, which preserve the combinatorial structure of the potential (6.126).

Beside the general form of Eq. (6.125), in the following we will also use the specific case for the GFT kinetic term introduced in Section 6.1, i.e.

$$K = \int \mathrm{d}\phi \int_{\mathrm{SU}(2)^4} (\mathrm{d}g)^4 \overline{\varphi}(g_\mathrm{I}, \phi) \mathcal{K}_{g_\mathrm{I}} \varphi(g_\mathrm{I}, \phi), \tag{6.127}$$

in the group representation, with

$$\mathcal{K}_{g_\mathrm{I}} = -\left( \tau \partial_\phi^2 + \sum_{i=1}^{4} \Delta_{g_i} \right) + m^2 \quad \tau, m^2 \in \mathbb{R}, \tag{6.128}$$

as studied in the context of GFT cosmology by the author in Refs. [457, 490] and used in the analysis of the first model in Section 6.2. It is motivated by the RG analysis of GFT models (see Ref. [393] and references therein). The same term can be given in the spin representation (using also the orthogonality of the intertwiners) as

$$
\begin{aligned}
K &= \int \mathrm{d}\phi \sum_{j_i, m_i, \iota_1, \iota_2} \overline{C}_{n_1 n_2 n_3 n_4}^{j_1 j_2 j_3 j_4 \iota_1} C_{n_1 n_2 n_3 n_4}^{j_1 j_2 j_3 j_4 \iota_2} \overline{\varphi}_{m_1 m_2 m_3 m_4}^{j_1 j_2 j_3 j_4 \iota_1} \hat{T}_{j_1 j_2 j_3 j_4} \varphi_{m_1 m_2 m_3 m_4}^{j_1 j_2 j_3 j_4 \iota_2} \\
&= \int \mathrm{d}\phi \sum_{j_i, m_i, \iota} \overline{\varphi}_{m_1 m_2 m_3 m_4}^{j_1 j_2 j_3 j_4 \iota} \hat{T}_{j_1 j_2 j_3 j_4} \varphi_{m_1 m_2 m_3 m_4}^{j_1 j_2 j_3 j_4 \iota},
\end{aligned}
\tag{6.129}
$$

with

$$\hat{T}_{j_1 j_2 j_3 j_4} = -\tau \partial_\phi^2 + \sum_{i=1}^{4} j_i (j_i + 1) + m^2. \tag{6.130}$$

Let us stress once more that the exact functional dependence on the discrete geometric data can be left more general for the EPRL model(s), since it is not uniquely fixed in the construction of the model, and it is only weakly constrained (mainly at large volumes) by the effective cosmological dynamics (which of course allows for the specific example above); the dependance on the scalar field variable $\phi$ is more important for obtaining the correct cosmological dynamics, at least in the isotropic case.

### 6.3.2 Emergent Friedmann dynamics

In this Section we derive the equations of motion of an isotropic cosmological background from the dynamics of the mean field $\sigma(g_\mathrm{I}; \phi)$ for this GFT model. We reproduce in more detail the analysis of Refs. [486, 487] and clearly spell out all the assumptions made in the derivation, including the necessary restrictions on the GFT field, such as isotropy (i.e. considering equilateral tetrahedra) and left invariance, which we now discuss.

#### 6.3.2.1 Left invariance and isotropic restriction

**(A) Left invariance**

As we have recalled in Section 5.1.2, the simplest effective cosmological dynamics are obtained as the mean field approximation of the full GFT quantum theory, for any specific model. The condensate states have a geometric interpretation as homogeneous continuum spatial geometries if they are left invariant under the diagonal group action. In this way, the domain becomes isomorphic to the minisuperspace of homogeneous geometries [480]. It is a property imposed on this specific class of states such that the above interpretation is guaranteed. With this, the field is decomposed as

$$\sigma_{m_1 m_2 m_3 m_4}^{j_1 j_2 j_3 j_4 \iota} = \sum_{\iota'} \sigma^{j_1 j_2 j_3 j_4 \iota \iota'} C_{m_1 m_2 m_3 m_4}^{j_1 j_2 j_3 j_4 \iota'}, \tag{6.131}$$

where $\iota'$ is another intertwiner label, independent from $\iota$. Using the assumption Eq. (6.131), Eq. (6.125) then becomes

$$
\begin{aligned}
K &= \int \mathrm{d}\phi \sum_{j_\mathrm{I}, m_\mathrm{I}, \iota} \overline{\sigma}_{m_1 m_2 m_3 m_4}^{j_1 j_2 j_3 j_4 \iota} \mathcal{K}_{m_1 m_2 m_3 m_4}^{j_1 j_2 j_3 j_4 \iota} \sigma_{m_1 m_2 m_3 m_4}^{j_1 j_2 j_3 j_4 \iota} \\
&= \int \mathrm{d}\phi \sum_{j_\mathrm{I}, \iota, \iota', \iota''} \overline{\sigma}^{j_1 j_2 j_3 j_4 \iota \iota'} \tilde{\mathcal{K}}^{j_1 j_2 j_3 j_4 \iota \iota' \iota''} \sigma^{j_1 j_2 j_3 j_4 \iota \iota''},
\end{aligned}
\tag{6.132}
$$

with

$$\tilde{\mathcal{K}}^{j_1 j_2 j_3 j_4 \iota \iota' \iota''} = \sum_{m_\mathrm{I}} \overline{C}_{m_1 m_2 m_3 m_4}^{j_1 j_2 j_3 j_4 \iota'} C_{m_1 m_2 m_3 m_4}^{j_1 j_2 j_3 j_4 \iota''} \mathcal{K}_{m_1 m_2 m_3 m_4}^{j_1 j_2 j_3 j_4 \iota} \quad . \tag{6.133}$$

When the kernel $\mathcal{K}_{m_1 m_2 m_3 m_4}^{j_1 j_2 j_3 j_4 \iota}$ does not depend on $\{m_1, \ldots, m_4\}$, Eq. (6.133) considerably simplifies to

$$\tilde{\mathcal{K}}^{j_1 j_2 j_3 j_4 \iota \iota' \iota''} = \left( \sum_{m_\mathrm{I}} \overline{C}_{m_1 m_2 m_3 m_4}^{j_1 j_2 j_3 j_4 \iota'} C_{m_1 m_2 m_3 m_4}^{j_1 j_2 j_3 j_4 \iota''} \right) \mathcal{K}^{j_1 j_2 j_3 j_4 \iota} = \delta^{\iota' \iota''} \mathcal{K}^{j_1 j_2 j_3 j_4 \iota}, \tag{6.134}$$

leading us to the following expression for the kinetic term

$$K = \int \mathrm{d}\phi \sum_{j_1, \iota, \iota'} \overline{\sigma}^{j_1 j_2 j_3 j_4 \iota \iota'} \mathcal{K}^{j_1 j_2 j_3 j_4 \iota} \sigma^{j_1 j_2 j_3 j_4 \iota \iota'}. \tag{6.135}$$

A particular example is given by Eq. (6.129) further simplifying to

$$\begin{aligned}
K &= \int \mathrm{d}\phi \sum_{j_1, m_1, \iota} \overline{\sigma}^{j_1 j_2 j_3 j_4 \iota}_{m_1 m_2 m_3 m_4} \hat{T}_{j_1 j_2 j_3 j_4} \sigma^{j_1 j_2 j_3 j_4 \iota}_{m_1 m_2 m_3 m_4} \\
&= \int \mathrm{d}\phi \sum_{j_1, \iota} \overline{\sigma}^{j_1 j_2 j_3 j_4 \iota} \hat{T}_{j_1 j_2 j_3 j_4} \sigma^{j_1 j_2 j_3 j_4 \iota}. \tag{6.136}
\end{aligned}$$

### (B) Isotropic restriction

Different definitions of "isotropy" are possible for GFT condensates, and have been used in the literature (cf. Refs. [457, 480, 486, 487, 490]), see also Section 6.2, depending on the chosen reconstruction procedure for the continuum geometry out of the discrete data associated to such GFT states. For all of them, though, the result is qualitatively similar, as it should be: the condensate wavefunction has to depend on a single degree of freedom, e.g. one single spin variable, corresponding to the volume information or the scale factor of the emergent universe. Also, we do not expect that these different definitions of isotropic wavefunction would result in very different cosmological dynamics, for any given GFT model, and in fact this seems to be confirmed so far in the literature.[29] In this Section, as in Refs. [486, 487], we adopt the simplest and most symmetric definition: we choose a condensate wavefunction such that the corresponding GFT quanta can be interpreted as equilateral tetrahedra. This implies that all of the spins labelling the quanta are equal $j_i = j, \ \forall i$, corresponding to tetrahedra with all triangle areas being equal. In this case, spin network vertices are said to be monochromatic. We further assume that the only non-vanishing coefficients for each $j$ are those which correspond to the largest eigenvalue of the volume and a fixed orientation of the vertex (which lifts the degeneracy of the volume eigenvalues). In this way, the label $\iota$ is uniquely determined in each intertwiner space following from right invariance (see Appendix E.2). We call this particular value $\iota^\star$. This means that we have fixed all the quantum numbers of a quantum tetrahedron. We are still left with the intertwiner label $\iota'$ following from left invariance. To fix this, we identify the two vectors in $\mathcal{H}_{\text{kin},4}$ determined by the decomposition of $\sigma$ by assuming that the only non-vanishing components $\sigma^{j_1 j_2 j_3 j_4 \iota \iota'}$ are such that $\iota = \iota'$, i.e.

---

[29]Notice that in the analysis of the first model in Section 6.2 we made use of a different notion of isotropy. Symmetry reductions onto the mean field there lead to configurations corresponding to trirectangular tetrahedra the three orthogonal faces of which are of equal size.

$$\sigma^{j_1 j_2 j_3 j_4 \iota \iota'} = \sigma^{j_1 j_2 j_3 j_4 \iota \iota} \delta^{\iota \iota'} \quad \text{(no sum)}. \tag{6.137}$$

Thus, also this extra label is fixed by the maximal volume requirement. The geometric interpretation of this further step is unclear, at present, but it is at least compatible with what we know about the (quantum) geometry of GFT states. Using Eqs. (6.131), (6.137), the expansion Eq. (6.121) simplifies to

$$\sigma(g_{\text{I}}; \phi) = \sum_{j_1, m_1, n_1, \iota} \sigma^{j_1 j_2 j_3 j_4 \iota \iota}(\phi) \, C^{j_1 j_2 j_3 j_4 \iota}_{m_1 m_2 m_3 m_4} C^{j_1 j_2 j_3 j_4 \iota}_{n_1 n_2 n_3 n_4} \prod_{i=1}^{4} d_{j_i} D^{j_i}_{m_i n_i}(g_i). \tag{6.138}$$

Bringing all these conditions together (and dropping unnecessary repeated intertwiner labels), we get for the kinetic term

$$K = \int \mathrm{d}\phi \, \sum_j \bar{\sigma}^{j \iota^\star} \tilde{\mathcal{K}}^{j \, \iota^\star} \sigma^{j \iota^\star} \,, \tag{6.139}$$

while the interaction term is given by

$$V_5 = \frac{1}{5} \int \mathrm{d}\phi \, \sum_j \left( \sigma^{j \iota^\star} \right)^5 \mathcal{V}_5''(j; \iota^\star). \tag{6.140}$$

In Eq. (6.140) we introduced the notation

$$\mathcal{V}_5''(j; \iota^\star) = \mathcal{V}_5'(\underbrace{j, j \ldots, j}_{10}; \underbrace{\iota^\star, \iota^\star \ldots \iota^\star}_{5}) = \mathcal{V}_5(j \ldots j; \iota^\star \ldots \iota^\star) \omega(j, \iota^\star), \tag{6.141}$$

with

$$\omega(j, \iota^\star) = \sum_{m_i} \prod_{i=1}^{10} (-1)^{j_i - m_i}$$
$$\times C^{j \iota^\star}_{m_1 m_2 m_3 m_4} C^{j \iota^\star}_{-m_4 m_5 m_6 m_7} C^{j \iota^\star}_{-m_7 - m_3 m_8 m_9} C^{j \iota^\star}_{-m_9 - m_6 - m_2 m_{10}} C^{j \iota^\star}_{-m_{10} - m_8 - m_5 - m_1}. \tag{6.142}$$

Thus, in the isotropic case, the effect of the interactions is contained in the diagonal of the potential and in the coefficient $\omega(j, \iota^\star)$ constructed out of the intertwiners. We also observe that distinct monochromatic components are decoupled and follow independent dynamics.

Eq. (6.142) can also be written in a different form, by expressing the intertwiner $C_{m_1 m_2 m_3 m_4}^{j \iota^\star}$ in terms of the intertwiners $\alpha_{m_1 m_2 m_3 m_4}^{jJ}$ defined in Appendix E.1 (see Eq. (E.8))

$$\omega(j, \iota^\star) = \sum_{J_k} \left( \prod_{k=1}^{5} c^{J_k \iota^\star} \right) \{15j\}_{J_k},$$ (6.143)

where we identified the contraction of five intertwiners $\alpha_{m_1 m_2 m_3 m_4}^{jJ}$ with a 15$j$-symbol of the first type

$$\{15j\}_{J_k} = \sum_{m_i} \prod_{i=1}^{10} (-1)^{j_i - m_i}$$

$$\times \alpha_{m_1 m_2 m_3 m_4}^{j J_1} \alpha_{-m_4 m_5 m_6 m_7}^{j J_2} \alpha_{-m_7 - m_3 m_8 m_9}^{j J_3} \alpha_{-m_9 - m_6 - m_2 m_{10}}^{j J_4} \alpha_{-m_{10} - m_8 - m_5 - m_1}^{j J_5}.$$ (6.144)

#### 6.3.2.2 Background equation

The equations of motion for the background can be found by varying the action Eq. (6.124). Using Eqs. (6.139), (6.141) we find (compare with Ref. [486])

$$\tilde{\mathcal{K}}^{j \iota^\star} \sigma^{j \iota^\star} + \mathcal{V}_5''(j; \iota^\star) \left( \bar{\sigma}^{j \iota^\star} \right)^4 = 0.$$ (6.145)

In the particular case given by Eq. (6.128) we can write

$$K = \int \mathrm{d}\phi \, \sum_j \bar{\sigma}^{j \iota^\star} \hat{T}_j \sigma^{j \iota^\star} \qquad \hat{T}_j = -\tau \partial_\phi^2 + 4j(j+1) + m^2.$$ (6.146)

For the purpose of studying a concrete example, from now on we consider the special case in which $j = \frac{1}{2}$ of SU(2). Using the definition Eq. (6.142), we then have

$$\omega \left( \frac{1}{2}, \pm \right) = \sum_{m_i} C_{m_1 m_2 m_3 m_4}^{\frac{1}{2} \iota_\pm} C_{m_4 m_5 m_6 m_7}^{\frac{1}{2} \iota_\pm} C_{m_7 m_3 m_8 m_9}^{\frac{1}{2} \iota_\pm} C_{m_9 m_6 m_2 m_{10}}^{\frac{1}{2} \iota_\pm} C_{m_{10} m_8 m_5 m_1}^{\frac{1}{2} \iota_\pm}$$ (6.147)

$$= \frac{3 \mp i\sqrt{3}}{18\sqrt{2}}.$$ (6.148)

$\iota^\star = \iota_\pm$ means that we are considering as an intertwiner the volume eigenvector corresponding to a positive (resp. negative) orientation, see Appendix E.2.

Thus, the equation of motion for the background in this special case reads

$$\left(-\tau\partial_\phi^2 + 3 + m^2\right)\sigma^{\frac{1}{2}\iota^\star} + \overline{\mathcal{V}''}_5\left(\frac{1}{2};\iota^\star\right)\left(\overline{\sigma}^{\frac{1}{2}\iota^\star}\right)^4 = 0, \tag{6.149}$$

with the coefficient of the interaction term given by Eqs. (6.141), (6.147). Notice that, under the assumption that $j = \frac{1}{2}$, used in addition to the isotropic reduction, even the more general form of the EPRL GFT model coupled to a massless free scalar field, Eqs. (6.125) and (6.126), as used in Refs. [486, 487] will collapse to the special case Eq. (6.149). Also, the dominance of a single (small) spin component in the cosmological dynamics of isotropic backgrounds can be shown to take place at late (relational) times [488], in the same way as presented for the other model in Section 6.2.4.1, and it can be expected to be a decent approximation at earlier ones. Thus, the special case we are considering is not too restrictive.

For completeness, we quickly discuss how the above-given effective cosmological dynamics are turned into an evolution equation (in relational time) for the volume of the universe, leading to an emergent Friedmann equation at late times with a generic quantum bounce replacing the big bang singularity [486, 487]. To see this, we consider the mesoscopic regime where we can neglect the interactions in Eq. (6.149) giving

$$\left(\partial_\phi^2 - \frac{3 + m^2}{\tau}\right)\sigma = 0 \tag{6.150}$$

and we neglected the indices in the mean field for convenience. Using the polar decomposition of $\sigma$, being

$$\sigma = \rho \, e^{i\theta}, \tag{6.151}$$

one obtains

$$\rho'' - \frac{Q^2}{\rho^3} - \frac{3 + m^2}{\tau}\rho = 0 \tag{6.152}$$

with the conserved quantities

$$Q = \rho^2\theta' \quad\text{and}\quad E = (\rho')^2 + \rho^2(\theta')^2 - \frac{3 + m^2}{\tau}\rho^2. \tag{6.153}$$

Equation (6.152) has the form of the equation of motion of a particle in a central potential. This potential diverges towards $\rho \to 0$ implying that $\rho$ can never vanish as long as $Q$ is non-zero. Hence, GFT predicts the occurrence of a bounce which replaces the initial spacetime singularity bedeviling classical cosmological models, as discussed in detail in Refs. [486, 487]. In this sense, the results of GFT condensate cosmology are reminiscent of those found in LQC where the resolution of the initial singularity was shown to be a robust feature [75, 76, 538].

The relation to classical cosmological dynamics follows immediately. To this aim, one expresses the first Friedmann equation in proper time $t$ in terms of relational time $\phi$ by

$$H^2 = \left(\frac{V'}{3V}\right)^2 \left(\frac{d\phi}{dt}\right)^2,$$ (6.154)

and does similarly for the second Friedmann equation. Using this together with the fact that the expectation value of the volume operator is directly related to the mean field via

$$V = V_{\frac{1}{2}}\rho^2,$$ (6.155)

one recovers the classical Friedmann equations in the semi-classical limit, i.e. for sufficiently large volumes and upon identifying $\frac{3+m^2}{\tau}$ with $3\pi G$ as

$$\left(\frac{V'}{3V}\right)^2 = \frac{4\pi G}{3} \quad \text{and} \quad \frac{V''}{V} = 12\pi G.$$ (6.156)

Notice that this amounts to a definition of Newton's constant from the fundamental parameters of the theory, see Refs. [486, 487, 536] for a detailed discussion of this matter.[30]

At a later stage of this Section, we will present how interactions, which become dominant away from the bounce, lead to a prolonged phase of accelerated expansion of the emergent space and a later recollapse of the universe to produce a cyclic evolution, as shown by the author in Ref. [489]. Before showing this, we focus our attention on perturbations around the isotropic case described here and governed by the above equations. This will further justify why the monochromatic background already captures the essential features of the emergent dynamics.

---

[30]When comparing with the model explored in Section 6.2, we see that the identification of Newton's constant in terms of the free parameters is in fact model dependent.

### 6.3.3 Non-monochromatic perturbations

We can now derive the equations of motion for perturbations around an isotropic background GFT field configuration $\varphi_0$ satisfying Eq. (6.149), i.e.

$$\sigma = \sigma_0 + \delta\sigma.^{[31]} \tag{6.157}$$

Let us start by writing down the more general equation of motion for the background, by relaxing the isotropy assumption while retaining the other hypotheses of Sections 6.3.2.1.

$$0 = \frac{\delta S[\sigma]}{\delta\overline{\sigma}^{abcd\,\tilde{\iota}}} = \frac{\delta K[\sigma]}{\delta\overline{\sigma}^{abcd\,\tilde{\iota}}} + \frac{\delta\overline{V}_5[\sigma]}{\delta\overline{\sigma}^{abcd\,\tilde{\iota}}} \tag{6.158}$$

Above, we explicitly wrote all of the spin labels $j_{\mathrm{I}} = (a, b, c, d)$. The first term is equal to

$$\frac{\delta K[\sigma]}{\delta\overline{\sigma}^{abcd\,\tilde{\iota}}} = \mathcal{K}^{abcd\,\tilde{\iota}}\sigma^{abcd\,\tilde{\iota}} \ , \tag{6.159}$$

while for the second term we have

$$
\begin{aligned}
\frac{\delta V_5[\varphi]}{\delta\sigma^{abcd\,\tilde{\iota}}} = \frac{1}{5}\sum \Big[ &\sigma^{d567\,\iota_2}\sigma^{7c89\,\iota_3}\sigma^{96b10\,\iota_4}\sigma^{1085a\,\iota_5}\mathcal{V}_5'\big(a,b,c,d,5,6,7,8,9,10;\tilde{\iota},\iota_2,\iota_3,\iota_4,\iota_5\big) + \\
&\sigma^{123a\,\iota_1}\sigma^{d389\,\iota_3}\sigma^{9c210\,\iota_4}\sigma^{108b1\,\iota_5}\mathcal{V}_5'\big(1,2,3,a,b,c,d,8,9,10;\iota_1,\tilde{\iota},\iota_3,\iota_4,\iota_5\big) + \\
&\sigma^{12b4\,\iota_1}\sigma^{456a\,\iota_2}\sigma^{d6210\,\iota_4}\sigma^{10c51\,\iota_5}\mathcal{V}_5'\big(1,2,b,4,5,6,a,c,d,10;\iota_1,\iota_2,\tilde{\iota},\iota_4,\iota_5\big) + \\
&\sigma^{1c34\,\iota_1}\sigma^{45b7\,\iota_2}\sigma^{738a\,\iota_3}\sigma^{d851\,\iota_5}\mathcal{V}_5'\big(1,c,3,4,5,b,7,8,a,d;\iota_1,\iota_2,\iota_3,\tilde{\iota},\iota_5\big) + \\
&\sigma^{d234\,\iota_1}\sigma^{4c67\,\iota_2}\sigma^{73b9\,\iota_3}\sigma^{962a\,\iota_4}\mathcal{V}_5'\big(d,2,3,4,c,6,7,b,9,a;\iota_1,\iota_2,\iota_3,\iota_4,\tilde{\iota}\big) \Big].
\end{aligned}
\tag{6.160}
$$

By just relabelling the indices, we obtain

$$
\begin{aligned}
\frac{\delta V_5[\sigma]}{\delta\sigma^{abcd\,\tilde{\iota}}} = \frac{1}{5}\sum &\sigma^{d567\,\iota_2}\sigma^{7c89\,\iota_3}\sigma^{96b10\,\iota_4}\sigma^{1085a\,\iota_5}\big[\mathcal{V}_5'\big(a,b,c,d,5,6,7,8,9,10;\tilde{\iota},\iota_2,\iota_3,\iota_4,\iota_5\big) + \\
&\mathcal{V}_5'\big(10,8,5,a,b,c,d,6,7,9;\iota_5,\tilde{\iota},\iota_2,\iota_3,\iota_4\big) + \mathcal{V}_5'\big(9,6,b,10,8,5,a,c,d,7;\iota_4,\iota_5,\tilde{\iota},\iota_2,\iota_3\big) + \\
&\mathcal{V}_5'\big(7,c,8,9,6,b,10,5,a,d;\iota_3,\iota_4,\iota_5,\tilde{\iota},\iota_2\big) + \mathcal{V}_5'\big(d,5,6,7,c,8,9,b,10,a;\iota_2,\iota_3,\iota_4,\iota_5,\tilde{\iota}\big)\big] \ ,
\end{aligned}
\tag{6.161}
$$

---

[31] Note that these perturbations are not to be confused with the fluctuations studied in Section 5.3.

which becomes, taking into account the discrete symmetries of the interaction kernel, the simpler expression

$$\frac{\delta V_5[\sigma]}{\delta\sigma^{abcd\,\bar{\iota}}} = \sum \sigma^{d567\,\iota_2}\sigma^{7c89\,\iota_3}\sigma^{96b10\,\iota_4}\sigma^{1085a\,\iota_5}\mathcal{V}'_5\big(a,b,c,d,5,6,7,8,9,10;\bar{\iota},\iota_2,\iota_3,\iota_4,\iota_5\big)\,. \tag{6.162}$$

Moreover, given the structure of the interaction term in Eq. (6.162), the only non-vanishing contributions to the first order dynamics of the perturbations around a monochromatic background come from terms having at least three identical spins among $(a,b,c,d)$. Therefore, depending on which of the four indices, labelled $j'$ is singled out to be different from the other three, labelled $j$, we obtain four independent equations

$$\mathcal{K}^{jjjj'\,\bar{\iota}}\delta\sigma^{jjjj'\,\bar{\iota}} + \sum_{\iota}\delta\overline{\sigma}^{j'jjj\,\iota}\left(\overline{\sigma_0}^{j\,\iota^{\star}}\right)^3\mathcal{V}'_5\big(j,j,j,j';j,j,j,j,j,j;\bar{\iota},\iota,\iota^{\star},\iota^{\star}\big) = 0\,. \tag{6.163}$$

$$\mathcal{K}^{jjj'j\,\bar{\iota}}\delta\sigma^{jjj'j\,\bar{\iota}} + \sum_{\iota}\delta\overline{\sigma}^{jj'jj\,\iota}\left(\overline{\sigma_0}^{j\,\iota^{\star}}\right)^3\mathcal{V}'_5\big(j,j,j',j;j,j,j,j,j,j;\bar{\iota},\iota^{\star},\iota,\iota^{\star}\big) = 0\,. \tag{6.164}$$

$$\mathcal{K}^{jj'jj\,\bar{\iota}}\delta\sigma^{jj'jj\,\bar{\iota}} + \sum_{\iota}\delta\overline{\sigma}^{jjj'j\,\iota}\left(\overline{\sigma_0}^{j\,\iota^{\star}}\right)^3\mathcal{V}'_5\big(j,j',j,j;j,j,j,j,j,j;\bar{\iota},\iota^{\star},\iota^{\star},\iota\big) = 0\,. \tag{6.165}$$

$$\mathcal{K}^{j'jjj\,\bar{\iota}}\delta\sigma^{j'jjj\,\bar{\iota}} + \sum_{\iota}\delta\overline{\sigma}^{jjjj'\,\iota}\left(\overline{\sigma_0}^{j\,\iota^{\star}}\right)^3\mathcal{V}'_5\big(j',j,j,j;j,j,j,j,j,j;\bar{\iota},\iota^{\star},\iota^{\star},\iota\big) = 0\,. \tag{6.166}$$

We define a new function

$$\mathcal{U}(j,j',\iota,\iota';n) \equiv \left(\overline{\sigma_0}^{j\,\iota^{\star}}\right)^3\mathcal{V}'_5\Big(\underbrace{j,...,j',...,j,j,j,j,j,j}_{n};\underbrace{\iota,...,\iota',...,\iota^{\star}}_{5-n}\Big)\,, \tag{6.167}$$

with $j'$ in the $n$-th position ($n = 1, 2, 3, 4$) and $\iota'$ appearing in position $5-n$ after $\iota$, which keeps the first place. For instance, one has for $n = 1$

$$\mathcal{U}(j,j',\iota^{\star},\iota,\iota';n) = \left(\overline{\sigma_0}^{j\,\iota^{\star}}\right)^3\mathcal{V}'_5\big(j',j,j,j,j,j,j,j,j,j;\iota,\iota^{\star},\iota^{\star},\iota'\big)\,. \tag{6.168}$$

Thus, the equations of motion for the perturbations can be more compactly rewritten as

$$\mathcal{K}^{j'jjj\,\iota}\delta\sigma^{j'jjj\,\iota} + \sum_{\iota'} \delta\bar{\sigma}^{jjjj'\,\iota'}\mathcal{U}(j,j',\iota^\star,\iota,\iota';1) = 0$$

$$\mathcal{K}^{jj'jj\,\iota}\delta\sigma^{jj'jj\,\iota} + \sum_{\iota'} \delta\bar{\sigma}^{jjj'j\,\iota'}\mathcal{U}(j,j',\iota^\star,\iota,\iota';2) = 0$$

$$\mathcal{K}^{jjj'j\,\iota}\delta\sigma^{jjj'j\,\iota} + \sum_{\iota'} \delta\bar{\sigma}^{jj'jj\,\iota'}\mathcal{U}(j,j',\iota^\star,\iota,\iota';3) = 0$$

$$\mathcal{K}^{jjjj'\,\iota}\delta\sigma^{jjjj'\,\iota} + \sum_{\iota'} \delta\bar{\sigma}^{j'jjj\,\iota'}\mathcal{U}(j,j',\iota^\star,\iota,\iota';4) = 0 \quad . \tag{6.169}$$

With the particular kinetic kernel (6.128), one has that the kinetic operator acting on the perturbation does not depend on the position of the perturbed index $j'$, neither it depends on the intertwiner label $\iota$. Hence, in that case we can define

$$\mathcal{K}' = \mathcal{K}^{j'jjj\,\iota} = \mathcal{K}^{jj'jj\,\iota} = \mathcal{K}^{jjj'j\,\iota} = \mathcal{K}^{jjjj'\,\iota}$$

$$= -\tau\partial_\phi^2 + \big(3j(j+1) + j'(j'+1)\big) + m^2. \tag{6.170}$$

The above equations are generic. However, recoupling theory imposes several restrictions on our perturbations, due to the conditions imposed on the fields: a) $j'$ is an integer (half-integer) if the background spin $j$ is an integer (half-integer); b) $j'$ cannot be arbitrarily large, since for $j' > 3j$ the closure (right invariance) condition would be violated; c) of course, the case $j' = j$ is uninteresting since such perturbations can be reabsorbed into the monochromatic background.

In the simplest example $j = \frac{1}{2}$ there is only one permitted value for the perturbed spin, namely $j' = \frac{3}{2}$, and the perturbation is identified with the state such that the total spin of a pair is $J = 1$. Any such state is trivially also a volume eigenstate since the volume operator is identically vanishing in such intertwiner space, as it is one-dimensional (see Appendix E.2, in particular the comment after Eq. (E.16)). For this reason, we will omit the indices $\iota$, $\iota'$ in the following.

Let us introduce some further notation for these specific perturbations. We define

$$\psi_1 = \delta\sigma^{\frac{3}{2}\frac{1}{2}\frac{1}{2}\frac{1}{2}}, \quad \psi_2 = \delta\sigma^{\frac{1}{2}\frac{3}{2}\frac{1}{2}\frac{1}{2}}, \quad \psi_3 = \delta\sigma^{\frac{1}{2}\frac{1}{2}\frac{3}{2}\frac{1}{2}}, \quad \psi_4 = \delta\sigma^{\frac{1}{2}\frac{1}{2}\frac{1}{2}\frac{3}{2}} \tag{6.171}$$

and similarly

$$\mathcal{K}_1 = \mathcal{K}^{\frac{3}{2}\frac{1}{2}\frac{1}{2}\frac{1}{2}}, \quad \mathcal{K}_2 = \mathcal{K}^{\frac{1}{2}\frac{3}{2}\frac{1}{2}\frac{1}{2}}, \quad \mathcal{K}_3 = \mathcal{K}^{\frac{1}{2}\frac{1}{2}\frac{3}{2}\frac{1}{2}}, \quad \mathcal{K}_4 = \mathcal{K}^{\frac{1}{2}\frac{1}{2}\frac{1}{2}\frac{3}{2}}. \tag{6.172}$$

Hence, it follows from Eq. (6.169) that the dynamics of the perturbations are governed (to first order) by the following equations (omitting the perturbation variables $j'$, $\iota$, $\iota'$ and the background spin $j = \frac{1}{2}$ in the argument of $\mathcal{U}$, Eq. (6.168)):

$$\mathcal{K}_1 \psi_1 + \mathcal{U}(\iota^\star; 1)\overline{\psi_4} = 0$$
$$\mathcal{K}_4 \psi_4 + \mathcal{U}(\iota^\star; 4)\overline{\psi_1} = 0$$
$$\mathcal{K}_2 \psi_2 + \mathcal{U}(\iota^\star; 2)\overline{\psi_3} = 0$$
$$\mathcal{K}_3 \psi_3 + \mathcal{U}(\iota^\star; 3)\overline{\psi_2} = 0 \quad . \tag{6.173}$$

The resulting equations for the perturbations are reasonably simple, mainly thanks to the isotropy assumption on the background, which considerably simplifies the contribution from the GFT interaction term $\mathcal{U}$. However, the simplified functional form in which the Lorentzian EPRL vertex amplitude appears in these equations remains unknown in exact analytic terms. The above equations would then have to be numerically studied or in more phenomenological approach, in which the exact function $\mathcal{U}$ is replaced by some simpler trial function, or several ones in different ranges of the variable $j$, approximating it. Luckily, for our present concerns, which relate to the behaviour of perturbations close to the cosmological bounce, these difficulties can be sidestepped since the interaction term is generically subdominant in that regime of the theory, mainly due to the smallness of the background condensate wavefunction (in turn related to the smallness of the universe 3-volume). This will allow us to perform a study of this dynamics in the following section.

Before we turn to such dynamics, let us notice that the equations (6.173) make manifest an asymmetry of the interaction terms of the GFT model we are considering, more specifically of the EPRL vertex amplitude, that is not apparent at first sight. The equations in fact couple perturbations in the first field argument with perturbations in the fourth, and perturbations in the second with perturbations in the third, with no other combination being present. This happens despite the isotropy assumption on the background and the other symmetries of the model. One can trace this asymmetry back to the combinatorial structure of the vertex amplitude itself: it corresponds to a 4-simplex as projected down to the plane but it is not symmetric with respect to the face pairings, if such faces are ordered in their planar projection: it only couples first and fourth faces across common tetrahedra sharing them, or second and third ones, i.e. exactly the type of asymmetry that is revealed in our perturbations equations. It is tempting to relate this asymmetry to an issue with orientability of the triangulations resulting from the Feynman expansion of the model, since the same type of issue has been identified in the Boulatov model for 3d gravity in Ref. [499]. It is unclear at this stage whether this is a problem or just a feature of the model; it is also unclear, in case one decides to remove such asymmetry, what is the best way to do so. The strategy followed in the 3d case Ref. [499], i.e. to maintain the ordering of the GFT field arguments but modify the combinatorics of the interaction vertex to ensure orientability, does not seem available in this 4d case. An easy solution would be to impose that the GFT fields themselves are invariant under (even) permutations of their arguments, which also ensure orientability of the resulting triangulations. We leave this point, not directly relevant for the analysis of the next section, for further study.

### 6.3.4   Dynamics of perturbations at the bounce

We now study the dynamics of the perturbations around a homogeneous and isotropic background solution of the condensate dynamics, in the mean field approximation. We focus on the bounce regime, since this is where typical bouncing models of the early universe have difficulties in controlling the dynamics of anisotropies. Luckily, as anticipated, this is also the regime where, in the GFT condensate cosmology framework we can have the best analytic control over the (quantum) dynamics of the theory, at least in the mean field approximation. In fact, the bouncing regime takes place, in the mean field approximation we are working in, for low densities, thus, intuitively, for low values of the modulus of the GFT mean field.

Considering the kernel of Eq. (6.128), assuming $j = \frac{1}{2}$ for the background, and neglecting the interaction term, the background equation reads as

$$\left(-\tau \partial_\phi^2 + 3 + m^2\right) \sigma^{\frac{1}{2}\iota^\star} \simeq 0. \tag{6.174}$$

On the other hand, to first order, perturbations satisfy the equation

$$\mathcal{K}'\psi \simeq 0, \tag{6.175}$$

where

$$\mathcal{K}' = -\tau \partial_\phi^2 + \big(3j(j+1) + j'(j'+1)\big) + m^2 = -\tau \partial_\phi^2 + 6 + m^2 \,, \qquad (6.176)$$

and we have indicated a generic perturbation by $\psi$, since there is no difference among them, in this approximation.

When the interaction term is no longer subdominant (i.e. after the universe exits the bouncing phase and after it has expanded enough), the dynamics of the perturbations is given by the systems of equations (6.173), which remain valid until $\psi \simeq \sigma^{\frac{1}{2}\iota^\star}$. At that point, higher order corrections are needed. On the other hand, it is important to stress that since the equations of motion become linear at the bounce, at that point we are no longer subject to the constraint that non-monochromatic components should be small. In other words, $\psi \simeq \sigma^{\frac{1}{2}\iota^\star}$ is allowed in that regime and perturbations can be large. This observation will be important in the following.

Using the analytic expression of the background solution, given in Ref. [536], we have

$$|\sigma^{\frac{1}{2}\iota^\star}| = \frac{\mathrm{e}^{\sqrt{\frac{3+m^2}{\tau}}(\Phi-\phi)}\sqrt{-2E\mathrm{e}^{2\sqrt{\frac{3+m^2}{\tau}}(\phi-\Phi)}\sqrt{\Omega} + \mathrm{e}^{4\sqrt{\frac{3+m^2}{\tau}}(\phi-\Phi)}\Omega + \Omega}}{2\sqrt{\frac{3+m^2}{\tau}}\sqrt[4]{\Omega}}, \qquad (6.177)$$

where

$$\Omega = E^2 + 4Q^2 \left(\frac{3+m^2}{\tau}\right). \qquad (6.178)$$

The two quantities $E$ and $Q$ are the conserved quantities of the monochromatic background, as discussed in Section 6.3.2.2.[32] Reality of the expression in Eq. (6.177) implies

$$\Omega \geq 0. \qquad (6.179)$$

In order to have the same dynamics for the background as in Refs. [486, 487], we demand that

$$\frac{3+m^2}{\tau} > 0. \qquad (6.180)$$

In this case, the modulus of the backround $|\sigma^{\frac{1}{2}\iota^\star}|$ has a unique global minimum at $\phi = \Phi$, corresponding to the quantum bounce. We will consider two possible cases in which

---

[32] The conservation of $Q$ is not exact, as it follows from an approximate $U(1)$-symmetry, which holds as long as interactions are negligible.

condition (6.180) is still satisfied, but different conditions are imposed on the parameters governing the dynamics of the perturbations. This qualitatively gives the same evolution of the background but two radically different pictures for the evolution of the perturbations:

- Case $i$) The first possibility is that $\tau,\ m^2 \geq 0$. In this case, also the perturbations satisfy an analogous condition

$$\frac{6+m^2}{\tau} > 0. \qquad (6.181)$$

  The analytic solution of the equation for the perturbations has the same form as Eq. (6.177)

$$|\psi| = \frac{e^{\sqrt{\frac{6+m^2}{\tau}}(\Phi_1-\phi)}\sqrt{-2E_1 e^{2\sqrt{\frac{6+m^2}{\tau}}(\phi-\Phi_1)}\sqrt{\Omega_1}+e^{4\sqrt{\frac{6+m^2}{\tau}}(\phi-\Phi_1)}\Omega_1+\Omega_1}}{2\sqrt{\frac{6+m^2}{\tau}}\sqrt[4]{\Omega_1}}. \qquad (6.182)$$

  We introduced the quantity $\Omega_1$, in analogy with Eq. (6.178)

$$\Omega_1 = E_1^2 + 4Q_1^2\left(\frac{6+m^2}{\tau}\right). \qquad (6.183)$$

  $E_1$ and $Q_1$ are two conserved quantities of the perturbations. Reality of Eq. (6.182) requires that $\Omega_1 \geq 0$. $|\psi|$ has a minimum at $\Phi_1$. From Eqs. (6.177), (6.182), we find in the limit of large $\phi$

$$\frac{|\psi|}{|\sigma^{\frac{1}{2}}\iota^\star|} \sim e^{\left(\sqrt{\frac{6+m^2}{\tau}}-\sqrt{\frac{3+m^2}{\tau}}\right)\phi}, \qquad (6.184)$$

  which means that perturbations cannot be neglected in this limit, i.e. away from the bounce occurring at $\phi = \Phi$ (see Eq. (6.177) and discussion below Eq. (6.180)). Therefore, when we are in this region of parameter space, they should be properly taken into account. Depending on the values of the parameters, they can become dominant already close to the bounce. At the same time, the value of relational time $\phi$ at which this approximation is usable cannot be too large, because then we expect the GFT interactions to grow in importance, breaking the approximation on the background and perturbation dynamics we have assumed to be valid so far.

- Case $ii$) A second possibility is represented by the case in which condition (6.180) is still satisfied while inequality (6.181) is not. This can be accomplished with $\tau < 0$ and $-6 < m^2 < -3$. In this case, the modulus of the perturbations oscillates around the minimum of a one-dimensional mechanical potential.

Writing $|\psi| = \rho$, its dynamics are given by (see Refs. [486, 487, 489])

$$\partial_\phi^2 \rho = -U_{,\rho}, \tag{6.185}$$

where

$$U(\rho) = \frac{Q_1^2}{2\rho^2} - \left(\frac{6 + m^2}{\tau}\right)\frac{\rho^2}{2}. \tag{6.186}$$

What happens in this case is that, away from the bounce, the perturbations are always dominated by the background.

In order to see this in a more quantitative way, we can make the simplifying assumption that the minimum of $U$ and the amplitude of the oscillations of $\rho$ are such that the interactions between quanta are always negligible for the perturbations. This can be realised by making an appropriate choice for the values of the parameters of the model.

In this case, Eq. (6.185) describes the evolution of the perturbations at all times. Their qualitative behaviour around the bounce is illustrated in Fig. 6.25. Non-monochromatic perturbations are relevant at the bounce but drop off quickly away from it.[33]

The behaviour of the perturbations is oscillatory, since $|\psi|$ is trapped in the potential well $U$ of Eq. (6.186). As a consequence of this, the number of non-monochromatic quanta $N_1 = |\psi|^2$ has an upper bound. Conversely, the number of quanta in the background grows unboundedly.

We conclude that, in this window of parameter space, perturbations can be relevant at the bounce but are negligible for large numbers of quanta in the background. For a suitable strength of the interactions, non-monochromatic perturbations can become completely irrelevant for the dynamics before interactions kick in.

### Measures of deviations from monochromatic GFT condensates

It is interesting to further explore the deviations from perfect monochromaticity, by computing some quantities which can characterise the dynamics of the perturbed condensate and distinguish it from the purely monochromatic case. We do so in the following. The quantities we compute do not have a clear cosmological meaning, and do not correspond to specific gauge-invariant observables characterising anisotropies in relativistic cosmology. They are however well-defined formal observables for GFT condensates.

---

[33]This is reminiscent of the results [490] obtained by the author in the context of the other model as presented in Section 6.2.6.1, also suggesting that such non-monochromatic modes are only relevant in the regime of small volumes.

The first one we consider is a surface-area-to-volume ratio. A first quantised area operator in GFT can be defined for a tetrahedron as in Ref. [482]: $\hat{A} = \kappa \sum_{i=1}^4 \sqrt{-\Delta_i}$, where the sum runs over all the faces of the tetrahedron, in analogy with the LQG area operator and $\kappa = 8\pi\gamma\ell_p^2$.

We have for its expectation value on a single monochromatic (equilateral) quantum:

$$A_0 = 2\kappa\sqrt{3} \qquad (6.187)$$

and for a perturbed non-monochromatic quantum:

$$A_1 = \frac{\kappa}{2}\sqrt{3}\left(3 + \sqrt{5}\right) \qquad . \qquad (6.188)$$

This operator can then be turned into a second quantised counterpart of the same (see e.g. Refs. [5, 31]), to be applied to ensembles of tetrahedra. One can then easily compute the expectation value of this operator, as well as the expectation value of the total volume operator, in both an unperturbed and in a perturbed condensate state (of the simplest type considered in this Section). The resulting quantity heuristically is the sum of the areas of the four faces of each tetrahedron times the number of tetrahedra with the same areas.

The area-to-volume ratio for the example considered can then be expressed as

$$\frac{A}{V} = \frac{A_0 N_0 + A_1 N_1}{V_0 N_0} = \frac{A_0}{V_0}\left(1 + \frac{A_1}{A_0}\,\frac{N_1}{N_0}\right). \qquad (6.189)$$

$A_0$ is the surface area of an unperturbed quantum and $A_1$ that of a perturbed one. $N_0$ and $N_1$ are the corresponding number of quanta, which can be computed using

$$N(\phi) = \int_{\mathrm{SU(2)}^4} (\mathrm{d}g)^4\, \bar{\sigma}(g_\mathrm{I}; \phi)\sigma(g_\mathrm{I}; \phi) \qquad (6.190)$$

thus leading to

$$N_0 = |\sigma^{\frac{1}{2}\iota^\star}|^2, \tag{6.191}$$

$$N_1 = |\psi|^2. \tag{6.192}$$

$V_0$ is the volume of a quantum of space in the background. We recall that the perturbed quanta considered in this example have *vanishing volume* (see Appendix E.2). This has significant consequences which we illustrate in the following. With the above, Eq. (6.189) leads to

$$\frac{A}{V} = \frac{A_0}{V_0} \left( 1 + \frac{3 + \sqrt{5}}{4} \frac{N_1}{N_0} \right). \tag{6.193}$$

Since $\frac{N_1}{N_0} \geq 0$, we have

$$\frac{A}{V} \geq \frac{A_0}{V_0}. \tag{6.194}$$

This inequality means that, for a given volume, quanta on average have more surface than they would in a purely mono-chromatic (isotropic) background.

The evolution of $\frac{A}{V}$ in case $ii$) is shown in Fig. 6.25. If the perturbations have minimal $E_1$ (as introduced in Eq. (6.183)), i.e. they sit at the minimum of $U$, $\frac{A}{V}$ drops off monotonically as we move away from the bounce, due to the growth of the background. One could say that anisotropies, to the extent in which they are captured by the non-monochromatic perturbations, are diluted away by the expansion of the isotropic background. The background value $\frac{A_0}{V_0}$ is a lower bound, which is asymptotically attained in the infinite volume limit (obviously, before too large volumes can be attained, one expects GFT interactions to kick in, breaking the approximation we have employed here). On the other hand, if $E_1$ of the perturbations is above the minimum of the potential $U$, the perturbations will start to oscillate around such minimum. Therefore, $\frac{A}{V}$ will oscillate as it drops off. The asymptotic properties are unchanged.

Another interesting quantity to compute is the effective volume per quantum, defined as

$$\frac{V}{N} = \frac{N_0 \, V_0}{N_0 + N_1} = \frac{V_0}{1 + \frac{N_1}{N_0}}, \tag{6.195}$$

where again all quantities entering the above formula are expectation values of 2nd quantised GFT observables in the (perturbed) GFT condensate state. It satisfies the bounds

$$0 \leq \frac{V}{N} \leq V_0. \tag{6.196}$$

Its profile for the example given above is shown in Fig. 6.26. The ratio $\frac{V}{N}$ represents the average volume of a quantum of space. Its value is generally lower than $V_0$, i.e. the volume of an equilateral quantum tetrahedron with minimal areas. In fact, zero volume quanta[34] can change the total number of quanta $N$, leaving $V$ unchanged. Explicit calculations show that, in the limit of large $N$, the ratio $\frac{V}{N}$ approaches the value $V_0$ (see Fig. 6.26). In Fig. 6.27 we show the plot relative to the case where perturbations do not reach their maximum amplitude at the bounce, resulting in a deformation of the profile of $\frac{A}{V}$. This corresponds to setting initial conditions for the microscopic anisotropies (non-monochromaticity) before the bounce.

To summarise, in the region of parameter space corresponding to Case ii) above, our results confirm that, from a bouncing phase, where the quantum geometry can be rather degenerate and anisotropies (encoded in non-monochromatic perturbations of the simplest GFT condensate state) quite large, a cosmological background emerges the dynamics of which can be cast into the form an effective Friedmann equation for a homogeneous, isotropic universe.

In the following, we will investigate the impact of effective interactions onto the background.

### 6.3.5   Non-linear dynamics of a GFT condensate

As shown in Ref. [486] the dynamics of an isotropic GFT condensate can be described by means of the effective action

$$S = \int \mathrm{d}\phi \; \left( A \, |\partial_\phi \sigma|^2 + V_{\mathrm{eff}} \right), \tag{6.197}$$

where $V_{\mathrm{eff}}$ subsumes the contributions of the mass, Laplacian and simplicial interaction terms.[35] Variation of this action with respect to $\sigma$ leads to the equation of motion of the isotropic mean field Eq. (6.145). As shown in Section 6.3.2, the macroscopic dynamics can be given in terms of effective Friedmann equations in relational form as

$$\frac{\partial_\phi V}{V} = 2\frac{\partial_\phi \rho}{\rho}, \tag{6.198}$$

$$\frac{\partial_\phi^2 V}{V} = 2\left[\frac{\partial_\phi^2 \rho}{\rho} + \left(\frac{\partial_\phi \rho}{\rho}\right)^2\right]. \tag{6.199}$$

where $\rho = |\sigma|$ and $V$ denotes the volume.

---

[34]See Appendix E.2.

[35]We note that there is an ambiguity in the choice of the sign of $A$, which is not fixed by the microscopic theory and will turn out to be particularly relevant for the cosmological applications of the model. In particular, it can be used to restrict the class of microscopic models by selecting only those that are phenomenologically viable. In fact, as we will show, only models entailing $A < 0$ are sensible from a phenomenological point of view since otherwise one would have faster than exponential expansion. This ambiguity has also been discussed earlier in Ref. [522] when exploring the possibility to embed LQC in GFT.

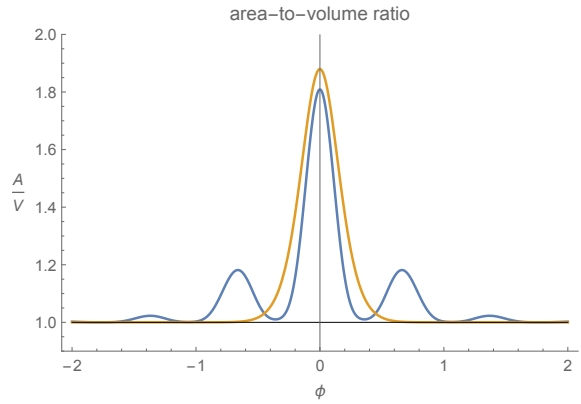

FIGURE 6.25:  Plot of the surface-area-to-volume ratio as a function of relational time $\phi$ in the case $\tau < 0$, $-6 < m^2 < -3$. The vertical axis is in units of $\frac{A_0}{V_0}$. The orange curve ($\tau = -1, m^2 = -42, Q_0 = 1, Q_1 = 1, E_0 = -70, E_1 = 3$) corresponds to perturbations sitting at the minimum of the potential $U$. Although the initial conditions can be chosen so that the surface-area-to-volume ratio $\frac{A}{V}$ is significantly different from its value for a single tetrahedron at the bounce, it decays exponentially away from it. The blue curve ($\mu = -24, Q_0 = 3, Q_1 = 1.5, E_0 = 2, E_1 = 14$) represents the case in which the the quantity $E$ of the perturbations is above the minimum of the potential, but the amplitude of the oscillations is small enough to justify the harmonic approximation. Initial conditions are chosen such that perturbations start oscillating with maximum amplitude at the bounce. The ratio $\frac{A}{V}$ undergoes damped oscillations away from the bounce. The value $\frac{A_0}{V_0}$ is always a lower bound, asymptotically attained. The qualitative behaviour represented by the blue curve is generic for any choice of parameters. When the "kinetic energy" of the perturbations is negligible compared to the potential, one obtains the behaviour represented by the orange curve.

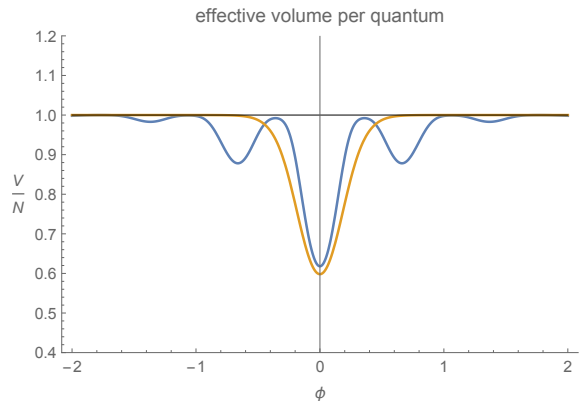

FIGURE 6.26: Evolution of the effective volume $\frac{V}{N}$ of a quantum over relational time for $\tau < 0$, $-6 < m^2 < -3$. The vertical axis is in units of $V_0$. The parameters chosen for the two curves correspond to those of Fig. 6.25. $\frac{V}{N}$ relaxes to the volume $V_0$ of a quantum in the backgound away from the bounce. However, at the bounce it can be significantly different from such value.

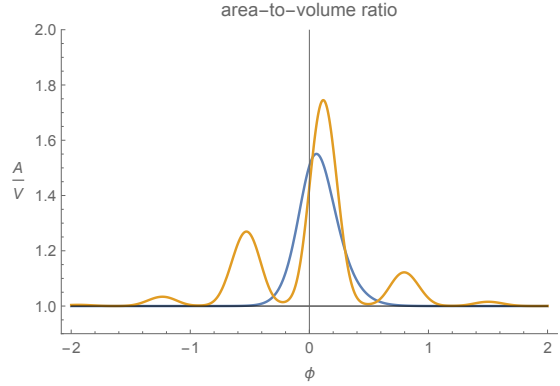

FIGURE 6.27: Plots of $\frac{A}{V}$ for the case of a non-maximal amplitude of the non-monochromatic perturbations at the bounce. The curves correspond to the same values considered in Fig. 6.25. An initial "velocity" $\partial_\phi|\psi|$ is given to the perturbations at the bounce $\phi = \Phi$, with the value 0.6 for the blue curve and 0.4 for the orange curve. The non-symmetric initial conditions results in a deformation of the profile of $\frac{A}{V}$ and is accompanied by a damping.

In the absence of a closed expression for the simplicial interaction term in EPRL form, we resort to phenomenological arguments and model interactions such that they at least capture the non-linearity of the standard GFT interactions. Thus, we consider an effective potential of the following form

$$V_{\text{eff}}[\sigma] = B|\sigma(\phi)|^2 + \frac{2}{n}w|\sigma|^n + \frac{2}{n'}w'|\sigma|^{n'}, \tag{6.200}$$

where we can assume $n' > n$ without loss of generality. Note that $B$ absorbs the contribution from the Laplacian and the mass term. The interaction terms appearing in GFT actions are usually defined in such a way that the perturbative expansion of the GFT partition function reproduces that of spin foam models. As noted above, spin foam models for $4d$ quantum gravity are mostly based on interaction terms of power 5, called simplicial. In the case that the GFT field is endowed by a particular tensorial transformation property, other classes of models can be obtained the interaction terms, called tensorial, of which are based on even powers of the modulus of the field. In this light, the particular type of interactions considered here can be understood as mimicking such types of interactions, which is the reason why we will refer to them as pseudosimplicial and pseudotensorial, respectively. In the following we will study their phenomenological consequences, and show how interesting physical effects are determined as a result of the interplay between two interactions of this type. The integer-valued powers $n$, $n'$ in the interactions will be kept unspecified throughout the rest of this Section, thus making our analysis retain its full generality. The particular values motivated by the above discussion can be retrieved as particular cases. In the following we will show how different ranges for such powers lead to phenomenologically interesting features of the model, most notably concerning an early era of accelerated expansion in Section 6.3.5.2.

For stability reasons, we require $V_{\text{eff}}[\sigma]$ to to be bounded from below, hence $w' > 0$. The equation of motion of the field $\sigma$ obtained from Eqs. (6.197), (6.200) is

$$-A\partial_\phi^2\sigma + B\sigma + w|\sigma|^{n-2}\sigma + w'|\sigma|^{n'-2}\sigma = 0. \qquad (6.201)$$

Writing the complex field $\sigma$ in polar form $\sigma = \rho\,\mathrm{e}^{i\theta}$ one finds (cf. Ref. [486]) that the equation of motion for the angular component leads to the conservation law

$$\partial_\phi Q = 0, \text{ with } Q \equiv \rho^2\partial_\phi\theta, \qquad (6.202)$$

while the radial component satisfies a second order ODE

$$\partial_\phi^2\rho - \frac{Q^2}{\rho^3} - \frac{B}{A}\rho - \frac{w}{A}\rho^{n-1} - \frac{w'}{A}\rho^{n'-1} = 0. \qquad (6.203)$$

The conserved charge $Q$ is proportional to the momentum of the scalar field $\pi_\phi = \hbar Q$ [486]. One immediately observes, that for large values of $\rho$ the term $\rho^{n'-1}$ becomes dominant. In order to ensure that Eq. (6.203) does not lead to drastic departures from standard cosmology at late times (Eq. (6.198)), the coefficient of such a term has to be positive

$$\mu \equiv -\frac{w'}{A} > 0, \qquad (6.204)$$

which implies, since $w' > 0$, that one must have $A < 0$. In fact, the opposite case $\mu < 0$ would lead to an open cosmology expanding at a faster than exponential rate, which relates to a Big Rip. Thus, considering $A < 0$, compatibility with the free case (see Ref. [486]) demands

$$m^2 \equiv \frac{B}{A} > 0, \qquad (6.205)$$

which in turn implies $B < 0$. The sign of $w$ is a priori not constrained, which leaves a considerable freedom in the model. Given the signs of the parameters $B$ and $w'$, the potential in Eq. (6.200) can be related to models with spontaneous symmetry breaking in statistical and quantum field theory. The sub-leading term in the potential plays an important role in determining an inflationary-like era, as shown below in Section 6.3.5.2. The connection to the theory of critical phenomena is to be expected from the conjecture that GFT condensates arise through a phase transition from a non-geometric to a geometric phase [461, 462], which could be a possible realisation of the geometrogenesis scenario [459, 460].

From Eqs. (6.204), (6.205) and defining

$$\lambda \equiv -\frac{w}{A}, \qquad (6.206)$$

we can rewrite Eq. (6.203) in the form

$$\partial_\phi^2 \rho - m^2 \rho - \frac{Q^2}{\rho^3} + \lambda \rho^{n-1} + \mu \rho^{n'-1} = 0, \qquad (6.207)$$

that will be used throughout the rest of this Section. The above equation has the form of the equation of motion of a classical point particle with potential (see the left hand side of Fig. 6.28)

$$U(\rho) = -\frac{1}{2}m^2\rho^2 + \frac{Q^2}{2\rho^2} + \frac{\lambda}{n}\rho^n + \frac{\mu}{n'}\rho^{n'}. \qquad (6.208)$$

Equation (6.207) leads to another conserved quantity, $E$, defined as

$$E = \frac{1}{2}(\partial_\phi \rho)^2 + U(\rho). \qquad (6.209)$$

As mentioned above, its physical meaning is yet to be clarified from a fundamental point of view [486].

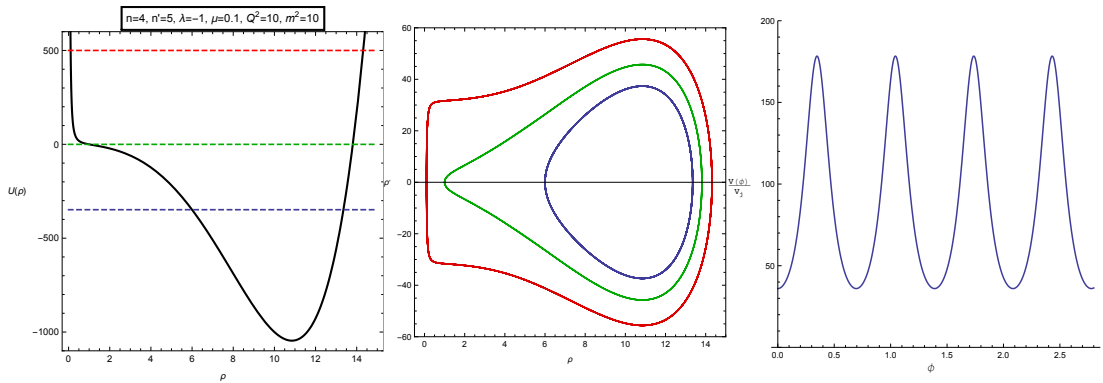

FIGURE 6.28: Left: Plot of the potential $U(\rho)$ (Eq. (6.208)) for the dynamical system described by Eq. (6.207) and a particular choice of parameters. The three horizontal curves correspond to different values of $E$, in turn corresponding to different choices of initial conditions for $\rho$, $\rho'$. The corresponding orbits in phase space are shown in the plot in the center. Recollapse is generic feature of the model and occurs for any values of the parameters, provided $\mu > 0$ and $Q \neq 0$. Middle: Phase portrait of the dynamical system given by Eq. (6.207). Orbits possess an $E$ given by the corresponding colour lines on the left hand side. Orbits are periodic and describe oscillations around the stable equilibrium point (center fixed point) given by the absolute minimum of the potential $U(\rho)$. This is a general feature of the model which does not depend on the particular choice of parameters, provided Eq. (6.204) is satisfied. Right: Plot of the volume of the universe as a function of relational time $\phi$ (in arbitrary units), corresponding to the blue orbit in the central plot. As a generic feature of the interacting model the universe undergoes a cyclic evolution and its volume has a positive minimum corresponding to a bounce.

### 6.3.5.1   Recollapsing universe

Some properties of the solutions of the model and its consequences for cosmology can already be drawn by means of a qualitative analysis of the solutions of the second order ODE in Eq. (6.208). In fact, solutions are confined to the positive half-line $\rho > 0$, given the infinite potential barrier at $\rho = 0$ for $Q \neq 0$. Moreover, since $\mu > 0$ the potential in Eq. (6.208) approaches infinity as $\rho$ takes arbitrarily large values. Therefore, provided that we fix $E$ at a value which is larger than both the absolute minimum and that of (possible) local maxima of the potential $U(\rho)$, the solutions of Eq. (6.207) turn out to be cyclic motions (see Fig. 6.28) describing oscillations around a stable equilibrium point. These, in turn, correspond via $V = V_j \rho^2$ to cyclic solutions for the dynamics of the universe Eqs. (6.198), (6.199) (see the right hand side of Fig. 6.28).

It is interesting to compare this result with what is known in the case where interactions are disregarded [486]. In that case one has that the universe expands indefinitely and in the limit $\phi \to \infty$ its dynamics follows the ordinary Friedmann equation for a flat universe, filled with a massless and minimally coupled scalar field. Therefore, we see that the given interactions in the GFT model induce a recollapse of the universe, corresponding to the turning point of the motion of $\rho$, as seen in Fig. 6.28.

It is well-known in classical cosmology that such a recollapse follows as a simple consequence of the closed topology of 3-space. In the condensate cosmology framework instead, the topology of space(time) is not fixed at the outset, but should rather be reconstructed from the behaviour of the system in the macroscopic limit. In other words, the simple condensate ansatz used here does not provide any information about the topology of spatial sections of the emergent spacetime which, as it is well-known, play an important role in the dynamics of classical cosmological models. Any topological information must therefore come from additional input. A possible strategy one could follow is to work with generalised condensates encoding such information [449]. Here, instead, we propose that the closedness of the reconstructed space need not be encoded in the condensate ansatz as an input, but is rather determined by the dynamics as a consequence of the GFT interactions. Hence, allowing only interactions that are compatible with reproducing a given spatial topology, one may recover the classical correspondence between closed spatial topology and having a finitely expanding universe.

### 6.3.5.2 Geometric inflation

Cosmology obtained from GFT displays a number of interesting features concerning the initial stage of the evolution of the universe, which mark a drastic departure from the standard Friedmann-Lemaître-Robertson-Walker (FLRW) cosmologies. In particular, the initial big-bang singularity is replaced by a regular bounce (see Refs. [486, 487]), followed by an era of accelerated expansion. Similar results were also obtained in the early LQC literature, see e.g. Refs. [538], [539]. However, it is not obvious a priori that they must hold for GFT as well. Nonetheless, it is remarkable that these two different approaches qualitatively yield similar results for the dynamics of the universe near the classical singularity, even though this fact by itself does not necessarily point at a deeper connection between the two.

In the model considered in this Section, our results have a purely quantum geometric origin and do not rely on the assumption of a specific potential for the minimally coupled scalar field [27].[36], which is taken to be massless and introduced for the sole purpose of having a relational clock. This is quite unlike inflation, which instead heavily relies on the choice of the potential and initial conditions for the inflaton in order to predict an era of accelerated expansion with the desired properties.

---

[36]This is also the case in LQC, see Ref. [539]. However, the number of e-folds computed in that framework turns out to be too small in order to supplant inflation [540].

In this Section we investigate under which conditions on the interaction potential of the GFT model it is possible to obtain an epoch of accelerated expansion that could last long enough, so as to account for the minimum number of e-folds required by standard arguments. The number of e-folds is given by

$$N = \frac{1}{3} \log \left( \frac{V_{\text{end}}}{V_{\text{bounce}}} \right), \tag{6.210}$$

where $V_{\text{bounce}}$ is the volume of the universe at the bounce and $V_{\text{end}}$ is its value at the end of the era of accelerated expansion. A necessary condition for it to be called an inflationary era is that the number of e-folds must be large enough, namely $N \gtrsim 60$.

Using $V = V_j \rho^2$, we rewrite Eq. (6.210) as

$$N = \frac{2}{3} \log \left( \frac{\rho_{\text{end}}}{\rho_{\text{bounce}}} \right), \tag{6.211}$$

with an obvious understanding of the notation. This formula is particularly useful since it allows us to derive the number of e-folds only by looking at the dynamics of $\rho$.

Since there is no notion of proper time in GFT, a sensible definition of acceleration can only be given in relational terms. In particular, we seek a definition that agrees with the standard one given in ordinary cosmology via the Raychaudhuri equation

$$\frac{\ddot{a}}{a} = \frac{1}{3} \left[ \frac{\ddot{V}}{V} - \frac{2}{3} \left( \frac{\dot{V}}{V} \right) \right]. \tag{6.212}$$

Since the momentum conjugate to the scalar field $\phi$ is given by $\pi_\phi = V \dot{\phi}$, together with $\dot{V} = \partial_\phi V \left( \frac{\pi_\phi}{V} \right)$, the Raychaudhuri equation rewrites as

$$\frac{\ddot{a}}{a} = \frac{1}{3} \left( \frac{\pi_\phi}{V} \right)^2 \left[ \frac{\partial_\phi^2 V}{V} - \frac{5}{3} \left( \frac{\partial_\phi V}{V} \right)^2 \right] \equiv \frac{1}{3} \left( \frac{\pi_\phi}{V} \right)^2 \mathfrak{a}(\rho). \tag{6.213}$$

We therefore define the acceleration as

$$\mathfrak{a}(\rho) \equiv \frac{\partial_\phi^2 V}{V} - \frac{5}{3}\left(\frac{\partial_\phi V}{V}\right)^2 .[37] \tag{6.214}$$

Hence, from Eqs. (6.198), (6.199) one gets the following expression for the acceleration $\mathfrak{a}$ as a function of $\rho$ for a generic potential

$$\mathfrak{a}(\rho) = -\frac{2}{\rho^2}\left\{\frac{(\partial_\phi U(\rho))\rho}{\rho'} + \frac{14}{3}\left[E - U(\rho)\right]\right\} . \tag{6.215}$$

Using Eq. (6.208) one finally has for our model

$$\mathfrak{a}(\rho) = -\frac{2}{\rho^2}\left[\frac{14}{3}E + \left(1 - \frac{14}{3n'}\right)\mu\rho^{n'} + \frac{4m^2\rho^2}{3} + \left(1 - \frac{14}{3n}\right)\lambda\rho^n - \frac{10Q^2}{3\rho^2}\right] . \tag{6.216}$$

Therefore, the sign of the acceleration is opposite to that of the polynomial

$$s(\rho) = P(\rho) + \left(3 - \frac{14}{n}\right)\lambda\rho^{n+2} + \left(3 - \frac{14}{n'}\right)\mu\rho^{n'+2}, \tag{6.217}$$

where we defined

$$P(\rho) = 4m^2\rho^4 + 14E\rho^2 - 10Q^2 . \tag{6.218}$$

In the following we will in detail study the properties of the era of accelerated expansion. The free case will be discussed in Section 6.3.5.2, whereas the role of interactions in allowing for an inflationary-like era will be the subject of Section 6.3.5.2.

### (A) The free case

In this case the acceleration is given by

$$\mathfrak{a}(\rho) = -\frac{2}{3\rho^4}P(\rho). \tag{6.219}$$

The bounce occurs when $\rho$ reaches its minimum value, i.e. when $U(\rho) = E$, leading to

$$\rho_{\text{bounce}}^2 = \frac{1}{m^2}\left(\sqrt{E^2 + m^2Q^2} - E\right). \tag{6.220}$$

---

[37]Note again, as in Section 6.2, that it is currently not possible to give an intrinsic derivation for the acceleration from within GFT condensate cosmology due to the lack of of more sophisticated observables.

A straightforward calculation shows that $\mathfrak{a}(\rho_{\text{bounce}}) > 0$ as expected. The era of accelerated expansion ends when $P(\rho)$ vanishes, which happens at a point $\rho_\star > \rho_{\text{bounce}}$, which is given by

$$\rho_\star = \frac{1}{4m^2} \left( \sqrt{49E^2 + 40m^2Q^2} - 7E \right).$$  (6.221)

We can then use Eqs. (6.211), (6.220), (6.221) to determine the conserved quantity $E$ as a function of the number of e-folds $N$. Reality of $E$ thus leads to the following bounds on $N$

$$\frac{1}{3} \log \left( \frac{10}{7} \right) \leq N \leq \frac{1}{3} \log \left( \frac{7}{4} \right),$$  (6.222)

that is

$$0.119 \lesssim N \lesssim 0.186.$$  (6.223)

Such tight bounds, holding for all values of the parameters $m^2$ and $Q^2$, rule out the free case as a candidate to replace the standard inflationary scenario in cosmology.

## (B) The interacting case

In the following we investigate the consequences of interactions for the evolution of the universe. In particular, we show how the interplay between the two interaction terms in the effective potential (Eq. 6.200) makes it possible to have an early epoch of accelerated expansion, which lasts as long as in inflationary models. Before studying their effect, we want to discuss how the occurrence of such interaction terms could be motivated from the GFT perspective. In principle, one could have infinitely many interaction terms given by some power of the GFT field. However, only a finite number of them will be of relevance at a specific scale, as dictated by the behaviour of the fundamental theory under the RG flow.

In a continuum and large scale limit new terms in the action could be generated, whereas others might become irrelevant. In this sense, one might speculate that, e.g., in addition to the five-valent simplicial interaction term the effective potential includes another term which becomes relevant on a larger scale. Ultimately, rigorous RG arguments will of course have the decisive word regarding the possibility to obtain such terms from the fundamental theory. Nevertheless, by studying the phenomenological features of such potentials and extracting physical consequences from the corresponding cosmological solutions, we aim at clarifying the map between the fundamental microscopic and effective macroscopic dynamics of the theory. At the same time, our results might help to shed some light onto the subtle issue of the physical meaning of such interaction terms.

Hereafter we assume the hierarchy $\mu \ll |\lambda|$, since otherwise an inflationary era cannot be easily accommodated. This means that the higher order term in the interaction potential $V_{\text{eff}}[\sigma]$ becomes relevant only for very large values of the condensate field $\sigma$, hence of the number of quanta representing the basic building blocks of quantum spacetime. Consequently, the dynamics in the immediate vicinity of the bounce is governed by the parameters of the free theory and the sub-leading interaction term.

To begin with, let us start by fixing the value of the conserved quantity $E$. We require the universe to have a Planckian volume at the bounce. Since the volume is given by $V = V_j \rho^2$, this is done by imposing $\rho_{\text{bounce}} = 1$. Such a condition also fixes the value of $E$ to

$$E = U \left( \rho_{\text{bounce}} = 1 \right). \tag{6.224}$$

In fact, we demand that $\rho_{\text{bounce}}$ is the minimal value of $\rho$ which is compatible with the the conserved quantity $E$ available to the system. Hence, we also have the condition

$$\partial_\rho U \left( \rho_{\text{bounce}} = 1 \right) \leq 0. \tag{6.225}$$

Notice that this is trivially satisfied in the free case. In the interacting case (holding the hierarchy $\mu \ll |\lambda|$) one can therefore use it to obtain a bound on $\lambda$

$$\lambda \leq m^2 + Q^2. \tag{6.226}$$

It is convenient for our purposes and in order to carry over our analysis in full generality, to introduce the definitions

$$\alpha \equiv \left( 3 - \frac{14}{n} \right) \lambda, \tag{6.227}$$

$$\beta \equiv \left( 3 - \frac{14}{n'} \right) \mu. \tag{6.228}$$

The acceleration Eq. (6.216) can thus be written as

$$\mathfrak{a}(\rho) = -\frac{2}{\rho^4} \left[ P(\rho) + \alpha \rho^{n+2} + \beta \rho^{n'+2} \right]. \tag{6.229}$$

As pointed out before, $\mathfrak{a} > 0$ is expected to hold at the bounce. The first thing to be observed is that $\alpha < 0$ is a necessary condition in order to have enough e-folds. In fact, if this were not the case, the bracket in Eq. (6.229) would have a zero at a point $\rho_{\text{end}} < \rho_{\star}$ (cf. Eq. (6.221)), thus leading to a number of e-folds which is even smaller than the corresponding one in the free case. Furthermore, it is possible to constrain the value of $\mu$ in a way that leads both to the aforementioned hierarchy and to the right value for $N$, which we consider as fixed at the outset. In order to do so, we solve Eq. (6.211) w.r.t. $\rho_{\text{end}}$, having fixed the bounce at $\rho_{\text{bounce}} = 1$

$$\rho_{\text{end}} = \rho_{\text{bounce}} \, \mathrm{e}^{\frac{3}{2}N}. \tag{6.230}$$

The end of inflation occurs when the polynomial in the bracket in Eq. (6.229) has a zero. Since $\rho_{\text{end}} \gg 1$, it is legitimate to determine this zero by taking only into account the two highest powers in the polynomial, with respect to which all of the other terms are negligible. We therefore have

$$\alpha \rho_{\text{end}}^{n+2} + \beta \rho_{\text{end}}^{n'+2} \approx 0, \tag{6.231}$$

which, using Eq. (6.230), leads to

$$\beta = -\alpha \mathrm{e}^{-\frac{3}{2}N(n'-n)}. \tag{6.232}$$

The last equation is consistent with the hierarchy $\mu \ll |\lambda|$ and actually fixes the value of $\mu$ once $\lambda$, $n$, $n'$ and $N$ are assigned. Furthermore, one has $\beta > 0$ which, together with Eqs. (6.204), (6.228), implies $n' > \frac{14}{3}$. Importantly, this means that $n' = 5$ is the lowest possible integer compatible with an inflationary-like era. This particular value is also interesting in another respect since in GFT typically only specific combinatorially non-local interactions minimally of such a power allow for an interpretation in terms of simplicial quantum gravity [36, 38, 433–436, 447].

Our considerations so far leave open two possibilities, viz.:

- $\lambda < 0$ and $n \geq 5$ ($n' > n$), which in the case of $n = 5$ could correspond to the just mentioned simplicial interaction term and the higher order $n'$-term could possibly be generated in the continuum and large scale limit of the theory and becomes dominant for very large $\rho$. For even $n'$ it mimics so-called tensorial interactions.

- $\lambda > 0$ and $2 < n < 5$ ($n' \geq 5$), which for $n' = 5$ could allow a connection to simplicial quantum gravity and would remain dominant for large $\rho$ over the $n$-term, which in the case $n = 4$ is reminiscent of an interaction of tensorial type.

However, this is not yet enough in order to guarantee an inflation-like era. In fact we have to make sure that there is no intermediate stage of deceleration occurring between the bounce at $\rho_b = 1$ and $\rho_{\text{end}}$, i.e., that $\mathfrak{a}(\rho)$ stays positive in the interval between these two points. In other words we want to make sure that $\rho_{\text{end}}$ is the only zero of the acceleration lying to the right of $\rho_b$. In fact $\mathfrak{a}(\rho)$ starts positive at the bounce and has a minimum when $P(\rho)$ becomes of the same order of magnitude of the term containing the power $\rho^{n+2}$ (see Eq. (6.229)). Thus we see that we have to require that the local minimum of $\mathfrak{a}(\rho)$ (i.e. the maximum of the poynomial in brackets in Eq. (6.229)) is positive (resp. negative). As $\rho$ increases further, the acceleration increases again until it reaches a maximum when the contribution coming from the term containing $\rho^{n'+2}$ becomes of the same order of magnitude of the other terms. Thereafter the acceleration turns into a decreasing function all the way until $\rho \to +\infty$ and therefore has a unique zero. Positivity of the local minimum of $\mathfrak{a}(\rho)$ translates into a further constraint on parameter space. By direct inspection, it is possible to see that the latter case listed above does not satisfy such condition for any value of the parameters of the model. Therefore we conclude that $\lambda$ must be negative if the acceleration is to keep the same sign throughout the inflationary era. The evolution of the acceleration as a function of relational time $\phi$ is shown in Fig. 6.29 for some specific choice of the parameters. It is worthwhile stressing that the behaviour of the model in the case $\lambda < 0$ is nevertheless generic and therefore does not rely on the specific choice of parameters. Furthermore, by adjusting the value of $N$ and the other parameters in Eq. (6.232), it is possible to achieve any desirable value of e-folds during inflation.

All we said in this Section applies to the model with effective potential Eq. (6.200) but does not hold in a model with only one interaction term. In fact in that case it is not possible to prevent the occurrence of an intermediate era of deceleration between $\rho_b$ and $\rho_{\text{end}}$, the latter giving the scale at which the higher order interaction term becomes relevant.

One last remark is in order: inflationary expansion was shown to be a feature of particular GFT models but only at the price of a fine-tuning in the value of the parameter $\mu$ (see Eq. (6.232)).

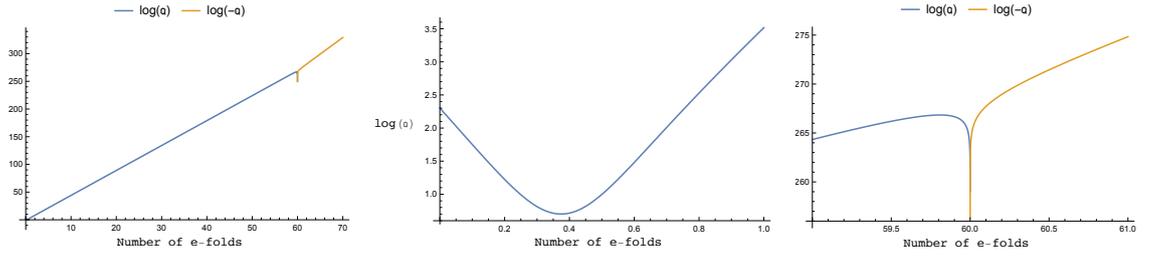

FIGURE 6.29: Inflationary era supported by GFT interactions in the model with two interaction terms. The blue (orange) curve represents the graph of the logarithm of the acceleration (minus the acceleration) as a function of the number of e-folds in the case $\lambda < 0$. The plot refers to the particular choice of parameters $n = 5$, $n' = 6$, $m = 1$, $Q = 1$, $\lambda = -3$. The value of $\mu$ is determined from Eq. (6.232) by requiring the number of e-folds to be $N = 60$. There is a logarithmic singularity at $N \simeq 60$, marking the end of the accelerated expansion. The figures to the center and right show the behaviour of the acceleration close to the bounce and at the end of inflation, respectively.

### 6.3.5.3 Interactions and the final fate of the universe

As we have seen above, it is possible to recast the dynamical equations for the volume of the universe in a form that bears a closer resemblance to the standard Friedmann equation, as shown in Ref. [486]. In fact, the Hubble expansion rate can be expressed as

$$H = \frac{1}{3} \frac{\partial_\phi V}{V^2} \pi_\phi,$$ (6.233)

where $\pi_\phi$ denotes the momentum conjugate to the scalar field $\phi$ defined by $\pi_\phi = V \dot{\phi}$. From Eq. (6.198) and the proportionality between the momentum of the scalar field and $Q$ we have

$$H^2 = \frac{4}{9} \frac{\hbar^2 Q^2}{V^2} \left( \frac{\partial_\phi \rho}{\rho} \right)^2.$$ (6.234)

The term in bracket can thus be interpreted as a dynamical effective gravitational constant, as in Ref. [536]. Alternatively, using Eqs. (6.209) and (6.208), the last equation Eq. (6.234) becomes (considering the case with only one interaction term, namely $\lambda = 0$)

$$H^2 = \frac{8\hbar^2 Q^2}{9} \left[ \frac{\varepsilon_m}{V^2} + \frac{\varepsilon_E}{V^3} + \frac{\varepsilon_Q}{V^4} + \frac{\varepsilon_\mu}{V^{3-n'/2}} \right],$$ (6.235)

where we defined

$$\varepsilon_E = V_j E, \qquad (6.236)$$

$$\varepsilon_m = \frac{m^2}{2}, \qquad (6.237)$$

$$\varepsilon_Q = -\frac{Q^2}{2} V_j^2, \qquad (6.238)$$

$$\varepsilon_\mu = -\frac{\mu}{n'} V_j^{1-n'/2} \qquad (6.239)$$

and $V_j$ is retrieved via $V = V_j \rho^2$. The exponents of the denominators in Eq. (6.235) can be related to the $w$ coefficients in the equation of state $p = w\varepsilon$ of some effective fluids, with energy density $\varepsilon$ and pressure $p$. Each term scales with the volume as $\propto V^{-(w+1)}$.

It is worth pointing out that Eq. (6.235) makes clearer the correspondence with the framework of ekpyrotic models[38], where one has the gravitational field coupled to matter fields with $w > 1$. Such models have been advocated as a possible alternative to inflation, see Ref. [541]. We observe that at early times (i.e. small volumes) the occurrence of the bounce is determined by the negative sign of $\varepsilon_Q$, which is also the term corresponding to the highest $w$. However, while this is sufficient to prevent the classical singularity, it is not enough to guarantee that the minimum number of e-folds is reached at the end of the accelerated expansion. In fact, the role of interactions is crucial in that respect, as our analysis in Section 6.3.5.2 has shown.

In the rest of this Section we focus instead on the consequences of having interactions in the GFT model for the evolution of the universe at late times. As we have already seen in Section 6.3.5.1, a positive $\mu$ entails a recollapsing universe. This should also be clear from Eq. (6.235). In particular, we notice that the corresponding term in the equation is an increasing function of the volume for $n' > 6$. This is quite an unusual feature for a cosmological model, where all energy components (with the exception of the cosmological constant) are diluted by the expansion of the universe. For $n' = 6$ one finds instead a cosmological constant term. It is also possible to have the interactions reproduce the classical curvature term $\propto \frac{\kappa}{V^{2/3}}$ by choosing $n' = \frac{14}{3}$, which is however not allowed if one restricts to integer powers in the interactions [36, 38, 433–436].

Our analysis shows that only $\lambda < 0$ leaves room for an era accelerated expansion analogous to that of inflationary models. In order for this to be possible, one must also have $n' > n \geq 5$. Moreover, if one rules out phantom energy (i.e. $w < -1$), there is only one case which is allowed, namely $n = 5$, $n' = 6$. Then during inflation the universe can be described as dominated by a fluid with equation of state $w = -\frac{1}{2}$. After the end of inflation its energy content also receives contribution from a negative cosmological constant, which eventually leads to a recollapse. It is remarkable that this particular case selects an interaction term which is in principle compatible with the simplicial interactions which have been extensively considered in the GFT approach. However, it must be pointed out that the realisation of the geometric inflation picture imposes strong restrictions also on the type of interactions one can consider, as well as on their relative strength. It also comes at the price of a resulting negative cosmological constant term, which seems empirically unfavourable.

---

[38]We are thankful to Martin Bojowald for this observation.

### 6.3.6 Discussion of the results of model 2

In this Section, we have developed further the application of the generalised version of the EPRL GFT model to quantum cosmology, as first done in Ref. [486], focusing on the dynamics of anisotropies close to a bouncing region of cosmological evolution and the effect of effective interactions onto the acceleration behaviour of the isotropic background for late relational times.

We reviewed the construction of the model in view of its cosmological application with particular regard to the left invariance and monochromaticity conditions imposed onto the mean field. For completeness, we presented the derivation of the effective Friedmann equation, as given in Ref. [486], which describes the evolution of the background. The most significant consequence of this equation is the resolution of the initial singularity of standard cosmology.

In the first part of this Section, we derived the effective dynamical equations for non-monochromatic perturbations of an isotropic (monochromatic) background, remaining at the mean field level. While a detailed characterisation of the anisotropic degrees of freedom in terms of GFT observables with a clear macroscopic, cosmological interpretation is lacking at present, we know that such anisotropic degrees of freedom are in fact encoded in the type of non-monochromatic perturbations we have analysed. Missing a clear geometric interpretation, however, we confined our analysis to a study of the non-monochromatic amplitudes, investigating in which region of parameter space they remain subdominant compared to the isotropic background. Specifically, we derived the evolution equations for such perturbations to first order in perturbation theory, in the general case, and then solved them exactly in a simpler case corresponding to a specific choice of EPRL-like GFT model, and for a simple background condensate in which a single spin component is excited and takes a fixed value. We have focused on the approximate regime of these equations corresponding to a cosmic bounce in the evolution of the background, replacing the initial big bang singularity. This is the most interesting regime from the cosmological point of view, where control over anisotropies is critical, but also the regime of GFT mean field dynamics that is technically simpler to study, since in this regime GFT interactions are expected to be subdominant compared to the free dynamics. Then, we determined different regions in the parameter space of the model where perturbations exhibit interesting behaviour; more precisely, we have identified in which region of parameter space, non-monochromatic perturbations decay rapidly away from the bounce, as the universe expands, even if significantly close to the bounce. Furthermore, for suitable values of the initial conditions and of the interaction strength, perturbations can become negligible before the interactions kick in. Hence, it is sufficient to consider the monochromatic background in the non-linear regime. Finally, we confirmed the behaviour of such perturbations by a quantitative study of some simple GFT observables: the surface-area-to-volume ratio and the effective 1-body volume. Although the relation between such quantities and physical observables with a cosmological interpretation is not clear, they are used in order to illustrate the departures from the case of an isotropic background of monochromatic tetrahedra, previously studied in the literature. Our analysis, therefore, strengthens the findings of Refs. [486, 487] that after a bouncing phase, where the quantum geometry can be rather degenerate, a cosmological background emerges the dynamics of which can be cast into the form an effective Friedmann equation.

It should be clear that this analysis, together with the results presented in Section 6.2.6.1, is only a first step towards a more comprehensive study of cosmological anisotropies in the emergent cosmological dynamics of GFT condensates. An immediate extension of our work would be to study non-monochromatic perturbations over a different, still isotropic condensate state, and still in a mean field approximation to confirm the general expectation that the ambiguities in associating a continuum geometry to GFT condensate wavefunctions do not drastically affect the effective cosmological dynamics. More importantly, we need to develop a precise characterisation of the anisotropic degrees of freedom encoded in non-monochromatic perturbations, in order to be able to describe their dynamics in more explicit, geometric terms. For this, it is necessary to identify suitable observables of clear geometric meaning of measuring cosmological anisotropies, i.e. gauge invariant combinations of the scale factors. In principle, simple condensate states like the ones we have used here are rich enough to capture such observables, at least at the kinematical level, since the domain of definition of the condensate wavefunction is isomorphic to the minisuperspace of generic anisotropic geometries. However, the construction of such suitable observables is far from trivial and has not been carried out so far; it could be that this construction is more naturally carried out by exploiting more involved condensate (or other many-body) states in the GFT Hilbert space, because it may be needed (or at least useful) to rely on the connectivity information present in generic states and absent in the simple coherent states we used. In any case, once a good definition of anisotropic observables is achieved, the effective dynamics of non-monochromatic perturbations should be translated into an effective dynamics for anisotropic geometries, and compared with those expected from classical GR, i.e. Bianchi models. This will be a crucial test of this approach, and at the same time a direction in which it could bring even more interesting fruits.

In the second part of this Section, we investigated the phenomenological consequences of simplified interactions for the dynamics of background. To solely consider their impact onto the background is in part motivated by the finding that condensate configurations quickly settle into a low-spin phase [457, 488, 490] and that anisotropies fade away quickly (in a region of parameter space) as shown just before.

Based on this, we considered an effective potential including two interaction terms besides the quadratic one, the latter being already present in the free theory. An ambiguity in the kinetic term, represented by the factor $A$, is fixed by requiring the expansion of the universe not to be faster than exponential at large volumes. A general prediction of the model is the occurrence of a recollapse when the higher order interaction term becomes codominant. Results that have already been obtained in the free theory [486] survive in the interacting case, in particular for what concerns the occurrence of a bounce. The former result, together with the recollapse induced by interactions, leads to cyclic cosmologies. A more detailed analysis of the latter, instead, leads to the conclusion that, in the free case, the era of accelerated expansion does not last for a number of e-folds which is at least as large as in inflationary models. This is instead made possible when suitable interaction terms are taken into account, as considered here. Indeed, we showed that one can attain an arbitrary number of e-folds as the universe accelerates after the bounce. Furthermore, having an inflationary-like expansion imposes a restriction on the class of viable models. In fact, this is possible only for $\lambda < 0$ and $5 \leq n < n'$ and when one has the hierarchy $\mu \ll |\lambda|$. Reasonable phenomenological arguments lead to select only the case $n = 5$, $n' = 6$ as physical. The two powers can be related, respectively, to simplicial interactions, commonly considered in the GFT framework, and to a negative cosmological constant. While the result is encouraging as a first step towards a quantum geometric description of the inflationary era, a few remarks are in order. In fact, it must be pointed out that it comes at the price of a fine tuning in the coupling constant of the higher order interaction term. From this point of view, it shares one of the major difficulties of ordinary inflationary models. Furthermore, an inflationary-like era does not seem to be a generic property of GFT models, but in fact requires interactions of a suitable form.

Future work must be devoted to studying the implications of interactions respecting the proper combinatorial structure which characterises GFTs with a geometric interpretation for the geometry of the emergent spacetime and its dynamics. As noticed above, this further step would also be required to properly take into account the effects of anisotropies.

To further corroborate the condensate cosmology approach, other degrees of freedom and inhomogeneities have to be incorporated in the description of the effective dynamics of the emergent spacetime. Their phenomenological signatures, in particular for what concerns the seeds for the growth of structures, are crucial in order to be able to give a definite answer to the problem of finding a valid alternative to the inflationary paradigm, which might come from quantum geometry. It is of course needless to say that going beyond the simple mean field approximation of the full quantum GFT dynamics, and to study the corresponding quantum-improved effective cosmological dynamics is an important and interesting task in itself.

It is the hope that the interplay between determining phenomenological constraints, as those obtained in this Chapter, and a fundamental approach involving, e.g. FRG arguments, might help to single out the correct microscopic theory (or even a family of such theories). Further work must come from both directions in a common effort to develop an appropriate framework for studying early universe cosmology, which correctly takes into account the quantum dynamics of all the relevant degrees of freedom of the gravitational field.

# Chapter 7

# Conclusion

<div style="text-align: right">

Dèyè mòn, gen mòn.

Behind mountains, more mountains.

—————————————————————

Haitian proverb.

</div>

In this thesis we studied the consequences of quantum gravitational effects for black hole physics and cosmology. This was done using the two related non-perturbative and background independent frameworks of LQG and GFT. More specifically, in the first thematic unit of this work (Chapters 1 and 2) we started by expounding the topic of the canonical quantisation of GR and then applied LQG to investigate details of the statistics of the black hole horizon quantum geometry. In the second part (Chapters 3, 4 and 5) we changed gears and motivated GFT via its relations to (and origins in) other approaches to the covariant quantisation of GR, then introduced the foundations of its condensate cosmology spin-off and employed it to various models and scenarios with particular regard to testing the condensate hypothesis and to studying the implications for the effective dynamics of the emergent cosmological geometries.

In the first chapter, we reviewed the reformulation of GR à la ADM and discussed essential features and problems of its canonical quantisation. This sets the incentive for the discussion of the second topic of this chapter: the construction of LQG at the classical level via the Holst action and aspects as well as peculiarities of its canonical quantisation which is most relevant for the ensuing chapters.

In the second chapter, we applied LQG to the horizon geometry defined via the isolated horizon boundary condition. We sidestepped the original formulation of the classical theory and focussed on its effective description where the horizon degrees of freedom are identified as punctures coupled to an SU(2) Chern-Simons theory living on a topological 2-sphere. We did so to have clearer access to the topological intricacies of the phase space needed later on. We then reviewed aspects of the quantisation of this configuration and followed the standard exposition of the entropy computation which leads to the area law. We then moved a step backwards and more thoroughly examined with tools from symplectic geometry the global obstructions (caused by the punctures) for symplectic vector fields on the isolated horizon to be Hamiltonian. Upon quantisation, we showed that this kinematical ambiguity leads to non-Abelian phases which give rise to non-Abelian anyonic statistics. In this way, our work represents a reinterpretation of the common reading of the statistics of these degrees of freedom. Such phases are unitary irreducible representations of the permutation group in $2d$, known as the braid group. Given the fact that this group is equivalent to the group of large diffeomorphisms, we clearly established a relation between this boundary symmetry group and the statistics of the model. We emphasised that this is a general result important also for other types of boundary surfaces obeying different boundary conditions. Finally, we showed that the imprint of the anyonic statistics would in principle be measurable for quasi-local observers in the vicinity of the horizon.

Future research could try to relate the standard statistical entropy computation where the boundary symmetry group is given by that of large diffeomorphisms explored here to recent investigations on entanglement entropy between local subsystems in gravity. There so-called edge states/soft modes appear as would-be gauge degrees of freedom on the separating boundary. These are needed for the (re-)construction of the full Hilbert space from the Hilbert spaces associated with the subsystems by means of an entangling/fusion product and encode entanglement between the latter. More precisely, the entangling product is a generalisation of the notion of the tensor product where states are restricted to be singlets under the boundary symmetry group. In its full generality, this philosophy has so far not been applied to the LQG black hole context and we may speculate that this could actually lead to the identification of both entropy notions therein, as is often suspected.

In the third chapter, we commenced with a survey of approaches to quantise gravity covariantly through the path integral. We reviewed aspects of the continuum path integral formulation and motivated through its flaws discrete attempts at its construction. We listed different discrete approaches (simplicial gravity, matrix/tensor models and spin foam theory) and explored to varying degree how macroscopic continuum geometries are recovered in these. This listing and exploration was done to motivate the GFT approach to quantum gravity in this chapter and its condensate cosmology offspring in the next one.

In the fourth chapter, we introduced the foundations of the GFT condensate cosmology framework which relies on the conjecture that a condensate phase of a suitable geometric GFT model correponds to a macroscopic continuum geometry. Based on this, the approach aims at deriving the effective dynamics for GFT condensate states directly from the microscopic GFT quantum dynamics (through approximations) and subsequently to extract a cosmological interpretation from them. It should be emphasised that the Hilbert/Fock space structure to formulate such states and the applicability of field theory methods successful in the context of real Bose-Einstein condensate systems is the major advantage of the condensate approach to extract continuum physics as compared to the other discrete approaches surveyed in the previous chapter. We elaborated on aspects of the above-mentioned conjecture by investigating solutions to the classical equations of motion of the dynamical Boulatov model for Euclidean quantum gravity in $3d$. We found under restrictions a sector of fully non-locally interacting solutions which minimise the classical action and can give rise to a large occupation number when the GFT field is interpreted as condensate field. It could be interesting to further analyse the properties of our solutions in the TFT limit where the occupation number blows up and to understand if this is potentially related to the triangulation invariance of the Ponzano-Regge model.

We then tested for various GFT models (with and without a geometric interpretation) whether a phase transition leading to a condensate can occur by means of Landau's mean field theory. For the Boulatov model we showed that mean field techniques cannot give a conclusive answer to this question and suggested that non-perturbative methods should be employed instead. Importantly, for a rank-1 toy model on $\mathrm{SL}(2, \mathbb{R})$ with local interaction we demonstrated that a phase transition can indeed take place. This might be taken as a hint that the non-compactness of the field domain in Lorentzian models will play a crucial role in establishing a condensate phase through a phase transition for a realistic GFT model for $4d$ spacetime.

In the fifth chapter, we studied two condensate models of rank-4 GFTs and their emergent geometries. The first model is based on a real-valued field and the second on a complex-valued field. These two different choices give rise to a different phenomenology: The former does not exhibit bouncing solutions (in the same way) as compared to the latter. Despite these subtle differences we may try to conclude on the results for both models in an integrated manner:

We analysed free and effectively interacting static and dynamic condensate configurations. For these we obtained non-vanishing condensate populations for which the expectation values of the volume and area operators imported from LQG are dominated by the lowest non-trivial configurations of the quantum geometry. The relative uncertainty of the geometric operators is shown to vanish quickly under evolution indicating the classicalisation of the quantum geometry. This analysis indicates that GFT condensates may consist of many smallest building blocks which may give rise to an effectively continuous emergent geometry.

Moreover, we explored the cosmological implications of effective interactions between the quantum geometric constituents of the condensate from a phenomenological perspective. We showed how such interactions can lead to a recollapse or infinite expansion of the emergent universe while preserving the bounce. It was then demonstrated how these interactions can lead to an early epoch of accelerated expansion purely of geometric origin, which can be fine-tuned to last for an arbitrarily large number of e-folds. (It should be highlighted and contrasted that the study of singularity resolution and accelerated expansion in the other discrete path integral approaches seems to be completely uncharted territory!)

Bouncing cosmologies are typically plagued by an uncontrolled growth of anisotropic stress in the contracting phase, which is the so-called Belinskii-Khalatnikov-Lifshitz (BKL) instability. On the other hand, any (quantum gravity inspired) model exhibiting a cosmological bounce has to be in agreement with the observed isotropy of our Universe at late times. Given this context, we commenced the investigation of anisotropies in the GFT condensate approach. In particular, we presented a simple mechanism by which an anisotropic condensate quickly isotropises under dynamical evolution. Subsequently, we examined the behaviour of perturbations of a perfectly isotropic background corresponding to microscopic anisotropies. Remarkably, it was found for a specific region of parameter space that these are under control at the bounce, become negligible away from it and are unimportant when interactions kick in, thus strengthening the findings of purely isotropic models. This work also represents a crucial step towards identifying anisotropic cosmologies, i.e. Bianchi models, within this approach.

It is the hope of the author that the research which led to this part of this manuscript will help to consolidate this promising approach to quantum cosmology and in particular to the extraction of continuum information from a discrete geometric setting. We believe that the field theoretic setting of GFT and specifically the use of field coherent states proves extremely useful and elegant to this aim, as compared to the techniques employed in EDT, CDT, tensor and spin foam models. Future work should aim at closing the main conceptual gaps of this approach and at rendering it more realistic. We list the (in our eyes) most important points in the following:

(1) The EPRL GFT model (or any related model) for $4d$ Lorentzian quantum gravity has to be spelled out in all its details. With this we want to point out that, so far, the GFT interaction term has not been put down as a function of its boundary data in an explicit manner. This is a serious limitation. Progress on the impact of realistic and not only phenomenologically motivated interactions can only then be accomplished. Furthermore, GFT interactions encode connectivity information between the building blocks of any extended quantum geometry, which at late times of the evolution cannot be neglected. Apart from its impact on the dynamics, this information will also be relevant to the construction of more involved observables, in particular in view of those capturing cosmological anisotropies and curvature. (In turn these will prove indispensable to classify and identify different emergent geometries from one another.) For such a model the condensate hypothesis then has to be probed to understand if it can truely exhibit a phase or phases which are related to $(3 + 1)$-dimensional Lorentzian continuum geometries. (In addition, notice that the models considered in GFT condensate cosmology are uncoloured. It can be expected that only coloured versions of these will lead to extended geometries free of pathologies!)

(2) At the current stage of research, it is not clear if higher order corrections to the so-far considered condensate equation of motion can be neglected or if they would have a drastic impact on the cosmological interpretation of this approach. At the same time, up to now only the dynamics of simple condensate states have been studied. It is questionable, if such a simple ansatz can be upheld close to the supposed phase transition and/or the bounce. Given that the vanishing of the order parameter in a phase transition (corresponding to a zero-volume state) would challenge the occurrence of a bounce, the clarification of this point is very important.

(3) Since the inception of the condensate approach, only homogeneous and isotropic models have been studied, apart from treating anisotropies (see above) and inhomogeneities as perturbations thereof. Modern Cosmology teaches us that today's cosmic structure can be related to the inhomogeneities of the very early Universe. Hence, the inclusion of other matter degrees of freedom (beyond the relational clock) as well as the identification and study of cosmological inhomogeneities in the condensate approach is mandatory to promote it to a realistic contestant theory of quantum cosmology.

# Epilogue

I.

It was six men of Indostan
To learning much inclined,
Who went to see the Elephant
(Though all of them were blind),
That each by observation
Might satisfy his mind.

II.

The First approached the Elephant,
And happening to fall
Against his broad and sturdy side,
At once began to bawl:
"God bless me!—but the Elephant
Is very like a wall!"

III.

The Second, feeling of the tusk,
Cried: "Ho!—what have we here
So very round and smooth and sharp?
To me 't is mighty clear
This wonder of an Elephant
Is very like a spear!"

IV.

The Third approached the animal,
And happening to take
The squirming trunk within his hands,
Thus boldly up and spake:
"I see," quoth he, the Elephant
Is very like a snake!

V.

The Fourth reached out his eager hand,
And felt about the knee.
"What most this wondrous beast is like
Is mighty plain," quoth he;
"'T is clear enough the Elephant
Is very like a tree!"

VI.

The Fifth, who chanced to touch the ear,
Said: "E'en the blindest man
Can tell what this resembles most;
Deny the fact who can,
This marvel of an Elephant
Is very like a fan!"

VII.

The Sixth no sooner had begun
About the beast to grope,
Than, seizing on the swinging tail
That fell within his scope,
"I see," quoth he, "the Elephant
Is very like a rope!"

VIII.

And so these men of Indostan
Disputed loud and long,
Each in his own opinion
Exceeding stiff and strong,
Though each was partly in the right,
And all were in the wrong!

So, oft in theologic wars
The disputants, I ween,
Rail on in utter ignorance
Of what each other mean,
And prate about an Elephant
Not one of them has seen!

J. G. Saxe,
The Blind Men and the Elephant (1872),
based on an Indian parable.

A poet once said, "The whole Universe is in a glass of wine." We will probably never know in what sense he meant that, for poets do not write to be understood. But it is true that if we look at a glass of wine closely enough we see the entire Universe. There are the things of physics: the twisting liquid which evaporates depending on the wind and weather, the reflections in the glass, and our imagination adds the atoms. The glass is a distillation of the Earth's rocks, and in its composition we see the secrets of the Universe's age, and the evolution of stars. What strange arrays of chemicals are in the wine? How did they come to be? There are the ferments, the enzymes, the substrates, and the products. There in wine is found the great generalisation: all life is fermentation. Nobody can discover the chemistry of wine without discovering, as did Louis Pasteur, the cause of much disease. How vivid is the claret, pressing its existence into the consciousness that watches it! If our small minds, for some convenience, divide this glass of wine, this Universe, into parts — physics, biology, geology, astronomy, psychology, and so on — remember that Nature does not know it! So let us put it all back together, not forgetting ultimately what it is for. Let it give us one more final pleasure: drink it and forget it all!

R. P. Feynman,

The Feynman Lectures on Physics (1964),

Volume I.

# Appendix A

# TQFT axiomatics, braid and mapping class group

## A.1   Atiyah's TQFT axiomatics

We briefly present the axiomatisation of Witten's notion of a topological quantum field theory (TQFT) [105–107] by Atiyah [542] to complement the content of Sections 3.1 and 3.2.

A $(2+1)$-dimensional TQFT $(Z, V)$ over $\mathbb{C}$ firstly consists of the association of a vector space $V(\Sigma)$ over $\mathbb{C}$ to every closed oriented smooth 2-dimensional manifold and secondly consists of the association of an element $Z(M) \in V(\partial M)$ to every compact oriented smooth 3-dimensional manifold $M$. These two associations are subject to the axioms:

1. $(Z, V)$ is functorial with respect to orientation preserving diffeomorphisms of $\Sigma$ and $M$: Let $\phi : \Sigma \to \Sigma'$ be such a diffeomorphism, then one associates to it a linear isomorphism $V(\phi) : V(\Sigma) \to V(\Sigma')$. For a composition of $\phi$ with $\chi : \Sigma' \to \Sigma''$ one has $V(\chi \circ \phi) = V(\chi) \circ V(\phi)$. If $\phi$ extends to an orientation preserving diffeomorphism $M \to M'$ with $\partial M = \Sigma$ and $\partial M' = \Sigma'$, one has $V(\phi)(Z(M)) = Z(M')$.

2. $(Z, V)$ is involutive, i.e. $V(-\Sigma) = V(\Sigma)^*$.

3. $(Z, V)$ is multiplicative.

4. If $\Sigma = \emptyset$ then one requires $V(\emptyset) = \mathbb{C}$ and if $M = \emptyset$ then $Z(\emptyset) = 1$. For generic $\Sigma$, the identity endomorphism of $V(\Sigma)$ reads: $Z(\Sigma \times \mathbb{1}) = \mathrm{id}_V(\Sigma)$ and crucially $\dim(V(\Sigma)) = \mathrm{tr}\, V(\mathrm{id}|_{V(\Sigma)}) = Z(\Sigma \times S^1)$ gives the dimension of the respective TQFT-vector space.

Endowed with additional structure, $V(\Sigma)$ turns into a Hilbert space $\mathcal{H}_\Sigma$.

**Example 1**: Let $M$ be a closed 3-manifold with $\partial M = \emptyset$. Then $Z(M) \in V(\emptyset) = \mathbb{C}$ is a constant and hence the theory produces numerical invariants of 3-manifolds. For the case of Chern-Simons (CS) theory, $Z(M)$ is just equal to Eq. (3.7). $Z_k(M)$ defines a topological invariant of the closed 3-manifold $M$, which is termed as the quantum $G$-invariant of $M$ at level $k$. A natural class of gauge invariant observables of CS-theory not requiring a choice of metric are the Wilson loop operators. Let $L$ be an oriented link embedded in $M = S^3$ with $N$ components $\{C_i\}_{i=1..N}$, each of them labelled with an irreducible representation $\rho_i$ of $G$. The expectation value of a product of Wilson loop operators $W(L) = \prod_{i=1}^{N} \text{tr}_{\rho_i}[P \, e^{i \oint_{C_i} A_i}]$ is

$$Z_k(M, L) = \langle W(L) \rangle = \frac{\int DA \, e^{iS_{\text{CS}}[A]} W(L)}{\int DA \, e^{iS_{\text{CS}}[A]}}. \tag{A.1}$$

Due to general covariance, this is invariant under smooth deformations of the (framed) link $L$. In $SU(2)_k$ CS-theory $\langle W(L) \rangle$ is equal to a corresponding evaluation of the Jones polynomial $J_L(q)$ with $q = e^{i\frac{2\pi}{k+2}}$ and it is a topological invariant of knot theory.

**Example 2**: Let $\partial M = \Sigma \neq \emptyset$, then the axioms assign to the boundary the physical Hilbert space $\mathcal{H}_\Sigma$ and to the 3-manifold $M$ the vector $Z_k(M) \in \mathcal{H}_\Sigma$, representing the time evolution of states.

The axioms imply how to yield representations of mapping class groups $\text{MCG}(\Sigma)$ of closed oriented surfaces $\Sigma$ from a $(2+1)$-dimensional TQFT. Let $\phi_t$ be the isotopy of an orientation preserving diffeomorphism $\Sigma \to \Sigma$, i.e. $\phi$ falls into one particular mapping class $[\phi]$, then

$$V(\phi) = \rho(\phi_t) : \mathcal{H}_\Sigma \to \mathcal{H}_\Sigma \tag{A.2}$$

is homotopically invariant. It is implied that

$$\rho : \text{MCG}(\Sigma) \to \text{End}(\mathcal{H}_\Sigma) \tag{A.3}$$

is a well-defined representation of $\text{MCG}(\Sigma) = \text{Diff}^+(\Sigma)/\text{Diff}_0(\Sigma)$, which acts as a symmetry on $\mathcal{H}_\Sigma$.

The axioms also imply how to obtain the dimension of $\mathcal{H}_\Sigma$ for $M \cong \Sigma \times S^1$. Coupling such a TQFT to a 1-dimensional one, corresponds to puncturing $\Sigma$ at the set of points $\{p_i\}$ by unknotted parallel circles labelled with their respective representations $\{\rho_i\}$. For

$\Sigma = S^2$ the application of the partition function to this configuration gives

$$Z(S^2 \times S^1; \{\rho_i\}) = \dim(\mathcal{H}_{S^2; \{\rho_i\}}) \tag{A.4}$$

and using techniques from CFT [88, 105–107, 140–142, 147–149] for a configuration of distinguishable punctures with occupation numbers $\{n_j\}$ one yields Eq. (3.27) [84–87, 99].

Notice that expression (A.3) also holds for the case of punctured surfaces [105–107, 140–142, 147–149]. The precise form of the mapping class group of the punctured sphere is recovered below.

## A.2 Braid group, symmetric group, pure braid group and their relations

Following Refs. [543–555], facts about the braid group are gathered.

**Definition 3.** The (Artin) braid group $B_N$ on $N$ strands is an infinite group, which has $N-1$ generators $\sigma_i$, with $(1 \leq i \leq N-1)$. The generators obey the following two relations

1. $\sigma_i \sigma_j = \sigma_j \sigma_i$, with $|i-j| \geq 2$,

2. the Yang-Baxter-relation

$$\sigma_i \sigma_{i+1} \sigma_i = \sigma_{i+1} \sigma_i \sigma_{i+1}, \quad i = 1, 2, \ldots, N-2 \ . \tag{A.5}$$

$(\sigma_i)^{-1}$ denotes the inverse and $e$ the identity. The generator $\sigma_i$ corresponds to the braiding of the $i$-th strand with the $i+1$-th strand in an anti-clockwise direction, where no other strands are enclosed. The multiplication of the generators is geometrically understood as a concatenation of braids. Fig. A.1 depicts an elementary braid. Taking the special case

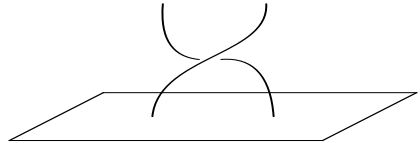

Figure A.1: Graphical representation of a braid.

where $\sigma_i^2 = e$ for $1 \leq i \leq N-1$, the braid group reduces to the permutation group $S_N$,

which is a finite subgroup of $B_N$. For braiding distinguishable strands, the pure braid group is introduced.

**Definition 4.** The pure braid group $PB_N$ is a normal subgroup of $B_N$ and has a presentation (Burau) with the generators

$$\gamma_{i,j} = \sigma_{j-1}\sigma_{j-2}\cdots\sigma_{i+1}\sigma_i^2\sigma_{i+1}^{-1}\cdots\sigma_{j-2}^{-1}\sigma_{j-1}^{-1}, \tag{A.6}$$

with $1 \leq i < j \leq n$ and the following relations

$$\gamma_{r,s}\gamma_{i,j}\gamma_{r,s}^{-1} =$$

$$\begin{cases} \gamma_{i,j} & s < i \text{ or } j < r \\ \gamma_{i,s}^{-1}\gamma_{i,j}\gamma_{i,s} & i < j = r < s \\ \gamma_{i,j}^{-1}\gamma_{i,r}^{-1}\gamma_{i,j}\gamma_{i,r}\gamma_{i,j} & i < r < j = s \\ \gamma_{i,s}^{-1}\gamma_{i,r}^{-1}\gamma_{i,s}\gamma_{i,r}\gamma_{i,j}\gamma_{i,r}^{-1}\gamma_{i,s}^{-1}\gamma_{i,r}\gamma_{i,s} & i < r < j < s. \end{cases} \tag{A.7}$$

The action of the generator $\gamma_{i,j}$ is illustrated in Fig. A.2. For pure braids the endpoints are kept fixed, whereas in $B_N$ they can be permuted. The kernel of the epimorphism $f : B_N \to S_N$ is $PB_N$, which can be compactly written as the short exact sequence

$$\{e\} \to PB_N \to B_N \to S_N \to \{e\}. \tag{A.8}$$

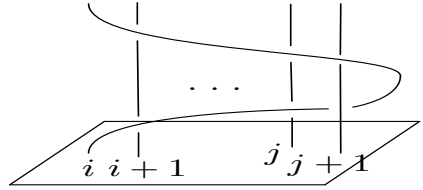

Figure A.2: Graphical representation of a pure braid.

## A.3 Topology of configuration spaces for (in)distinguishable particles

Let the configuration space of one particle be denoted by $\mathcal{F} = X$. For $N$ indistinguishable particles one cannot make a distinction between points in $\mathcal{F}_N = X^N$ differing by the order of the particle coordinates. Let $x = (x_1, \ldots, x_N) \in X^N$ and a different point $x' \in X^N$ with $x' = P(x) = (x_{P^{-1}(1)}, \ldots, x_{P^{-1}(N)})$, where $P \in S_N$. Physically equivalent configurations are thus orbits of points in $X^N$ with respect to $S_N$. The configuration space is $Q_N \equiv X^N/S_N$.

More formally, let $M$ be a connected manifold of dimension $d = 2$ or higher. Let $N$ be a positive integer, denoting the total particle number. Define Faddell's configuration space of a set of N ordered points in $M$ to be

$$\mathcal{F}_N(M) = \{(x_1, \ldots, x_N) \in M \times \cdots \times M | x_i \neq x_j \; for \; i \neq j\}. \tag{A.9}$$

In the physical context the ordered points are distinguishable particles. In contrast,

$$Q_N(M) \equiv \mathcal{F}_N(M)/S_N \tag{A.10}$$

is the configuration space of a set of $N$ unordered points in $M$, representing indistinguishable particles.

A particle exchange by means of an adiabatic transport in $d = 2$ spatial dimensions is different from $d = 3$. In $3d$ paths can be continuously deformed, whereas in $2d$ the topology of the configuration space allows for an oriented winding by an arbitrary number of times around other particles. Mathematically, these properties of the transport paths are captured by the first homotopy group of the configuration space. For indistinguishable particles it is given as:

$$\pi_1\Big(Q_N(M)\Big) \cong S_N \;(d=3); \; B_N(M) \;(d=2). \tag{A.11}$$

There are only two one-dimensional representations of $S_N$, namely the identical ($\sigma_i = 1$) and the alternating one ($\sigma_i = -1$), giving in the corresponding quantum theory rise to bosonic and fermionic statistics. Quantum states for $N$ indistinguishable particles in $2d$ are elements of a Hilbert space which transforms unitarily under representations of $B_N$. If the wave functions are multiplets, one deals with higher-dimensional representations of $B_N$. These depict non-Abelian anyons, giving rise to non-Abelian braiding statistics, introduced in Ref. [135]. The representation

$$\rho : B_N(M) \to U(\mathcal{H}_{M;N}), \tag{A.12}$$

maps into the unitary transformations of the Hilbert space $\mathcal{H}_{M,N}$, being in accordance with expression (A.3). An element of $B_N$ acts on states as

$$\rho(\sigma_i) \, |\psi\rangle = |\psi'\rangle. \tag{A.13}$$

The non-Abelian character is due to

$$[\rho(\sigma_i), \rho(\sigma_j)] \neq 0. \tag{A.14}$$

In contrast to the above discussion, one has for distinguishable particles/punctures

$$\pi_1\Big(\mathcal{F}_N(M)\Big) \cong PB_N \ (d=2), \tag{A.15}$$

whereas for $d = 3$ the fundamental group is just $e \in S_N$.

If a $N$-particle system consists of a variety of distinct and thus distinguishable species, one has $n_j$ particles of species $j$ with $N = \sum_j^{j_{\max}} n_j$. The configuration space is

$$Q_N = \mathcal{F}_N(M)/S_{n_1} \times \cdots \times S_{n_{j_{\max}}} \tag{A.16}$$

and its first homotopy group is

$$\pi_1(Q_N) = B_{n_1,\dots,n_{j_{\max}}}(M). \tag{A.17}$$

It generalises the braid group $B_N(M)$ to $j_{\max}$ distinguishable strand species. Eq. (A.17) is an extension of $PB_{n_1+\cdots+n_{j_{\max}}}$ by $S_{n_1} \times \cdots \times S_{n_{j_{\max}}}$ and one has the short exact sequence

$$\{e\} \to PB_{n_1+\cdots} \to B_{n_1,\dots,n_{j_{\max}}} \to S_{n_1} \times \cdots \to \{e\}. \tag{A.18}$$

## A.4   Spherical braid and pure braid group

A braid on $M = S^2$ has the following geometric picture. One can draw two spheres with different radii around the same center point. Moving a point on the first sphere to another position is kept track of by a strand, connecting both spheres. The according braid groups are $\pi_1(\mathcal{F}_N(S^2)) = PB_N(S^2)$ and $\pi_1(Q_N(S^2)) = B_N(S^2)$, respectively. The generators of $B_N(S^2)$ are those of $B_N$ supplemented by

$$\sigma_1\sigma_2\cdots\sigma_{N-1}^2\cdots\sigma_2\sigma_1 = 1. \tag{A.19}$$

This constraint reflects that a closed loop can be continuously deformed and shrunk to a point on the back of the sphere due to its compactness [543–555].

The spherical pure braid group $PB_N(S^2)$ needs apart from the upper presentation for the $\gamma_{i,j}$'s the conditions 1.)   $\gamma_{i,j} = \gamma_{j,i}$ for $i < j \leq N$, 2.)   $\gamma_{i,i} = 1$ and 3.) $\gamma_{i,i+1}\gamma_{i,i+2}\cdots\gamma_{i,i+N-1} = 1$ for $i \leq N$, where the indices in the latter are considered to run $\mathrm{mod}\,N$.

Finally, for $j_{\max}$ species of punctures distributed on $S^2$ together with expression (A.18) the braid group reads

$$B_{n_1,\ldots,n_{j_{\max}}}(S^2). \tag{A.20}$$

## A.5   Mapping class group and braid group on the sphere

Consider $S_{g,b,N}$ to be an oriented surface of genus $g$, with $b$ boundary components and a set of $N$ marked points/punctures in the surface, following Refs. [543–555]. $\mathrm{Homeo}^+(S_{g,b,N})$ is the group of orientation preserving self-homeomorphisms of $S_{g,b,N}$. These point-wisely fix the boundary if $b > 0$ and they map the set of $N$ marked points into itself. $\mathrm{Homeo}_0(S_{g,b,N})$ is its normal subgroup and its elements are isotopic to the identity. It is a fact, that homotopic homeomorphisms of the compact surface $S$ (even with a finite number of marked points) are isotopic, as long as $S$ is not the disc or the annulus. Additionally, one can improve homeomorphisms of this $S$ to diffeomorphisms. Then isotopies are replaced by smooth isotopies. The mapping class group $M_{g,b,N}$, is defined as

$$M_{g,b,N} \equiv \pi_0(\mathrm{Homeo}^+(S_{g,b,N})) = \mathrm{Homeo}^+(S_{g,b,N})/\mathrm{Homeo}_0(S_{g,b,N}). \tag{A.21}$$

With the given facts this can be restated as

$$M_{g,b,N} \equiv \pi_0(\mathrm{Diff}^+(S_{g,b,N})) = \mathrm{Diff}^+(S_{g,b,N})/\mathrm{Diff}_0(S_{g,b,N}), \qquad (A.22)$$

also denoted as $\mathrm{MCG}(S)$ or $\Gamma_{g,N}$. $\mathrm{Diff}^+(S_{g,b,N})$ is the group of orientation preserving diffeomorphisms of $S_{g,b,N}$, that are the identity on the boundary and that non-trivially act on the punctures. They are also called "large diffeomorphisms". On the other hand, $\mathrm{Diff}_0(S_{g,b,N})$ is the group of small diffeomorphisms. Alltogether, $M_{g,b,N}$ is the group of diffeomorphisms of $S$, which leave the set of punctures invariant, modulo isotopies, which leave the set of punctures invariant. It is the space of path components or isotopy classes of $\mathrm{Diff}^+(S_{g,b,N})$. However, this allows the diffeomorphisms in $\mathrm{Diff}^+(S_{g,b,N})$ to permute the $N$ punctures. In contrast, for an ordered set of $N$ punctures, indicated by $\widehat{N}$, one has $\mathrm{Diff}^+(S_{g,b,\widehat{N}})$. Due to the ordering, different orderings are discernible and the punctures are thus distinguishable. The according pure mapping class group constitutes itself through the isotopy classes of diffeomorphisms, which preserve the punctures point-wisely. It is defined as

$$PM_{g,b,N} = \mathrm{Diff}^+(S_{g,b,\widehat{N}})/\mathrm{Diff}_0(S_{g,b,\widehat{N}}). \qquad (A.23)$$

There is a natural epimorphism $f : M_{g,b,N} \to S_N$, whose kernel is precisely $PM_{g,b,N}$ and one is lead to the short exact sequence

$$\{e\} \to PM_{g,b,N} \to M_{g,b,N} \to S_N \to \{e\}. \qquad (A.24)$$

Importantly, these groups are closely related to braid groups. In Appendices A.3 and A.4 $\pi_1(Q_N(S^2)) = B_N(S^2)$ was recovered. In Ref. [556] it was shown that $\pi_1(SO(3)) = \pi_1(\mathrm{Diff}^+(S^2)) = \mathbb{Z}_2$. When $N \geq 2$, this group maps non-trivially onto $\pi_1(\mathrm{Diff}^+(S^2))$. The short exact sequence

$$\{e\} \to \pi_1(\mathrm{Diff}^+(S^2)) \to \pi_1(Q_N(S^2)) \to M_N(S^2) \to \{e\} \qquad (A.25)$$

is equivalent to

$$\{e\} \to \mathbb{Z}_2 \to B_N(S^2) \to M_N(S^2) \to \{e\}. \qquad (A.26)$$

From this one finds

$$M_N(S^2) \cong B_N(S^2)/\mathbb{Z}_2. \qquad (A.27)$$

$M_N(S^2)$ has the same generators as $B_N(S^2)$ but is supplemented by an additional condition generating the occuring $\mathbb{Z}_2$, namely

$$[\sigma_1 \cdots \sigma_{N-1}]^N = 1. \tag{A.28}$$

This is equivalent to $[\sigma_1...\sigma_{N-1}\sigma_1...\sigma_{N-2}...\sigma_1\sigma_2\sigma_1]^2 = 1$ when using the definition of $B_N$. Elements which obey Eq. (A.28) correspond to those of $B_N(S^2)$, where the $N$ strands are rotated by a $2\pi$ twist. This twist can be untangled when applying it twice, also known as Dirac's belt trick. In contrast to this, one has $M_{0,1,N} \cong B_N(D^2)(\cong B_N(\mathbb{R}^2))$ for the disc. For the pure case one has

$$PM_N(S^2) \cong PB_N(S^2)/\mathbb{Z}_2. \tag{A.29}$$

Analogously to expression (A.20), the generalisation for $n_{j_{\max}}$-species leads to

$$M_{n_1,...,n_{j_{\max}}}(S^2) \cong B_{n_1,...,n_{j_{\max}}}(S^2)/\mathbb{Z}_2. \tag{A.30}$$

# Appendix B

# Non-Fock coherent states

The notion of non-Fock coherent states in GFT was discussed for the first time in some detail in Ref. [457]. It follows largely from the established literature on optical coherence given in axiomatic form in Refs. [450–456, 532, 557–565].

From an algebraic point of view, a GFT quantum system can be defined by its algebra of observables $\mathcal{A}$ which is a unital $C^*$-algebra. In this language, a GFT state is a linear functional $\omega : \mathcal{A} \to \mathbb{C}$ which is positive (i.e. $\omega(a^\dagger a) \geq 0 \ \forall a \in \mathcal{A}$) and normalised (i.e. $\omega(\mathbb{1}) = 1$) with $\omega(A) = \langle A \rangle$. Without proof let us assume that for each such $\omega$ there is a GNS triple (determined up to unitary transformations), $(\mathcal{F}_\omega, \pi_\omega, \psi_\omega)$, where $\mathcal{F}_\omega$ is the bosonic Fock space (as introduced in Section 4.2.4.3), $\pi_\omega$ is a unit-preserving representation of $\mathcal{A}$ in terms of linear operators over $\mathcal{F}_\omega$ and $\psi_\omega \in \mathcal{F}_\omega$ is cyclic, that means $\pi_\omega(\mathcal{A})\psi_\omega$ is dense in $\mathcal{F}_\omega$. Using the scalar product in $\mathcal{F}_\omega$, $\langle \psi_\omega | \pi_\omega(a)\psi_\omega \rangle = \omega(a)$ holds for all $a \in \mathcal{A}$. Using this language with Section 4.2.4.3, one can write for example for the number opertor $\langle \hat{N}_i \rangle = \omega(\hat{c}_i^\dagger \hat{c}_j) = N_i$.

In the following, let $M$ the domain of the left and/or right invariant GFT fields built on SU(2)-valued domains. The measure on $M$ is denoted by $\mathrm{d}h$. Using the distributional character of the field operators, we smear the creation and annihilation operators with the real functions $f_i \in C_0^\infty(M)$ which form an orthonormal set $\{f_i\}$, giving e.g. $\hat{c}(f_i) = \hat{\psi}(f_i) = \int_M \mathrm{d}h \ \hat{\psi}(g_\mathrm{I})f_i(g_\mathrm{I})$.

Using the above, a state $\omega$ is called (fully) coherent if it possesses a factorisation property of the correlation functions in the sense that with a linear form, the so-called coherence function, $L : C_0^\infty(M) \to \mathbb{C}$ one has

$$\omega(\hat{c}^\dagger(f_1) \cdots \hat{c}^\dagger(f_k)\hat{c}(\tilde{f}_1) \cdots \hat{c}(\tilde{f}_k)) = L(f_1) \cdots L(f_k)\bar{L}(\tilde{f}_1) \cdots \bar{L}(\tilde{f}_l) \tag{B.1}$$

for all $k$, $l \in \mathbb{N}_0$ with $k = l$ and for all $\{f_k\}$ and $\{\tilde{f}_l\} \in C_0^\infty(M)$. In particular, $\omega(\hat{c}^\dagger(f_i)\hat{c}(f_i)) = |L(f_i)|^2 \overset{!}{=} N_i$ holds. One calls the coherence function $L$ bounded, if there exists a constant $c_L \geq 0$ with $|L(f)| \leq c_L \|f\|$. Otherwise $L$ is unbounded. In the GFT condensate cosmology context, $f$ is strictly related to the mean field $\sigma$.

With this one can make the following statements. A coherent state $\omega$ in GFT is normal to the Fock representation, if and only if $L$ is bounded, that means the state is given by a unique density operator in Fock space. For unbounded $L$ the state $\omega$ is not representable by a density operator in Fock space, i.e., $\omega$ is disjoint from the Fock sector. This implies that the set of all occupation numbers is unbounded. Suppose now, that $\omega$ is a coherent state of GFT in the above sense. For bounded $L$ one calls $\omega$ a Fock coherent or a microscopic coherent state. In contradistinction to that one calls $\omega$ a non-Fock coherent or a macroscopic coherent state if the coherence function $L$ is unbounded.[12]

---

[1] In the context of local QFTs one can show that an unbounded $L$ exhibits specific classical features, such as a collective phase and amplitude which means that it acquires the status of a classical field due to the ordering effect of the present phase correlations. Furthermore, the unboundedness of $L$ leads to a finite particle density at infinite volume in contrast to a vanishing particle density for bounded $L$ in the same limit [532, 557–565].

[2] We refer to Ref. [458] where inequivalent representations of GFT were studied in detail and the notion of non-Fock coherent state was rigorously explored.

# Appendix C

# Harmonic analysis on $\mathrm{SU}(2)$ and $\mathrm{SL}(2, \mathbb{R})$

## C.1   Harmonic analysis on Lie Groups

Fourier transformations on flat space can be generalised to semi-simple compact Lie groups and to some extent also to non-compact ones. One can use irreducible unitary representations $\pi$ to define a transform of a $L^2$-function on the Lie group $G$ to a function $\hat{f}$ on representation space,

$$\hat{f}(\pi) = \int_G \mathrm{d}g f(g) \pi_{g^{-1}} \tag{C.1}$$

in terms of the Haar measure $\mathrm{d}g$. If available, the Plancherel inversion formula describes the decomposition of $f$ into such modes,

$$f(g) = \int_{\hat{G}} \mathrm{d}\mu(\pi^\lambda) \operatorname{tr}\left(\hat{f}(\pi^\lambda)\pi_g^\lambda\right) \tag{C.2}$$

where $\hat{G}$ is the unitary dual of $G$, *i.e.*, $\hat{G}$ is the set of all equivalence classes of irreducible unitary representations of $G$. One can choose a representation $\pi^\lambda$ for each class $\lambda$ in $\hat{G}$. The Plancherel measure is denoted by $\mathrm{d}\mu(\pi^\lambda)$, see Refs. [566–568] for details.

Accordingly, the Plancherel theorem for $L^2$-functions on $G$ is

$$\int_G \mathrm{d}g |f(g)|^2 = \int_{\hat{G}} \mathrm{d}\mu(\pi^\lambda) ||\hat{f}(\pi^\lambda)||_{HS}^2 \tag{C.3}$$

with $||\hat{f}(\pi^\lambda)||_{HS}^2 = \operatorname{tr}\left(\hat{f}(\pi^\lambda)\hat{f}(\pi^\lambda)^*\right)$ the Hilbert-Schmidt norm. The direct-integral decomposition of the regular representation $R_g$ for $g \in G$ into the sum of primary components

is

$$R_g \simeq \oint_{\hat{G}} \mathrm{d}\mu(\pi^\lambda) \pi_g^\lambda \otimes \mathbb{1}_{d(\pi^\lambda)}, \tag{C.4}$$

where $\mathbb{1}_d$ is the identity on the vector space of dimension given by multiplicity $d = d(\pi^\lambda)$ which may be finite or infinite [567].

## C.2  Harmonic analysis on SU(2)

### C.2.1  General features

On a semi-simple compact Lie group unitary irreducible representations act on finite vector spaces and representations have matrix coefficients [523, 566, 569]. In particular, on $G = \mathrm{SU}(2)$ unitary irreducible representations are defined by the Wigner matrices $D^j(g)$, labelled by half integers $j \in \left\{0, \frac{1}{2}, 1, \dots\right\}$ and the representation spaces have dimension $d_j = 2j + 1$. For these the matrix coefficients are given by $D_{mn}^j(g)$ with $m, n \in \{-j, \dots, j\}$.

Thus, the Plancherel inversion formula for an $L^2$-function $f$ on SU(2) takes the form

$$f(g) = \sum_j \mu(\pi^j) \mathrm{tr}\left(\hat{f}(\pi^j)\pi_g^j\right) \tag{C.5}$$

$$= \sum_j d_j \sum_{m,n=-j}^j f_{mn}^j D_{mn}^j(g) \tag{C.6}$$

where $f_{mn}^j$ are the coefficents of the transforms $\hat{f}(\pi^j)$. One refers to this expression also as the Peter-Weyl formula.

Notice that there is a stronger fomulation of it for $\mathcal{C}_0^\infty$-functions [500, 523]. To see this, let $\mathcal{C}^\infty(\mathrm{SU}(2))$ the space of smooth functions $f$ on SU(2) which is equipped with the topology given by semi-norms

$$\|f\|_n = \sup_{g \in \mathrm{SU}(2)} |\Delta^n f(g)|, \tag{C.7}$$

with $(-)\Delta$ the Laplace-Beltrami operator and $n \in \mathbb{N}$. For any $f \in \mathcal{C}^\infty(\mathrm{SU}(2))$ there exists a sequence of complex numbers $\left(f_{mn}^j\right)$ with $j \in \frac{\mathbb{N}}{2}$ and $m, n \in \{-j, \dots, j\}$ and $D_{mn}^j(x)$ denote the Wigner matrix coefficients with $d_j = 2j + 1$ such that

$$\lim_{N \to \infty} \sum_{j=0}^N \sum_{m,n=-j}^j d_j f_{mn}^j D_{mn}^j = f, \tag{C.8}$$

in the above topology. The sequence of Fourier coefficients $\left( f_{mn}^j \right)$ is rapidly decreasing, i.e. for any $K \in \mathbb{N}$

$$\sup_j \left| j^K \sum_{m,n=-j}^{j} \bar{f}_{mn}^j f_{mn}^j \right| < \infty. \tag{C.9}$$

If we call the space of rapidly decreasing sequences $\mathcal{S}(\mathbb{N})$, then Eq. C.9 defines a family of semi-norms on $\mathcal{S}(\mathbb{N})$ and in the corresponding topology it becomes a Fréchet space. Then the Peter-Weyl transform $\mathcal{F} : \mathcal{C}^\infty(\mathrm{SU}(2)) \to \mathcal{S}(\mathbb{N})$ is a topological isomorphism between the space of smooth functions and the space of rapidly decreasing sequences [500].

Tensor product representations are easily obtained from this [569]. We would like to remark that depending on the source, one finds different conventions on whether the Plancherel measure $d_j$ is fully spelled out in the inversion formula or is in part or fully absorbed in the Fourier coefficients. When in this thesis different conventions are used (e.g. $d_j$ vs. $\sqrt{d_j}$), this has no impact onto the final results of calculations since redundant factors can always be reabsorbed into the Fourier coefficients.

As a special example, the $\delta$-distribution with transforms $\hat{\delta}(\pi^j) = \mathbb{1}_{d_j}$ for all $j$ is given by

$$\delta(g) = \sum_j d_j \chi^j(g) \tag{C.10}$$

in terms of characters $\chi^j(g) \equiv \mathrm{tr}\, D^j(g)$.

In the following, we give some relevant properties of Wigner matrices and characters:

1. Under complex conjugation one has

$$\overline{D_{mn}^j} = (-1)^{2j+m+n} D_{-m-n}^j. \tag{C.11}$$

2. The Wigner-matrix coefficients form an orthogonal basis of the Hilbert space $L^2(\mathrm{SU}(2), \mathrm{d}g)$ with

$$\int \mathrm{d}g\, D_{m_1 n_1}^{j_1}(g)\, \overline{D_{m_2 n_2}^{j_2}}(g) = \frac{1}{d_{j_1}} \delta^{j_1 j_2} \delta_{m_1 m_2} \delta_{n_1 n_2}. \tag{C.12}$$

The volume element $\mathrm{d}g$ defines the Haar measure as the unique measure (up to rescalings) which is invariant under right and left action of the group onto itself.

3. These coefficients form a basis of eigenfunctions for the Laplace-Beltrami operator $-\Delta$ (defined with the canonical metric), i.e.

$$-\Delta D^j_{mn}(g) = j(j+1) D^j_{mn}(g).$$  (C.13)

Note that some sources in the literature adopt the convention of multiplying the eigenvalues by a factor 4.

4. The characters are smooth real-valued functions satisfying $\chi^j(g) = \chi^j(g^{-1})$.

5. For $g_1, g_2 \in$ SU(2) one has the convolution relation

$$\int dh \; \chi^j(hg_1)\chi^l(g_2h) = \frac{\delta_{jl}}{d_j}\chi^j(g_2g_1^{-1})$$  (C.14)

from which the orthogonality relation $\int dh \chi^j(h)\chi^l(h) = \delta_{jl}$ is retrieved.

6. The Wigner $6j$-symbol can be defined in terms of characters as [569],

$$\begin{Bmatrix} l_{01} & l_{02} & l_{03} \\ l_{23} & l_{13} & l_{12} \end{Bmatrix}^2 = \int (dh)^4 \prod_{i<j}^3 \chi^{l_{ij}}(h_j h_i^{-1}).$$  (C.15)

## C.2.2   Basis for left and right invariant functions

In the above notations, the left and right invariant functions on SU(2)$^3$ are given by group averaging, such that for any $f \in C^\infty\left(\text{SU}(2)^3\right)$ and any $(g_1, g_2, g_3) \in$ SU(2)$^3$, $\int dl dr \; f(lg_1r, lg_2r, lg_3r)$, with $l, r \in$ SU(2). In the Peter-Weyl decomposition a left and right invariant function $f$ assumes the form

$$f(g_1, g_2, g_3) = \sum_J f^J d_{j_1} d_{j_2} d_{j_3} \int dh \; \chi^{j_1}(g_1h)\chi^{j_2}(g_2h)\chi^{j_3}(g_3h),$$  (C.16)

where $J = (j_1, j_2, j_3) \in (\frac{\mathbb{N}}{2})^3$. We denote the integral of the product of three characters by

$$\mathcal{X}^J(g_1, g_2, g_3) \equiv d_{j_1} d_{j_2} d_{j_3} \int dh \; \chi^{j_1}(g_1h)\chi^{j_2}(g_2h)\chi^{j_3}(g_3h).$$  (C.17)

It can be easily checked that $\mathcal{X}^J$ has the following properties:

1. Using the orthogonality of characters, $\int dh \; \chi^j(hg_1)\chi^l(g_2h) = \frac{\delta_{jl}}{d_j}\chi^j(g_2g_1^{-1})$, and reality of characters, the $\mathcal{X}^J$'s are real valued and form an orthonormal family with respect to the $L^2 \left( \mathrm{SU}(2)^3, (\mathrm{d}g)^3 \right)$ scalar product:

$$\int (\mathrm{d}g)^3 \, \mathcal{X}^J \left( g_1, g_2, g_3 \right) \mathcal{X}^K \left( g_1, g_2, g_3 \right) = \delta_{J,K}; \tag{C.18}$$

2. $\mathcal{X}^J$ is proportional to the Wigner-3$J$ symbol with three equal $j$'s and sum over the magnetic indices and hence vanishes if $j$ is not an integer;

3. Using Eq. (C.16), the family of $\mathcal{X}^J$'s is dense in the space of left and right invariant functions, such that any left and right invariant function $f$ can be written as

$$f \left( g_1, g_2, g_3 \right) = \sum_{J \in \mathfrak{J}} f^J \, \mathcal{X}^J \left( g_1, g_2, g_3 \right); \tag{C.19}$$

4. Using Eq. C.15, the 6$j$-symbol is given by a $\mathcal{X}^J$ integral as

$$\delta_{j_1 k_1} \delta_{j_2 l_1} \delta_{j_3 q_1} \delta_{q_2 l_3} \delta_{k_2 q_3} \delta_{k_2 l_3} \begin{Bmatrix} j_1 & j_2 & j_3 \\ q_2 & l_2 & k_2 \end{Bmatrix}^2 = $$
$$\int (\mathrm{d}g)^6 \mathcal{X}^J \left( g_1, g_2, g_3 \right) \mathcal{X}^K \left( g_1, g_4, g_5 \right) \mathcal{X}^L \left( g_2, g_5, g_6 \right) \mathcal{X}^Q \left( g_3, g_6, g_4 \right), \tag{C.20}$$

with $J = (j_1, j_2, j_3)$, $K = (k_1, k_2, k_3)$, $L = (l_1, l_2, l_3)$, $Q = (q_1, q_2, q_3)$.

Since we are interested in functions that are invariant under cyclic permutation we need to symmetrise the characters $\mathcal{X}^J \left( g_1, g_2, g_3 \right)$. To achieve this, we introduce the symmetrisation operator

$$P \mathcal{X}^J \left( g_1, g_2, g_3 \right) = \frac{1}{3} \sum_{\sigma \in \mathrm{Cyc}} \mathcal{X}^{\left( j_{\sigma(1)}, j_{\sigma(2)}, j_{\sigma(3)} \right)} \left( g_1, g_2, g_3 \right), \tag{C.21}$$

where Cyc denotes the set of cyclic permutations of $\{1, 2, 3\}$. All aforementioned properties of $\mathcal{X}^J$ can be adapted to $P \mathcal{X}^J \left( g_1, g_2, g_3 \right)$ by including a normalised sum over cyclic permutations of indices. Since for the equilateral case we have

$$P \mathcal{X}^J \left( g_1, g_2, g_3 \right) = \mathcal{X}^J \left( g_1, g_2, g_3 \right), \tag{C.22}$$

we simply use the notation $\mathcal{X}^J \left( g_1, g_2, g_3 \right)$ for symmetric characters on $\mathcal{S}_{\mathrm{EL}}$ and on $\mathcal{S}$ in Section 5.2.

## C.3 Harmonic analysis on SL(2, ℝ)

### C.3.1 Group structure of SL(2, ℝ)

The non-compact, simple and multiply connected Lie group $G = \mathrm{SL}(2, \mathbb{R})$ is the group of $2 \times 2$ real matrices of determinant 1, i.e.,

$$\mathrm{SL}(2, \mathbb{R}) = \left\{ \begin{pmatrix} a & b \\ c & d \end{pmatrix}, \; a, b, c, d \in \mathbb{R}, \; ad - bc = 1 \right\}. \tag{C.23}$$

Its largest normal subgroup is its centre $Z = \{\pm \mathbb{1}\}$. The group acts by linear transformation on $\mathbb{R}^2$ while preserving oriented area. The eigenvalues of a matrix $g \in \mathrm{SL}(2, \mathbb{R})$ are

$$\lambda_g^{\pm} = \frac{\mathrm{tr}\,(g) \pm \sqrt{(\mathrm{tr}\,(g))^2 - 4}}{2}. \tag{C.24}$$

such that elements in $SL(2, \mathbb{R})$ are classified according to the following scheme:

- If $|\mathrm{tr}\,(g)| < 2$, $g$ is called elliptic,

- if $|\mathrm{tr}\,(g)| = 2$, $g$ is called parabolic,

- if $|\mathrm{tr}\,(g)| > 2$, $g$ is called hyperbolic.

There are different ways to decompose $G$ in terms of three special subgroups:

- the maximal compact subgroup (isomorphic to SO(2))

$$H_0 = \left\{ u = u_\theta = \begin{pmatrix} \cos(\theta) & \sin(\theta) \\ -\sin(\theta) & \cos(\theta) \end{pmatrix}, \; 0 \leq \theta \leq 2\pi \right\}, \tag{C.25}$$

- the upper/lower unipotent subgroups

$$N = N_\pm = \left\{ n = \begin{pmatrix} 1 & \nu \\ 0 & 1 \end{pmatrix} \text{ and } \begin{pmatrix} 1 & 0 \\ \nu & 1 \end{pmatrix}, \; \nu \in \mathbb{R} \right\} \tag{C.26}$$

- and the diagonal group

$$H_1 = \left\{ a = \pm a_t = \begin{pmatrix} \pm e^t & 0 \\ 0 & \pm e^{-t} \end{pmatrix}, \; t \in \mathbb{R} \right\} \tag{C.27}$$

with its positive part, the semigroup

$$H_{1,+} = \{a_t : t > 0\}. \tag{C.28}$$

With these one can give the so-called Iwasawa decomposition $G = H_0 N_- H_1$ or $G = H_1 N_+ H_0$ and the Cartan decomposition $G = H_0 H_{1,+} H_0$.

The Lie algebra of SL(2, ℝ) consists of the traceless $2 \times 2$ real matrices, i.e.

$$\mathfrak{sl}(2, \mathbb{R}) = \{g \in \text{Mat}(2, \mathbb{R}) : \text{tr}(g) = 0\} \tag{C.29}$$

with the commutator acting as the Lie bracket. A basis of the three dimensional vector space $\mathfrak{sl}(2, \mathbb{R})$ shall be given by $\{h, x, y\}$. The structure of the Lie algebra is then encoded by the commutator relations

$$[h, x] = 2x, \ [h, y] = -2y \text{ and } [x, y] = h. \tag{C.30}$$

In its fundamental representation the generators can be represented by

$$h = \begin{pmatrix} 1 & 0 \\ 0 & -1 \end{pmatrix}, \ x = \begin{pmatrix} 0 & 1 \\ 0 & 0 \end{pmatrix}, \text{ and } y = \begin{pmatrix} 0 & 0 \\ 1 & 0 \end{pmatrix}. \tag{C.31}$$

$\mathfrak{sl}(2, \mathbb{R})$ is a simple, particularly a semi-simple Lie algebra. Remarkably, it has two non-conjugated Cartan subalgebras, generated by $x - y$ and $h$ [570]. Thus, SL(2, ℝ) has two Cartan subgroups. (This is to be contrasted to the case of SL(2, ℂ) which has only one Cartan subgroup.) One of them is compact, given by $H_0$, see Eq. C.25, while the other is non-compact, given by $H_1$, see Eq. C.27.

Elements which can be conjugated to a Cartan subgroup are called regular. They form a set which decomposes into the conjugacy classes, specifically

(i) the *elliptic* classes

$$G_0 = \bigcup_{0 < \theta < \pi} \{g u_\theta g^{-1} : \ g \in G\} \tag{C.32}$$

and the

(ii) *hyperbolic* classes

$$G_\pm = \bigcup_{t > 0} \{g (\pm a_t) g^{-1} : \ g \in G\}. \tag{C.33}$$

The Haar measure on $\mathrm{SL}(2, \mathbb{R})$ can then be desintegrated into invariant measures on these classes. Together with the Weyl integration formula, the averaging of a $C_0^\infty$-function $f$ over $G$ leads to

$$\int_G \mathrm{d}g\, f(g) = \alpha_0 \int_0^\pi \mathrm{d}\theta \sin^2\theta\, f_0(\theta) + \alpha_1 \int_0^\infty \mathrm{d}t \sinh^2 t\, f_1(t), \tag{C.34}$$

with $f_1(t) \equiv f_1(\pm a_t)$ and $(\alpha_0 = 1, \alpha_1 = 1)$, see e.g. Ref. [514]. The functions $f_0$ and $f_1$ denote the averaging of $f$ over the corresponding elliptic and hyperbolic conjugacy classes, that is

$$f_0(\theta) = \int_{G/H_0} \mathrm{d}g\, f\left(g u_\theta g^{-1}\right) \tag{C.35}$$

and

$$f_1(t) = \int_{G/H_1} \mathrm{d}g\, f\left(g(\pm a_t)g^{-1}\right). \tag{C.36}$$

However, using group averaging arguments, the only way to consistently define an $\mathrm{Ad}(G)$-invariant function $f$ through averaging, is given by the two choices $(\alpha_0 = 1, \alpha_1 = 0)$ or $(\alpha_0 = 0, \alpha_1 = 1)$. For $L^2$-functions, this amounts to defining two Hilbert spaces $\mathcal{H}_0$ for functions with support on $G_0$ with $(\alpha_0 = 1, \alpha_1 = 0)$ and $\mathcal{H}_1$ for functions with support on $G_\pm$ and $(\alpha_0 = 0, \alpha_1 = 1)$. For a detailed discussion of this point, we refer to Ref. [514]. Notice that these two sectors cannot be mapped into one another. This can be interpreted as a superselection rule [515–517]. In the following subsection we discuss the Fourier decomposition for functions on $G_0$ and $G_\pm$ of the type $f_0(\theta)$ and $f_1(t)$.

### C.3.2   Harmonic analysis on $\mathrm{SL}(2, \mathbb{R})$

Here we collect some facts regarding the harmonic analysis on $\mathrm{SL}(2, \mathbb{R})$ to supplement the main body of this thesis focusing on the characters and the Plancherel formula. We closely follow Refs. [566–568, 570–575].

**Characters of $\mathrm{SL}(2, \mathbb{R})$**

All unitary irreducible representations of $\mathrm{SL}(2, \mathbb{R})$ are exhausted by the three series: principal, complementary and discrete. In the following, we give the characters of these and refer to Refs. [566–568, 575] for their derivation from the respective representations.

**1.)** The characters of the principal series representation labelled by $s \in \mathbb{R}^+$ are

$$\chi_s^\pm(g) = \begin{cases} \frac{\cos(st)}{|\sinh t|} \epsilon_\pm(\lambda_g), & \text{for } g \text{ hyperbolic,} \\ \\ 0 & , \text{ for } g \text{ elliptic,} \end{cases} \tag{C.37}$$

where $\epsilon_+(\lambda_g) = 1$ and $\epsilon_-(\lambda_g) = \text{sgn}(\lambda_g)$, depending on the eigenvalues $\lambda_g$, Eq. C.24.

**2.)** The characters of the complementary series $\chi_\rho(g)$ take the same form, only that for these $is$ is replaced by $\rho \in [-1, 1]$. Importantly, the complementary series does not contribute to the Plancherel formula (for distributions or $L^2$-functions on $G$) [566, 567, 570, 574].

**3.)** The characters of the discrete series are

$$\chi_n^\pm(g) = \begin{cases} \frac{e^{-n|t|}}{2|\sinh t|} \epsilon_\pm(\lambda_g), & \text{for } g \text{ hyperbolic,} \\ \\ \mp \frac{e^{\pm in\theta}}{2i \sin \theta} & , \text{ for } g \text{ elliptic,} \end{cases} \tag{C.38}$$

labelled by $n = 1, 2, \dots$.

The characters are eigenfunctions of the Laplacian with spectrum [567]

$$\frac{1 + s^2}{4} \text{ for } \chi_s^\pm \quad \text{and} \quad \frac{1 - n^2}{4} \text{ for } \chi_n^\pm. \tag{C.39}$$

The individual parts of the Laplacian act on averaged functions $f_0(\theta)$ and $f_1(t)$ in the standard way.

**Plancherel formula for** SL(2, ℝ)

In view of Appendix C.1, in the case of SL(2, ℝ) the inversion formula [566–568, 571–573, 575] reads

$$\begin{aligned} f(g) \ = \ & \sum_{n=1}^\infty \frac{n}{4\pi} \left( \text{tr} \left( f^+(n) \pi_g^{n,+} \right) + \text{tr} \left( f^-(n) \pi_g^{n,-} \right) \right) \\ & + \int_0^\infty \frac{\mathrm{d}s}{4\pi} \frac{s}{2} \tanh\left(\frac{\pi s}{2}\right) \ \text{tr} \left( f^+(s) \, \pi_g^{s,+} \right) \\ & + \int_0^\infty \frac{\mathrm{d}s}{4\pi} \frac{s}{2} \coth\left(\frac{\pi s}{2}\right) \ \text{tr} \left( f^-(s) \, \pi_g^{s,-} \right) \end{aligned} \tag{C.40}$$

with the Fourier coefficients given by

$$f^{\pm}(s) = \int_G \mathrm{d}g f(g) \pi^{s,\pm}_{g^{-1}} \tag{C.41}$$

and

$$f^{\pm}(n) = \int_G \mathrm{d}g f(g) \pi^{n,\pm}_{g^{-1}}. \tag{C.42}$$

where $s_{\pm}$ and $n_{\pm}$ label the positive and negative branches of the principal and discrete series respectively. The expression of the inversion formula is due to Harish-Chandra, building on foundational work of Bargmann [566, 570–573]. The first term stems from the discrete series and encapsulates both the contributions coming from the compact and non-compact directions. The second and third terms stem from the continuous series contribution originating from the non-compact directions. In particular, this decomposition can be applied to the $\delta$-distribution on $G$ [574, 575], which is simply

$$\begin{aligned}
\delta(g) &= \delta_0(\theta) + \delta_1(t) \\
&= \sum_{n=1}^{\infty} \frac{n}{4\pi} \left( \chi^+_n(\theta) + \chi^-_n(\theta) + \chi^+_n(t) + \chi^-_n(t) \right) \\
&\quad + \int_0^{\infty} \frac{\mathrm{d}s}{4\pi} \frac{s}{2} \left( \tanh \frac{\pi s}{2} \chi^+_s(t) + \coth \frac{\pi s}{2} \chi^-_s(t) \right)
\end{aligned} \tag{C.43}$$

with $\delta_0(\theta) \equiv \delta_0(u_\theta)$, $\delta_1(t) \equiv \delta_1(\pm a_t)$ and the characters are taken as in Appendix C.3.2. One observes the structural similarities with the case of SU(2) where the $\delta$-distribution is expanded in terms of characters, see Appendix C.2.

For functions $f_0(\theta)$ and $f_1(t)$ as given in Appendix C.3.1, we can use a similar decomposition as Eq. (C.43), namely

$$f_0(\theta) = \sum_{n=1}^{\infty} \frac{n}{4\pi} \left( f^+(n) \chi^+_n(\theta) + f^-(n) \chi^-_n(\theta) \right) \tag{C.44}$$

and

$$\begin{aligned}
f_1(t) &= \sum_{n=1}^{\infty} \frac{n}{4\pi} \left( f^+(n) \chi^+_n(t) + f^-(n) \chi^-_n(t) \right) \\
&\quad + \int_0^{\infty} \frac{\mathrm{d}s}{4\pi} \frac{s}{2} \tanh \frac{\pi s}{2} f^+(s) \chi^+_s(t) \\
&\quad + \int_0^{\infty} \frac{\mathrm{d}s}{4\pi} \frac{s}{2} \coth \frac{\pi s}{2} f^-(s) \chi^-_s(t)
\end{aligned} \tag{C.45}$$

with Fourier coefficients $f^{\pm}(n)$ and $f^{\pm}(s)$ for the respective series.

# Appendix D

# Proofs for the Section "Solving the dynamical Boulatov model"

## D.1    Proofs

### D.1.1    Proof of proposition 1

Consider the action $S_{m,\lambda}$ Eq. (5.13) and $S'_{m,\lambda}$ Eq. (5.16); $\mathcal{S}_{(EL)}$ means either $\mathcal{S}$ (space of right invariant functions) or $\mathcal{S}_{EL}$ (space of left and right invariant and equilateral functions). The following statement holds:

**Lemma 1.** *The field $\varphi \in \mathcal{S}_{(EL)}$ is an extremum of $S_{m,\lambda}$ iff*

$$S'_{m,\lambda}[\varphi, \mathcal{X}^J] = 0 \tag{D.1}$$

*for all $J \in \mathfrak{J}_{(EL)}$.*

*Proof.* Let $\varphi$ be an extremum of $S_{m,\lambda}$, then the "only if" direction is obvious since for any $J \in \mathfrak{J}_{(EL)}$ the functions $\mathcal{X}^J$ are in $\mathcal{S}_{(EL)}$.

For the "if" direction we observe the following: since the set $\left\{\mathcal{X}^J\right\}_{J \in \mathfrak{J}_{(EL)}}$ is dense in $\mathcal{S}_{(EL)}$, for any $f \in \mathcal{S}_{(EL)}$ there exists a family of real numbers $\left\{f^J\right\}_{J \in \mathfrak{J}_{(EL)}}$ such that the sequence of functions given for all $N \in \mathbb{N}$ as

$$f_N(g_1, g_2, g_3) = \sum_{\substack{J \in \mathfrak{J}_{(EL)} \\ |J| < N}}^{|J| < N} f^J \mathcal{X}^J(g_1, g_2, g_3), \tag{D.2}$$

converges to $f$. Then $c = \sup_{(g_1, g_2, g_3) \in \mathrm{SU}(2)^3} \sup_{N \in \mathbb{N}} |f_N(g_1, g_2, g_3)|$, exists and dominates each $f_N$ such that, $|f_N| \leq c$. Moreover, $c$, seen as a constant function on $\mathrm{SU}(2)^3$, is integrable since $\mathrm{SU}(2)^3$ is compact.

For any $f \in \mathcal{S}_{(\mathrm{EL})}$ the extremal condition for the action $S_{m,\lambda}$ reads as

$$
\begin{aligned}
S'_{m,\lambda}[\varphi, f] = & \int (\mathrm{d}g)^3 \, f(g_1, g_2, g_3) \left( -\Delta + m^2 \right) \varphi(g_1, g_2, g_3) \\
& + \frac{\lambda}{3!} \int (\mathrm{d}g)^6 f(g_1, g_2, g_3) \varphi(g_1, g_4, g_5) \varphi(g_2, g_5, g_6) \varphi(g_3, g_6, g_4).
\end{aligned} \tag{D.3}
$$

Using the Peter-Weyl decomposition for $f$, we can interchange the limit and the integral by the dominant convergence theorem (using the bound $c$) and obtain

$$
S'_{m,\lambda}[\varphi, f] = \lim_{N \to \infty} \sum_J f^J \, S'_{m,\lambda}[\varphi, \mathcal{X}^J] = 0,
$$

for any $f \in \mathcal{S}_{(EL)}$, from which the statement follows. $\qquad \square$

**Corollary.** *$\varphi \in \mathcal{S}_{(EL)}$ is an extremum of $S$ if and only if the Peter-Weyl coefficients of $\varphi$ — denoted by $\varphi^J$ — satisfy for any $J \in \mathfrak{J}_{(EL)}$,*

$$
(J^2 + m^2)\varphi^J + \frac{\lambda}{3!} \sum_{k_i} \varphi^{j_1 k_2 k_3} \varphi^{j_2 k_3 k_1} \varphi^{j_3 k_1 k_2} \begin{Bmatrix} j_1 & j_2 & j_3 \\ k_1 & k_2 & k_3 \end{Bmatrix}^2 = 0. \tag{D.4}
$$

*Proof.* From lemma 1 the extremal condition is given by the variation in the basis direction $\mathcal{X}^J$ for any $J \in \mathfrak{J}_{(\mathrm{EL})}$. Inserting the Peter-Weyl decomposition of $\varphi$ in the action $S_{m,\lambda}(\varphi)$, interchanging the limit with the integral by the dominant convergence theorem and using the relation in Eq. (C.20) we obtain the desired statement. $\qquad \square$

### D.1.2 Proof of theorem 1

**Theorem.** *For any $C \in \mathcal{E}_{m,\lambda}$ the field $\varphi \in \mathcal{S}_{EL}$*

$$
\varphi(g_1, g_2, g_3) = \sum_{J \in \mathfrak{J}_{EL}} C^J \, \mathcal{X}^J(g_1, g_2, g_3) \tag{D.5}
$$

*is an extremum of the action $S_{m,\lambda}$. Moreover, every equilateral extremum of $S_{m,\lambda}$ in $\mathcal{S}_{EL}$ is of the above form.*

*Proof.* To show that $\varphi$ solves the extremal condition we need to show, by proposition 1, that each $C^J$ satisfies Eq. (5.18), which follows by direct calculation.

Conversely, every equilateral function can be written as

$$f(g_1, g_2, g_3) = \sum_{J \in \mathfrak{J}_{\mathrm{EL}}} A_J \, \mathcal{X}^J (g_1, g_2, g_3),$$ (D.6)

with $\left(A^J\right)_{J \in \mathfrak{J}_{\mathrm{EL}}}$ being a rapidly decreasing sequence [500]. Using proposition 1, we find that the extremal solutions have coefficients $A^J$ which satisfy

$$A^J \in \left\{ \pm \frac{1}{|\{6j\}|} \sqrt{-\frac{3!}{\lambda} (J^2 + m^2)}, 0 \right\},$$ (D.7)

or $A^J \in \mathbb{R}$ for $J \in \mathfrak{J}_{\mathrm{EL}}/\mathfrak{J}_{\mathrm{EL}}^S$ with $J^2 + m^2 = 0$. If $A^J$ is not trivial we can estimate its growth using the asymptotic behaviour of $6j$-symbols [576] as

$$A^J \sim \frac{j}{|\{6j\}|} \sim j^{\frac{5}{2}}.$$ (D.8)

However, for $\left(A^J\right)_{J \in \mathfrak{J}_{\mathrm{EL}}}$ to be a rapidly decreasing sequence, the coefficients have to satisfy for any $n \in \mathbb{N}$,

$$\lim_{j \to \infty} |j|^n \left| A^J \right| \to 0.$$ (D.9)

This is only possible if $A^J = 0$ *for all but finitely many* $J \in \mathfrak{J}_{\mathrm{EL}}$. □

**Corollary.** *The space* $\tilde{\mathcal{E}}_{m,\lambda}$ *is a vector space over the discrete algebraic field* $(\mathbb{Z}_3, +, \cdot)$.

*Proof.* Denote the space of sequences with finitely many non-zero elements over $\mathbb{Z}_3$ by $c_{00}(\mathbb{Z}_3)$. Clearly, it is a vector space over $\mathbb{Z}_3$. Consider the map

$$\mathcal{I} : \tilde{\mathcal{E}}_{m,\lambda} \to c_{00}(\mathbb{Z}_3)$$

$$\varphi \mapsto \left( \mathrm{sgn}\left(C^1\right), \mathrm{sgn}\left(C^2\right), \dots \right),$$

with the convention $\mathrm{sgn}(0) = 0$. $\mathcal{I}$ is one-to-one on its image, however, it may not be onto $c_{00}(\mathbb{Z}_3)$ simply because the nontrivial zeros of the $6j$-symbol are not fully characterised. Nevertheless, the image of $\mathcal{I}$ is algebraically closed and forms a subspace of $c_{00}(\mathbb{Z}_3)$. For any $s = (s_0, s_1, \ldots) \in \mathcal{I}\left(\tilde{\mathcal{E}}_{m,\lambda}\right)$, the inverse mapping is given by

$$\mathcal{I}^{-1} : s \mapsto [\mathcal{I}^{-1}s](g_1, g_2, g_3) = \sum_{j \in \mathbb{N}} \operatorname{sgn}(s_j) \left|C^{J_j}\right| \mathcal{X}^{J_j}(g_1, g_2, g_3),$$

where $J_j = (j, j, j)$, $j \in \mathbb{N}$, with

$$\left|C^J\right| = \frac{1}{|\{6j\}|} \left|\sqrt{-\frac{3!}{\lambda}\left(J^2 + m^2\right)}\right|. \tag{D.10}$$

Since there are only finitely many non-zero coefficients, $s_j \neq 0$, the sum trivially converges in $\mathcal{S}_{\mathrm{EL}}$. Since $\mathcal{I}$ is linear it is an isomorphism between $\tilde{\mathcal{E}}_{m,\lambda}$ and $\mathcal{I}(c_{00}(\mathbb{Z}_3))$.

We define the sum on $\tilde{\mathcal{E}}_{m,\lambda}$ by

$$\varphi_1 +_{\mathbb{Z}_3} \varphi_2 \equiv \mathcal{I}^{-1}\left(\mathcal{I}(\varphi_1) + \mathcal{I}(\varphi_2)\right). \tag{D.11}$$

<div align="right">□</div>

### D.1.3  Proof of theorem 2

*Proof of theorem 2.* In the following, let $\varphi(g_1, g_2, g_3)$ denote an extremum and let $f \in \mathcal{S}_{\mathrm{EL}}$ be a generic function with the Peter-Weyl decomposition given by $f(g_1, g_2, g_3) = \sum_{J \in \mathfrak{J}_{\mathrm{EL}}} f^J \mathcal{X}^J(g_1, g_2, g_3)$. We remind here that a necessary condition for an extremum $\varphi(g_1, g_2, g_3)$ to be a minimiser (maximiser, respectively) is given by

$$S''_{m,\lambda}[\varphi, f] \geq 0 \qquad \left(S''_{m,\lambda}[\varphi, f] \leq 0, \text{ resp.}\right), \tag{D.12}$$

for any $f \in \mathcal{S}_{EL}$. In the Peter-Weyl decomposition the second variation recasts as

$$S''_{m,\lambda}[\varphi, f] = \sum_{J \in \mathfrak{J}_{\mathrm{EL}}} \left(f^J\right)^2 \left(\left(J^2 + m^2\right) - \frac{\lambda}{2} \sum_{K \in \mathfrak{J}_{\mathrm{EL}}^S} \delta_{J,K} \, \varphi^K \, \{6K\}^2\right), \tag{D.13}$$

where $\varphi^K$ is the Peter-Weyl coefficient of the extremum $\varphi$. The above condition is necessary but not sufficient, nevertheless, it turns out to be useful to exclude some extrema.

*Case (a)* $(m^2 \leq 0, \ \lambda \leq 0)$: By theorem 1, extremal solutions only contain finitely many non-zero Fourier coefficients. Therefore, it is possible to find $J_> \in \mathfrak{J}_{EL}$ such that $J_>^2 - |m|^2 > 0$ and $\varphi^{J_>} = 0$. Choosing $f_>(g_1, g_2, g_3) \equiv f^{J_>} \mathcal{X}^{J_>}(g_1, g_2, g_3)$ the second

variation gives

$$S_{m,\lambda}''[\varphi, f_>] = \left(f^{J_>}\right)^2 \left(J_>^2 - \left|m^2\right|\right) > 0, \tag{D.14}$$

which violates the maximiser condition.

To see that the minimiser condition is also violated, choose

$$f_< (g_1, g_2, g_3) \equiv f^{J_<} \mathcal{X}^{J_<} (g_1, g_2, g_3) \tag{D.15}$$

such that $J_<^2 - \left|m^2\right| < 0$. Then the second variation is written as

$$S_{m,\lambda}''[\varphi, f_<] = \left(f^{J_<}\right)^2 \left(J_<^2 - \left|m^2\right|\right) \leq 0. \tag{D.16}$$

Hence, each extremum in this parameter region violates the minimiser and the maximiser condition and therefore is a saddle point.

*Case (b)* ($m^2 \geq 0$, $\lambda \leq 0$): For the nontrivial minimiser the above argument can also be applied in this case. Choosing the functions $f_> (g_1, g_2, g_3)$ and $f_< (g_1, g_2, g_3)$ as above, we find

$$S_{m,\lambda}''[\varphi, f_>] = \left(f^{J_>}\right)^2 \left(J_>^2 + \left|m^2\right|\right) > 0, \quad \text{and}$$
$$S_{m,\lambda}''[\varphi, f_<] = -2 \left(f^{J_<}\right)^2 \left(J_<^2 + \left|m^2\right|\right) < 0. \tag{D.17}$$

Hence, nontrivial extrema are saddle points. For the trivial extremum the second variation of $S_{m,\lambda}$ reads for any $f \in \mathcal{S}_{\text{EL}}$

$$S_{m,\lambda}''[0, f] = \sum_{J \in \mathfrak{J}_{\text{EL}}} \left(f^J\right)^2 \left(J^2 + \left|m^2\right|\right) \geq 0,$$

and the necessary condition is satisfied. Indeed, the trivial extremum is a local minimum. To prove this, we first notice that the Peter-Weyl transform is a topological isomorphism from $\mathcal{S}_{\text{EL}}$ to the space of rapidly decreasing sequences $\mathcal{S}(\mathbb{N})$ with topology given by the family of semi-norms [500, theorem 4],

$$\| \left(f^J\right)_{J \in \mathfrak{J}_{\text{EL}}} \|_n = \sup_{J \in \mathfrak{J}_{\text{EL}}} \left|J^n f^J\right|. \tag{D.18}$$

The action evaluated at $f$ becomes

$$S_{m,\lambda}[f] = \sum_{J \in \mathfrak{J}_{\mathrm{EL}}} (f^J)^2 (J^2 + m^2) - \frac{\lambda}{4!} \sum_{J \in \mathfrak{J}_{\mathrm{EL}}} (f^J)^4 \{6j\}^2 \qquad (\mathrm{D}.19)$$

Since the Wigner-6$j$-symbol is upper-bounded by 1, we can estimate

$$\begin{aligned} S_{m,\lambda}[f] &\geq \sum_{J \in \mathfrak{J}_{\mathrm{EL}}} (f^J)^2 \left( (J^2 + |m^2|) - \frac{\lambda}{4!} (f^J)^2 \right) \\ &\geq \sum_{J \in \mathfrak{J}_{\mathrm{EL}}} (f^J)^2 \left( m^2 - \frac{\lambda}{4!} (f^J)^2 \right). \end{aligned} \qquad (\mathrm{D}.20)$$

Since the Peter-Weyl transform is a topological isomorphism, we get for any $f \in \mathcal{S}_{\mathrm{EL}}$ with $\|f\|_0 \leq \sqrt{\frac{4! m^2}{|\lambda|}}$, an estimate on the Fourier coefficients

$$|f^J| \leq \| (f^J)_{J \in \mathfrak{J}_{\mathrm{EL}}} \|_0 \leq \sqrt{\frac{4! m^2}{|\lambda|}}. \qquad (\mathrm{D}.21)$$

Inserting this bound in Eq. (D.20), we obtain $S_{m,\lambda}[f] \geq 0 = S_{m,\lambda}[0]$. Hence, in the neighborhood $N_{\epsilon,0} \cap \mathcal{S}_{\mathrm{EL}}$ with $\epsilon = \sqrt{\frac{4! m^2}{|\lambda|}}$ the trivial extremum is a minimiser.

*Case (c)* ($m^2 > 0$, $\lambda > 0$): In this case the space of extremal sequences only contains the zero-sequence, procuring the trivial extremum $\varphi(g_1, g_2, g_3) = 0$. Denoting the quadratic part of the action in Eq. (5.13) by $Q_m[f]$ and the interaction part by $\lambda I[f]$ such that

$$S_{m,\lambda}[f] = Q_m[f] + \lambda I[f], \qquad (\mathrm{D}.22)$$

we have for any $f \in \mathcal{S}_{\mathrm{EL}}$

$$Q_m[f] = \sum_{J \in \mathfrak{J}_{\mathrm{EL}}} (f^J)^2 (J^2 + m^2) \geq 0, \qquad \lambda I[f] = \frac{\lambda}{4!} \sum_{J \in \mathfrak{J}_{\mathrm{EL}}} (f^J)^4 \{6j\}^2 \geq 0. \qquad (\mathrm{D}.23)$$

Hence, $S_{m,\lambda}[0] = 0 \leq S_{m,\lambda}[f]$, $\forall f \in \mathcal{S}_{\mathrm{EL}}$. We obtain a global minimiser, since the minimal condition is satisfied on the whole $\mathcal{S}_{\mathrm{EL}}$.

*Case (d)* ($m^2 < 0$, $\lambda > 0$): For any $f \in \mathcal{S}_{\mathrm{EL}}$ the action evaluated at $f$ gives

$$S_{m,\lambda}[f] = \frac{1}{2} \sum_{J \in \mathfrak{J}_{\mathrm{EL}}} (f^J)^2 (J^2 - |m^2|) + \frac{\lambda}{4!} \sum_{J \in \mathfrak{J}_{\mathrm{EL}}} (f^J)^4 \{6j\}^2. \qquad (\mathrm{D}.24)$$

Splitting $f$ such that $f(g_1, g_2, g_3) = f^-(g_1, g_2, g_3) + f^+(g_1, g_2, g_3)$ with

$$f^-(g_1, g_2, g_3) = \sum_{J \in \mathfrak{J}_{\mathrm{EL}}}^{|J| \leq 3j_{\max}} f^J \mathcal{X}^J(g_1, g_2, g_3)$$

$$f^+(g_1, g_2, g_3) = \sum_{J \in \mathfrak{J}_{\mathrm{EL}}}^{|J| > 3j_{\max}} f^J \mathcal{X}^J(g_1, g_2, g_3), \qquad (\mathrm{D.25})$$

we have $S_{m,\lambda}[f] = S_{m,\lambda}[f^- + f^+] \geq S_{m,\lambda}[f^-]$. Hence, verifying the minimiser condition, it is enough to show that $S_{m,\lambda}[\varphi] \leq S_{m,\lambda}[f^-]$. The space of functions of the form $f^-$ is finite-dimensional and we can use the usual minimisation procedure for functions. More specifically, let $s_J : \mathbb{R} \to \mathbb{R}$ a function such that

$$s_J(f^J) = (f^J)^2 \left[ \frac{1}{2}(J^2 - |m^2|) + \frac{\lambda}{4!}(f^J)^2 \{6j\}^2 \right].$$

The action $S_{m,\lambda}[f^-]$ is smallest when each $s_J$ is minimal on $\mathbb{R}$ for each $J \leq J_{\max}$. Taking the first and second derivative of $s_J$, we see that the minimum is reached by the coefficients $C^J$ from Eq. (5.20). Hence, an extremum given by an extremal sequence of maximal length is a global minimiser on the whole $\mathcal{S}_{\mathrm{EL}}$.

If $\varphi$ is given by an extremal sequence $C$ of length $\ell(C) < j_{\max}$, then there exists a $\mathcal{X}^{J_0}$ with $J_0 \leq J_{\max}$ and $\varphi^{J_0} = 0$. For $\delta \in \mathbb{R}$ define the field $v(g_1, g_2, g_3) = \varphi(g_1, g_2, g_3) + \delta \cdot \mathcal{X}^{J_0}(g_1, g_2, g_3)$. Inserting $v$ into the action we get

$$S_{m,\lambda}[v] = S_{m,\lambda}[\varphi] + \delta^2 \left[ \frac{1}{2}(J_0^2 - |m^2|) + \frac{\lambda}{4!}\delta^2 \{6j_0\}^2 \right]. \qquad (\mathrm{D.26})$$

If $\delta^2$ is in the range $0 < \delta < 2C^{J_0}$ the square bracket is negative and it follows $S_{m,\lambda}[v] \leq S_{m,\lambda}[\varphi]$. Moreover, for any $\epsilon > 0$ and $\delta < \frac{\epsilon}{J_0^{2n}}$ we have

$$\begin{aligned} \|v - \varphi\|_n &= \delta \sup_{(g_1, g_2, g_3) \in \mathrm{SU}(2)^3} \left| \Delta^n \mathcal{X}^{J_0}(g_1, g_2, g_3) \right| \\ &= \delta \sup_{(g_1, g_2, g_3) \in \mathrm{SU}(2)^3} \left| J_0^{2n} \mathcal{X}^{J_0}(g_1, g_2, g_3) \right| < \epsilon, \qquad (\mathrm{D.27}) \end{aligned}$$

since the characters are bounded by one, $\left| \mathcal{X}^{J_0}(g_1, g_2, g_3) \right| \leq 1$. Hence, $v \in N_{\epsilon,n}(\varphi)$. For any $\epsilon > 0$ choosing $\delta < \min\left( \frac{\epsilon}{J_0^{2n}}, C^{J_0} \right)$ we get $S_{m,\lambda}[f] < S_{m,\lambda}[\varphi]$. This shows that we can find a function $g$ in any neighborhood of $\varphi$ that decreases the value of the action, and hence, $\varphi$ is not a minimiser. $\qquad \square$

# Appendix E

# Intertwiner space and volume operator

## E.1 Intertwiner space of a four-valent spin network node

In the following, we review the construction of the intertwiner space of a four-valent spin network node. With this we explicitly construct the intertwiners of volume eigenstates in a simple example in the subsequent Section E.2.

Consider four links to meet at a spin network node as in Fig. E.1 each supporting a holonomy $g_i$ labelled with an irreducible representation $D^{j_i}(g_i)$ of the Lie group SU(2) living in the Hilbert space $\mathcal{H}_{j_i}$. The intertwiner space $\mathcal{H}_{\text{kin},4}$ is defined as the subspace of the tensor product $\mathcal{H}^{j_1} \otimes \mathcal{H}^{j_2} \otimes \mathcal{H}^{j_3} \otimes \mathcal{H}^{j_4}$ whose elements are invariant under the (right) diagonal action of SU(2), i.e. we define it as the space of invariant tensors [577]

$$\mathcal{H}_{\text{kin},4} = \text{Inv}_{\text{SU}(2)} \left[ \mathcal{H}^{j_1} \otimes \mathcal{H}^{j_2} \otimes \mathcal{H}^{j_3} \otimes \mathcal{H}^{j_4} \right]. \tag{E.1}$$

Thus, $\mathcal{H}_{\text{kin},4}$ is the space of singlets that can be constructed out of four spins. It can be interpreted as the Hilbert space of a quantum tetrahedron [5, 578].

We construct a basis in $\mathcal{H}_{\text{kin},4}$ starting with dividing the spins in two pairs $(j_1, j_2)$ and $(j_3, j_4)$, corresponding to the recoupling channel $\mathcal{H}^{j_1} \otimes \mathcal{H}^{j_2}$ [577]. The total spin of a pair is labelled by the quantum number $J$ which is the same for each of the two pairs since they sum to give a singlet. Basis vectors can thus be expressed in terms of the tensor product basis as

$$|j_1, j_3, j_3, j_4; J\rangle = \sum_{m_1, \ldots, m_4} \alpha^{j_1 j_2 j_3 j_4, J}_{m_1 m_2 m_3 m_4} |j_1, m_1\rangle |j_2, m_3\rangle |j_3, m_3\rangle |j_4, m_4\rangle. \tag{E.2}$$

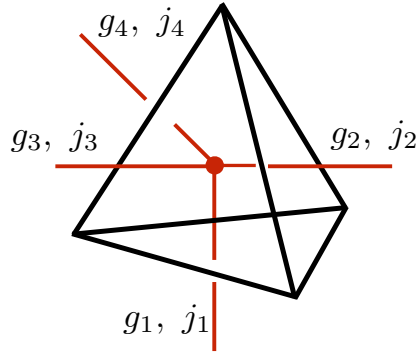

Figure E.1: An open spin network node which corresponds to an elementary excitation over the GFT vacuum. Each link supports a holonomy $g_i$ which is an element of SU(2). Dual to the four-valent node is a tetrahedron whose faces are labelled by $\mathfrak{su}(2)$ representations with spin $j_i$. The four spins satisfy the closure condition Eq. (E.13).

The coefficients $\alpha^{j_1 j_2 j_3 j_4, J}_{m_1 m_2 m_3 m_4}$ are the elements of a unitary matrix, which implements the change of basis from the tensor product basis

$$\{|j_1, m_1\rangle |j_2, m_3\rangle |j_3, m_3\rangle |j_4, m_4\rangle\} \quad \text{to} \quad \{|j_1, j_3, j_3, j_4; J\rangle\} \tag{E.3}$$

in the space of singlets $\mathcal{H}_{\mathrm{kin},4}$.[1] $\alpha^{j_1 j_2 j_3 j_4, J}_{m_1 m_2 m_3 m_4}$ is an invariant tensor, i.e. all of its components are invariant under SU(2). The quantum number $J$ satisfies the inequalities

$$\max\{|j_1 - j_2|, |j_3 - j_4|\} \leq J \leq \min\{j_1 + j_2, j_3 + j_4\}. \tag{E.4}$$

Moreover, in order to get a singlet one must have

$$m_1 + m_2 + m_3 + m_4 = 0. \tag{E.5}$$

If Eqs. (E.4), (E.5) are not satisfied for certain values of $J$ and $\{m_1, \ldots, m_4\}$, $\alpha^{j_1 j_2 j_3 j_4, J}_{m_1 m_2 m_3 m_4}$ vanishes and the corresponding term gives no contribution to Eq. (E.2).

We can express the coefficients of the decomposition in Eq. (E.2) in terms of Clebsch-Gordan coefficients as

$$\alpha^{j_1 j_2 j_3 j_4, J}_{m_1 m_2 m_3 m_4} = \eta \, \frac{(-1)^{J-M}}{\sqrt{d_J}} C^{j_1 j_2 \, J}_{m_1 m_2 \, M} C^{j_3 j_4 \, J}_{m_3 m_4 \, -M}, \tag{E.6}$$

---

[1]Notice that it is not a unitary matrix over the whole Hilbert space $\mathcal{H}^{j_1} \otimes \mathcal{H}^{j_2} \otimes \mathcal{H}^{j_3} \otimes \mathcal{H}^{j_4}$, since $\alpha^{j_1 j_2 j_3 j_4, J}_{m_1 m_2 m_3 m_4}$ vanishes when the set $\{m_1, \ldots, m_4\}$ fails to satisfy Eq. (E.5)

where we defined $M = m_1 + m_2 = -(m_3 + m_4)$ and $\eta$ is a phase factor. The latter can depend on $J$ as well as on the fixed values of the four spins $\{j_1, \ldots, j_4\}$. We omit the functional dependence to avoid confusion with tensor indices. The value of $\eta$ does not affect the unitarity relation satisfied by the coefficients defined in Eq. (E.6)

$$\sum_{m_1, \ldots, m_4} \overline{\alpha}_{m_1 m_2 m_3 m_4}^{j_1 j_2 j_3 j_4, J} \alpha_{m_1 m_2 m_3 m_4}^{j_1 j_2 j_3 j_4, J'} = \delta^{JJ'}. \tag{E.7}$$

We choose the value of the phase $\eta$ such that

$$\alpha_{m_1 m_2 m_3 m_4}^{j_1 j_2 j_3 j_4, J} = (-1)^{J-M} \sqrt{d_J} \begin{pmatrix} j_1 & j_2 & J \\ m_1 & m_2 & -M \end{pmatrix} \begin{pmatrix} J & j_4 & j_3 \\ M & m_4 & m_3 \end{pmatrix}. \tag{E.8}$$

It is convenient to choose $\eta$ in this way so that the contraction of five four-valent intertwiners coincides with the definition of the $\{15j\}$-symbol, see Ref. [579].

All intertwiners, i.e. elements of $\mathcal{H}_{\mathrm{kin},4}$, can be expressed as linear combinations of the above-given coefficients $\alpha_{m_1 m_2 m_3 m_4}^{j_1 j_2 j_3 j_4, J}$ as

$$C_{m_1 m_2 m_3 m_4}^{j_1 j_2 j_3 j_4, \iota} = \sum_J c^{J_\iota} \alpha_{m_1 m_2 m_3 m_4}^{j_1 j_2 j_3 j_4, J}. \tag{E.9}$$

Thus, $\iota$ labels any linear subspace in the intertwiner space $\mathcal{H}_{\mathrm{kin},4}$. Different choices correspond to different physical properties of the quanta of geometry.

## E.2 Volume operator

In the LQG literature there are several different definitions of the volume operator available, see Refs. [512, 580, 581]. Importantly, they all agree in the case of a four-valent node [577] and match the operator introduced in Ref. [578]. In the following, we will largely follow Ref. [537] for the definition of the volume operator and the derivation of its spectrum.

The volume operator acting on a spin network node (embedded in a differentiable manifold) is defined as

$$\hat{V} = \sqrt{\left| \sum_{I < J < K} \epsilon(e_I, e_J, e_K) \epsilon_{ijk} J_I^i J_J^j J_K^k \right|} = \sqrt{\left| \frac{i}{4} \sum_{I < J < K} \epsilon(e_I, e_J, e_K) \hat{q}_{IJK} \right|}, \tag{E.10}$$

where $(e_I, e_J, e_K)$ is a triple of links adjacent to the node. Their orientation $\epsilon(e_I, e_J, e_K)$ is given by the triple product of the vectors tangent to the links. There is one spin degree of freedom attached to each link. Angular momentum operators corresponding to distinct links commute, i.e.

$$\left[J_I^i, J_J^j\right] = i\delta_{IJ}\epsilon^{ijk}J_I^k. \tag{E.11}$$

In Eq. (E.10) we also introduced the operator

$$\hat{q}_{IJK} = \left(\frac{2}{i}\right)^3 \epsilon_{ijk}J_I^i J_J^j J_K^j. \tag{E.12}$$

Spin network nodes are gauge-invariant, i.e. the angular momenta carried by the links entering a node satisfy a closure condition. For a four-valent node, as depicted by Fig. E.1, the closure condition is given by

$$\vec{J}_1 + \vec{J}_2 + \vec{J}_3 + \vec{J}_4 = \vec{0}. \tag{E.13}$$

Hence, the Hilbert space of the node is that of Eq. (6.122). The closure condition yields the following simplification in the evaluation of the sum in Eq. (E.10)

$$\sum_{I<J<K} \epsilon(e_I, e_J, e_K)\hat{q}_{IJK} = 2\,\hat{q}_{123}. \tag{E.14}$$

With this, the squared volume operator rewrittes as

$$\hat{V}^2 = \left|\frac{i}{2}\hat{q}_{123}\right|. \tag{E.15}$$

Observe that, while the definition (E.10) makes explicit reference to the embedding map, the final expression (E.15) does not depend on it.

By means of the recoupling channel $\mathcal{H}^{j_1} \otimes \mathcal{H}^{j_2}$ as in Appendix E.1 and labelling with $J$ the eigenvalue of $(\vec{J}_1 + \vec{J}_2)^2$, one finds the non-vanishing matrix elements

$$
\begin{aligned}
\langle J|\hat{q}_{123}|J-1\rangle = \frac{1}{\sqrt{4J^2-1}} \times \\
\Big[ (j_1+j_2+J+1)(-j_1+j_2+J)(j_1-j_2+J)(j_1+j_2-J+1) \times \\
(j_3+j_4+J+1)(-j_3+j_4+J)(j_3-j_4+J)(j_3+j_4-J+1) \Big]^{\frac{1}{2}} \\
= -\langle J-1|\hat{q}_{123}|J\rangle
\end{aligned}
\tag{E.16}
$$

in the recoupling basis [537]. The eigenvalues of $\hat{q}_{123}$ are non-degenerate. Moreover, if $\hat{q}_{123}$ has a non-vanishing eigenvalue $a$, also $-a$ is an eigenvalue, where the sign corresponds to the orientation of the node. If the dimension of the intertwiner space is odd, $\hat{q}_{123}$ has a non-degenerate zero eigenvalue.

**Monochromatic node**

Following the convention of Ref. [537], we call a node monochromatic if the four incident spins are all identical ($j_1 = j_2 = j_3 = j_4 = j$). This simplifies Eq. (E.16) to

$$
\langle J|\hat{q}_{123}|J-1\rangle = \frac{1}{\sqrt{4J^2-1}} J^2(d_j^2 - J^2),
\tag{E.17}
$$

where $d_j = 2j+1$ denotes the dimension of the irreducible representation with spin $j$.

In view of the work presented in Sections 6.3.2, 6.3.3, we consider the fundamental representation $j = \frac{1}{2}$. In this case the intertwiner space is just two-dimensional, with a basis given by $\{|0\rangle, |1\rangle\}$, i.e. the four-valent gauge-invariant node is constructed using two singlets and two triplets, respectively. Using Eq. (E.17), the squared volume operator is rewritten in this basis as

$$
\hat{V}^2 = \frac{\sqrt{3}}{2} \left| \begin{pmatrix} 0 & -i \\ i & 0 \end{pmatrix} \right| = \left| \hat{Q} \right|,
\tag{E.18}
$$

where we introduced a new matrix $\hat{Q} = \frac{\sqrt{3}}{2}\sigma_2$, which is equal to $\hat{V}^2$ up to a sign. The sign of the eigenvalues of $\hat{Q}$ gives the orientation of the node. Its normalised eigenvectors are

$$
|+\rangle = \frac{1}{\sqrt{2}} \begin{pmatrix} 1 \\ i \end{pmatrix}, \quad |-\rangle = \frac{1}{\sqrt{2}} \begin{pmatrix} 1 \\ -i \end{pmatrix},
\tag{E.19}
$$

with eigenvalues $\pm\frac{\sqrt{3}}{2}$. We decompose the volume eigenstates $|\pm\rangle$ in the tensor product basis of $\mathcal{H}^{\frac{1}{2}} \otimes \mathcal{H}^{\frac{1}{2}} \otimes \mathcal{H}^{\frac{1}{2}} \otimes \mathcal{H}^{\frac{1}{2}}$ as

$$|\pm\rangle = \sum_J c^{J\pm} |J\rangle = \sum_{m_1,\dots,m_4,J} c^{J\pm} \alpha^{\frac{1}{2}\,J}_{m_1 m_2 m_3 m_4} |\tfrac{1}{2}, m_1\rangle |\tfrac{1}{2}, m_2\rangle |\tfrac{1}{2}, m_3\rangle |\tfrac{1}{2}, m_4\rangle. \qquad (E.20)$$

Finally, we can define the intertwiners corresponding to the volume eigenstates $|\pm\rangle$ as

$$C^{\frac{1}{2}\,\pm}_{m_1 m_2 m_3 m_4} = \sum_J c^{J\pm} \alpha^{\frac{1}{2}\,J}_{m_1 m_2 m_3 m_4} = \frac{1}{\sqrt{2}} \left( \alpha^{\frac{1}{2}\,0}_{m_1 m_2 m_3 m_4} \pm i\, \alpha^{\frac{1}{2}\,1}_{m_1 m_2 m_3 m_4} \right). \qquad (E.21)$$

Hence, we write

$$|\pm\rangle = \sum_{m_1,\dots,m_4} C^{\frac{1}{2}\,\pm}_{m_1 m_2 m_3 m_4} |\tfrac{1}{2}, m_1\rangle |\tfrac{1}{2}, m_2\rangle |\tfrac{1}{2}, m_3\rangle |\tfrac{1}{2}, m_4\rangle. \qquad (E.22)$$